\documentclass[twocolumn,preprintnumbers,showpacs,nofootinbib,amsmath,amssymb,rmp,floatfix]{revtex4}
\usepackage{natbib,graphicx,journals}% Include figure files
\usepackage{dcolumn}% Align table columns on decimal point
\usepackage{bm}% bold math

\newcommand\ms{M$_{\odot}$}
\newcommand\msy{M$_{\odot} \ \rm {yr^{-1}} $}
\newcommand\mspcb{M$_{\odot} \ \rm {pc^{-2}}$}

\newcommand\zs{Z$_{\odot}$}
\newcommand\Ls{L$_{\odot}$}
\def\apjs{Astrophys. J. Suppl. Ser.}
\def\araa{Annu. Rev. Astron. Astrophys.}
\def\mnras{Mon. Not. R. Astron. Soc.}
\def\pasj{Publ. Astron. Soc. Japan}
\newcommand{\Al}{$^{26}$Al}

\newcommand{\degree}{$^{\circ}$}

\newcommand{\fluxrad}{ph~cm$^{-2}$s$^{-1}$rad$^{-1}$\ }
\newcommand{\pcmq}{\mbox{cm$^{-2}$}}
\newcommand{\pcmc}{\mbox{cm$^{-3}$}}
\newcommand{\psec}{\mbox{s$^{-1}$}}

\newcommand{\funit}{\mbox{ph \pcmq \psec}}

\def\dg{\ensuremath{^\circ}}

\def\degree{\hbox{$^o$}}
\def\lsim{\;\raise0.3ex\hbox{$<$\kern-0.75em\raise-1.1ex\hbox{$\sim$}}\;}
\def\gsim{\;\raise0.3ex\hbox{$>$\kern-0.75em\raise-1.1ex\hbox{$\sim$}}\;}

\begin{document}

%\preprint{APS/123-QED}

\title{The  511 keV emission from positron annihilation in the Galaxy}% Force line breaks with \\

\author{N. Prantzos}
\affiliation{CNRS, UMR7095, UMPC and  Institut d'Astrophysique de Paris, F-75014, Paris, France}

\author{C. Boehm}
\affiliation{LAPP - 9 Chemin de Bellevue - BP 110 F-74941 Annecy-le-Vieux , France}

\author{A. M. Bykov}
\affiliation{A.F. Ioffe Institute of Physics and Technology, Russian Academy of Sciences, 194021, St. Petersburg, Russia}

\author{R. Diehl}
\affiliation{Max Planck Institut f\"ur Extraterrestrische Physik, D-85741 Garching, Germany}

\author{K. Ferri\`ere}
\affiliation{Laboratoire d'Astrophysique de Toulouse-Tarbes,
  Universit\'e de Toulouse, CNRS,   14 avenue Edouard Belin, F-31400 Toulouse, France}

\author{N. Guessoum}
\affiliation{American University of Sharjah, College of Arts \& Sciences / Physics Dpt., P.O Box 26666, Sharjah, UAE}

\author{P. Jean}
\affiliation{CESR, 9, Av. du Colonel Roche, B.P. 4346, 31028 Toulouse Cedex 4, France}

\author{J. Knoedlseder}
\affiliation{CESR, 9, Av. du Colonel Roche, B.P. 4346, 31028 Toulouse Cedex 4, France}

\author{A. Marcowith}
\affiliation{L.P.T.A., Universit\'e Montpellier II,  CNRS, Place Eug\`ene Bataillon, 34095 Montpellier, France}

\author{I. V. Moskalenko}
\affiliation{Hansen Experimental Physics Laboratory and Kavli Institute for Particle Astrophysics and Cosmology, \\
                 Stanford University, Stanford, CA 94305, U.S.A.}

\author{A. Strong}
\affiliation{Max Planck Institut f\"ur Extraterrestrische Physik, D-85741, Garching, Germany}

\author{G. Weidenspointner}
\affiliation{Max Planck Institut fur Extraterrestrische Physik, Garching, Germany, {\rm and} \\
                  MPI Halbleiterlabor, Otto-Hahn-Ring 6,   81739 Muenchen,   Germany
}

\date{\today}% It is always \today, today,
             %  but any date may be explicitly specified

\begin{abstract}
The first  gamma-ray line originating from outside the solar system that was ever detected is
the 511 keV  emission from positron annihilation in the Galaxy. 
%is the first gamma-ray line ever detected from outside the solar system. 
Despite 30 years of intense theoretical and observational investigation, the main sources of positrons  have not been identified up to now. 
Observations in the 1990's with OSSE/CGRO showed that the emission  is strongly concentrated towards the Galactic bulge. In the 2000's, the SPI instrument aboard ESA's INTEGRAL 
$\gamma$-ray observatory  allowed scientists to measure that emission across the entire Galaxy, revealing that  the bulge/disk luminosity ratio  is
larger than observed in any other wavelength. This mapping prompted a number of novel explanations, including rather ``exotic'' ones (e.g. dark matter annihilation). However, conventional astrophysical sources, like type Ia supernovae, microquasars or X-ray binaries, are still plausible candidates for a large fraction of the observed total 511 keV emission of the bulge. A closer study of the subject reveals new layers of complexity,
since positrons may propagate far away from their production sites,
making it difficult to infer the underlying source distribution from the observed map of 511 keV emission.
However, contrary to the rather well understood propagation of  high energy ($>$GeV) particles of Galactic cosmic rays, understanding the propagation
of low energy ($\sim$MeV) positrons in the turbulent, magnetized interstellar medium, still remains a formidable challenge.
We review the spectral and imaging properties  of the observed 511 keV  emission 
and we critically discuss  candidate positron sources and models of positron propagation in the
Galaxy.

\end{abstract}

% PACS, the Physics and Astronomy
                             % Classification Scheme.

%\keywords{Suggested keywords}%Use showkeys class option if keyword
                              %display desired
\maketitle

\tableofcontents

%.........................................................................................................................................................................................
\section{ Introduction}
\label{sec:Intro}

The existence of a particle with equal mass but opposite charge to that of the electron
 was predicted by \citet{Dirac31}, who named it "anti-electron". 
Unaware of Dirac's prediction,  \citet{Anderson32} found 
the first experimental hints for such a particle in cloud-chamber photographs of cosmic rays, and he named it
{\it positron}. His finding was confirmed the following year by \citet{BO33}, who identified it
with Dirac's anti-electron.
One  year later, \citet{KC34} detected the characteristic $\gamma$-ray line at 511 keV resulting from
e$^-$-e$^+$ annihilation, a convincing proof that positrons are indeed electron's antiparticles. That same year,
Croatian physicist \citet{M34} predicted the existence of a bound system
composed of an electron and a positron (analogous to the hydrogen atom, but with the proton
replaced by a positron), which he called "electrum". This state was experimentally found by \citet{Deutsch51}
at MIT and became known as {\it positronium}.

For about 30 years after their discovery, all detected positrons were of terrestrial origin. Those detected
by  \citet{Anderson32} and   \citet{BO33} were created by cosmic ray interactions with molecules in
Earth's atmosphere. \citet{JC34} identified another positron producing process: $\beta^+$
radioactivity of artificially created unstable nuclei. The first positrons of extraterrestrial origin were reported
by \citet{dSHM64}, who loaded a spark chamber on a stratospheric balloon to detect  positrons within the cosmic rays.
\citet{Ginzburg56} had already suggested that high energy p-p interactions in cosmic rays would produce pions $\pi^+$,
which would decay to positrons (via $\mu$ decays) . The production rate of those pions was evaluated by \citet{PF63} who 
predicted a $\gamma$-ray flux from the  Galaxy at 511 keV of $\sim$10$^{-3}$ cm$^{-2}$ s$^{-1}$.

The properties of e$^-$-e$^+$ annihilation were explored in the 1940's. 
Direct e$^-$-e$^+$ annihilation produces a single $\gamma$-ray line at 511 keV, while the annihilation of
positronium produces a composite spectrum with a lower energy continuum and a 511 keV line
(\onlinecite{Ore-Powell:1949} and Sec.~\ref{subsec:positronium}). 
\citet{Stecker69} was the first to point out that in the conditions of the
interstellar medium, most positrons would annihilate after positronium formation; 
this would reduce the
511 keV flux  from cosmic rays to values lower than evaluated by \citet{PF63}.

The 511 keV emission of e$^+$ annihilation was first detected from the general 
direction of the Galactic center in the early 1970's, by balloon-borne 
instruments of low energy resolution  \citep{Johnson:1972}. It was unambiguously
identified a few years later with high resolution Ge detectors \citep{Leventhal:1978}.
It is the first and most intense 
$\gamma$-ray line originating  from outside the solar system that was ever detected. 
Its flux on Earth ($\sim$10$^{-3}$
cm$^{-2}$ s$^{-1}$), combined with the distance to the Galactic center 
($\sim$8 kpc\footnote{1 pc(parsec)=3.26 light years = 3.09 10$^{18}$ cm.}), implies the
annihilation of $\sim$2  10$^{43}$ e$^+$ s$^{-1}$ (Sec.~\ref{subsub:imaging}), releasing a  power of
$\sim$10$^{37}$ erg s$^{-1}$ or $\sim$10$^4$ \Ls~ in $\gamma$-rays.
Assuming a steady state, i.e. equality between production and annihilation rates of positrons,
one should then look for a source (or sources) able to provide $\sim$2  10$^{43}$ e$^+$ s$^{-1}$. If the
activity of that site were maintained to the same level during the $\sim$10$^{10}$ yr of Galaxy's
lifetime, a total amount of positrons equivalent to $\sim$3 \ms~ would have been annihilated.

Several years earlier, the Sun had already become the first astrophysical laboratory for the study of 
positron annihilation~\cite{Crannell:1976}. The solar annihilation $\gamma$-ray line had been detected 
with a simple NaI instrument aboard the OSO-7 satellite~\cite{1975IAUS...68..341C}. 
The Solar Maximum Mission (SMM), designed for solar flare observations and launched in 1980, 
featured a $\gamma$-ray spectrometer with exceptional stability. 
Based on detailed measurements with SMM, positrons in solar flares were 
found to originate from  flare-accelerated particles when hitting the upper photosphere. 
Nuclear interactions of flare-accelerated protons and ions with atomic nuclei of the 
photosphere produce radioactive nuclei and pions that decay by emission of positrons, which annihilate locally \cite{1983SoPh...86..395R,2005AGUSMSP21A..09M}.

Imaging the Galaxy in annihilation $\gamma$-rays was considered to be the exclusive
way to identify the cosmic e$^+$ sources (assuming that the spatial morphology of the 
$\gamma$-ray emission reflects the spatial distribution of the sources, i.e. that positrons
annihilate close to their production sites). Because of the difficulties of imaging in the MeV region,
progress was extremely slow in that field: only in the 1990s 
were the first constraints on the spatial distribution    of the 511 keV emission  in the inner Galaxy
 obtained by the OSSE instrument aboard the Compton Gamma Ray Observatory 
(CGRO, \onlinecite{Cheng:1997}; \onlinecite{Purcell:1997}).
The most reliable imaging of the 511 keV emission was obtained by the SPI 
instrument aboard ESA's INTEGRAL Gamma Ray Observatory: the emission is strongly
concentrated in the inner   Galaxy (the bulge, \onlinecite{Knodlseder:2005}) 
and a weaker emission is detected from the Galactic
disk \citep{Weidenspointner+08a}, unlike the situation in any other wavelength. 

Several candidate sources of positrons were proposed over the years: radioactivity from $\beta^+$
 decay of unstable nuclei produced in stellar explosions, high energy interactions 
 occurring in cosmic rays or near
 compact objects (like pulsars and  X-ray binaries) or the supermassive black hole in the Galactic center, etc.
 For a long time, radioactivity from $^{56}$Co produced in thermonuclear supernovae (SNIa)
 appeared as the most promising candidate, provided that just a few per cent of the released positrons could escape the
 supernova remnant and annihilate in the interstellar medium. However, none of the candidate sources
 has a spatial pattern resembling the one of the detected $\gamma$-ray emission. In particular, the
 release of the first year of SPI data, revealing the bulge but not yet the disk, 
 prompted a series of "exotic"  explanations involving dark matter particles, superconducting
 cosmic strings, etc. The confirmation of disk emission a few years later,
 made such explanations lose much of their interest,
 without completely eliminating them up to now.
 
The spectral analysis of the 511 keV emission  established already at the late 1970's that
most of the positrons annihilate after positronium formation
\citep{Bussard:1979}. This result  constitutes an
important diagnostic tool for the physical properties of the annihilation medium, as analyzed
in Guessoum et al. (1991). Only recently,
i.e. in the 2000's, was it realized that the spectral analysis may also provide important hints on
the e$^+$ source(s). In particular, positrons appear to annihilate at 
low energies, while in most candidate sources
they are produced at relativistic energies; during the long period of their slowing down, positrons
may travel far away from their sources, making the detected $\gamma$-ray emission  useless as a tracer
of their production sites. Unfortunately, propagation of low energy positrons in the turbulent, magnetized 
interstellar plasma of the Galaxy, is poorly understood at present. 

In this paper we present a synthetic view of the various facets of this complex issue, concerning the
production, propagation and annihilation of positrons in the Galaxy, in relation with the characteristic
signature of that annihilation, namely the 511 keV emission. The  paper is structured as follows:

In Sec.~\ref{sec:obs} we first present  a historical account of observations of the 511 keV emission,  
which illustrates the difficulties of $\gamma$-ray line astronomy in the MeV range. We also provide a 
summary of the latest SPI/INTEGRAL data analysis, of relevant observations in other wavelengths and of the constraints
imposed on the e$^+$ sources.

Sec. III provides astrophysical background material concerning the stellar and supernova populations
of the Milky Way, as well as the properties of the 
various phases of the interstellar medium (ISM) and of the
Galactic magnetic fields in which positrons propagate. This material can be skipped at a first reading by astronomers, but it contains
important and updated information, which is used in all other sections.

In Sec. IV we discuss the physical processes and candidate sources of positron production 
(radioactivity from stars  and supernovae, high energy
processes in cosmic rays, compact objects and the central supermassive black hole, 
dark matter, other "exotica", etc.) and, in some cases, 
we present new estimates of their e$^+$ yields. We discuss critically the properties of those sources, in 
the light of the observational constraints presented in Sec. II.

Sec. V summarizes the various physical processes of e$^+$ slowing down and annihilation, taking into account
the properties of the ISM, as well as the corresponding $\gamma$-ray spectral signature. The spectral analysis 
of the SPI data is  then used to constrain the energy of the emitted positrons (thus eliminating some of the 
candidate sources) and to constrain the properties of the annihilation medium. The results of this 
analysis offer some hints for e$^+$ propagation away from the sources.

The intricacies of low energy positron propagation are discussed in some depth in Sec. VI, in the light of recent work.
The implications of e$^+$ propagation for the 511 keV emission are also discussed.
This is   not a mature topic yet, and our poor knowledge of the plasma 
properties in the inner Galaxy prevents any definitive conclusions. A synthetic summary of the  subject and directions
for future research are presented in the last section.

% Sec. II Observations   .....................................................................................................................................................
%\input{SecII_v3}
%==================================================================
%  Section I: OBSERVATIONS
%==================================================================
%
% revised by Roland Diehl 091002 (V4): change eps to pdf files;
%          revised intro to annihilation states and photons
%          revise wordings in other Sections describing the observations and data analyses
%%%%%%%%%%%%%%%%%%%%%%%%%%%%%%%%%%%%%%%%%%%%%%%%%%%
\section{ Observations}
\label{sec:obs}
Gamma-ray observations in the MeV domain (from $\sim$100 keV up to a few MeV) provide access 
to Galactic positrons through three main windows:

i) Emission from e$^+$ annihilation at sub-relativistic energies, with its prominent 511~keV line and the associated
three-photon continuum from positronium annihilation;

ii) Continuum emission at energies E$>$0.5 MeV from energetic positrons propagating 
through interstellar space and annihilating "in-flight";

iii) Emission of characteristic gamma-ray lines from radioactive  nuclei, such as $^{26}$Al
and $^{44}$Ti, which also produce positrons by $\beta^+$-decay.

In this section, we describe the relevant observations and the constraints that they impose
on our understanding of Galactic positrons (see also \onlinecite{2009arXiv0906.1503D}). We start with a brief description of the radiative signatures of positron annihilation.

\subsection{Radiative signatures of positron  annihilation} 
\label{subsec:positronium}

The annihilation of a positron with an electron releases a total (rest-mass) energy of 1022 keV in the form of
 two or more photons.
Direct annihilation of a e$^-$-e$^+$ pair at rest produces two photons of 511 keV each. The situation is more complex in the case of  positronium (Ps). 
%Its state  can be formally treated as the one of hydrogen atom. The corresponding Schroedinger equations
%are identical, with the reduced mass being half of the electron mass in the case of Ps. 
%Because of that, the frequencies of
%the de-excitation spectral lines are roughly half of those of the H atom. %% deleted RoD: misleading
The ground state of positronium has two total spin states, 
depending on the relative orientations of 
the spins of the electron and the positron.
The singlet state has antiparallel spins, total spin S=0, is denoted as $^1S_0$ and  is known as {\it para-positronium} (p-Ps).
The triplet state has parallel spins, total spin S=1, is denoted as $^3S_1$ and  is known as {\it ortho-positronium} (o-Ps). From the
(2S+1) spin degeneracy, it follows that Ps will be formed 1/4 of the time in the p-Ps state and 3/4 of the time in the o-Ps state.
\footnote{The energy difference between the two spin states ("hyperfine splitting") is 8.4 10$^{-4}$ eV. Transitions between these states  similar to the spin-flip transition in hydrogen, which produces the astrophysically-important 21~cm line, are unimportant due to the short Ps lifetimes (see text).}

\begin{figure}
\includegraphics[angle=270,width=0.49\textwidth]{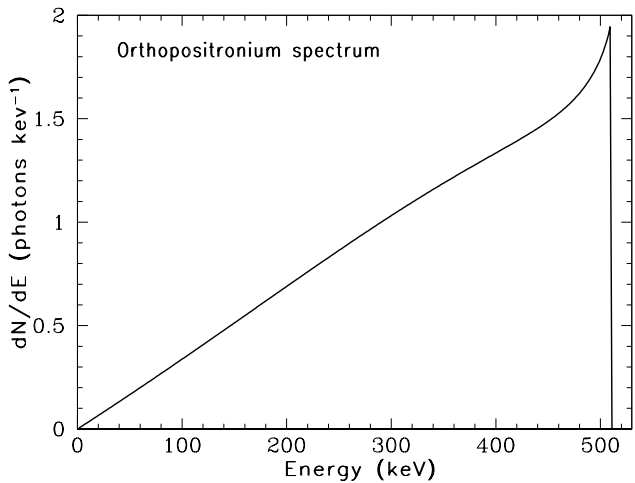}
\caption{Spectrum of ortho-positronium annihilation with the  three-photon continuum
(from \onlinecite{Ore-Powell:1949}).
\label{fig:pos_ann}}
\end{figure}

Spin and momentum conservation constrain the release of annihilation energy in the form of photons. Para-positronium annihilation releases
two photons of 511 keV each in opposite directions 
(as in the case of direct e$^-$-e$^+$ annihilation). 
Ortho-positronium annihilation requires a final state with more than two photons from spin conservation; momentum conservation distributes the total energy of 1022~keV among three  photons\footnote{
Annihilation into a larger number of photons (an even number for para-positronium, an odd number for ortho-positronium) is possible, but the corresponding
branching ratios are negligible ($\sim$10$^{-6}$ for four photons in the case of para-Positronium or five photons in the case of ortho-positronium). Even lower
are the branching ratios for annihilation into neutrino-antineutrino pairs. Finally, a single-photon annihilation is also possible, provided
momentum conservation is obtained through positronium being bound to another particle, such as a dust grain. None of those cases
is important in astrophysical settings.} producing a continuum of energies up to 511~keV (Fig.~\ref{fig:pos_ann}).
The corresponding lifetimes before annihilation (in vacuum) are 1.2 10$^{-10}$ s for p-Ps and 1.4 10$^{-7}$ s for o-Ps.

If a fraction $f_{Ps}$ of the positrons annihilate via positronium formation, then the 3-photon $\gamma$-ray continuum of ortho-positronium will  have an integrated intensity of
\begin{equation}
I_{3\gamma} \ \propto \ \frac{3}{4} \ 3 \ f_{Ps}
\label{eq:3gamma}
\end{equation}
The remaining fraction $1-f_{Ps}$ will annihilate directly to 2 photons of 511 keV each, to which should be added the 2-photon contribution of the para-Positronium state; thus, the 2-photon (511 eV line) intensity will be:
\begin{equation}
I_{2\gamma} \ \propto \ 2 (1-f_{Ps}) \ + \ \frac{1}{4} \ 2 \ f_{Ps} \ = \ 2 \ - \ 1.5 \ f_{Ps}
\label{eq:2gamma}
\end{equation}
By measuring the intensities of the 511 keV line and of the Ps continuum one can then derive the positronium fraction
\begin{equation}
f_{Ps} \ = \  \frac{8 \ I_{3\gamma}/I_{2\gamma}}{9+6I_{3\gamma}/I_{2\gamma}}
\label{eq:fPos}
\end{equation}
which offers a valuable diagnostic of the physical conditions of the ISM where positrons annihilate, 
(see discussion in Sec. V.E).

The state of Ps  can be formally treated as the one of hydrogen atom. The corresponding Schroedinger equations
are identical, with the reduced mass being half of the electron mass in the case of Ps. Because of that, the frequencies of
the de-excitation spectral lines of are roughly half of those of the H atom \citep{CMB75}. Radiative recombination
of ortho-Ps could, under certain circumstances, be observed in the near infra-red by next
generation instruments (see \onlinecite{EB09}).
%%%%%%%%%%%%%%%%%%%%%%%%%%%%%%%%%%%%%%%%%%%%%%%%%%
%%%%%%%%%%%%%%%%%%%%%%%%%%%%%%%%%%%%%%%%%%%%%%%%%%

\subsection{Observations of $\sim$ MeV emission from e$^+$ annihilation}
%__________________________________________________________________  
\subsubsection{Early balloon and satellite observations}
\label{subsec:earlyobs}

The first evidence for Galactic 511 keV emission was obtained in the early 1970's,
through a series of balloon flights by teams from Rice University.
 \citet{Johnson:1972} announced the first detection of a celestial 
gamma-ray line originating  from outside the solar system.
Using a sodium iodine (NaI) scintillation detector they
observed a spectral excess with a flux of $1.8 \times 10^{-3}$ \funit\ at 
an energy of $473 \pm 30$ keV during a balloon flight in 1970.
A second balloon flight of the same team in 1971 confirmed this signal 
%reality of the excess 
\citep{Johnson:1973}.
Although the authors mentioned e$^+$ annihilation as a possible origin
of the observed feature, the significant offset between observed (473~keV)
and expected (511~keV) line centroid led them to conclude that the
feature was, perhaps, due to radioactive decay of unknown origin.

\citet{Leventhal:1973} proposed an interesting alternative explanation that
could reconcile observations with a positron origin of the feature.
As e$^+$  annihilation in the interstellar medium occurs mainly via
positronium, a superposition of a narrow line at
511~keV and a  continuum emission below 511~keV is expected (Fig. \ref{fig:pos_ann}).
Because of the poor spectral resolution of the NaI detector 
and the low signal/noise ratio of  these balloon
flights these two contributions  could not be disentangled, 
     and were reported as  a
single emission peak at an  energy of $\sim490$~keV,
in reasonable agreement with the observations.

It took five years before a group from Bell-Sandia \citep{Leventhal:1978} could confirm  this
conjecture. With their high-resolution balloon-borne germanium (Ge) detector they
could separate the line and continuum components of the
emission. \citet{Leventhal:1978} located the narrow component (FWHM=3.2 keV) at an energy
of $510.7 \pm 0.5$ keV, consistent with the expectations for e$^+$
annihilation at rest.
The observed line flux of $(1.2 \pm 0.2)$ 10$^{-3}$ \funit\ was below
the value reported by \citet{Johnson:1972}, as expected if the earlier measurements
were a superposition of two components.
\citet{Leventhal:1978} also detected the positronium continuum component,
and the comparison of line and continuum intensities implied that 92\%
of the annihilations occurred after the formation of  a positronium atom.

Subsequent observations of the Galactic center by different balloon-borne instruments in the 
1970's found a surprising variability of the 511~keV line flux
\citep{Haymes:1975,Leventhal:1980,Albernhe:1981,Gardner:1982}.
But \citet{Albernhe:1981} recognized that the flux measured by the various
balloon experiments increased with increasing size of the detector's field
of view, which could mean that the annihilation emission was extended along the
Galactic plane. 
\citet{Riegler:1981} proposed
a different scenario, based on analysis of their HEAO-3 satellite data.
These data showed a decline by almost a factor three of the 511~keV flux  between fall 1979 
and spring 1980, suggesting that positron annihilation was
variable in time.  From the $\Delta t\sim$6 months interval between the observations they inferred a maximum 
size of $\Delta r \sim c \Delta t \sim$ 0.3 pc
of the annihilation site, which implies gas densities for the annihilation medium
of $10^4$ - $10^6$ \pcmc.
These extreme conditions suggest that the positrons were produced by a compact source such as
a massive black hole within 4\degree \ of the Galactic center
\citep{Lingenfelter:1981}.
%__________________________________________________________________  
%\subsubsection{Contradictory observations}  

While balloon-borne experiments seemed to establish the variability of the 511~keV
emission \citep{Leventhal:1982,Paciesas:1982,Leventhal:1986}, contemporaneous
observations by the Solar Maximum Mission (SMM) 
satellite did not confirm  such a trend.
SMM carried a NaI-detector 
with a large field of view (130$^o$) and provided the first long-term monitoring of the
inner Galaxy (1980-1987). The variability of the 511 keV emission was constrained to be less than 30\%
\citep{Share:1988,Share:1990}.
The apparent disagreement between the balloon and SMM observations
could still be understood by assuming an extended distribution of 511~keV emission
along the Galactic plane; however, in order to reconcile 
the observations with a time variable source one had to adopt rather complex scenarios. 
For example, \citet{Lingenfelter:1989} suggested the combination of a steady, 
extended 511~keV emission along the Galactic plane and a compact variable source 
at the Galactic center (assumed to be active from 1974 through 1979) in order 
to explain all
data available at that time. Such a scenario could not be ruled out,
since no imaging of the 511~keV emission had  been achieved and
the morphology and spatial extent of the positron annihilation emission
were essentially unconstrained.

The hypothesis of a time variable central Galactic positron source was revitalized in the
early 1990s by the observation of transient $\gamma$-ray line features 
with the SIGMA telescope. The french coded mask imager SIGMA was launched
in 1989 on board the soviet GRANAT satellite.  It was the first imaging $\gamma$-ray
instrument, with an angular resolution of 15 arcmin and it used NaI detectors covering
the energy range 35-1300 keV.
In October 1991, an unusual spectrum was observed
from 1E 1740.7-2942, during an outbreak of this hard X-ray source which lasted $\sim$17 hours
\citep{Bouchet:1991,Sunyaev:1991}.
Superimposed on a candidate typical black hole
continuum spectrum, there appeared a strong (flux 
$F\sim10^{-2}$~ph~cm$^{-2}$~s$^{-1}$) and broad (FWHM $\sim 200$~keV)
emission line centered at about 440~keV. If interpreted as broadened
and redshifted annihilation line, this observation seemed to make
1E~1740.7-2942 be the long sought compact and variable source of
positrons. Follow-up observations led to the classification of
1E~1740.7-2942 as the first micro-quasar \citep{Mirabel+92}: a binary system
involving a compact object (neutron star or black hole, see sec. IV.B.3) accreting material
from its companion and emitting part of the accreted energy in the form of jets.  It
was therefore proposed that 1E~1740.7-2942 would occasionally emit jets of
positrons (produced in e$^-$-e$^+$  pairs), some of which would annihilate in the
inner edge of the accretion disk as presumably observed by SIGMA; the
remaining positrons would eventually lose their energy and give rise
to time-variable narrow 511~keV line emission. 
Different SIGMA teams also reported
narrow and/or broad $\gamma$-ray lines near 511~keV, lasting for a day or so, 
from the transient X-ray source ``Nova Muscae'' \citep{Sunyaev:1992,Goldwurm:1992}
and the Crab nebula \citep{Gilfanov:1994}.
Another transient $\gamma$-ray line source was discovered from archival HEAO 1 data by
\citet{Briggs:1995}.

However, the line feature seen by SIGMA was not seen in
simultaneous observations of 1E~1740.7-2942 performed
with the OSSE \citep{Jung:1995} and BATSE \citep{Smith:1996a} instruments aboard the 
NASA Compton Gamma Ray Observatory (CGRO, launched in 1991).
Besides, BATSE data 
did not confirm the transient event seen by SIGMA from the Crab nebula \citep{Smith:1996a}.
Moreover, the search on 6 years of BATSE data did not reveal any transient line feature from
any direction of the sky \citep{Smith:1996b,Cheng:1998}.
Similarly, 9 years of SMM data did not show any transient event from the
Galactic center direction \citep{Harris:1994a} or the Crab nebula
\citep{Harris:1994b}.
A reanalysis of HEAO 3 data then revealed that the drop in 511~keV flux
reported earlier by \citet{Riegler:1981} was not significant \citep{Mahoney:1994}.
Thus, the idea of a steady 511~keV Galactic emission was gradually established.

The contradictory results obtained during the 1980s and early 1990s
provide a dramatic illustration of  the difficulties affecting the
 analysis of $\gamma$-ray line data.
In this domain, astrophysical signals rarely exceed the instrumental
background by more than a few percent and any systematic
uncertainty in the treatment of the background immediately disturbs the analysis.
In particular, the time variability of the instrumental background (due to
changing radiation environments along the orbital trajectory, or due to
solar activity) can easily fake time variable signals.
In addition, hard X-ray sources often exhibit highly variable continuum
emission components that may further  affect the data analysis and require their proper modeling; 
this concerns,  in particular, the
densely populated regions towards the Galactic center, which were not spatially resolved by 
older instruments.

%__________________________________________________________________  
\subsubsection{Early mapping of the spatial  distribution of e$^+$ annihilation}  

%To unveil the elusive nature of the galactic positron source, imaging the 511~keV
%annihilation emission became more and more pressing.
Before the launch of CGRO 
in 1991 with its OSSE collimated (11.4\degree~x~3.8\degree) spectrometer, the spatial
distribution of 511~keV line emission was only poorly constrained.
Hypotheses on a possible extent of the emission were mainly 
based on theoretical expectations \citep{1987ApJ...316..801K},  
on the different fluxes received by detectors with different fields of view
\citep{Albernhe:1981,Dunphy:1983}, 
and on a marginal detection of the 511~keV line near $\sim25\dg$ Galactic 
longitude with the balloon-borne  GRIS telescope
\citep{Leventhal:1989,Gehrels:1991}.
Nine years of OSSE observations drastically improved this situation.

\begin{figure}
\includegraphics[width=0.49\textwidth]{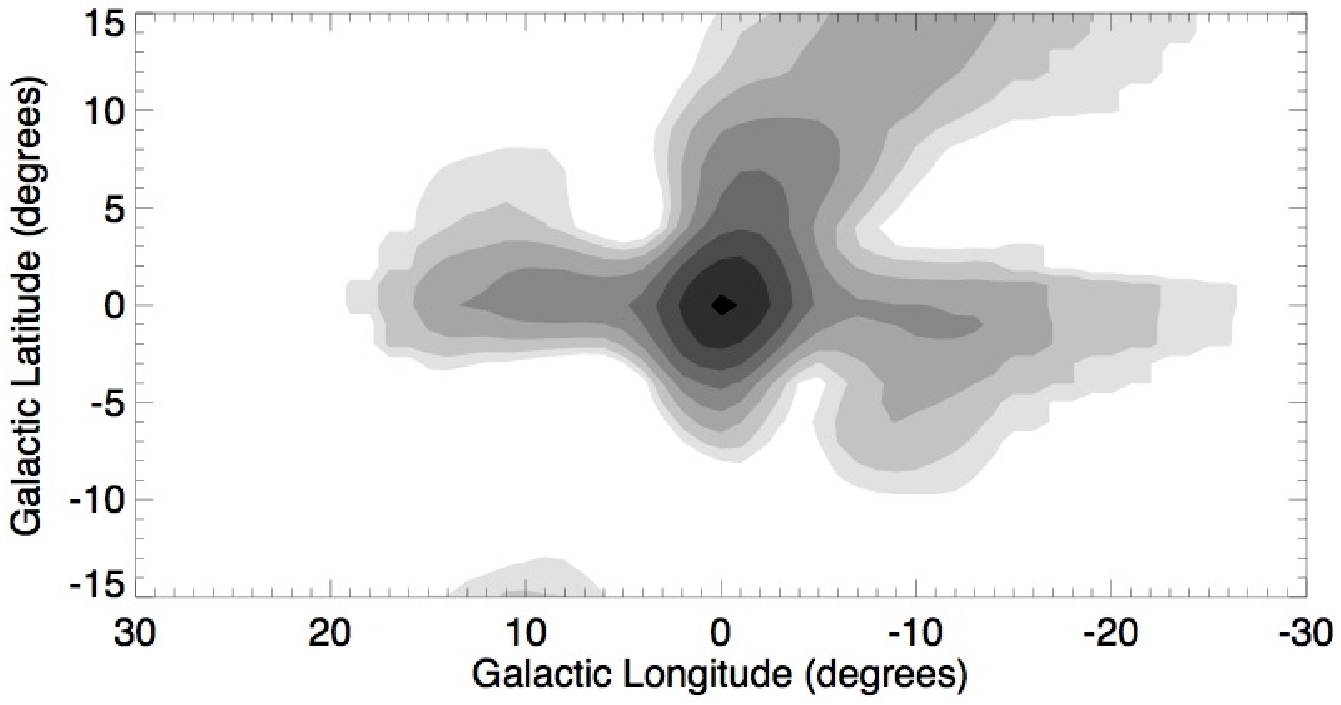}
\includegraphics[width=0.49\textwidth]{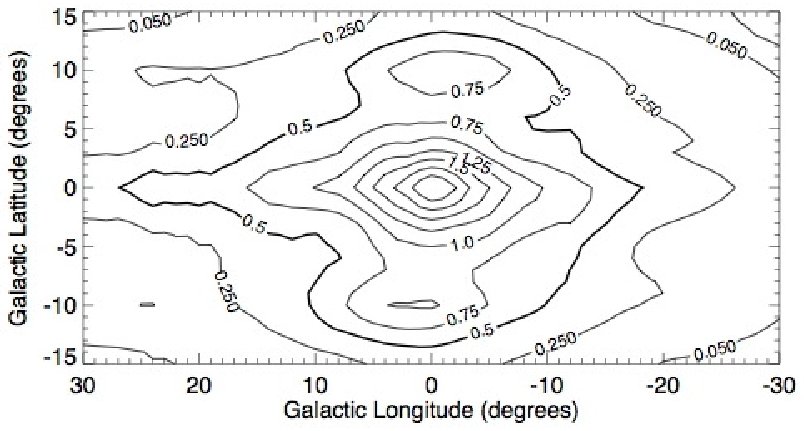}
\caption{OSSE 511~keV line map of the galactic center region ({\it top}) and
corresponding exposure map ({\it bottom}) (from \onlinecite{Purcell:1997}).
\label{fig:ossemap}}
\end{figure}

OSSE data  could clearly exclude a single point source as the
origin of observed 511~keV line emission \citep{Purcell:1994}.
The data were best understood in terms of an extended source consisting of a symmetrical
bulge (centered on the Galactic center) and emission from the
Galactic plane.
\citet{Cheng:1997} and \citet{Purcell:1997} established the first 511~keV
line emission map of the central Galactic ridge (Fig.~\ref{fig:ossemap}).
Beyond the aforementioned  components, there was hint of a third  component located
at Galactic coordinates of longitude $l \sim -2\dg$ and latitude 
$b \sim 12\dg$, dubbed the {\em Positive
Latitude Enhancement} ({\em PLE}).
However, the intensity and morphology of this feature were only weakly
constrained by the data \citep{Milne:2001a}, and the non-uniform exposure
of the sky may have biased the sky maps \citep{VonBallmoos:2003}.
\citet{Kinzer:1996}, \citet{Kinzer:2001} and \citet{Milne:1998} studied the spatial distribution
of the continuum emission from  positronium annihilation and concluded that it follows closely the
distribution of the 511~keV  line. However,
no {\em PLE} was visible in the continuum emission image \citep{Milne:2001b}.

%%%%%%%%%%%%%%%%%%%%%%%%%%%%%%%%%%%%%%%%%%%%%%%%%%
%%%%%%%%%%%%%%%%%%%%%%%%%%%%%%%%%%%%%%%%%%%%%%%%%%
At this point one should mention that images in the hard X-ray and soft $\gamma$-ray domains
are obtained through complex non-linear iterative deconvolution techniques, and they generally
represent only a family of solutions, which 
explains the observed data within the given    statistical and convergence constraints.
 The reader should be aware of this particularly important point when
inspecting all images in this paper. For instance, other OSSE images of the Galactic 511 keV
line emission are presented in \citep{Milne:2001b} or \citep{Milne:2002}.

Several models have been proposed to describe the spatial distribution of the
annihilation emission observed by OSSE
\citep{Prantzos93,Purcell:1994,Kinzer:1996,Purcell:1997,Milne:2000,Kinzer:2001,Milne:2001a}.
They all had in common a two-component emission from a
spheroid located in the inner Galaxy and from the extended Galactic disk (see Sec. III.A
for a detailed discussion of the Galaxy's morphology). However,
both morphology and relative intensity of these two components were only
poorly constrained by the data; depending on the adopted model, the spheroidal component
was claimed to be dominant or sub-dominant, i.e
%While models using a halo type spheroidal component suggested a halo dominated
%morphology, models using a Gaussian type spheroidal suggested a disk dominated 
%morphology. Consequently, 
the Galactic spheroidal-to-disk  flux ratio was only 
constrained in the broad interval $F_{spher}/F_{disk}\sim 0.2 - 3.3$.
The uncertainty on the total Galactic 511~keV line flux was also rather large, spanning the
range 1-3~10$^{-3}$ \funit.

Despite the considerable progress achieved by OSSE observations, the
origin of Galactic positrons remained unclear.
The data did not constrain  the morphology of the 511~keV emission enough  to
clarify the underlying source population(s).
Yet, the strong concentration of the 511~keV emission towards the Galactic
bulge led several authors to suggest that the $\beta^+$ decay of radioactive $^{56}$Co,
produced by Galactic Type Ia supernovae (SNIa), should be the dominant Galactic
positron source \citep{Kinzer:1996,Kinzer:2001,Milne:2002}.
The emission from the Galactic disk was generally attributed to radioactive $\beta^+$ decays
of  $^{26}$Al, $^{56}$Co and $^{44}$Ti produced by a variety of stellar sources
\citep{Purcell:1994,Kinzer:1996,Purcell:1997}. In fact,  $^{26}$Al had already been detected from the
inner region of the Galaxy through its characteristic $\gamma$-ray line at 1809~keV  in the 1979/1980 HEAO-C data \cite{1982ApJ...262..742M},  
and its contribution to Galactic-disk e$^+$ production was established (Sec. II.B.2 and IV.A.2). 
%Contributions of radioactive $\beta^+$decay nuclei from SNIa and core-collapse supernovae were plausible but less clear (see Sec. IV.A.4).

%__________________________________________________________________  
\subsubsection{Imaging with INTEGRAL/SPI}
\label{subsub:imaging}

With the launch of  ESA's INTEGRAL observatory \citep{Winkler2003} in 2002 for a multi-year mission,
a new opportunity became available for the study of Galactic e$^+$
annihilation. The SPI imaging spectrometer \citep{Vedrenne2003} combines for the first time imaging with high-resolution spectroscopy. 
The spatial resolution of $3\dg$ (FWHM) of SPI, 
though inferior to telescopes optimized for slightly lower energies (SIGMA, IBIS), 
is superior to that of OSSE; 
its energy coverage and sensitivity around  the annihilation line and its large field of view allow an
 improved study of the 511~keV emission morphology.
The spectral resolution of $\sim2.1$~keV (FWHM, at 0.5~MeV) is comparable to that of other
Ge detectors employed on balloons or the HEAO~3 satellite, allowing for
a spatially resolved fine spectroscopy of the signal (including the
underlying continuum emission).

%%%%%%%%%%%%%%%%%%%%%%%%%%%%%%%%%%%%%%%%%%%%%%%%%%
\begin{figure}
\includegraphics[width=0.49\textwidth]{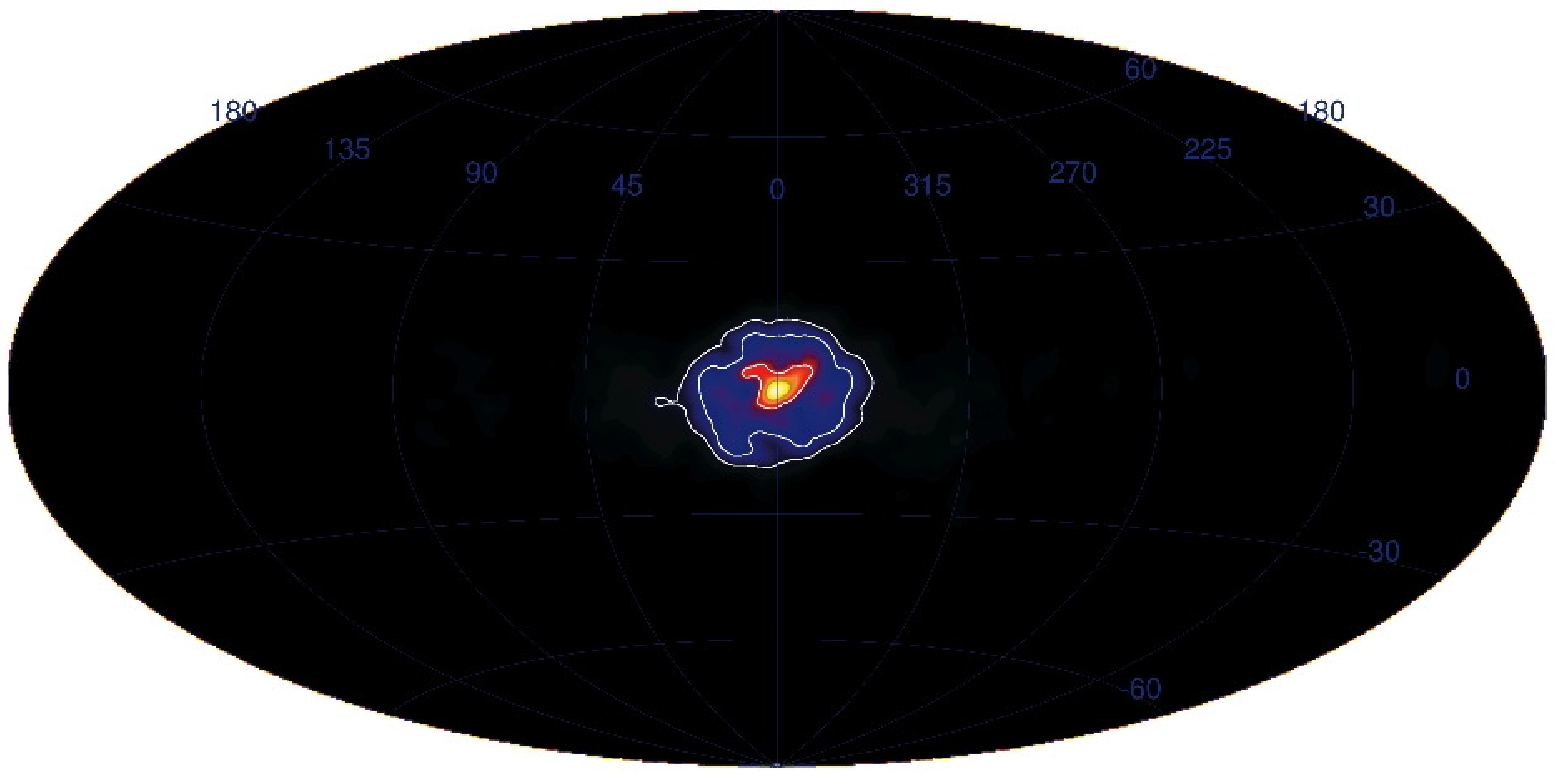}
\includegraphics[width=0.49\textwidth]{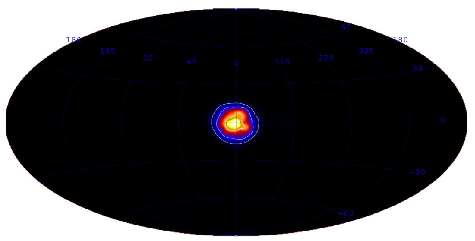}
\caption{511~keV line map ({\it top}) and positronium continuum map
({\it bottom}) derived from one year of INTEGRAL/SPI data
(from \onlinecite{Knodlseder:2005} and \onlinecite{Weidenspointner:2006}, respectively).
\label{fig:spimap}}
\end{figure}
%%%%%%%%%%%%%%%%%%%%%%%%%%%%%%%%%%%%%%%%%%%%%%%%%%

The first 511~keV line and positronium continuum all-sky maps have been presented
by \citet{Knodlseder:2005} and \citet{Weidenspointner:2006}, respectively, 
based on approximately one year of SPI data (Fig.~\ref{fig:spimap}).
The two maps are compatible with each other (within their uncertainties), 
suggesting that the positronium fraction does not vary over the sky.
The images illustrate the remarkable predominance of the spheroidal component.
In contrast to  OSSE data, which suggested a relatively strong disk component,
the Galactic disk seemed to be completely absent in the first year SPI images.
Model fitting indicated only a marginal signal from
the Galactic disk, corresponding to a bulge-to-disk flux ratio $>1$
\citep{Knodlseder:2005}.
This strong predominance of the Galactic bulge, unseen in any other wavelength, stimulated
"unconventional" models involving dark matter (Sec. IV.C).
However, \citet{Prantzos06} pointed out that the data could not exclude the
presence of disk emission of a larger latitudinal extent (resulting from positrons propagating
far away from their sources), which  could be rather luminous and still undetectable
by SPI, because of its low surface brightness.

%%%%%%%%%%%%%%%%%%%%%%%%%%%%%%%%%%%%%%%%%%%%%%%%%%
\begin{figure}
\includegraphics[width=0.49\textwidth]{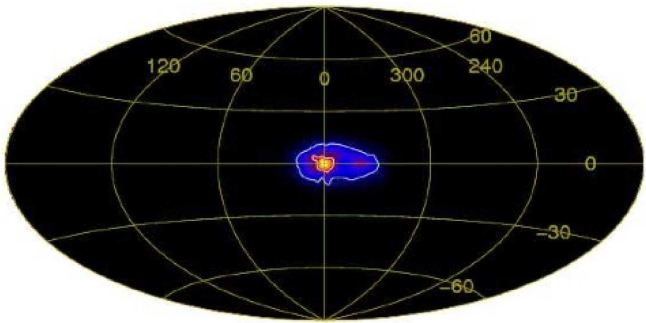}
\caption{511~keV line map derived from 5 years of INTEGRAL/SPI data
(from \onlinecite{Weidenspointner+08a}).
\label{fig:spiasmap}}
\end{figure}
%%%%%%%%%%%%%%%%%%%%%%%%%%%%%%%%%%%%%%%%%%%%%%%%%%

After accumulating 5 years of INTEGRAL/SPI data the 511~keV line emission all-sky image 
revealed also fainter emission extending along the Galactic plane (Fig.~\ref{fig:spiasmap}).
With a much improved exposure with respect to the first year  (in particular along the 
Galactic plane), 511~keV emission from the Galactic disk is now clearly detected
\citep{Weidenspointner+08a}. 
However, the detailed quantitative characterization of components of 511 keV emission requires
parameterizing these in the form of (necessarily idealized) spatial emission models fitted to the data.
No unique description emerges at present, since both the spheroid and the disk may have
faint extensions contributing substantially to their total $\gamma$-ray emissivities.
It turns out that the bulge emission is best described by combining  a narrow and a broad Gaussian,
with widths (FWHM, projected onto the sky) of 3\degree \ and 11\degree, respectively. Another, more extended 
component is needed to fit the data,  a rather
thick disk of vertical extent 7\degree \ (FWHM projected on the sky). 
The model implies a total e$^+$ annihilation rate of 2 10$^{43}$ e$^+$ s$^{-1}$ and a
spheroid/disk  ratio of 1.4  (Table~\ref{tab:MorphGeorg}).  It should be noted, however, that alternative
models, involving extended components of low surface brightness (thus far undetected by SPI) are
also possible. One such  alternative \citep{Weidenspointner+08b}
involves a centrally condensed
but very extended halo and a thinner disk (projected vertical extent of 4\degree), with a spheroid/disk ratio of 6 
(Table~\ref{tab:MorphGeorg}).
 
With more SPI data, it was possible to proceed to more detailed constraints on the morphology of
the disk emission. The flux in the disk component remains concentrated to 
longitudes $|l| < 50\dg$; no significant 511~keV
line emission has been detected from beyond this interval so far.
The accumulated SPI data yield a flux from negative longitudes of the Galactic disk that is
twice as large as the flux from an equivalent region at positive longitudes.
The significance of this asymmetry is still rather low, about $\sim4\sigma$.  
Indications for such an asymmetry were already noticed in the OSSE data
(M. Leising, private communication). It should be noted, however, that
a different analysis of the same SPI data finds no evidence for a disk asymmetry
\citep{Bouchet+08,BRJ10}, although it cannot exclude it, either. Clearly, clarifying the asymmetric or symmetric nature of the disk profile
should be a major aim of the 511 keV studies in the years to come\footnote{INTEGRAL will
continue operations until 2012, at least.}.

\begin{table}
     \caption{Two model fits of the Galactic 511 keV emission 
     (from \onlinecite{Weidenspointner+08b}): fluxes, photon
     emissivities and e$^+$ annihilation rates (computed for a positronium 
     fraction of $f_{ps}$=0.967, see Sec. II.B.4). Notice that "thin" and "thick" 
    disks have not the same meaning as in Sec. III.
     \label{tab:MorphGeorg}}
     \begin{tabular}[b]{llll}
\noalign{\smallskip}
\hline
\hline
\noalign{\smallskip}
     &    $F_{511}$  &  $L_{511}$       &  $\dot{N}_{e^+}$      \\
     &     ($10^{-4}$ cm$^{-2}$ s$^{-1}$) &   (10$^{42}$ s$^{-1}$)       &  (10$^{42}$ s$^{-1}$ )     \\
\hline
\noalign{\smallskip}
{\it Bulge + thick disk}  &   &   & \\
\noalign{\smallskip}
Narrow bulge   & 2.7$^{+0.9}_{-0.4}$   &   2.3$^{+0.8}_{-0.7}$  & 4.1$^{+1.5}_{-1.2}$  \\
Broad  bulge   & 4.8$^{+0.7}_{-0.4}$   &   4.1$^{+0.6}_{-0.4}$  & 7.4$^{+1.0}_{-0.8}$  \\
Thick disk     & 9.4$^{+1.8}_{-1.4}$   &   4.5$^{+0.8}_{-0.7}$  & 8.1$^{+1.5}_{-1.4}$  \\
Total          & 17.1                  &   10.9                 & 19.6                 \\
Bulge/Disk  &     0.8             &    1.4             &     1.4                 \\
\noalign{\smallskip}
\noalign{\smallskip}
{\it Halo + thin disk}  &   &   & \\

\noalign{\smallskip}
Halo   & 21.4$^{+1.1}_{-1.2}$   &   17.4$^{+0.9}_{-1.1}$  & 31.3$^{+2.2}_{-2.6}$  \\
Disk   & 7.3$^{+2.6}_{-1.9}$   &   2.9$^{+0.6}_{-0.6}$  & 5.2$^{+1.1}_{-1.1}$  \\
Total          & 28.7                  &   20.3                 & 36.5                 \\
Halo/Disk  &   2.9                    &    6                  &      6                \\
\noalign{\smallskip}
  \hline
\end{tabular}
\end{table}

%__________________________________________________________________  
\subsubsection{Spectroscopy with INTEGRAL/SPI}

Before INTEGRAL, the spectral shape of the positron annihilation emission
was only poorly constrained by observations.
All high-resolution observations suggested a modest line broadening 
of FWHM$\sim$ 2 keV  \citep{Leventhal:1993,Smith:1993,Mahoney:1994,Harris:1998}. 
The excellent spectral resolution of SPI allows now for the first time to study
the spectrum of this  emission in great detail and for different regions.

\begin{figure}
\includegraphics[width=0.49\textwidth]{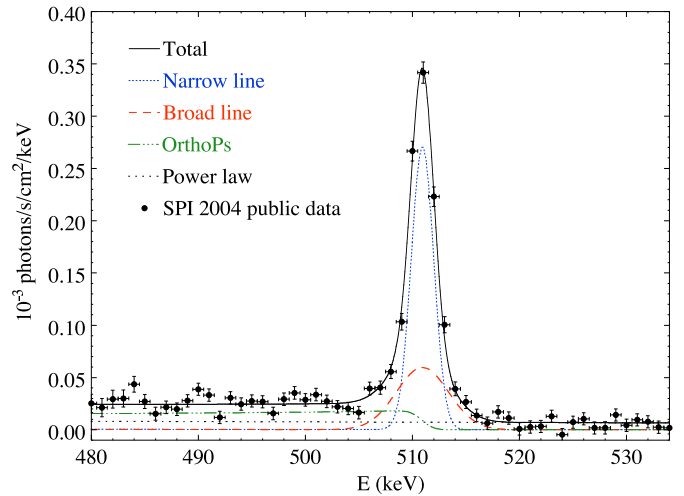}
\caption{Fit of the spectrum of the annihilation 
emission measured by SPI with narrow and broad Gaussian lines and an 
ortho-positronium continuum. The power-law account for the Galactic 
diffuse continuum emission \citep{Jean:2006}.
\label{fit:spispec}}
\end{figure}

Spectral results for the Galactic spheroidal emission were presented by \citet{Churazov:2005} 
and \citet{Jean:2006}, based on the first year of SPI data.  
The line displays no spectral shift, i.e. it has an energy $E$=511$\pm$0.08 keV \citep{Churazov:2005} and
it  is composed of
two spectral components (assumed to be represented by Gaussians):
a narrow line with a width of  FWHM=$1.3 \pm 0.4$ keV and a broad
component with a width of FWHM=$5.4 \pm 1.2$ keV
(Fig.~\ref{fit:spispec}). 
The width of the broad line is in agreement with the broadening expected
from positronium annihilation via charge exchange with hydrogen atoms
(see section V.B.2). The narrow line component contains $\sim2/3$ 
of the total annihilation line flux while the broad one makes up the remaining $\sim1/3$ 
of the flux. Table \ref{tab:specJean} summarizes the results of the spectral analysis
of the Galactic 511 keV emission after the first year of SPI data.

SPI also clearly detected the ortho-positronium continuum with an intensity
that corresponds to a positronium fraction of $f_{Ps}$=97$\pm$2 \%
(\onlinecite{Jean:2006}; see Eq. \ref{eq:fPos}).
This value is in good agreement with earlier measurements obtained by
OSSE (97$\pm$3 \%, \onlinecite{Kinzer:1996}) 
and TGRS (94$\pm$4 \%, \onlinecite{Harris:1998}).

The shape of the annihilation line and the relative intensity of the 
ortho-positronium continuum are closely related to the physical conditions
such as temperature, ionisation stage and chemical abundances of the interstellar medium in 
which positrons annihilate. These conditions, obtained from the analysis of the  measured 
spectrum, are presented and discussed in Sec. V.E.
Important complementary information on the energies of the annihilating positrons
 is obtained from the analysis of the observed continuum emission at somewhat higher
energies (above 511 keV and into the MeV region), as discussed in the next section and Sec. V.B.

%%%%%%%%%%%%%%%%%%%%%%%%%%%%%%%%%%%%%%%%%%%%%%%%%%

\begin{table}
     \caption{Results of spectral analysis of Galactic 511 keV emission from the region within
     8\degree from the Galactic center.  $I_{n}$, $\Gamma_{n}$, 
     $I_{b}$ and $\Gamma_{b}$ are the flux and width (FWHM) of the narrow and broad 
     lines, respectively, $I_{3\gamma}$ is the flux of the ortho-positronium 
     continuum and $A_{c}$ is the amplitude of the Galactic continuum at 511~keV. 
     The first set of error bars refers to  1$\sigma$ 
     statistical errors and the second set to  systematic errors (from \onlinecite{Jean:2006}). 
     \label{tab:specJean}}
     \begin{tabular}[b]{ll}
\noalign{\smallskip}
\hline
\hline
\noalign{\smallskip}
\mbox{Parameters}  & \mbox{Measured values} \\
\hline
\noalign{\smallskip}
I$_{n}$ \; \mbox{(10$^{-3}$ s$^{-1}$ cm$^{-2}$)} & 0.72 $\pm$ 0.12 $\pm$ 0.02  \\
\noalign{\smallskip}
$\Gamma_{n}$ \; \mbox{(keV)} & 1.32 $\pm$ 0.35 $\pm$ 0.05 \\
\noalign{\smallskip}
I$_{b}$ \; \mbox{(10$^{-3}$ s$^{-1}$ cm$^{-2}$)} & 0.35 $\pm$ 0.11 $\pm$ 0.02 \\
\noalign{\smallskip}
$\Gamma_{b}$ \; \mbox{(keV)} & 5.36 $\pm$ 1.22 $\pm$ 0.06 \\
\noalign{\smallskip}
I$_{3\gamma}$ \; \mbox{(10$^{-3}$ s$^{-1}$ cm$^{-2}$)} & 4.23 $\pm$ 0.32 $\pm$ 0.03 \\
\noalign{\smallskip}
A$_{c}$ \; \mbox{(10$^{-6}$ s$^{-1}$ cm$^{-2}$ keV$^{-1}$)} & 7.17 $\pm$ 0.80 $\pm$ 0.06 \\
\noalign{\smallskip}
  \hline
\end{tabular}
\end{table}

\subsection{Relevant observations at MeV energies}

\subsubsection{The MeV continuum}

Positrons are typically emitted at relativistic energies, in some cases even far above 1 MeV (Sec. IV). 
They behave essentially like relativistic electrons of cosmic rays, by producing bremsstrahlung and inverse-Compton emission while slowing down to the thermal energies (eV) of the interstellar medium, where they eventually annihilate. But positrons may also annihilate "in flight" while still having relativistic energies, 
giving rise to a unique $\gamma$-ray continuum  signature at energies above 511 keV (as the center-of-mass energy is transferred to annihilation photons; Sec. V). The shape and amplitude of this $\gamma$-ray emission depend on the  injection spectrum of positrons and the corresponding total annihilation rate. For positrons injected at low energies (of the order of $\sim$MeV, such as those released by radioactivity), the amplitude of the in-flight annihilation continuum above 1 MeV is quite small, while for  sources injecting positrons at much higher energy (such as cosmic-ray  positrons from pion decay), the annihilation $\gamma$-ray spectrum would extend up to GeV energies and include a considerable $\gamma$-ray flux. The high energy $\gamma$-ray continuum above 1 MeV therefore constrains the energy and the annihilation rate of relativistic positrons, when all other sources of such high energy emission are properly accounted for.

Diffuse Galactic continuum emission has been well-measured at least in the inner part of the Galactic disk (longitudes $-30^\circ < l < 30^\circ$) in the hard-X-ray through $\gamma$-ray regime by INTEGRAL, OSSE, COMPTEL, and EGRET \cite{Bouchet+08,1999ApJ...515..215K,1994A&A...292...82S}. The spectrum of the underlying continuum emission in the 511 keV region  is best represented as a power-law with index 1.55 \cite{Bouchet+08} and is mostly due to emission from
cosmic-ray electrons and positrons (Sec. \ref{subsubsec:GalaxyCR}). The corresponding emission processes  are modelled in detail in e.g. the GALPROP code  \citep{SMP07}, which 
includes 3D distributions of interstellar gas and photon fields, as well as all relevant interaction cross sections and constraints from near-Earth observations of cosmic-ray fluxes and spectra.  This model reproduces well  the entire range of $\gamma$-ray observations,
 from keV  to GeV energies; however, there exist tantalizing hints of residual emission in the MeV region,  when comparison is made to COMPTEL measurements
 \footnote{ The CGRO/COMPTEL data points have an uncertainty of up to a factor 2 due
to the difficulty of producing skymaps with a Compton telescope with high
background. The most reliable COMPTEL values come from a maximum-entropy imaging
analysis and are model-independent, but 
the zero level is uncertain and contributes to the systematic error.
The enormous gap in sensitivity between the current Fermi
mission ($>$30 MeV) and the 1-30 MeV range (sensitivity factor $\sim$100 !)
highlights the urgent need for new experiments in the MeV range.}: 
 the data points appear to lie on the high side of model predictions. In view of the possible systematic uncertainties of such measurements, but also of the parameters of GALPROP code and the possible
 contributions of unresolved X-ray binaries,
some (though little) room is still left for a contribution of in-flight e$^+$ annihilation to the MeV continuum.

%%%%%%%%%%%%%%%%%%%%% spectra vs. phases %%%%%%%%%%%%%%%%%%%%%%%%
\begin{figure}
\includegraphics[width=0.47\textwidth, height=0.35\textwidth]{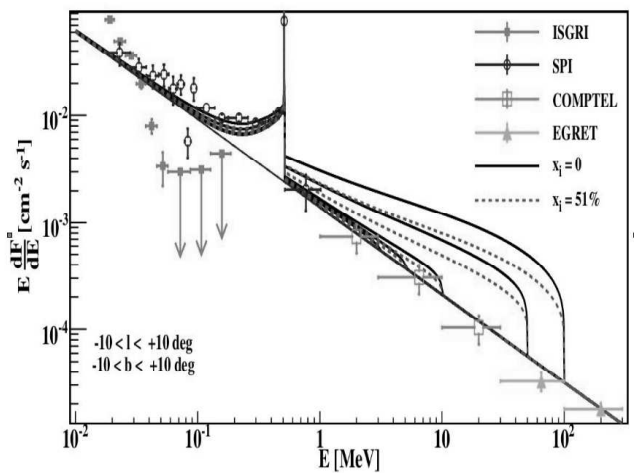}
\caption{\label{fig:sizun} Spectrum of the inner Galaxy as measured by various instruments, compared to various theoretical estimates made under the assumption that positrons are injected at high energy: the four pairs of curves result from positrons
injected at  100, 30, 10 and 3 MeV (from top to bottom) and correspond to positrons propagating in neutral ({\it solid}) or
50\% ionized ({\it dotted}) media (from  \onlinecite{Sizun:2006}). This constrains the injected positron energy (or,
equivalently, the mass of decaying/annihilating dark matter particles) to a few MeV.}
\end{figure}
%%%%%%%%%%%%%%%%%%%%%%%%%%%%%%%%%%%%%%%%%%%%%%%%%%%%%%%%%%%%%%%%%%

The physics of the in-flight annihilation of positrons will be analyzed in Sec. V. Here we simply note that
the corresponding constraints on the injection energy of positrons wee
 pointed out many years ago by
\citet{Agaronyan-Atoyan:1981}.  They showed that the positrons which are 
responsible for the Galactic 511~keV line cannot be produced in a 
steady state by the decay of the $\pi^{+}$ created in proton-proton 
collisions or else the in-flight annihilation emission should have been 
detected. A similar argument was used by \citet{Beacom-Yuksel:2006} 
and \citet{Sizun:2006} to constrain the mass of the dark matter particle 
which could be the source of positrons in the Galactic spheroid
(see section IV.C). If such particles produce positrons (in their decay or 
annihilation) at a rate which corresponds to the 
observed 511 keV  emission, then their mass should be 
less than a few  MeV, otherwise the kinetic energy of the created positrons would 
have been sufficiently high to produce a measurable $\gamma$-ray continuum 
emission in the 1-30 MeV range (Fig.  \ref{fig:sizun}).
 The same argument allows one to constrain
the initial kinetic energy of positrons and thus to eliminate several classes of candidate
sources, like e.g. pulsars, millisecond pulsars, magnetars, cosmic rays etc., as major positron producers
(Sec. IV. D).

\subsubsection{\label{sec2:ObsAl26}Gamma-rays and positrons from radioactive $^{26}$Al}
% revised RD 04Oct09

\Al \ is a long-lived (half-life $\tau_{1/2}$=7.4 10$^5$ yr) radioactive isotope. It decays by emitting a positron, while the de-excitation of daughter nucleus  $^{26}$Mg emits a characteristic $\gamma$-ray line at 1808.63~keV.  Based on predictions of nucleosynthesis calculations in the 1970's,
\citet{Arnett77} and  \citet{RL77} suggested that its $\gamma$-ray emission should be detectable by forthcoming space instruments. 
The detection of the 1809~keV line from the inner Galaxy with the HEAO-C germanium spectrometer \cite{1982ApJ...262..742M} came 
as a small surprise \citep{Clayton84} because of its unexpectedly high flux ($\sim$4~10$^{-4}$ cm$^{-2}$ s$^{-1}$).
Being the first radioactivity ever detected through its $\gamma$-ray line signature, it
provided direct proof of ongoing nucleosynthesis in our Galaxy (see review by \onlinecite{1996PhR...267....1P}).

\begin{figure}
\includegraphics[angle=-90,width=0.49\textwidth]{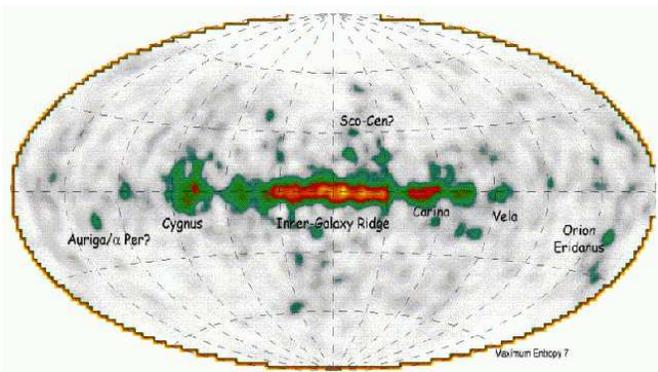}
\caption{Map of Galactic $^{26}$Al $\gamma$-ray emission after 9-year observations with COMPTEL/CGRO (from \onlinecite{Pluschke+01}).
}
\label{fig:Al26_map}
\end{figure}

Several balloon experiments, and in particular the Gamma-Ray Spectrometer (GRS)
aboard the Solar Maximum Mission (SMM)  rapidly confirmed the HEAO-C finding  \citep{1985ApJ...292L..61S}. 
Early experiments  had large fields of view (130\degree\ for SMM, 42\degree\ for HEAO-C) with no or modest imaging  capabilities. The first map of Galactic 1.8 MeV emission was
obtained with the COMPTEL instrument aboard CGRO \citep{Diehl+95}, which had a spatial resolution of 3.8\degree \  (FWHM) within a field of view of 30\degree. The sky map derived from the 9-year survey of  COMPTEL is shown in Fig. \ref{fig:Al26_map}.
Unlike the 511 keV maps of Fig. \ref{fig:spimap} and \ref{fig:spiasmap} the \Al \ emission is 
concentrated along the Galactic plane (brightest within the inner Galactic radian) and is irregular, with emission maxima aligned with spiral-arm tangents. 
The Cygnus region stands out as a significant and bright emission region.
The "patchy" nature of the \Al $\gamma$-ray emission suggests that massive stars are the most important contributors to Galactic \Al, 
as  suggested  at a time when the morphology of \Al \ emission was
unknown (\onlinecite{Prantzos91} and Sec. IV.A.2). It is consistent with the (statistically significant) similarity to the
Galactic free-free emission map, which reflects  electron radiation from HII regions ionized from
the same massive stars that eventually release \Al   \citep{Knodlseder99}.

The total flux of \Al \ $\gamma$-rays depends slightly on the measuring instrument.
In terms of statistical precision, the SMM result of 4.0$\pm$0.4~10$^{-4}$~\fluxrad has been considered the canonical value. Imaging instruments, however, have consistently reported lower flux values of 2.6$\pm$0.8~10$^{-4}$~\fluxrad (COMPTEL) and 3.1$\pm$0.4~10$^{-4}$~\fluxrad (SPI), respectively. 
%This has been attributed to either instrumental biases from the treatment of an underlying background line, or  to some large-scale extended low surface-brightness emission. The latter would be missed by the imaging instruments (which normalize their background mostly at high-latitude exposures  by assuming that these are free of celestial \Al  emission), while SMM had employed Earth occultation analysis to determine its background level. Thus, each of the reported  flux values for the  inner Galactic radian includes, to some degree, observational uncertainties and biases, as well as a dependency on the 3D source distribution model used to convert measured flux into \Al abundance. Analysis with the SPI/INTEGRAL has shown that the latter amounts to values differing by almost 30\% even among plausible models (the statistical variance being only \about~4\%, though). 
The latest SPI value is compatible with the full  range of measured values by other instruments (within statistical uncertainties), and we adopt it here.
The detected flux translates into a decay rate of \Al \  which depends slightly on the adopted
3D distribution of \Al \ in the Galaxy \citep{Diehl+06}. The most recent analysis of SPI data results in a rate of $\dot{N}_{26}$=
4.3 10$^{42}$ s$^{-1}$ or 2.7 \ms/Myr \citep{Wang+09}. Assuming a steady state, i.e. equality between production and decay rates, this is also the present production rate of \Al \ in the Galaxy;  recent models of massive star nucleosynthesis can readily explain such a production rate (\onlinecite{Diehl+06} and Sec. IV.A.2).

Being predominantly a $\beta^+$-emitter (with a branching ratio of $f_{e^+,26}$=82\%, see Table \ref{tab:Emitters})  \Al  \ is 
itself a source of positrons. The corresponding Galactic e$^+$ production rate is  $\dot{N}_{e^+,26}$=
$f_{e^+,26} \dot{N}_{26}\sim$ 3.5 10$^{42}$ s$^{-1}$ . This constitutes a significant contribution
to the total Galactic e$^+$ production rate  (Sec. II.A.3 and Table \ref{tab:MorphGeorg}): 17\% of the total  e$^+$ annihilation rate and almost half  of the (thick) disk in the double bulge+thick disk model, or 10\% of the total and 70\% of the thin disk in the Halo+thin disk model. We shall see in Sec. IV that positrons from other $\beta^+$-decaying nuclei can readily explain the remaining disk emissivity, while the bulge emissivity remains hard to explain.

%Assuming prompt or steady-state annihilations of \Al positrons, this would contribute with (1.58$\pm$0.2)~10$^{-4}$\fluxrad in the inner Galaxy region (-28\degree$\leq l \leq$28\degree) to the observable 511~keV line emission. This would account for \about~17\% of the 511~keV flux, determined for the ``bulge region'' as \about (9.35$\pm$0.54)~10$^{-4}$~\flux. But we must account for the different origins of positrons, the \Al positrons deriving exclusively from the Galactic disk with no bulge component. Therefore, this \Al related annihilation flux is the {\it background} underlying some kind of bulge-origin-dominated positron annihilation source. From the disk brightness in 511~keV  of (5.5$\pm$2.5)~10$^{-4}$~\flux \cite{2005A&A...441..513K} and 1808.65~keV (11.1~10$^{-4}$~\flux all-sky, \cite{2001egru.conf...55P}) photons, respectively, we see that \Al decay may readily explain the {\it complete} 511~keV emission {\it in the region outside the bright bulge}, within uncertainties. Therefore, from another point of view, uncertainties in annihilation originating in the Galactic disk and in \Al gamma-ray brightness would allow for only modest other contributions to the integrated 511~keV disk emission, constrained to \about~2--3~10$^{-4}$~\flux at most.

\subsection{Summary of observational constraints}
\label{sec:Summary}
The results of the analysis of Galactic $\gamma$-ray emissions in the MeV range  can be summarized as follows:

1) {\it Intensity}: The total rate of positron annihilation observed in $\gamma$-rays is at least $L_{e^+}$=2 10$^{43}$ s$^{-1}$, depending
on the adopted source configuration. Most of it comes from the bulge (unless there is important emission from an extended,
low surface brightness, disk).

2) {\it Morphology:} The bulge/disk ratio of e$^+$  annihilation rates
is $B/D\sim$1.4; however, substantially different ratios cannot be excluded {\it if} there is important emission
of low surface brightness (currently undetectable by SPI) either from the disk or the spheroid.
About half of the disk emission can be explained by the observed radioactivity
of \Al \ (provided its positrons annihilate in the disk). 
There are hints for an asymmetric disk emission with flux ratio $F(l<$0\degree)/$F(l>$0\degree)$\sim$1.8,
which has yet to be confirmed.

3) {\it Spectroscopy:} The ratio of the 511 keV line to the E$<$511 keV continuum suggests a positronium fraction 
of 97$\pm$2 \% and constrain the physical conditions in the annihilation region.
The observed continuum at $\sim$MeV
energies can  be mostly explained with standard inverse Compton
emission from cosmic ray electrons. A contribution from unresolved compact
sources is possible, while a (small) contribution
from high-energy ($>$MeV) positrons annihilating in flight cannot be excluded.
 
These are the key observational constraints that should be satisfied by the source(s) 
and annihilation site(s) of
Galactic positrons. We shall reassess them in the light of theoretical analysis in the end of Sec. IV and V.

% Sec. III The Galaxy   ........................................................................................................................................................

\section{ The Galaxy}
\label{sec0:Galaxy}

The expected spatial distribution and intensity 
of the positron annihilation emission obviously depends on the corresponding distribution of the potential e$^+$ sources, as well as on the properties of the ISM in which positrons first slow down and then annihilate. One may distinguish two types of e$^+$ sources, depending on whether their lifetimes ($\tau_S$) are shorter or longer than the lifetime of positrons in the ISM ($\tau_{e^+}$). Calculation of the total e$^+$ production rate requires in the former case  ($\tau_S < \tau_{e^+}$) an estimate of (i) the Galactic birthrate $R_S$ of the sources and (ii) the individual e$^+$ yields $n_{e^+}$ (i.e. the average amount of positrons released by each source). In the latter case ($\tau_S > \tau_{e^+}$), the total number of such sources in the Galaxy $N_S$ is required, as well as the individual e$^+$ production rate $\dot{n}_{e^+}$ of each source. In the former class belong  supernovae or novae and the corresponding positron production rate is $\dot{N}_{e^+}=R_S n_{e^+}$; in the latter class belong e.g. low mass XRBs or millisecond  pulsars, and the corresponding positron production rate is
$\dot{N}_{e^+}=N_S \dot{n}_{e^+}$.

The galactic distribution of any kind of stellar source of positrons is somewhat related to the distribution of stars in the Milky Way. Similarly, the birthrate of any kind of positron source is somewhat related to the Galactic star formation rate. In this section we present a summary of current knowledge about the stellar populations of the Milky Way and  their spatial distribution and we 
discuss the birthrates of stars and supernovae. Since the slowing down and annihilation of positrons depend on the properties of the ISM, we present a brief overview of the  ISM in the bulge and the disk of the Milky Way. Positron propagation depends  also on the properties of the Galactic magnetic field, which are reviewed in Sec. III.D. Finally, the main properties of the Milky Way's dark matter halo are presented in Sec. III.E.

\subsection{\label{sec:GalaxyStars} Stellar populations }

The Milky Way is a typical spiral galaxy, with a total baryonic mass of $\sim$5 10$^{10}$ \ms, of which more than 80\% is in the form of stars. Stars are found in three main components: the central bulge,  the disk and the halo, while the gas is found essentially in the plane of the disk.
Because of its low mass, estimated to 4 10$^8$ \ms \ i.e. less than 1\% of the total, (\onlinecite{Bell+07}), the Galactic halo  plays no significant role in the positron production.  The bulge contains $\sim$1/3 of the total mass and an old stellar population (age$>$10 Gyr). The dominant component of the Milky Way is the so-called {\it thin disk}, a rotationally supported structure composed of stars of all ages (0-10 Gyr). A  non negligible, contribution is brought by the {\it thick disk}, an old ($>$10 Gyr) and kinematically distinct entity   identified by \citet{GR83}.

%\subsubsection{\label{sec:GalaxyBulgeStars} The Bulge and the Galactic center}

To a first approximation, and by analogy with external galaxies, the bulge of the Milky Way can be considered as spherical, with a density profile either  exponentially decreasing
with radius $r$  or of Einasto-type ($\rho(r)\propto$exp(-A $r^\alpha$).
Measurements in the near infrared (NIR), concerning either integrated starlight observations or  star counts  revealed that the bulge is not spherical, but elongated. Recent  models suggest a tri-axial ellipsoid, but its exact shape is difficult to determine 
( \onlinecite{LCG05}; \onlinecite{RMSS07}) because of the presence of a Galactic bar. 
%The distinction between the triaxial bulge and the bar is not yet clear. An operational definition
%sometimes adopted is that the bulge is included within Galactic longitudes $|l|<$10\degree \ (radius $R<$1.5 kpc) and the bar is its triaxial component \citep{VGG09} extending up to $\sim$3 kpc from the center.
The mass of the bulge lies in the range 1-2 10$^{10}$ \ms. 
%Integrated light analysis leads to lower values ($\sim$1.3 10$^{10}$ \ms, 
(\onlinecite{Dwek+95};
%) than the star count analysis (2$\pm$0.26 10$^{10}$ \ms, 
\onlinecite{RRDP03}).
%; note, however, that the latter study includes the contribution of white dwarfs, estimated to be of $\sim$26\%. Removing that contribution leaves a luminous stellar mass of $\sim$1.5 10$^{10}$ \ms.
%The metallicities
%\footnote{The term {\it metallicity} in astrophysics indicates the sum of the abundances of all elements heavier than He; in the case of the Sun, the metallicity is Z$_{\odot}\sim$0.014 by mass fraction, very close to the one of nearby young stars and the local interstellar medium. Metallicities are usually expressed in logarithmic scale, i.e. for an element X [X/H]=log($\frac{({\rm X/H}_*)}{({\rm X/H})_{\odot}}$), where X/H is the number ratio of X with respect to hydrogen. In observational stellar astrophysics, metallicity usually means {\it iron abundance}, i.e. [Fe/H]. } of stars in the bulge span a large range, from less than 1/10 solar to almost 3 times solar. Their mean metallicity is slightly higher than solar,
%[Fe/H]$\sim$0.15, as obtained from both photometric \cite{Zoccali+03} and
%spectroscopic (\onlinecite{Zoccali+07}, \onlinecite{Lecureur+07}) determinations. 
By comparing colour-magnitude diagrams of stars in
the bulge and in metal-rich globular clusters, \citet{Zoccali+03} find
that the populations of the two systems are co-eval, with an age of
$\sim$10 Gyr. 
%Such a large age is confirmed by the deepest CMD of the bulge today \cite{Sahu+06}. 
%Traces of younger populations, such as
%OH/IR stars, bright AGB variables etc. appear to be confined to the
%Galactic plane, but the bulk of the bulge is certainly old. Such an
%old age is supported by the observed behaviour of the [O/Fe] ratio as
%a function of [Fe/H] (\onlinecite{Zoccali+07}, \onlinecite{Lecureur+07}), which
%requires a rapid {\it and early} evolution of the system.

The innermost regions of the bulge, within a few hundred pc, are dominated by a distinct, disk-like component, called the Nuclear Bulge
%. Revealed in IRAS and COBE-DIRBE NIR data, this component 
which contains about 10\% of the bulge stellar population ($\sim$1.5 10$^9$ \ms) within a flattened region of radius 230$\pm$20 pc and scaleheight 45$\pm$5 pc \cite{LZM02}.
% Such a small region, with projected surface aerea smaller than 1 square degree on the sky, is %barely resolved by SPI/INTEGRAL observations. 
%Its stellar population is dominated by two young, massive
%clusters, the Arches and Quintuplet clusters (\onlinecite{Stolte+05}; \onlinecite{Figer08}),
% of ages 2 Myr and 4 Myr, respectively,  
%each one containg more than 10$^4$ \ms \ and more than a  hundred O-type stars. 
%The central few pc %which are not resolved by COBE, 
%are dominated by the Nuclear Stellar Cluster (NSC), a spherical configuration with a power-law density profile in the K-band $\propto(R/R_0)^{-2}$, with core radius $R_0$=0.22 pc and %central density $\rho_0$=3.3. 10$^6$ \mspcc ; its 
%total stellar mass of 3$\pm$1.5 10$^7$ \ms, of which 7 10$^5$ \ms \ are found within the central pc$^3$ \cite{LZM02}. The massive star populations in the central 0.5 pc have been resolved by    \citet{Paumard+06}, who find two sub-clusters of total mass slightly smaller than 10$^4$ \ms \ each, with ages evaluated from 4 to 8 Myr. 
It is dominated by three massive stellar clusters  (Nuclear Stellar Cluster or NSC in the innermost 5 pc, Arches and  Quintuplex), which have a mass distribution substantially flatter than the classical Salpeter IMF \footnote{Stars are born with a mass distribution called  Initial mass function (IMF). Observed IMFs of young stellar clusters in the Milky Way and other galaxies have similar IMFs, with the  upper part (M$>$1 \ms) described by a power-law ($dN/dM \propto M^{-(1+X)}$, where $X$ is the slope of the IMF; in most cases, $X$=1.35 (as determined by \onlinecite{Salpeter55}, for the local IMF.}\cite{Figer08}. Finally, in the center of the Milky Way, at the position of SgrA$^*$ source,  lies the supermassive Galactic black hole (SMBH) with a total mass of 
$\sim$ 4 10$^6$ \ms \ \citep{Gillessen+08}.

%\subsubsection{\label{sec:GalaxyDiskStars} The Disk}

The Sun is located in the thin disk of the Milky way, at a distance of $R_{\odot}\sim$8 kpc from the Galactic center; a recent evaluation, based on Cepheids, gives $R_{\odot}$=7.94$\pm$0.37(statistical)$\pm$0.26(systematic) kpc (\onlinecite{GUB08} and references therein). Furthermore, the Sun is not located exactly on the plane, but at a distance from it  $z_{\odot}\sim$25 pc, as evaluated from the recent analysis of  the  Sloan Digital Sky Survey (SDSS) data \citep{Juric+08}.

In studies of the Milky Way the solar neighborhood plays a pivotal role, since local properties can, in general, be measured with greater accuracy than global ones.
%An accurate determination of distances to nearby stars was made possible in the 90's by the HIPPARCOS satellite \citep{Perryman+97}. This allowed to establish a rather complete census of stars within a few tens of pc from the Sun. The resulting volume luminosity function is transformed into a surface luminosity function by taking into account the observed relationship between stellar luminosity and stellar velocity dispersion perpendicularly to the galactic plane \footnote{Bright stars stars are younger and lie closer to the disk plane (i.e. the have smaller vertical velocity dispersion) than faint stars, which are, in general, older and have been submitted for a longer time to ``disk heating''; this effect is generally attributed  to star encounters with giant molecular clouds, although supplementary mechanisms appear to be required to exlain quantitatively the observations (e.g. \citealp{Kroupa02})}. 
%\citet{Flynn+06}  find a total stellar surface density in the ``solar cylinder''\footnote{The solar cylinder is defined as a cylinder of radius 500 pc centered on Sun's position and extending perpendicularly to the Galactic plane up to several kpc.} of $\Sigma_*$=35.5 \mspcb
%Stellar remnants (mostly white dwarfs) contribute for 5.4 \mspcb \  and brown dwarfs for about 1.8 \mspcb \ to this amount; the %remaining 28.2 \mspcb \ are live stars, contributed by to which the thick disk and the halo contribute 7 \mspcb \ and 0.8 \mspcb, respectively,  while the remaining belongs to the thin disk.
%By adding 13 \mspcb \ for the interstellar gas (see next section), one obtains 
The total baryonic surface density of the solar cylinder\footnote{The solar cylinder is defined as a cylinder of radius 500 pc centered on Sun's position and extending perpendicularly to the Galactic plane up to several kpc.} is estimated to $\Sigma_T$=48.8 \mspcb \citep{Flynn+06},  with $\sim$13 \mspcb \ belonging to the gas (see Sec. III.B). This falls on the lower end of the dynamical mass surface density estimates( from kinematics of stars perpendicularly to the plane) which amount to $\Sigma_D$=50-62 \mspcb \ \citep{HF04} or 57-66 \mspcb \citep{Bienayme+06}. 
%The difference between $\Sigma_T$ and $\Sigma_D$  may be attributed either totally to dark matter (which is not dominant locally, in any case) or, at least partially, to invisible baryons, in the form of gas, low mass stars or stellar remnants. 
Thus, the values for the baryon content of the solar cylinder, summarized in Table~\ref{Tabstars}, should be considered rather as lower limits \citep{Flynn+06}: the total stellar surface density could be as high as 40 \mspcb.  

The density profiles of the stellar thin and thick disks can be satisfactorily fit with exponential functions, both in the radial direction and perpendicularly to the Galactic plane. 
%In the case of the thin disk, the scalelength depends sensitively on the wavelength, with larger scalelengths obtained typically for shorter wavelengths. This is due to the colour gradient of the Milky Way, which is attributed to the inside-out formation of the disk: the inner disk is, on average, older and redder than the outer one. Most studies in the NIR (which traces the bulk stellar popultion) converge to scalength values of $L\sim$2-3 kpc, while in the $B$-band (tracing the young population), values as large as $L$=5.5 kpc have been suggested \citep{Hammer+07} and references therein).
%The situation appears to be less clear for the thick disk, for which a scalelength smilar to the one of %the thin disk is sometimes adopted (e.g. \citealp{RRDP03}).
The recent SDSS data analysis of star counts, with no {\it a priori} assumptions as to the functional form of the density profiles finds exponential disks with scalelengths
as displayed in Table \ref{Tabstars} (from  \onlinecite{Juric+08}).
% $L_{thin}$=2.6 kpc for the thin and $L_{thick}$=3.6 kpc for the thick disk, respectively \citep{Juric+08}. The same study finds for the corresponding scaleheights at $R_{\odot}$=8 kpc $H_{thin}$=300 pc and  $H_{thick}$=900 pc, i.e. close to the average values found in other works. \citet{Juric+08}  find that the local thick/disk volume density normalisation is $\rho_{0,thick}/\rho_{0,thin}$=0.12, which translates into a substantial contribution of the thick disk to the local surface density (of the order of 25\%),  in agreement with the independent evaluation of \citet{Flynn+06}. The scaleheights of the disks increase, in principle, with galactocentric distance $R$, because of the radial variation of the gravitational potential, but that variation turns out to be small inside the solar circle (e.g. \citealp{NJ02}). 
%On the other hand, observations suggest a rarefaction of the inner stellar disk near the plane (for galactic latitudes $|b|<$ 3$^o$) when compared to the predictions of a simple exponential density profile \citep{LCGG04}; this is attributed to the presence of the bar, which contributes to dynamical heating of the inner stellar disk.
The thin and thick disks cannot extend all the way to the Galactic center, since dynamical arguments constrain the spatial co-existence of such rotationally supported structures with the pressure-supported bulge. The exact shape of the ``central hole'' of the disks is poorly known (see, e.g. \onlinecite{Freudenreich98}; \onlinecite{RRDP03},  for parametrizations), but for most practical purposes (i.e. estimate of the total disk mass) the hole can be considered as trully void of stars for disk radius $R<$2 kpc.

The data presented in this section (as summarized in Table~ \ref{Tabstars}), allow one to estimate the total mass of the thin and thick disks as $M_{D,thin}$=2.3 10$^{10}$ \ms \ and  $M_{D,thick}$=0.5 10$^{10}$ \ms, respectively, in the galactocentric distance range 2-15 kpc. 
%As argued in \citet{Flynn+06} a 10\% increase should be  adopted in the case of the thin disk, to account for density enhancements due to spiral arms (since the solar cylinder, which is used for the overall normalisation, is found outside such an arm). 
Overall, the disk of the MW is twice as massive as the bulge.
% Including $M_{Gas}\sim$9 10$^9$ \ms \  for the gas (see next section) leads to a total baryonic disk mass of $M_{Disk}\sim$3.7 10$^{10}$ \ms.

\begin{table}
\caption{Properties of the stellar populations of the thin and thick disk$^a$.
\label{tab:TableStars}}
\begin {center}
\begin{tabular}{lclcc}
\hline \hline
\noalign{\smallskip}
 & & & {Thin } & {Thick} \\
 \noalign{\smallskip}
\hline
 \noalign{\smallskip}
 Mass density &$\rho_{0,\odot}$ & (M$_{\odot}$ pc$^{-3}$)    & 4.5 10$^{-2}$ &  5.3 10$^{-3}$   \\
 Surface density & $\Sigma_{\odot}$ & (M$_{\odot}$ pc$^{-2}$)    & 28.5      &    7          \\
 Scaleheight & $H_{\odot}$ & (pc)  &    300    &   900           \\
 Scalelength & $L$ & (pc)  &    2600    &  3600            \\
 Star mass & $M_D$ & (10$^{10}$ M$_{\odot}$)   & 2.3      &    0.53           \\
 $\langle {\rm Age} \rangle_{\odot}$ & $\langle {\rm A} \rangle_{\odot}$ & (Gyr)   &   5     &   10           \\
$\langle {\rm  Metallicity}  \rangle_{\odot}$ & $\langle {\rm [Fe/H]} \rangle_{\odot}$ & (dex)    &   -0.1     & -0.7      \\
\hline 
\end{tabular}
\end{center}
$a$: The indice $\odot$ here denotes quantities measured at Galactocentric distance $R_{\odot}$=8 kpc. Average 
quantities are given  within $\langle  \ \rangle$.
\label{Tabstars}
\end{table}

\subsection{Interstellar matter}
\label{sec1:GalaxyISM}

Interstellar matter is primarily composed of hydrogen, 
but it also contains helium ($\simeq 9~\%$ by number or $28~\%$ by mass) 
and heavier elements, called ``metals''  ($\simeq$  0.12~\% 
by number or $1.5~\%$ by mass in the solar neighborhood).
All the hydrogen, all the helium, and approximately half the metals
 exist in the form of gas; 
the other half of the metals is locked up in small solid grains of dust.
Interstellar dust manifests itself through its selective absorption of
starlight (leading to extinction, reddening, and polarization
of starlight) and through its thermal infrared emission.
Dust grains cover a whole range of radii, from $a_{\rm min} \lesssim 100$~\AA \ to 
$a_{\rm max} \gtrsim 0.25~\mu$m, as implied from the overall shape of extinction curves 
which can be reproduced with a power-law 
distribution in radius, 
$N(a) \, da \propto a^{-3.5} \, da$ \citep{mrn77, kmh94, bcj00}.
Overall, gas and dust appear to be spatially well correlated
\citep{bp88, babbdhlp96}.

Interstellar gas can be found in molecular, atomic (cold or warm) and
ionized (warm or hot) forms. 
The physical properties of the different gas components
in the Galactic disk were reviewed by \citet{ferriere01} and are summarized 
in Table ~\ref{tab:TableISM}.
The gas properties in the Galactic bulge are less well established,
but on the whole, all gas components appear to be hotter and denser 
in the bulge than in the disk \citep{fgj07}.

Spatially, the molecular gas is confined to discrete clouds, 
which are roundish, gravitationally bound, and organized hierarchically 
from large complexes (size $\sim 20 - 80$~pc, 
mass $\sim 10^5 - 2  \ 10^6~M_\odot$) 
down to small clumps (size $\lesssim 0.5$~pc, 
mass $\lesssim 10^3~M_\odot$) \citep{goldsmith87}.
The cold atomic gas is confined to more diffuse clouds,
which often appear sheet-like or filamentary, 
cover a wide range of sizes (from a few pc up to $\sim 2$~kpc),
and have random motions with typical velocities of a few km~s$^{-1}$ 
\citep{kh87}.
The warm and hot components are more widespread and they
form the intercloud medium.

\begin{figure}
\includegraphics[width=0.60\textwidth]{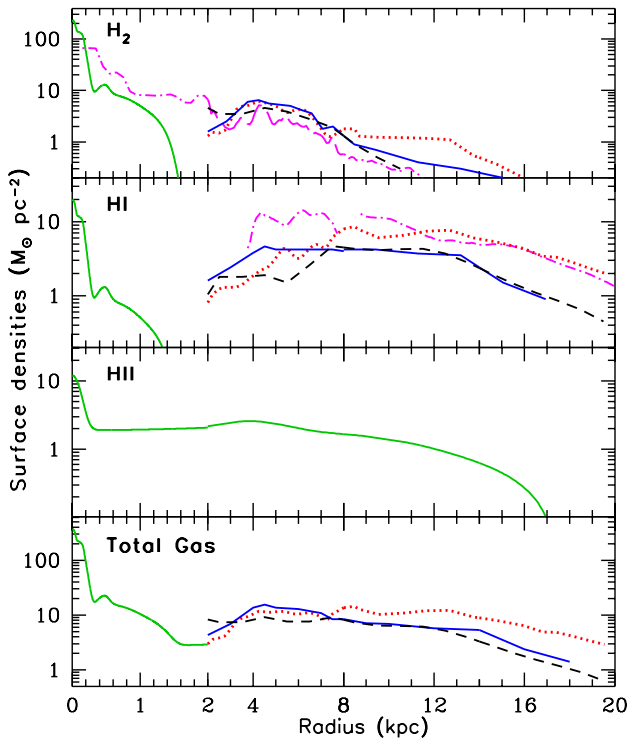}% Here is how to import EPS art
\caption{Azimuthally-averaged surface densities
of interstellar atomic, molecular and ionized hydrogen as functions of Galactic radius. 
The total gas ({\it bottom}) includes a  40\% contribution by helium.  Notice the change of scale at $R$=2 kpc.
For $R<$2 kpc (bulge) data are
derived by \citet{fgj07}, based on a compilation of earlier works:
\citet{shhc04} for the molecular gas in the Central Molecular Zone,
\citet{lb80} for the neutral gas in the tilted disk,
and \citet{cl02} for the ionized gas. 
In all panels, disk data ($R>$2 kpc) are from:  \citet{Dame93} ({\it solid});   
\citet{om01} ({\it dotted}) ;  \citet{ns06} for H{\sc I} and \citet{ns03} for H$_2$,
respectively ({\it dashed}); and  \citet{kd08} for  H{\sc I} and and \citet{peb08} for $H_2$, respectively ({\it dot-dashed}).
The curve in the H{\sc II} panel is from the NE2001 free-electron 
density model of \citet{cl02} (for simplicity, we identified the H density 
with the free-electron density, i.e., we neglected the contribution 
of free electrons originating from He in the HIM). 
}
\label{fig:Gas_SurfDens}
\end{figure}

\begin{figure}
\includegraphics[width=0.49\textwidth]{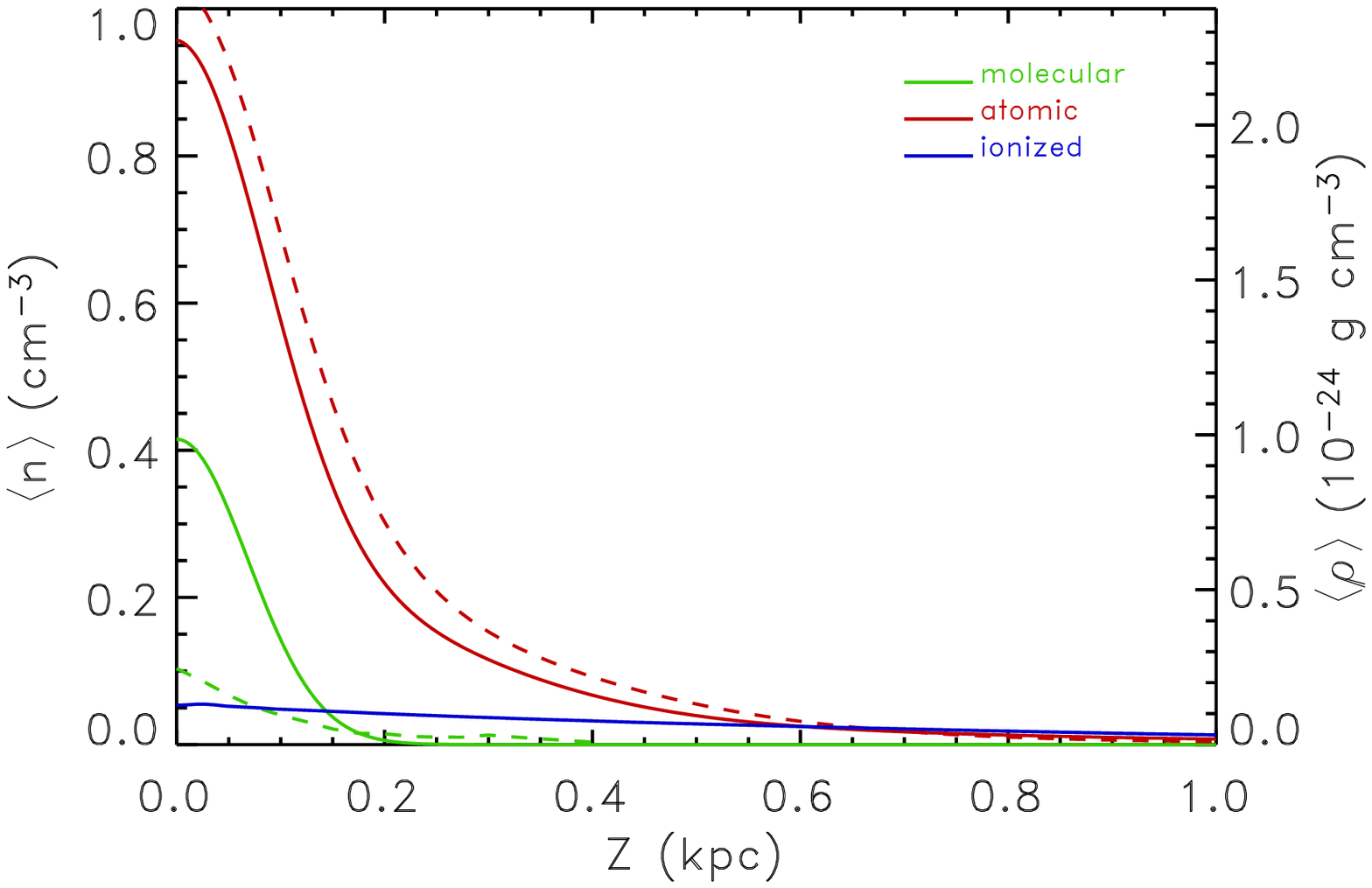}
\caption{Space-averaged volume densities of interstellar 
H$_2$, H{\sc i} and H{\sc II},
averaged along the solar circle ($R = R_\odot$),
as functions of distance from the Galactic plane $Z$. Data: H$_2$  from \citet{bcamt88},
rescaled to $X_{\rm CO} = 2.3 \times 10^{20}~{\rm cm^{-2}~K^{-1}~km^{-1}~s}$,
as in \citet{om01}; H{\sc i} from \citet{dk90}, scaled up by a factor 1.58 such as to match
the H{\sc i} surface density of \citet{om01}.
All the other curves are from the same sources 
as in Fig.~\ref{fig:Gas_SurfDens}.
}
\label{Gas_Vert_Profile}
\end{figure}

The different gas components also differ by their spatial
distributions at large scales.
The observational situation was reviewed by \citet{ferriere01}
for the Galactic disk and by \citet{fgj07} for the Galactic bulge.
Fig.~\ref{fig:Gas_SurfDens} gives the radial variation of 
the azimuthally-averaged surface densities of H$_2$, H{\sc i}, H{\sc ii} 
and the total gas (accounting for a 28\% contribution from He),
while Fig.~\ref{Gas_Vert_Profile} gives the vertical variation of
their respective space-averaged volume densities averaged along 
the solar circle (at $R = R_\odot$).
The total interstellar masses  of the three
gas components in the Galactic disk are uncertain and their sum 
in the Galactic disk, i.e. between 2 and 20 kpc,
is probably comprised between $\sim 0.9~10^{10}~M_\odot$ and 
$\sim 1.5~10^{10}~M_\odot$.
%  representing $\sim 25\%-35\%$  of the total mass of the disk.

\begin{table}
\caption{Physical properties (typical temperatures, hydrogen densities 
and ionization fractions) of the different ISM phases in the Galactic disk.
\label{tab:TableISM}}
\begin {center}
\begin{tabular}{llccc}
\hline \hline
\noalign{\smallskip}
Phase & & $T$ (K) & $n_{\rm H}$ (cm$^{-3}$) & $x_{\rm ion}$ \\
\noalign{\smallskip}
\hline
\noalign{\smallskip}
Molecular & (MM) & $10 - 20$ & $10^2 - 10^6$ & $\lesssim 10^{-4}$ \\
Cold neutral & (CNM) & $20 - 100$ & $20 - 100$ & $4 \ 10^{-4} - 10^{-3}$ \\
Warm neutral & (WNM) & $10^3 - 10^4$ & $0.2 - 2$ & $0.007 - 0.05$ \\
Warm ionized & (WIM) & $\sim 8000$ & $0.1 - 0.3$ & $0.6 - 0.9$ \\
Hot ionized  & (HIM) & $\sim 10^6$ & $0.003 - 0.01$ & $1$ \\
\noalign{\smallskip}
\hline
\end{tabular}
\end{center}
% MM: molecular medium,  
%CNM: cold neutral medium, WNM: warm neutral medium, 
%WIM: warm ionized medium, HIM: hot ionized medium 
%\label{TabISMproperties}
\end{table}

%at $\sim 0.9~10^9~M_\odot$ \citep[from][]{peb08} to 
%$\sim 2.5~10^9~M_\odot$ \citep[from][]{om01} for the molecular component,
%$\sim 6.5~10^9~M_\odot$ \citep[from][]{om01} to 
%$\sim 1.1~10^{10}~M_\odot$ \citep[from][]{kd08} for the atomic component,
%and $\sim 1.5~10^9~M_\odot$ \citep[from][]{cl02} for the ionized component.

Since most of the transport and annihilation of positrons takes place
in the Galactic bulge, its gas distribution deserves a more detailed
description.
In the bulge, the interstellar gas is roughly equally divided between
the neutral (molecular + atomic) and ionized components,
and the neutral component is $\sim 90\%$ molecular.
The molecular gas tends to concentrate in the so-called
central molecular zone \citep[CMZ;][]{morris&s_96},
a thin sheet parallel to the Galactic plane,
which, in the plane of the sky, extends out to $R \sim 250$~pc
at longitudes $l>0^\circ$ and $R \sim 150$~pc at $l<0^\circ$
and has a FWHM thickness $\sim 30$~pc.
Projected onto the Galactic plane, the CMZ would appear as
a $500~{\rm pc} \times 200~{\rm pc}$ ellipse inclined (clockwise)
by $\sim 70^\circ$ to the line of sight \citep{shhc04}.
Outside the CMZ, the molecular gas is contained in a significantly
tilted disk \citep{liszt&b_78,burton&l_92}, extending in the plane
of the sky out to $R \sim 1.3$~kpc on each side of the GC
and having a FWHM thickness $\sim 70$~pc.
According to \citet{lb80}, the tilted disk has the shape of
a $3.2~{\rm kpc} \times 1.0~{\rm kpc}$ ellipse, which is tilted
(counterclockwise) by $\sim 13\fdg5$ out of the Galactic plane
and inclined (near-side down) by $\sim 70^\circ$ to the plane of the sky.
The tilted disk is also believed to feature an elliptical hole
in the middle, just large enough to enclose the CMZ.
The spatial distribution of the atomic gas is arguably similar to that
of the molecular gas (\citet{burton&l_92}, but see also \citet{combes_91}
for another point of view), with this difference that the atomic
layer is about three times thicker than the molecular layer,
both in the CMZ and in the tilted disk.
The ionized gas, for its part, is not confined to either the CMZ or
the tilted disk; it appears to fill the entire bulge and to connect
with the ionized gas present in the disk.

 The dramatic density and temperature contrasts between the different 
 ISM phases as well as the supersonic random motions observed in all 
 of them bear witness to a highly turbulent state.
 Mainly responsible for this turbulence are the powerful winds
 and terminal supernova explosions of the most massive stars.
 Interstellar turbulence manifests itself over a huge range of spatial
 scales, from $\lesssim 10^{10}$~cm up to $\gtrsim 10^{20}$~cm;
 throughout this range, the power spectrum of the free-electron density 
 in the local ISM is consistent with a Kolmogorov-like power law 
 \citep{ars95}.

\subsection{ Star formation and supernova rates}
\label{sec1:GalaxySFSN}

%\subsubsection{\label{sec:GalaxySF}Star formation}

\begin{figure}
\includegraphics[width=0.49\textwidth]{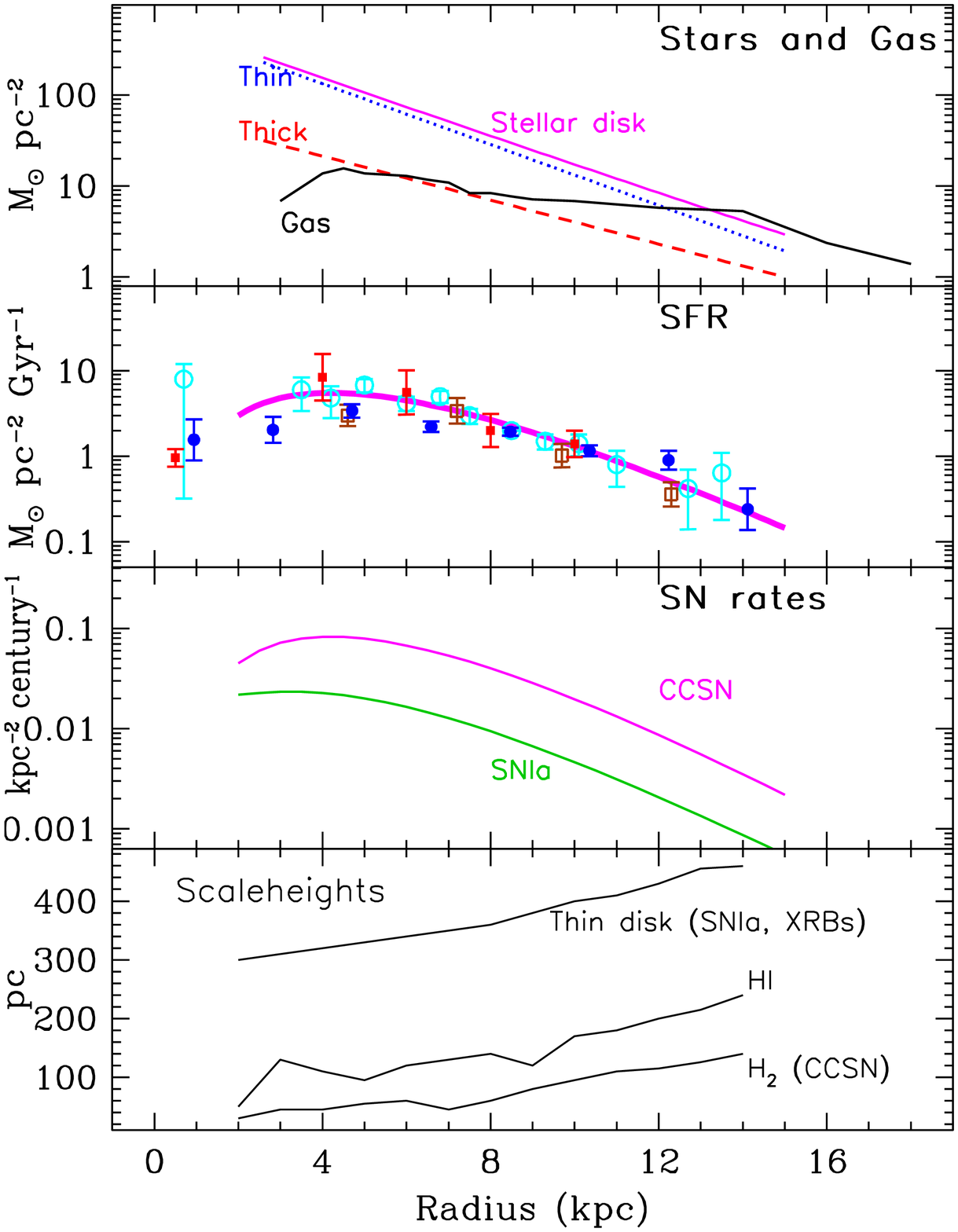}% Here is how to import EPS art
\caption{\label{fig:epsart}Surface densities of stars+gas, SFR, SN rates and scaleheights of gas and stars 
as a function of Galactocentric distance.
Star profiles are from data of Table II and gas profile is the one of \citet{Dame93} (bottom pane of Fig. 6). 
Data for SFR are from :      
\citet{LMT85} ({\it open cicles}); \citet{CB98} ({\it filled circles}); \citet{MW97} ({\it open squares}); \citet{GLV78}
({\it filled squares}).
 The solid curve is an approximate fit, normalized to 2 \ms yr$^{-1}$ for the whole Galaxy.
The same curve is used for the CCSN rate profile (third panel), normalized to 2 CCSN/century; the SNIa rate profile 
is calculated by Eq. ~\ref{eq:SNrate} \  and normalized to 0.5 SNIa/century (Table~\ref {tab:SNrates}).
 }
\label{SFRdens}
\end{figure}

Determination of absolute values of star formation rates (SFR,  in \msy) constitutes one of the most challenging tasks in modern astrophysics
 \citep{Kennicutt98}. 
%Each one of the tracers used is sensitive to only some part of the stellar initial mass function (IMF) and has its own advantages and drawbacks (e.g. \onlinecite{Kennicutt98} ). 
%The UV continuum (optimal range: 1250 - 2500 A), which is often used in extragalactic studies, is useless in the cas of the Galaxy because of the huge extinction it suffers by gas and dust along the Galactic plane. Near infrared emission suffers little extinction, but results partly from ageing stars (red giants) and cannot be used to measure recent star formation. Far infrared (FIR) and radio emission are used then, both of them resulting from reprocessed light from massive stars. Thermal radio emission (``free-free'' emission) originates from recombination lines emitted by regions surrounding massive stars (HII regions), which are ionized by the Lyman continuum stellar photons. The FIR continuum is emitted by interstellar dust, which is heated by absorption of stellar light. However, that dust may either surround young massive star forming regions or be heated by the general interstellar radiation field; in the former case (which describes well  dusty circumnuclear starbursts) FIR emission measures really the SFR, but in the latter (which corresponds to normal disk galaxies) the situation is less clear.
In the case of the Milky Way, methods  based  on counts of various short-lived objects (with lifetimes less than a few Myr, like e.g. pulsars, SN remnants or OB associations) are used.
%these include e.g. counts of supernova remnants;
% (detected through their synchrotron radio-emission, resulting from electrons accelerated by their shock waves); 
%pulsars; 
%(detected through synchrotron emission from electrons, accelerated in the pulsar magnetospheres, e.g. \onlinecite{LMT85}; \onlinecite{CB98}) ; 
%or OB associations.
% (\onlinecite{MW97}). 
%Distance determinations to those objects are crucial for an accurate evaluation of the Galactic profile of their  surface density. 
Those methods establish in fact the {\it relative } star formation rate across the Galactic disk.
%i.e. with respect to its value at some Galactocenric distance (for instance, the solar neighborhood). 
%The derivation of an {\it absolute } value (i.e. the ``calibration'' of the SFR profile) remains a delicate enterprise.
Surface density profiles of various stellar tracers appear in Fig. \ref{SFRdens}. 
%They all increase towards the inner disk and pass through a broad maximum at the distance of  4 kpc (the ``molecular ring'').
% Those profiles can be compared to the star and gas  profiles ($\Sigma_*(R)$ and $\Sigma_G(R)$, respectively in panels a and b of the same figure), through the so-called Schmidt-Kennicutt law (SFR$\propto \Sigma_G^n$), modified 
%Schmidt-Kennicutt law ( SFR$\propto \Sigma_G^n \Omega$, with $\Omega$ being the rotation period), or Dopita-Ryder law ( SFR$\propto \Sigma_G^n \Sigma_*^m$). As discussed in \citet{BPBG03}, all those empirically and/or theoritically motivated laws provide satisfactory fits to the data of the MW disk (taking into account the large uncertainties in the data) and have been extensively used in models of the chemical evolution of the Galaxy.
For the calibration of the SFR profile one needs to know either the total SFR of the MW disk or the local one in the solar neighborhood. A ``ball-park'' estimate of the former value is obtained by noting that the late spectral type (Sbc) of the MW  suggests a slow formation at a relatively steady rate $\langle SFR \rangle$ over the past $\Delta T\sim$10 Gyr, leading to $\langle SFR \rangle$=M$_{D,thin}$/$\Delta T \ \sim$2.3 \msy. Most empirical estimates of the present-day total Galactic SFR, based on the aforementioned tracers (and {\it assumptions} on the
IMF)  produce values within a factor of two of the $\langle SFR \rangle$ (e.g. \onlinecite{MW97}; \onlinecite{RW10}, and references therein).

In the context of Galactic positrons, special attention should be paid to the star formation activity in the central regions of the bulge. The massive star population of the  three major star clusters inside the Nuclear Bulge clearly indicate important recent star formation, obviously fed from the gas of the Central Molecular Zone (CMZ). Deep field observations of late-type stars with the {\it NICMOS/HUBBLE} \cite{Figer+04} and with {\it SINFONI/VLT} \cite{Maness+07} suggest that the star forming activity in that region has proceeded at a relatively steady rate, of the order of a few 10$^{-2}$ \ms/yr, over the past $\sim$10 Gyr.

%\subsubsection{\label{sec:GalaxySN}Supernova rates}

%From the observational point of view, supernovae (SN) are classified in two major types, depending on the absence (SNI) or presence (SNII) of hydrogen in their spectra (e.g \onlinecite{Turatto03}). In the 80ies, the class of SNI was subdivided into SNIa (presence of Si lines, no He) and SNIb (absence of Si, presence of He), whereas a little later the subtype of SNIc (absence of both Si and He) was added. That classification, although suggestive (few objects in the Universe lack H or He, which are the most abundant elements), concerns the envelope of the exploding star and reveals nothing about the explosion mechanism itself.

\begin{table*}
\caption{Supernova rates in the Milky Way
\label{tab:SNrates}}
\begin {center}
\begin{tabular}{lcccccc}
\hline \hline
\multicolumn{3}{c}{} &  \multicolumn{2}{c}{\bf SNIa } & \multicolumn{2}{c}{\bf Core collapse SN} \\
\hline

 &Stellar mass$^a$ &Spectral type &Specific rate$^b$ & Rate & Specific rate$^b$ & Rate \\

 &10$^{10}$ M$_{\odot}$  &  & SNuM  & century$^{-1}$  & SNuM  & century$^{-1}$  \\

\hline

Bulge         & 1.4  & E0  & 0.044 &0.062 & -  & -  \\

Nuclear Bulge & 0.15 & Sbc/d-Irr$^c$ & 0.17-0.77   &0.025-0.115 &0.86-2.24  &0.13-0.33  \\

Thin disk     & 2.3  & Sbc & 0.17  &0.4   &  0.86  & 2  \\

Thick disk    & 0.5  & E0  & 0.044 &0.022 & -   &  \\

\hline

Total bulge   & 1.5  &     &       &0.087-0.18  &  &0.13-0.33  \\

Total disk    & 2.8  &     &       &0.42  &  &2  \\

\hline

Total Milky Way &4.3 &     &       &0.5-0.6   &  &2.13-2.33  \\

\hline

Bulge/Disk   ratio    & $<$0.5  & & & 0.21-0.43 & & 0.06-0.15 \\
\hline \hline

\end{tabular}
\end{center}

\noindent SNuM: = 1 SN per 10$^{10}$ M$_{\odot}$ per  century;  $a$: See Sec. III.A for references;   $b$ Mannucci et al. (2005);
 \noindent $c$ Very uncertain, in view of uncertainties in star formation efficiency  and slope of IMF (see text).
\label{tab:SNrates} 
\end{table*}

From the theoretical point of view, SN are now classified mainly in {\it thermonuclear} supernovae (the explosion energy being due to the thermonuclear disruption of a white dwarf accreting matter in a  binary system) and {\it core collapse} supernovae (CCSN, where the energy originates from the gravitational collapse of the iron core of a massive star having exhausted all its nuclear fuel). Thermonuclear supernovae are identified with SNIa (lacking hydrogen in their spectra)
and are observed in all types of galaxies: old ellipticals with no current star formation, but also young, star forming, spiral and irregular galaxies. All other supernova types (SNII, SNIb, SNIc) are exclusively observed in the star forming regions of spirals (i.e. inside spiral arms) and irregulars\footnote{The degree of mass loss suffered by the massive star prior to the explosion determines the appearance of the core collapse supernova as SNII (little H lost), SNIb (all H and little He lost) or SNIc (all H and most He lost).}.

No supernovae have been observed in the Galaxy in the past four centuries, and the handful of so called ``historical supernovae'' offers a very biased estimate of the Galactic SN frequency \cite{TLS94}. All methods  used to determine the Galactic SN rate which are based  exclusively   on Galactic data 
%(e.g. counts of SN remnants, the present-day SFR, the present-day mass of $^{26}$Al) and are, in principle, able to evaluate solely the core-collapse SN rate ; all those methods 
suffer from various systematic uncertainties 
%(e.g. shape of the adopted massive star IMF, yields of $^{26}$Al etc.) 
and converge to a value of $R_{CCSN}$= a few per century (\onlinecite{Diehl+06} and references therein).
The most accurate way to evaluate the Galactic SN rate is, probably,  through statistics of SN rates in external galaxies. The work of \citet{Mannucci+05}, corrected for various observational biases, offers a valuable database for such an estimate and can be used, along with the stellar masses of the various Galactic components
(Sec. III.A) , to derive the Galactic rate of the main SN types
(Table~\ref{tab:SNrates}). 

The spatial distribution of core collapse SN in the MW should obviously follow the one of the SFR (Fig. ~\ref{SFRdens}).
 Such an azimuthally averaged surface density masks the fact that CCSN are exclusively concentrated inside spiral arms. 
%Also, the ratio of SNIb,c/SNII seems to increase towards the central regions of external galaxies \cite{Hakobyan08}, and this should be also the case for the Milky Way. 
The scaleheight of core collapse SN should be comparable to the scaleheight of the molecular gas, i.e. less than 100 pc, and little varying with Galactocentric distance.
More difficult is the evaluation of the radial profile of SNIa, since the progenitor white dwarfs may originate from stars of a wide variety of stellar masses (1-8 \ms) and corresponding lifetimes (10-0.05 Gyr). 
%Furthermore, it is not yet clear whether SNIa result from the accretion of matter from a companion star onto the white dwarf (``single degenerate'' scenario) or from the collision and merging of two white dwarfs (``double degenerate'' scenario); in fact, it may well be that several channels contribute to the overall SNIa rate. 
Various models have been developed in order to calculate the SNIa  rate (e.g. \onlinecite{Greggio05}). A useful empirical approach is the one adopted in \citet{SB05}, where the SNIa rate is calculated as the sum of two terms: one depending on the stellar mass $M_*$ and one on the SFR of the system, i.e.
\begin{equation}
{{R_{SNIa}}\over{\rm century}} \ = \ A \ {{M_*}\over{10^{10} \ {\rm M_{\odot}}}} \ + \ B \ {{SFR} \over {\rm M_{\odot} \ yr^{-1}}}
\label{eq:SNrate}
\end{equation}
with parameters $A$ and $B$ empirically determined (\onlinecite{SB05}; \onlinecite{Sullivan+06}).
%Recent studies converge on $A\sim$0.04 - 0.05 (\onlinecite{SB05}; \onlinecite{Sullivan+06}), but give discrepant  values for $B$. Adopting $A$=0.05 and using the data of Table ~\ref{tab:SNrates} leads to $B$=0.14.  
The  parametrized SNIa profile in the Milky Way disk appears in the third panel of Fig. \ref{SFRdens}.
Taking into account the nature of the SNIa progenitors, it is expected that the distribution  of SNIa vertically to the disk plane will follow the corresponding distribution of the thin disk, i.e. with a scaleheight of 300 pc (an insignificant contribution from the thick disk is also expected).

\subsection{Interstellar magnetic fields}
\label{sec1:GalaxyMagnField}

The magnetic field strength, $B$, in cold, dense regions of interstellar space
can be inferred from the Zeeman splitting of the 21-cm line of H{\sc i}
(in atomic clouds) and centimeter lines of OH and other molecules 
(in molecular clouds).
In practice, it is the line-of-sight component of the magnetic field,
$B_\parallel$, that is measured.
With appropriate statistical corrections for projection effects,
it is found that in atomic clouds, $B$ is typically a few $\mu$G, 
with a slight tendency to increase with increasing density 
\citep{th86, ht05}, 
while in molecular clouds, $B$ increases approximately 
as the square root of density, from $\sim 10~\mu$G to $\sim 3\,000~\mu$G 
\citep{crutcher99, crutcher07}.

The interstellar magnetic field, ${\bf B}$, in the ionized medium 
is generally probed with Faraday rotation measures of Galactic pulsars 
and extragalactic radio sources.
Here, too, the quantity that is actually measured is $B_\parallel$.
Faraday rotation studies have provided the following interesting pieces
of information.

1) ${\bf B}$ has a uniform (or regular) component, ${\bf B}_{\rm u}$, 
and a random (or turbulent) component, ${\bf B}_{\rm r}$.
Near the Sun, $B_{\rm u} \simeq 1.5~\mu$G and $B_{\rm r} \sim 5~\mu$G 
\citep{rk89}.
Away from the Sun, $B_{\rm u}$ increases toward the GC, 
to $\gtrsim 3~\mu$G at $R = 3$~kpc \citep{hmlqv06}, 
i.e., with an exponential scale length $\lesssim 7.2$~kpc.
In addition, $B_{\rm u}$ decreases away from the midplane, albeit 
at a very uncertain rate; for reference, the exponential scale height 
inferred from the rotation measures of extragalactic sources 
is $\sim 1.4$~kpc \citep{it81}.

2) In the Galactic disk, ${\bf B}_{\rm u}$ is nearly horizontal
and generally dominated by its azimuthal component.
It is now widely accepted that ${\bf B}_{\rm u}$ reverses several times 
with decreasing radius, but the number and radial locations of 
the reversals are still highly controversial 
\citep{rl94, hmq99, vallee05, hmlqv06, bhgtb07}.
These reversals have often been interpreted as evidence that
${\bf B}_{\rm u}$ is bisymmetric (azimuthal wavenumber $m \!=\! 1$),
while an axisymmetric ($m \!=\! 0$) field would be expected
from dynamo theory.
In reality, \citet{mfh08} showed that neither the axisymmetric
nor the bisymmetric picture is consistent with the existing pulsar 
rotation measures, and they concluded that ${\bf B}_{\rm u}$ must have 
a more complex pattern.

3) In the Galactic halo, ${\bf B}_{\rm u}$ could have a significant
vertical component. 
For the local halo, \citet{taylor&ss_09} obtained
$(B_{\rm u})_Z \sim -0.14~\mu$G above the midplane ($Z > 0$)
and $(B_{\rm u})_Z \sim +0.30~\mu$G below the midplane ($Z < 0$),
whereas \citet{mao&ghz_10} obtained 
$(B_{\rm u})_Z \sim 0.00~\mu$G toward the north Galactic pole 
and $(B_{\rm u})_Z \sim +0.31~\mu$G toward the south Galactic pole.
In contrast to the situation in the Galactic disk, the azimuthal 
component of ${\bf B}_{\rm u}$ shows no sign of reversal with 
decreasing radius.

4) At low latitudes (basically, in the disk), ${\bf B}_{\rm u}$ 
appears to be roughly symmetric in $Z$ \citep{rl94, frick&sss_01}, 
while at high latitudes (in the halo), 
%%the rotation-measure sky exhibits a rather striking antisymmetry/symmetry 
%%in $Z$ in the inner/outer Galactic quadrants 
%%\citep{oren&w_95, hmbb97, hmq99}, which suggests that
${\bf B}_{\rm u}$ appears to be 
roughly antisymmetric/symmetric in $Z$ inside/outside the solar circle
\citep{hmbb97, hmq99}.
Finding ${\bf B}_{\rm u}$ to be symmetric in the disk and antisymmetric
in the inner halo is consistent with the predictions of dynamo theory 
and with the results of galactic dynamo calculations 
\citep[e.g.,][]{ruzmaikin&ss_88, moss&s_08}.
However, there is no reason to believe that ${\bf B}_{\rm u}$ is simply
a pure quadrupole (in the disk) and a pure dipole (in the inner halo),
sheared out in the azimuthal direction by the Galactic differential rotation.
In this respect, one should emphasize that the picture of 
an azimuthally-sheared pure dipole, originally proposed by \citet{han_02} 
and often used in the cosmic-ray propagation community
\citep[e.g.,][]{alvarez&es_02,Pv03}, is supported
neither by numerical simulations of galactic dynamos nor by observations
of external edge-on galaxies, which generally reveal X-shaped field
patterns \citep{beck08}.

A more global method to map out the spatial distribution of ${\bf B}$ 
rests on the observed Galactic synchrotron emission. 
Relying on the synchrotron map of \citet{bkb85} and assuming equipartition
between magnetic fields and cosmic rays, \citet{ferriere98}
found that the total magnetic field has a local value $\simeq 5.1~\mu$G,
a radial scale length $\simeq 12$~kpc,
and a local vertical scale height $\simeq 4.5$~kpc.
Besides, synchrotron polarimetry indicates that the local ratio of
ordered (regular + anisotropic random) to total fields is $\simeq 0.6$
\citep{beck_01}, implying an ordered field $\simeq 3~\mu$G near the Sun.

In the vicinity of the GC, the interstellar magnetic field has completely 
different properties from those prevailing in the Galactic disk.
In that region, systems of nonthermal radio filaments were discovered,
which run nearly perpendicular to the Galactic plane and pass through it
with little or no distortion \citep{ymc84, liszt85}.
The morphology of the filaments strongly suggests that they follow 
magnetic field lines, and radio polarization measurements (assuming 
synchrotron emission) confirm that the magnetic field in the filaments 
is oriented along their long axis \citep{tih85,reich_94}.
From this, it has naturally been concluded that the interstellar
magnetic field near the GC is approximately vertical, at least close to 
the midplane.
Farther from the midplane, the filaments tend to lean somewhat 
outwards, consistent with the interstellar magnetic field having 
an overall poloidal geometry \citep{morris90}.

The radio filaments have equipartition or minimum-energy field strengths 
$\sim 50 - 200~\mu$G \citep[and references therein]{ape91,larosa&nlk_04}.
On the other hand, the fact that the filaments remain nearly straight 
all along their length suggests that their magnetic pressure is stronger 
than the ram pressure of the ambient interstellar clouds, or, equivalently,
that their field strength is $B \gtrsim 1$~mG \citep{ym87}.

Low-frequency radio observations of diffuse nonthermal
(supposedly synchrotron) emission from a $6^\circ \times 2^\circ$ region
centered on the GC imply that the diffuse ISM near the GC has 
a minimum-energy field strength $\gtrsim 6~\mu$G -- possibly up to 
$\sim 80~\mu$G if the cosmic-ray proton-to-electron energy ratio
is as high as 100 and the filling factor of the synchrotron-emitting gas
is as low as 0.01 \citep{larosa&bsl_05}.
A more reliable estimation of the general field strength in the GC
region emerges from the recent analysis of \citet{crocker&jmo_10}, 
which combines radio and $\gamma$-ray data and comes to the conclusion 
that $B \gtrsim 50~\mu$G.

Far-infrared/submillimeter polarization studies of dust thermal emission 
from the GC region indicate that the magnetic field inside GC molecular 
clouds is roughly parallel to the Galactic plane \citep{ncrgn03}.
More precisely, the field direction appears to depend on the molecular
gas density, being nearly parallel to the plane in high-density regions 
and nearly perpendicular to it in low-density regions \citep{cdddh03}.
Near-infrared polarization studies of starlight absorption by dust
also find the magnetic field inside GC molecular clouds to be roughly
horizontal, although without any obvious correlation between field 
direction and gas density \citep{nishiyama&thk_09}.

Zeeman splitting measurements have yielded mixed results.
In the circumnuclear disk, the innermost molecular region with
radius $\lesssim 7$~pc, \citet{klc92} and \citet{plc95} derived
line-of-sight magnetic fields $|B_\parallel| \simeq 2$~mG and 
$|B_\parallel| \simeq 0.6 -4.7$~mG, respectively.
Farther from the GC, \citet{crmt96} measured values of $|B_\parallel|$ 
ranging between $\simeq 0.1$ and $0.8$~mG
toward the Main and North cores of Sgr~B2.
In contrast, \citet{ug95} reported only non-detections, 
with $3\sigma$ upper limits to $|B_\parallel|$ $\sim 0.1 - 1$~mG, 
toward 13 selected positions within a few degrees of the GC 
(including Sgr~B2).

Faraday rotation measures have also yielded somewhat disparate 
results. The disparity lies not so much in the absolute value of
$B_\parallel$, which is generally estimated at a few $\mu$G
\citep[e.g.,][]{tih85,ym87,gray&ceg_91},
but more in the $(l,b)$-dependence of its sign.
\citet{ncrgn03}, who collected all the available rotation measures 
toward synchrotron sources within $1^\circ$ of the GC, found that 
$B_\parallel$ reverses sign both across the rotation axis and
across the midplane.
A different pattern was uncovered by \citet{roy&rs_05}, who derived 
the rotation measures of 60 background extragalactic sources 
through the region $(|l|<6^\circ,|b|<2^\circ)$ and obtained 
mostly positive values, with no evidence for a sign reversal 
either with $l$ or with $b$.

The properties of the turbulent magnetic field are not well established.
 \citet{rk89} provided a first rough estimate for the typical spatial 
 scale of magnetic fluctuations, $\sim 55$~pc, although they recognized 
 that the turbulent field cannot be characterized by a single scale.
 Later, \citet{ms96} presented a careful derivation of the power spectrum 
 of magnetic fluctuations over the spatial range $\sim (0.01 - 100)$~pc; 
 they obtained a Kolmogorov spectrum below $\sim 4$~pc and a flatter 
 spectrum consistent with 2D turbulence above this scale.
 In a complementary study, \citet{hfm04} examined magnetic fluctuations
 at larger scales, ranging from $\sim 0.5$ to 15~kpc; at these scales,
 they found a nearly flat magnetic spectrum, with a 1D power-law index 
 $\sim -0.37$.

\begin{figure}
\includegraphics[width=0.49\textwidth]{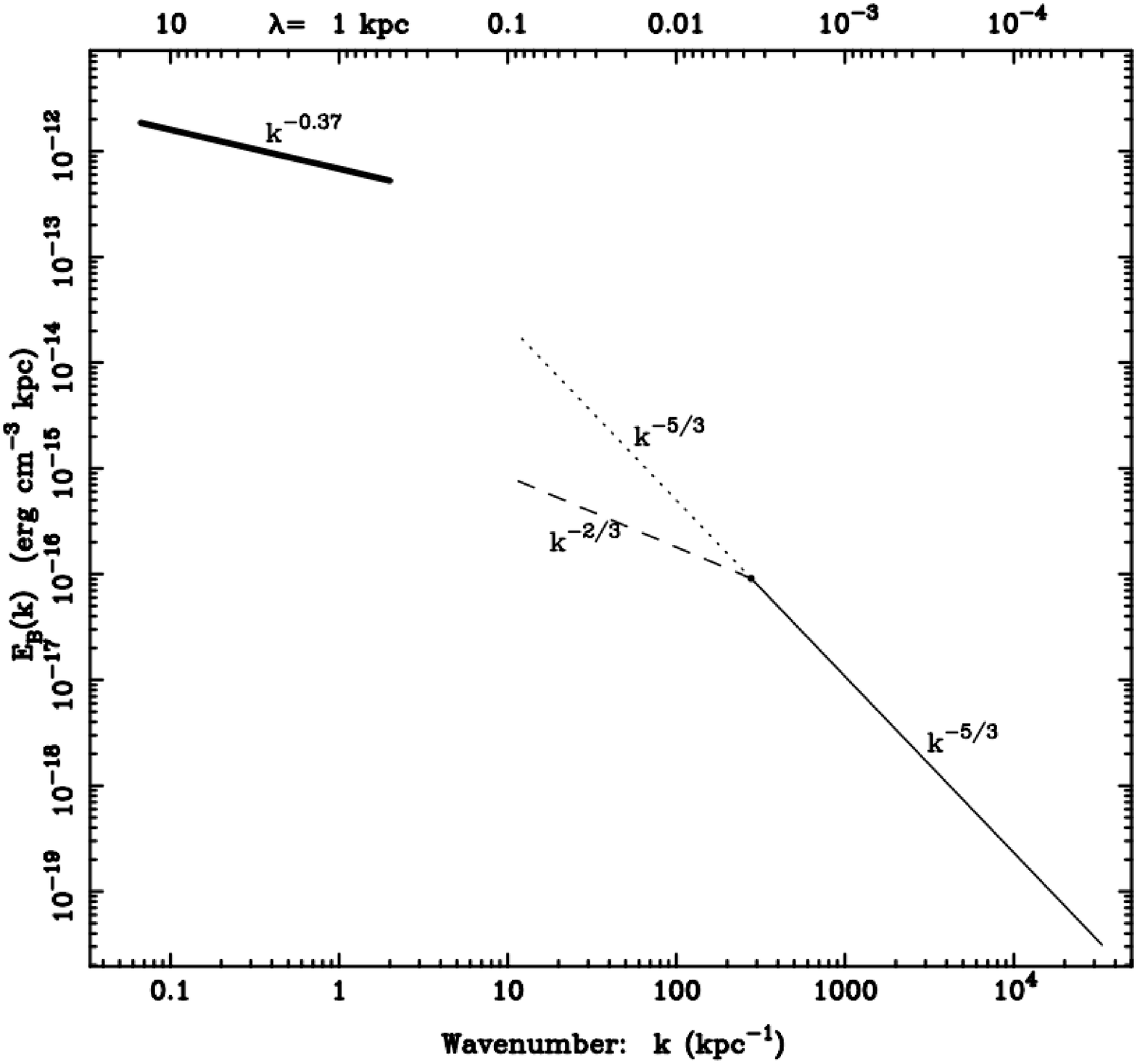}
\caption{Composite magnetic energy spectrum in our Galaxy. The {\it  thick
solid} line is the large-scale spectrum. The {\it thin solid} and
{\it dashed/dotted} lines give the Kolmogorov and two-dimensional turbulence
spectra, respectively, inferred from the Minter and  Spangler (1996) study.
The two-dimensional turbulence spectrum is uncertain; it 
probably lies between the {\it dashed} [$E_B(k) \propto k^{-2/3}$ ] 
and {\it dotted} [$E_B(k) \propto k^{-5/3}$] lines (from \onlinecite{hfm04}).
}
\label{fig:Turb_spec}
\end{figure}

The, poorly understood at present, properties of the turbulent Galactic magnetic field
as well as its overall configuration,
are extremely important for understanding positron propagation in the Milky Way (Sec. VI).

\subsection{The dark matter halo  }
\label{sec:GalaxyDM}

A large body of observational data on the extragalactic Universe suggests that its
mass is dominated by non-baryonic dark matter.
%\footnote{Alternatively, the theory of general relaticity may fail at small accelerations, e.g. }.
 In the presently widely accepted
"standard" cosmological model ($\Lambda$CDM, for Cold Dark Matter with 
cosmological constant $\Lambda$) dark matter accounts for a fraction
$\Omega_{DM}\sim$24\% of the overall matter/energy budget of the Universe, baryons
for $\sim$4 \% and dark energy - or cosmological constant - for the remaining
$\sim$72\% \citep{Bartelmann09}.

The presence of dark matter in spiral galaxies  is deduced from the fact that their rotation curves
beyond a radius of $\sim$ 3 scalelengths do not fall off as rapidly as expected from their baryonic
content. In the case of the Milky Way, the rotation curve is poorly determined beyond
the Sun's location (R$_{\odot}$=8 kpc). It is then assumed, rather than directly inferred from observations,
that the MW is found inside a dark matter halo with a density profile $\rho_{DM}(r)$ similar
to those found in numerical simulations of structure formation in a $\Lambda$CDM universe (e.g.
\onlinecite{Navarro:1996gj}).
In the absence of baryons such simulations predict approximately universal density profiles 
$\rho_{DM}(r) \propto r^{-k}$, with $k$ being itself a positive function of radius $r$: $k(r) \propto r^s$
("Einasto profile"). Because of finite numerical resolution, values of $k$ cannot yet be reliably
determined in the inner halo. Some simulations find $k$=1.5 in the inner galaxy \citep{Moore:1999nt},
but the analysis of one of the largest simulations so far \citep{Navarro+08} suggests
that $k$=0.9$\pm$0.1, i.e. a value compatible with the value of $k$=1 in the
classical NFW profile \citep{Navarro:1996gj}. For values of $k\geq$1 mass diverges as $r\longrightarrow$0
({\it cuspy} profiles). Including interactions of dark matter with baryons \citep{Blumenthal:1985qy}
or with a central black hole \citep{Gondolo:1999ef} generically tend to enhance the cusp 
(e.g. \onlinecite{SM05}).

\begin{figure}
\includegraphics[width=0.49\textwidth]{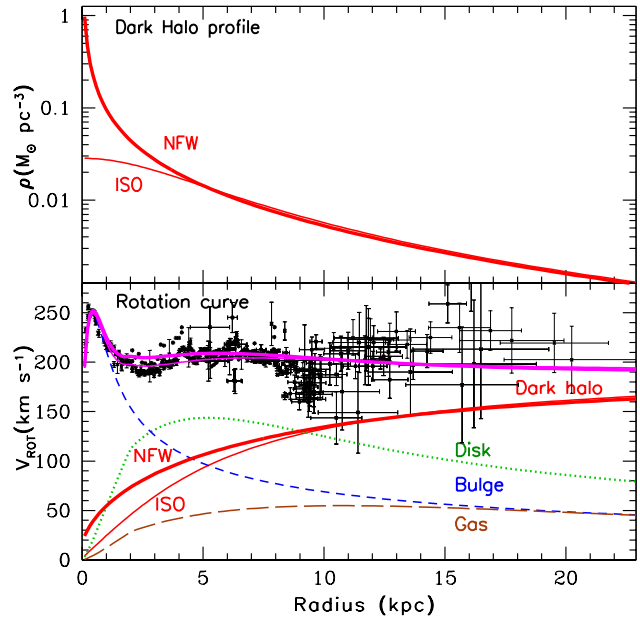}
\caption{Dark matter density profile ({\it top}) and rotational velocity ({\it bottom})
of the Milky Way; the various components (bulge, stellar disk, gas and dark halo) 
contributing to the latter are also indicated. In both panels {\it thick} and {\it thin solid} curves correspond to
NFW and isothermal ("ISO") dark halo profiles, respectively. Data points are from \citet{SHO08}.
}
\label{fig:Rot_Curve}
\end{figure}

The shape of the dark matter density profile in the inner Galaxy is obviously crucial for the
corresponding profile of the putative positrons released from dark matter
decay, annihilation or de-excitation (see Sec. \ref{sec1:DM}). Since dark matter is sub-dominant
in the inner Galaxy (see Fig.~\ref{fig:Rot_Curve}) , observations of the rotation curve
 cannot help to determine its density profile. Analyzing observations of the optical depth
of the inner Galaxy to microlensing events (which are affected only by the baryonic mater)  
\citet{Binney:2001wu} find $k\sim$0.3. On the other hand, rotation curves of dwarf
galaxies (which are dominated by dark matter) systematically suggest flat  profiles
(\onlinecite{GTS07}; \onlinecite{Spano+08}) with $k\sim$0, such as those obtained in the case
of cored isothermal dark halo (see also \onlinecite{Merritt10}). A useful parametrization of the density profiles is
 
\begin{equation}
\rho(r) = \frac{\rho_0(r_0)}{ (r/r_0)^{\gamma} \ [1 +
(r/r_0)^\alpha]^{(\beta-\gamma)/\alpha}},
\end{equation}
where $\rho_0$ and $r_0$ are, respectively, the characteristic mass/energy density and radius of the halo and
$\alpha, \beta, \gamma$ are parameters with values (found, either from simulations or from
observations), reported in Table~\ref{tab:DMHaloProfile}.

\begin{table}%[htbp]
\caption{Models for the Milky Way dark halo profile.}
\begin{center}
\begin{tabular}{lcccc}
\hline
\hline
DM profile: &  ISO  & BE & NFW  & M99 \\
\hline

$\alpha$ & 2 & 1 & 1 & 1.5 \\

$\beta$ & 2& 3 & 3 & 3 \\

$\gamma$ & 0 & 0.3 & 1 & 1.5 \\

$r_0$ (kpc) & 5& 10 & 20 & 30 \\

$\rho_0$ \rm{(M$_{\odot}$  pc$^{-3}$)} & 5. 10$^{-2}$ & 7. 10$^{-2}$   & 1.0  10$^{-2}$  & 1.7 10$^{-3}$  \\

$\rho_0$ \rm{(GeV cm$^{-3}$)}  & 1.89  &2.65  &0.38 &0.065 \\

\hline			
\end{tabular}
\end{center}
ISO: Isothermal; BE: Binney and Evans (2001); NFW (Navarro et al. 1997); M99 (Moore et al. 1999)

\label{tab:DMHaloProfile}
\end{table}

The shape of the dark halo profile may deviate from spherical symmetry. A triaxial shape arises naturally 
from the fact that gravitational collapse of the halo starts first (and proceeds more rapidly) in one direction.
However, other processes may subsequently erase it (e.g. gas cooling, \onlinecite{Kazantzidis:2004vu}).
Various observations in the Milky Way have been interpreted as suggesting a spherical \citep{MartinezDelgado:2003qy},
oblate \citep{MartinezDelgado:2003qy} or prolate \citep{Helmi:2004id} dark halo, but in any case, deviations
from spherical symmetry appear to be small.

Structure formation in the $\Lambda$CDM model leads to a hierarchy of dark haloes embedded within the main halo
of a galaxy. Since smaller galaxies are more dark matter dominated than larger ones, the strongest signal from
dark matter annihilation may not arise from the main halo, but from satellite galaxies. This important issue
has been extensively studied recently. Analyzing one of the largest "Milky Way size" simulations so far, \citet{Springel+08} find
that the most intense emission is expected to arise from the main halo. We shall further discuss this point in Sec. IV.D.3.

%  Sec. IV Production / Sources ..........................................................................................................................................

%%%% RMP Positron Paper Chapter IV: Positron Production %%%%%%%%%%%%%%%%%%%
%
% revised by Roland Diehl 091014 ->V4: change eps to pdf files; edit text (various), up to 'novae' subsection
% revised by Roland Diehl 091018 ->V5: edit text (various), from "XRBs and mircorquasars" subsection onward
%%%%%%%%%%%%%%%%%%%%%%%%%%%%%%%%%%%%%%%%%%%%%%%%%%%

\section{ Positron production: processes and sources}
\label{sec0:Processes}

\subsection{Radioactivity  from stellar nucleosynthesis}
\label{sec1:Nucleosynthesis}

\subsubsection{Radioactivity  }
\label{sec:Radioactivity}

Positrons are emitted by the $\beta^+$-decay of unstable nuclei which
turns a proton into a neutron, provided the mass difference between
parent and daughter nucleus is $\Delta M>m_e c^2$ (where $m_e$ is electron's mass and $c$ the
light velocity). 
$\beta^+$-decay of unstable nuclei produced in stellar explosions was one of the earliest candidates 
proposed to explain the Galactic 511 keV emission \citep{Clayton73}. 

Astrophysically important e$^+$ emitters are produced in proton-rich environments, either
hydrostatically (e.g. in massive star cores) or explosively (in novae or SN explosions); in both cases,
proton captures  occur on shorter timescales than the corresponding lifetimes of $\beta^+$-decaying
nuclei along the nucleosynthesis path.
Important e$^+$ emitters are also found in the Fe-peak region and are produced in the
so-called {\it Nuclear Statistical Equilibrium} (NSE) regime, at temperatures
$T>4 \ 10^9$ K.  In the short timescale of the
explosion ($\tau \sim$1 s in SNIa and in the inner layers of CCSN) weak 
interactions can hardly operate and material is nuclearly processed under the effect of strong interactions alone,
moving in the neutron vs. proton ($N-Z$) plane along a $N/Z\sim const$  trajectory. The original
stellar material is essentially composed either of $^{28}$Si ($N=Z=14$) in the case of CCSN 
or of $^{12}$C ($N=Z=6$) and $^{16}$O ($N=Z=8$) in the case of the white dwarf progenitors
of SNIa; this  $N/Z\sim1$ ratio is mostly preserved during the explosion. Since the last
stable nucleus with $N=Z$ is $^{40}$Ca ($N=Z=20$), NSE reactions produce mostly unstable Fe-peak nuclei,
which decay later back to the nuclear stability valley by electron captures (EC) or e$^+$
emission. This is typically the case of the most abundant Fe-peak nucleus $^{56}$Fe ($Z=26, N=30$),  which
is produced as $^{56}$Ni ($Z=N=28$) through the decay chain  
$^{56}$Ni $\longrightarrow$ $^{56}$Co $\longrightarrow$ $^{56}$Fe; the first decay
proceeds by EC and the second one by both EC and e$^+$ emission, with branching ratios
of 81\% and 19\%, respectively \citep{Nadyozhin94}.

\begin{table*}
\caption{Astrophysically important positron-emitting radioactivities
\label{tab:Emitters}}
\begin {center}
\begin{tabular}{lccccccc}
\hline \hline
 Nuclide & Decay chain  & Decay mode  & Lifetime & Associated $\gamma$-ray lines  & Endpoint e$^+$  &   Mean e$^+$   &Sources  \\
               &            &   and e$^+$  BR$^a$        &    & Energy in keV (BR$^a$)  &energy (keV)  &energy (keV)    & \\
\hline
 & &  & & &  &  &\\
$^{56}$ Ni  &   $^{56}$ Ni $\longrightarrow$ $^{56}$ Co$^*$ & EC$^b$ & 6.073 d & 158(0.99), 812(0.86) & & &SNIa \\
                     & $^{56}$ Co $\longrightarrow$ $^{56}$ Fe$^*$ & e$^+$ (0.19)  & 77.2 d & 2598(0.17), 1771(0.15) &1458.9  & 610  & \\
 & &  & & & & &\\
$^{22}$ Na  &   $^{22}$ Na $\longrightarrow$ $^{22}$ Ne$^*$ & e$^+$ (0.90)   & 2.61 y & 1275(1) &1820.2  & 215.9  &Novae \\
& &  & & & & &\\
$^{44}$ Ti  &   $^{44}$ Ti $\longrightarrow$ $^{44}$ Sc$^*$ & EC$^b$ & 59.0 y & 68(0.94), 78(0.96)  & &  & Supernovae\\
                     & $^{44}$ Sc $\longrightarrow$ $^{44}$ Ca$^*$ & e$^+$ (0.94)  & 3.97 h & 1157(1) &1474.2 & 632.  & \\
& &  & &  &  & \\
$^{26}$ Al  &   $^{26}$ Al $\longrightarrow$ $^{26}$ Mg$^*$ & e$^+$ (0.82)   & 7.4 10$^5$ y & 1809(1) & 1117.35 & 543.3 & Massive stars \\

\hline \hline

\end{tabular}
\end{center}
($a$) BR:Branching Ratio  (in parenthesis);  ($b$) EC: Electron capture
\label{tab:Emitters}
\end{table*}

Other important astrophysical e$^+$ emitters are displayed in 
Table~\ref{tab:Emitters}, along with various relevant data.
An important feature of $\beta^+$-decay is that positrons are released 
with energies in the MeV range, i.e. they naturally satisfy the constraint imposed by the
continuum observations of the inner Galaxy in that energy range (see Sec. II.B.1).

Contrary to all other e$^+$ sources presented in this section, it is well established that stellar radioactivities contribute at a non-negligible level to the e$^+$ production rate, because of the
observed presence of $^{26}$Al in the disk (see Sec. II.B.2). The uncertainties related to their
overall contribution stem from two factors:

i) In the case of short-lived radioactivities (i.e. with lifetimes short compared to the characteristic timescales of SN expansion)  positrons are released in high density
environments and in magnetic fields of unknown configuration. Those conditions render difficult
the evaluation of the fraction of e$^+$ escaping to environments of sufficiently low density  for their annihilation photons to be detectable. This is the case of $^{56}$Co in SNIa.

ii) In some cases of long-live radioactivities, the corresponding stellar yields and/or
the frequencies of the nucleosynthesis sites are quite uncertain. Indirect methods should then 
be used to evaluate their contribution to the Galactic e$^+$ production. This is the case of $^{44}$Ti.

\subsubsection{Massive stars: $^{26}$Al and $^{44}$Ti}

The observed irregularities in the \Al \ $\gamma$-ray emission along the plane of the Galaxy as shown in the COMPTEL map \cite{Diehl+95} suggest that massive stars are the dominant source, as these are the only candidate \Al \ sources clustered along spiral arms (see \onlinecite{1996PhR...267....1P}, and Sec. II.B).
 \Al \ is produced in such stars both hydrostatically (during H-burning) and explosively (in the C-Ne-O) layers;  it is ejected by the Wolf-Rayet stellar winds in the former case and by the supernova explosion in the latter.   \citet{LC06} find that in their $Z=$\zs \ models explosive nucleosynthesis is always dominant; however,  models with rotation and at $Z>$\zs \ (appropriate for the inner Galaxy) may modify this conclusion somewhat. 
 Stellar nucleosynthesis models find typical yields of $\sim$10$^{-4}$ \ms \  of \Al \ per star, which combined with the derived CCSN frequency in the Galaxy (Table \ref{tab:SNrates}) results in a production rate comparable to the observed one of $\sim$2.7\ms/Myr; thus, the nucleosynthesis of \Al \ is
considered to be rather well understood quantitatively (within a factor of 2).
Independently of theoretical considerations, however, the observed Galactic decay rate of \Al \  corresponds to a production rate of $\dot{N}_{e+,26}\sim4 \ 10^{42}$ s$^{-1}$ in the Galactic disk. 

$^{44}$Ti is produced in the innermost layers of the supernova, in the "$\alpha$-rich freeze-out"
regime of NSE (\onlinecite{Meyer93}, \onlinecite{TNH96}). Its yields are much more uncertain than the ones of \Al \ because of uncertainties either in the nuclear reaction rates (which affect its yields by a factor of 2, \onlinecite{The+06};
\onlinecite{Magkotsios+08}) or, most importantly,
in the explosion mechanism itself
%The proximity of the $^{44}$Ti production layers  to the nascent compact object (neutron star or black hole) make its yields very sensitive to the details of the explosion (energetics, reverse shock, etc.,  
(\onlinecite{WW95}; \onlinecite{TWHH96}).
Moreover, asphericity effects (due e.g. to rotation)
appear to be critical, leading to the production of substantially higher $^{44}$Ti yields 
(and $^{44}$Ti/$^{56}$Ni ratios) than in the case of spherically symmetric models \citep{Nagataki+98}.
 
Observations offer little help in this case. $^{44}$Ti has been directly  detected in the Cassiopeia A (CasA) SN remnant, through its $\gamma$-ray lines, both with COMPTEL/CGRO \citep{Iyudin+94} and with SPI/INTEGRAL \citep{Renaud+06}. Its presence is also indirectly derived in SN1987A, the closest observed supernova in the past four centuries, since it is required to explain the late
 lightcurve \citep{MotKu04}. In both cases the derived $^{44}$Ti yield is $Y_{44}\sim2 \ 10^{-4}$ \ms, substantially larger than predictions of spherically symmetric models, but comparable to predictions of aspherical models. Asphericity is also favoured for CasA and SN1987A on the basis of other observables (\onlinecite{Prantzos04}, and references therein). Does this mean that typical sources of $^{44}$Ti  are aspherical and have the aforementioned yield? 

CasA is found at a distance of $\sim$3 kpc from the Earth in the outer Galaxy (outside 
the active star forming regions of the inner Galaxy)  and its age is estimated to 300 yrs (much larger
than the $^{44}$Ti lifetime). That a  supernova with such properties is the only one detected
so far through its $^{44}$Ti lines, despite the sensitivity of  COMPTEL/CGRO and SPI/INTEGRAL Galactic surveys, appears to be statistically improbable (\onlinecite{The+06}; \onlinecite{Renaud+06}). It may imply
that typical $^{44}$Ti sources are rare, i.e. with frequencies much lower than the CCSN frequencies of Table~\ref{tab:SNrates}, and, consequently, much larger  yields.  Sub-Chandrasekhar mass SNIa (i.e.
thermonuclear SN  induced by surface He-detonation, see Sec.  IV.4) are  potential candidates, since they produce 10--20 times more $^{44}$Ti than a typical massive star explosion \citep{WW94};
but, provided that such objects exist and have the required yields,  their frequencies are totally unknown.

In those conditions, the only way to evaluate the Galactic $^{44}$Ti production rate is through a nucleosynthesis argument, based on i) the solar ($^{44}$Ca/$^{56}$Fe)$_{\odot}=1.2 \ 10^{-3}$ ratio \citep{Lodders03} , i.e the ratio of the stable products of $^{44}$Ti and $^{56}$Ni decays and ii) the knowledge of the current production rate
of $^{56}$Fe, based on disk SN frequencies of Table  \ref{tab:SNrates} and on presumably well-known typical yields of $^{56}$Fe: $Y_{56}^{SNIa}\sim$0.7 \ms \ (see Sec. IV.A.4) and $Y_{56}^{CCSN}\sim$0.07 \ms \ (from the observed lightcurve of SN1987A, \onlinecite{ABKW89}). The production rate of $^{44}$Ti is then:
\begin{equation}
\dot{M}_{44} =  \left({{^{44}Ca}\over{^{56}Fe}}\right)_{\odot} (R_{SNIa}Y_{56}^{SNIa}  + 
\ R_{CCSN}Y_{56}^{CCSN})
\end{equation}
and the corresponding e$^+$ production rate is  
$\dot{N}_{e+,44}\sim3\times 10^{42}$ e$^+$ s$^{-1}$, i.e. 
comparable to the one of \Al. Thus, the two long-lived radioactivities together may account for most, if not all,
of the disk production rate of positrons, as revealed by the SPI/INTEGRAL analysis. The same analysis,
applied to the bulge (and assuming the bulge $^{44}$Ca/$^{56}$Fe ratio to be solar) leads to
a e$^+$ production rate three  times smaller, i.e. an insignificant fraction of the obervationally
required rate for that region.

\subsubsection{Hypernovae and $\gamma$-ray bursts}

Hypernovae are very energetic supernova explosions, with typical observed kinetic energies $>$10$^{52}$ ergs (i.e. about ten times larger than normal supernovae) and ejected $^{56}$Ni masses of $\sim$0.5 \ms \ (e.g. \onlinecite{Nomoto+07}).
Their properties are usually interpreted in terms of aspherical explosions of rotating massive stars (with mass $>$30 \ms).
The rotating Fe core implodes to a black hole, around which  the surrounding material forms a short-lived ($\sim$0.1 s) 
accretion disk. The gravitational energy of accretion is partially transferred (by some still unclear mechanism) to two jets along the rotation axis, which launch the supernova explosion. Heavy nuclei (among which $^{56}$Ni)  are formed in the hot basis of the jet and ejected in the ISM.  This model was originally proposed to account for the phenomenon of Gamma-Ray Bursts (GRB), the most powerful electromagnetic beacons in the Universe, releasing $\sim$10$^{51}$ erg in short flashes of  $\gamma$-rays beamed along the jet direction (the "collapsar" model of  \onlinecite{Woosley93}). Observed metallicities of GRB host galaxies are typically  few times lower than solar \citep{SGLB08};  such low metallicities prevent substantial losses of mass and angular momentum and allow for a rapid rotation of the core at the moment of the explosion, a crucial ingredient of the collapsar model.

Hypernovae/GRBs have been suggested as potential sources of the Galactic positrons, produced either from the $^{56}$Ni decay (\onlinecite{Nomoto+01}; \onlinecite{CCPS04}) or from pair creation, as photons backscattered from the ionized medium ahead of the jet interact with the GRB $\gamma$-ray photons ({\onlinecite{PCLP05};  \onlinecite{Bertone+06}).  Because of the complex (and still very uncertain) nature of those objects, the corresponding positron yield is virtually unknown. In the light of the observational (and theoretically motivated) constraint of low metallicity for the progenitor stars, the existence of such objects 
in the metal-rich bulge (see Sec. III.A.1) should be excluded. Besides, a small bulge/disk ratio would be logically expected in that
case, contrary to observations.

\subsubsection{Thermonuclear supernovae (SNIa)}

SNIa  display a remarkable uniformity in  their properties, like
e.g. the peak luminosity, which is attributed to the power input of 
$\sim$0.7 \ms \ of radioactive $^{56}$Ni \citep{Arnett82}\footnote{In fact, 
the $^{56}$Ni mass may vary by a factor of $\sim$10, as shown by \citet{SMSB06},
who  find values in the range of 0.1-0.9 \ms \ for a sample of seventeen well observed SNIa.
However, obervations indicate that the shape of the SNIa light curves is 
associated to the $^{56}$Ni mass (with brighter SNIa fading more slowly) and
after correction is made for that effect \citep{Phillips93} SNIa can 
indeed be used as ``standard candles'' for the determination of cosmological 
distances (see \onlinecite{Leibundgut01} for a review).}.
There is general agreement that SNIa result from the thermonuclear 
disruption of a white dwarf, igniting explosively its carbon. The thermonuclear flame may propagate
either subsonically (deflagration) or supersonically (detonation) inside the white dwarf;
\citet{2007Sci...315..825M} show that the SNIa variety   can be understood within a single, combined, model, 
involving both deflagration and detonation. 
There are  two main scenarios for the precursors of SNIa: the {\it single degenerate}
(SD) model, in which accretion is made from a main sequence or red giant companion 
\citep{WI73}; and the {\it double degenerate} (DD) model, which involves 
the merging of two white dwarfs in a close binary system (\onlinecite{Webbink84}; \onlinecite{IT84}).
\citet{PBJB07} discuss all available observational evidence
%\footnote{Only recently it was made possible to obtain observational arguments for either one of these scenarios. (\onlinecite{Napiwotzki+04},; \onlinecite{Patat+07}).} 
 and find  that the SD channel is by far the dominant one (but see \onlinecite{GB10} for a different view).

Most studies of SNIa were made in the framework of the SD scenario and, up to the late 1990s with one-dimensional (1D) models. Detailed 1D models exploring the various  possibilities
%\footnote{ It was realised quite early that, after carbon ignition in degenerate conditions, the thermonuclear flame propagates subsonically ({\it deflagration}), turning the dense ($\sim$10$^9$ g cm$^{-3}$) interior of the white dwarf into iron peak nuclei, mostly $^{56}$Ni \citep{NSN76}. It is not yet clear whether a transition to a supersonic flame ({\it delayed detonation}) takes place at some point, producing some more $^{56}$Ni in the outer layers \citep{Khokhlov91}.  Those two cases belong to the class of "Chandrasekhar mass" SD models for SNIa.  A third class of models("sub-Chandrasekhar mass") involves lower mass white dwarfs accreting helium; a sufficiently thick helium shell detonates at the surface and triggers central carbon ignition  (see e.g. \onlinecite{HN00} for a review).} 
(and the corresponding parameter space) have been developed over the years.
% making predictions on the density, velocity and composition profiles of the SNIa ejecta. Based on those predictions, synthetic spectra at different epochs and light curves in different wavelengths   have been calculated and compared to observations. 
Perhaps the most successful 1D model developed so far is the so-called W7 model \citep{NTY84}, the  physics of which has been updated  in \citet{Iwamoto+99}; it is a deflagration model producing in its inner layers $\sim$0.7 \ms \ of $^{56}$Ni and negligible amounts of other positron emitters.  A more accurate description of reality is pursued by the upcoming generation of multi-dimensional models
(\onlinecite{THRT04}, \onlinecite{BG06}, \onlinecite{SN06}, \onlinecite{Ropke+07}, \onlinecite{RN07}).
Preliminary results show interesting features for the stratification of radioactivities, in particular the presence of substantial $^{56}$Ni amounts within outer, high velocity, layers (Fig. ~\ref{SNIa_ejecta}).
%(despite the fact  that it is a pure deflagration model).

\begin{figure}
\includegraphics[width=0.49\textwidth]{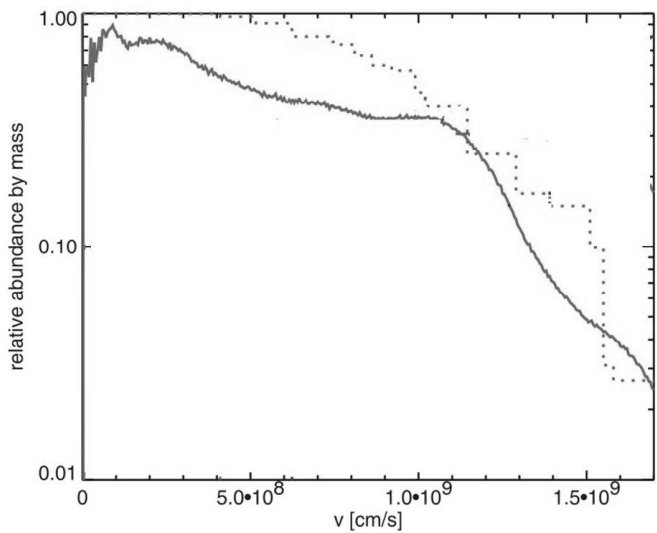}% Here is how to import EPS art
\caption{\label{fig:epsart} Mass fraction of $^{56}$Ni as a function of ejecta velocity after a SNIa explosion. The space-averaged profile of  3-D mode ({\it solid curve} from \citet{Ropke+07}) is  compared to observational data ({\it dotted curve}).
Both theory and observations find non-negligible amounts of $^{56}$Ni in the outer (high velocity) ejecta. Figure from 
 \citet{Ropke+07}.
  }
\label{SNIa_ejecta}
\end{figure}

The fate of the $\beta^+$-decay products ($\gamma$-rays and positrons) in the expanding SNIa ejecta has been extensively studied in 1D (\onlinecite{GIJ98}, \onlinecite{Milne+04}), and more recently, in 3D models (\onlinecite{IBH07}, 
  \onlinecite{SM08}).
%Indeed, one of the aims of astrophysical $\gamma$-ray spectroscopy was to provide accurate diagnostics %of SN interiors through the study of the evolving spectral profiles of $\beta^+$-decay $\gamma$-ray %lines (Clayton et al. 1967, Clayton 1982). 
Generically, before peak luminosity, the SNIa envelope is opaque, and both the energy of 
the explosion and of $\beta^+$-decay are deposited in and diffuse outwards through the ejecta. After the peak of the bolometric lightcurve ($\sim$20 days after the explosion) the luminosity evolves from radioactive energy deposits and increasing energy leakage in a way following (surprisingly closely, given the interplay of these processes) the decay rate of $^{56}$Co. About 6 months later, the ejecta are completely transparent to $\gamma$-rays and the SNIa luminosity results almost exclusively from energy deposited by positrons from ongoing radioactive decays. If positrons are trapped (escape) a flattening (steepening) of the light curve results. How many positrons ultimately escape to the ISM depends on the distribution of the parent radioactivities within the supernova, the evolution of its density, temperature and ionization profiles and, most importantly for the late phases,  on the unknown configuration of its magnetic field. Progenitor white dwarfs have field strengths of 10$^5$-10$^9$ G. \citet{CL93} found that, in the case of radially combed magnetic fields and fully mixed ejecta a substantial fraction of $^{56}$Co positrons ($>$10\%) may escape. Building on the same ideas, \citet{MTL99}  compared  SNIa models to observations of late lightcurves of a dozen SNIa (mostly in B and V bands) and concluded that, typically, a few \% of positrons finally escape the ejecta; the average positronic  yield of a SNIa is $n_{e^+}$(SNIa)$\sim$8 10$^{52}$ (corresponding to an escape fraction of $f_{esc}\sim$0.03). They also found that the mean energy of escaping positrons is  $\sim$0.5 MeV (Fig. ~\ref{SNIa_PosEscape_Spectrum}).

\begin{figure}
\includegraphics[width=0.49\textwidth]{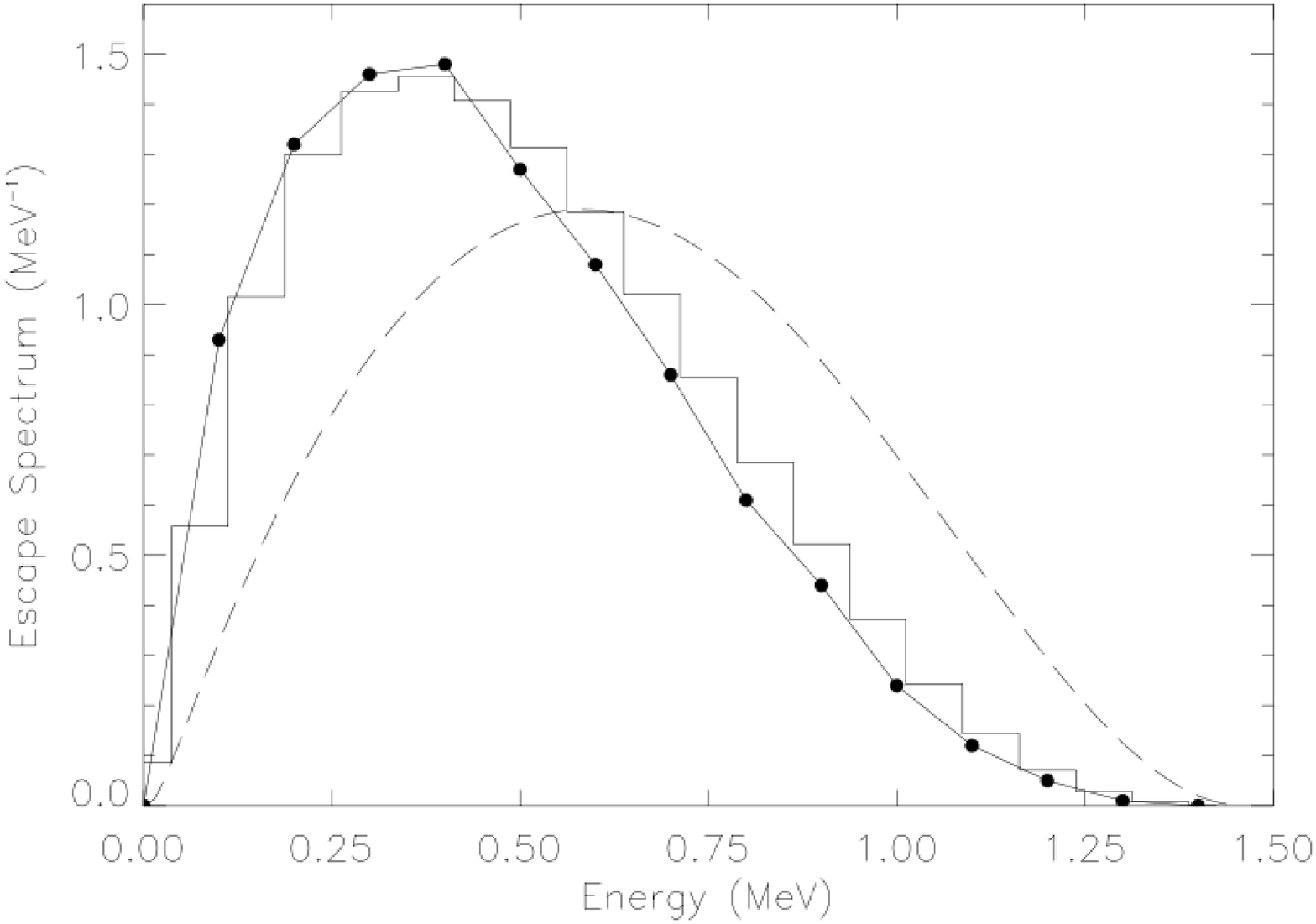}
\caption{ Distribution of emitted positron kinetic energies as estimated
by Segre (1977, dashed curve) compared to the spectrum of escaping positrons
from W7, as estimated by Chan and Lingenfelter (1993, {\it solid curve}) and Milne {\it et al.} (1999, {\it solid histogram}). The slowing of the positrons leads to the mean energy shifting from 632 to 494 keV . Figure from Milne {\it et al}. (1999).
}
\label{SNIa_PosEscape_Spectrum}
\end{figure}

The corresponding Galactic positron yield is then estimated as
\begin{equation}
\dot{N}_{e+,SNIa}  \ = \ n_{e^+}(SNIa)  \ R_{SNIa} \ \sim \ {\rm 1.6 \ 10^{43} \ s^{-1} }
\end{equation} 
where $R_{SNIa}$ is the SNIa frequency in Table~\ref{tab:SNrates}. The total e$^+$ yield is comparable to the observed Galactic one, but the bulge/disk positron emissivity ratio is B/D$\sim$0.4, considerably less than derived from observations. 

This simplified picture may not apply to SNIa in general, though.
Bolometric observations (including the near IR) of the late lightcurves  of SN 2000cx (a rather peculiar at early times SNIa) and of SN 2001el and SN2003hv (two typical SNIa), interpreted in the framework of 1D models, suggest that no positrons escape (\onlinecite{Sollerman+04}, \onlinecite{SS07}, \onlinecite{Leloudas+09}); in that case, despite their large $\beta^+$-decay yields,  the SNIa would be insignificant  e$^+$ producers. 

However, 3D effects may considerably alter the stratification of radioactivities inside the SNIa \citep{Blinnikov+06}, allowing for substantial amounts of $^{56}$Ni to be mixed out to the surface  (Fig.~\ref{SNIa_ejecta}) and for the released positrons to escape at early times (i.e. when the lightcurve is dominated by $\gamma$-rays, and not yet e$^+$  deposition) without being noticed. Studying the very early optical spectra of six SNIa, \citet{Tanaka+08} find indeed indications for asphericity and substantial amounts of $^{56}$Ni present in the high velocity ejecta ($v\sim$10 000 - 15 000 km/s). Positrons produced by the subsequent decay of $^{56}$Co may escape  the ejecta if the magnetic field of the supernova is radially combed.
Observations indicate that this may be the dominant configuration of magnetic fields in young SN remnants
(e.g. \onlinecite{MSH93}; \onlinecite{KR01}) although the origin of such a configuration is not yet clearly understood 
(\onlinecite{JN96}, \onlinecite{SVAK08}). On the other hand, the late lightcurve of SNIa may also be (at least partially) 
powered by internal conversion and Auger electrons released from the decay of $^{57}$Co \citep{STS09}, thus allowing for some $^{56}$Co 
positrons to escape.
Thus, the issue of the positron yield of SNIa 
is not settled yet: $n_{e+}$ may well be  as high as envisioned by \citet{MTL99} (albeit for different reasons), but also much lower.

When the SN becomes sufficiently diluted, the annihilation $\gamma$-ray photons may be directly observed. \citet{KBMR06} find no such signal in observations with SPI/INTEGRAL  of the SN remnant SN1006. They exclude then SNIa as major e$^+$ producers in the Galaxy under the assumption that the e$^+$ lifetime is $\tau_{e^+}<$10$^5$ yr. However, even the low energy positrons of $\beta^+$-decay  may live much longer before annihilation and then escape the SN remnant, especially in the case of a radially combed magnetic field.

\subsubsection {Novae}

Novae result from explosive H-burning on the surfaces of white dwarfs in binary systems. Accretion of material from the companion star increases the density, pressure and temperature at the base of the white dwarf envelope, up to the point where hydrogen ignites in degenerate conditions and burns explosively at peak temperatures of several 10$^8$ K. The ejected mass is $\sim$10$^{-4}$ \ms \ and is substantially enriched with the material of the white dwarf, leading to CO or to ONe novae (see, e.g. \onlinecite{Hernanz05} for a review).

Major positron producers are  $^{13}$N and $^{18}$F  (produced in the hot-CNO cycle) and, in the case of  ONe novae,  $^{22}$Na (produced in the hot-NeNa cycle). The short lifetimes of $^{13}$N and $^{18}$F ($\tau$=862 s and 158 min, respectively)  make unlikely a substantial  escape of their positrons from the nova ejecta. Positrons from $^{22}$Na decay certainly escape and recent calculations suggest that up to 10$^{-8}$ \ms \ of this nucleus may be  produced in ONe novae \citep{HJ06}, releasing up to $n_{e^+,nova}$=10$^{48}$ e$^+$.

The novae frequency in the Galaxy is estimated to $R_{nova}\sim$35 yr$^{-1}$ ( (\onlinecite{Shafter97}; \onlinecite{Darnley+06}). About 1/3 of those may originate from ONe white dwarfs \citep{GilPons+03}, leading to a Galactic e$^+$ production rate of $\dot{N}_{e+,Novae}= R_{nova}n_{e^+,nova}\sim$1.5 10$^{41}$ s$^{-1}$, ie. smaller by two orders of magnitude than the observed rate in the bulge or in the disk. 
It should be noticed that ONe novae appear mostly close to the Galactic plane \citep{dVL98}.
 %% RoD: So? Would we exclude them being associated wth massive-star regions? Would we be able to constrain their vicinity, and be able to predict a scale height of annihilation different from SNIa or from massive-star positrons? I think that would be difficult. If discussed, it should be at a place where we discussed source scale heights altogether... RoD 14 Oct 09

%% RoD: only changes made up to here in V4 of this file; next section (3) unchanged; Oct 14 %%%

\subsection{High energy processes in cosmic rays and compact objects}
\label{sec1:Compact}

\subsubsection{High energy processes}\label{HEprocesses}

\noindent{\it 1a. Inelastic p-p collisions }

\medskip

Relativistic protons and heavier nuclei are present in many astrophysical 
environments in the Galaxy. Their inelastic interactions with interstellar gas produce 
secondary particles including numerous neutral and charged pions and kaons
$pp\to\pi+X$, $pp\to K+X$. In turn, decay of positively charged mesons
produces secondary positrons. The dominant channel is pion decay 
$\pi^+\to\mu^+\nu_\mu$, $\mu^+\to\tilde{\nu}_\mu\nu_ee^+$, though  
a non-negligible contribution comes from the charged kaon decays.
The two main kaon decay modes contributing to
the secondary $e^\pm$ spectrum are $K^\pm \to \mu \nu_\mu$ (63.5\%)
and $K^\pm \to \pi^0 \pi^\pm$ (21.2\%). The processes as the source of secondary
cosmic ray positrons and diffuse $\gamma$ ray emission have been thoroughly
studied (e.g., 
%\citet{Stecker70}, \citet{Orth76}, \citet{Proth82}, \citet{Dermer86a}, \citet{MDR87},
%\citet{MS1998}, \citet{Kelner2006}, \citet{Kamae2006,Kamae2007}, \citet{Porter2008}, 
\onlinecite{SMP07} and references therein). A review of the
experimental data for pion production in proton-proton collisions
and relevant cross section parameterizations $<$50 GeV were presented by
\citet{Blattingea01}. New parameterizations of neutral
and charged pion cross sections which provide an accurate
description of the experimental data in a wide energy range from the pion production
threshold up to $10^5$ TeV
are discussed in \citet{Kelner2006}, \citet{Kamae2006,Kamae2007}.

The energy spectra of positrons from the decay of $\pi^+$
mesons produced in collisions of isotropic monoenergetic protons
with protons at rest are shown in Fig.~\ref{Fig_pp}; they typically present a maximum
at $E\sim30-40$ MeV.

%Positrons are copiously produced in
%inelastic collisions of energetic protons with the interstellar gas.
%In many astrophysical sites of interest,  sources of energetic relativistic protons are
%present. Pions and kaons are produced in the inelastic proton-proton
%collisions $pp\to\pi^+X$,~ $pp\to K^+X$ and also in proton-nuclei
%interactions. Then most of the positrons are created in charged pion
%decays through muon, though a non-negligible contribution comes from
%the charged kaon decays.  The two main kaon decay modes contributing to
%the secondary $e^\pm$ spectrum are $K^\pm \to \mu \nu_\mu$ (63.5\%)
%and $K^\pm \to \pi^0 \pi^\pm$ (21.2\%). The processes as the source of secondary
%cosmic ray positrons and diffuse gamma ray emission have been
%studied in detail in many papers(e.g., \citet{Stecker70},
%\citet{Orth76}, \citet{Proth82}, \citet{Dermer86a}, \citet{MDR87},
%\citet{MS1998}, \citet{SMP07} and reference therein). A review of the
%experimental data for pion production in proton-proton collisions
%and relevant cross section parameterizations were presented by
%\citet{Blattingea01}. They discussed parameterizations of neutral
%and charged pion cross sections which provide an accurate
%description of the experimental data.

\begin{figure}
\includegraphics[width=0.49\textwidth]{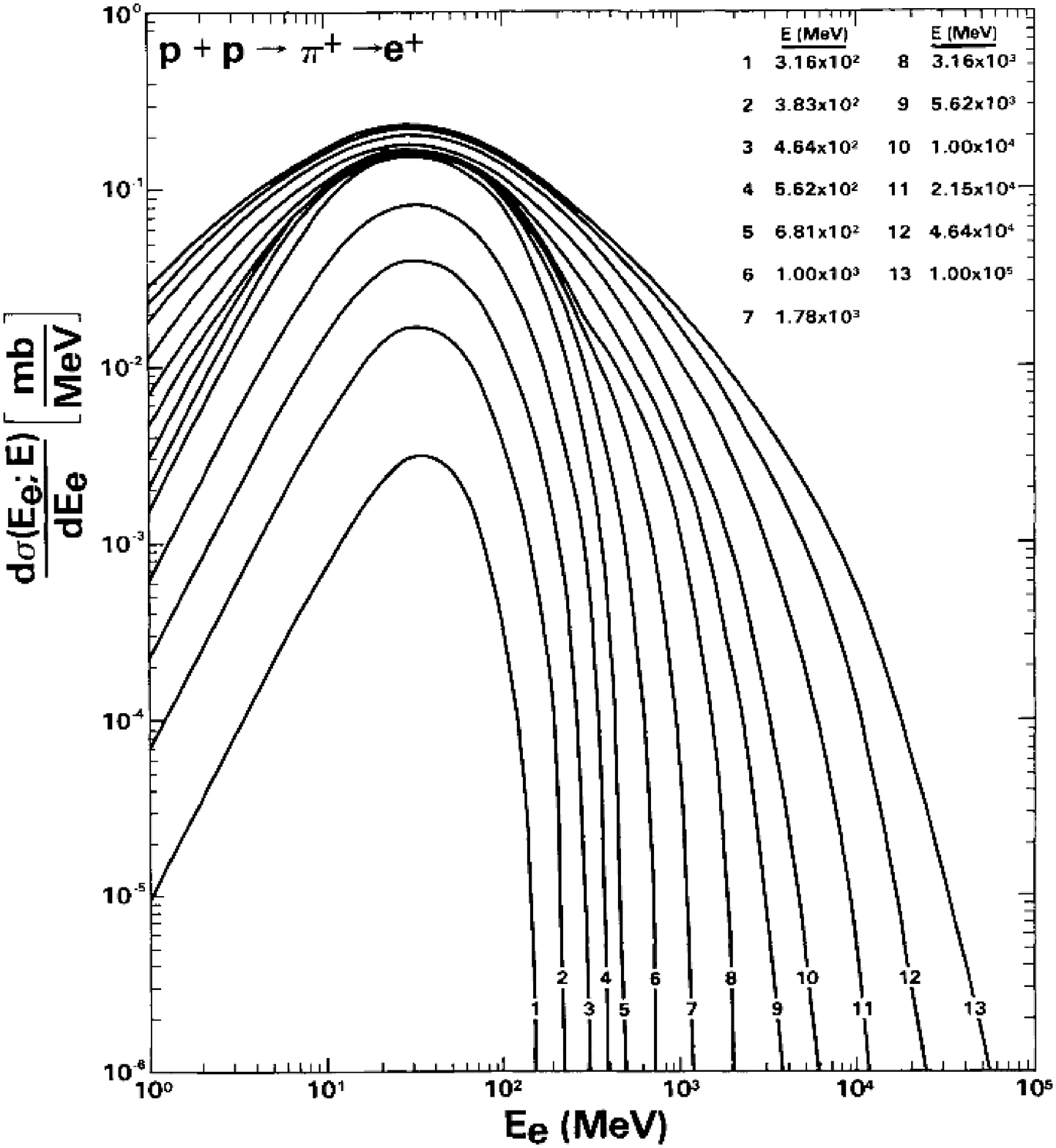}
%\includegraphics*[scale=0.9]{fig3a.eps}}
%\plotone{fig3a.eps}
 \caption{
The energy spectra of positrons from the decay of $\pi^+$ produced in collisions
of isotropic monoenergetic protons with protons at rest for various
proton kinetic energies (from bottom to top):
0.316, 0.383, 0.464, 0.562, 0.681, 1.0, 1.78, 3.16, 10.0, 100.0 GeV \citep{MDR87}.}
 \label{Fig_pp}
\end{figure}

\bigskip

\noindent {\it 1b. $\gamma-\gamma$ pair production}}

\medskip

Positrons can also be produced in photon-photon interactions
when the product of their energies is $> 2 m_e^2c^4/(1-\cos\theta)$ where
$\theta$ is the angle between the photon directions. 
The total unpolarized cross section for the creation of $ e^{\pm}$
by two photons $\gamma \gamma \rightarrow e^{+} e^{-}$
can be expressed as a function of a dimensionless velocity $\beta$
of the produced particles in the center of mass frame:

\begin{equation}
\sigma_{\gamma \gamma} =  \frac{3 \sigma_{\rm T}}{16}\,(1 - \beta^2)
\left[(3 - \beta^2) \ln\frac{1 + \beta}{1 - \beta} - 2\beta (2 -
\beta^2)\right],
\end{equation}
where  $\sigma_{\rm T} \approx 6.65  \ 10^{-25}$ cm$^{2}$ is the
Thomson scattering cross section \citep{bw34, greiner03}.

\begin{figure}
\includegraphics*[width=0.49\textwidth]{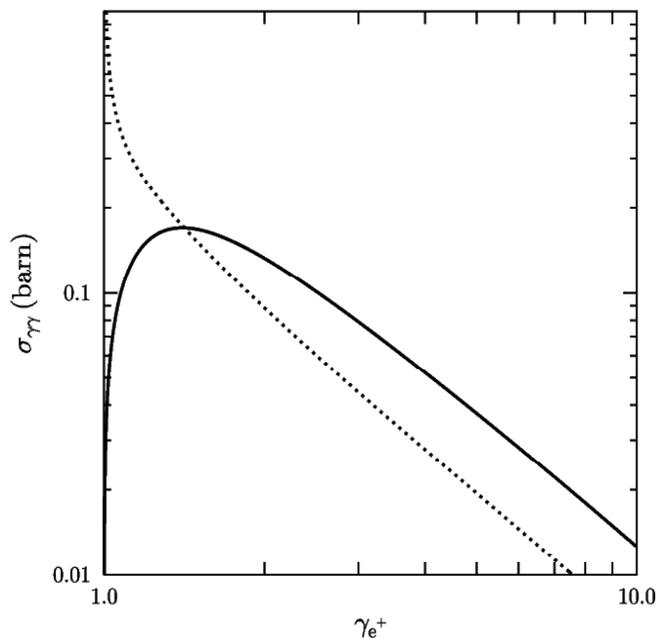}
%\plotone{fig3a.eps}
\caption{
The total cross section of $\gamma+\gamma \rightarrow
e^{+}+e^{-}$ reaction as a function of the Lorentz factor of created
positron in the center of mass frame ({\it solid}). The
total cross section $\sigma_{a}$ for two-photon pair annihilation
reaction $e^{+}+e^{-}\rightarrow \gamma + \gamma$ (unpolarized) is
indicated as a {\it dotted} curve.
 \label{pair1}}
\end{figure}

In Fig.~\ref{pair1} the cross section is presented as a function of
the Lorentz factor of created positron in the center of mass frame.
The positron production due to photon-photon collisions is
suppressed at the threshold and reaches a maximum at $\beta \approx
0.7$. 

%In real applications one should have in mind that the
%probability of pair annihilation at the low energy cut-off region is
%large. The total crossection for two-photon pair annihilation
%reaction $e^{+}+e^{-}\rightarrow \gamma + \gamma$ (unpolarized) can
%be expressed as
%\begin{equation}
%\sigma_{a} = \frac{1}{2 \beta^2} \sigma_{\gamma\gamma}.
%\end{equation}

The production of a $e^+e^-$-pair by a single photon is possible in magnetic 
fields $B\ga$10$^{12}$ G observed in highly magnetized objects such as pulsars and magnetars
\citep{Klepikov1954,DH1983}.
This occurs with significant probability when the photon energy
is $\simeq 3/(B_{12} \sin\theta)$ MeV, where $B_{12}$
is the external magnetic field strength in units $10^{12}$ Gauss and $\theta$
is the angle between the photon direction and the magnetic field. 

%The critical energy of a photon to produce a pair is $E_{\rm crit} \simeq 3/B_{12}$ MeV, where $B_{12}$
%is the external magnetic field strength in units $10^{12}$ Gauss. 

%A specific pair creation process may occur in highly magnetized objects (pulsars, magnetars): 
%the {\it magnetic pair creation} which  involves the interaction with a magnetic photon
%$\gamma + \gamma_{\rm B} \rightarrow e^{\pm} + \gamma_{\rm B}$. 
%The critical energy of a photon to produce an 
%electron-positron pair with the {\it local} magnetic field $B_{\rm 12}$ (in 10$^{12}$ Gauss) is $E_{\rm crit} \simeq$ 3 B$_{\rm 12}^{-1}$ MeV. 

\subsubsection{Galactic cosmic rays  }
\label{subsubsec:GalaxyCR}

The majority of positrons in cosmic rays (CR) are believed to be secondaries
produced by interactions of relativistic particles with interstellar gas;
however recent measurements of positron fraction in cosmic rays $e^+/(e^-+e^+)$
by PAMELA \citep{Adriani2009} indicate that there may be another component at 
high energies. If produced by CR interactions, the positron 
fraction is expected to decrease with energy \citep{MS1998} while the
PAMELA data show it rises above $\sim$10 GeV. 
The origin of this additional component is intensively debated.
%\footnote{ At the time when this paper was in preparation, the PAMELA positron fraction 
%measurement had a citation count $\sim$350 and it was rapidly increasing.}. 
The ideas proposed can be roughly divided into two broad classes: conventional sources,
such as SNR or nearby pulsars 
%\citep[e.g.,][]{Hooper2009,Yuksel2009,Blasi2009,Fujita2009,Shaviv2009,Zhang2001,Coutu1999}, 
(e.g. \onlinecite{Blasi2009} and references therein)
and exotic sources such as WIMP annihilation or decay 
%\citep[e.g.,][]{Arvanitaki2009,Arkani2009,Bergstrom2009a,Bergstrom2009b,Bi2009,Finkbeiner2009,Regis2009}. 
(e.g. \onlinecite{Arvanitaki2009} and references therein).
The predicted behaviour of the positron fraction at energies higher
than currently measured ($\sim$100 GeV) depends on the model and can be used to
distinguish between different possibilities. 

In this section we will discuss
positrons produced by conventional CR interactions with interstellar gas.
The most important factors 
are the energetics of cosmic rays and their diffusion
in the interstellar medium \citep[for more details see, e.g.,][]{SMP07}. 

The major CR sources  are believed to be supernovae (SN) and their remnants (SNR) with some
fraction coming from pulsars, compact objects in close binary
systems, and stellar winds. Recent observations of X-ray and $\gamma$-ray
emission from SNRs \cite{2003ApJ...593..377P,2006A&A...449..223A}
%(Pannuti et al. 2003; Aharonian et al. 2006) 
reveal the presence of energetic electrons, thus
testifying to efficient acceleration processes near these objects.
The total power of
Galactic CR sources needed to sustain the observed CR
density is estimated at  $\sim$10$^{41}$ erg s$^{-1}$ which corresponds
to approximately $10^{50}$ erg per SN, if the SN rate in the Galaxy is 1 every 30 years
(Table~\ref{tab:SNrates}). 
This value is $\sim$10\% of the corresponding total
kinetic power of the SN ejecta, an efficiency which is in agreement with the predictions
of diffusive shock acceleration theory \citep{Blandford1987,Jones1991}.
After injection into the interstellar medium (ISM),
 cosmic rays remain contained in the gaseous disk for $\sim$15 myr and
 in the Galaxy for $\sim$100 Myr
 before escaping into intergalactic space \citep{berea90}.
Note that the latter value is much larger than estimates based on the so-called
leaky-box model \citep{Yanasak+01};  see \citet{SMP07}for a full discussion of this point.

\iffalse%##############################################################
The wealth of information contained in the CR isotopic abundances makes
it possible to study various aspects of their acceleration and
propagation in the interstellar medium as well as the source
composition. Stable secondary nuclei tell us about the diffusion
coefficient and Galactic winds (convection) and/or re-acceleration in
the interstellar medium, while long-lived radioactive secondaries provide
constraints on global Galactic properties such as, e.g., the Galactic halo
size. Abundances of K-capture isotopes, 
which decay via electron K-capture after attaching an electron from
the ISM, can be used to
probe the gas density and acceleration time scale. Details of the
Galactic structure, such as the non-uniform gas distribution (spiral
arms, the Local Bubble), radiation field and magnetic field (regular
and random) distributions, and close SNRs may also affect the local
fluxes of CR particles,
%
especially CR leptons (electrons and positrons) whose energy losses 
are large at both low and high energies (due to ionization, 
Coulomb scattering, bremssrahlung, inverse Compton, and synchrotron emission). 
%
Heliospheric influence (modulation) changes
the spectra of CR particles below $\sim$10--20 GeV/nucleon as they
propagate from the boundaries of the solar system toward the
orbits of the inner planets; these distorted spectra are finally
measured by balloon-borne and satellite instruments (Parker 1965; Gleeson \& Axford 1968).

\fi%###################################################################

Propagation of cosmic rays in the ISM is usually modelled as diffusion, where 
the energetic particles scatter on
iregularities (fluctuations) of the turbulent Galactic magnetic field (see Sec. III.D).
The diffusion equation may include stochastic reacceleration in the ISM,
convection by the Galactic wind, continuous and catastrophic energy 
losses, nuclei fragmentation, radioactive decay, and production of secondary
particles and isotopes \citep[for a recent review of cosmic ray transport, see][]{SMP07}.
Isotopes of light elements (Li, Be, B) in cosmic rays are almost all secondaries 
produced in spallations of heavier (CNO) nuclei  during CR propagation.
If the diffusion is fast (slow), the secondary nuclei are present,
after propagation,  in small (large) amounts;
therefore, the relative abundances of secondary and primary nuclei can be used to 
determine the propagation parameters\footnote{
The stable secondary/primary ratio does not allow one to derive a unique set of
propagation parameters. The radioactive isotope abundances
are then used to break the degeneracy. Four
radioactive isotopes, $^{10}$Be, $^{26}$Al, $^{36}$Cl, and $^{54}$Mn,
are commonly used to probe the effective Galactic volume filled with
cosmic rays and derive the CR confinement time  in the Galaxy. Their
half-lives range from 3.07 10$^5$ yr ($^{36}$Cl) to
1.60 10$^6$ yr ($^{10}$Be) with the shortest half-life being most
sensitive to the local structure.}.
The derived propagation parameters (timescale of CR confinement, diffusion coefficient, etc.)
are model dependent and can vary significantly \citep[e.g.,][]{Ptuskin2006}. 

%###################################################################

%Secondary particles, such as positrons and antiprotons, are produced
%in energetic interactions of cosmic ray protons and heavier nuclei with
%interstellar gas (see Section \ref{HEprocesses}). 
The production spectra of secondary particles are 
determined by the kinematics of the collision and depend on the 
ambient spectrum of cosmic rays while their propagation is governed by 
the same propagation equation as for other cosmic ray species. 
The production rate of secondary positrons slightly depends on the 
assumed propagation model and is about $(1-2) 10^{42}$ s$^{-1}$
\citep{Porter2008}, i.e.  $\sim$5-10\% of the Galactic e$^+$ annihilation
rate.  A cosmic ray origin of the positrons annihilating at the Galactic center 
can still be reconciled
with the production rate if cosmic ray intensities were significantly higher 
in the past (Sec. IV.B.5 for an analogous situation for the Galactic supermassive black hole).

Heliospheric influence (modulation) changes
the spectra of cosmic ray particles below $\sim$10--20 GeV/nucleon as they
propagate from the boundaries of the solar system toward the
orbits of the inner planets \cite{1965P&SS...13....9P,1968ApJ...154.1011G}.
%(Parker 1965; Gleeson \& Axford 1968); 
The heliospheric modulation is a combination of effects of convection by the solar wind, 
diffusion, adiabatic cooling, drifts, and diffusive acceleration 
\citep[e.g.,][]{Potgieter1998}. 
%Drift models predict a clear charge-sign dependence 
%for modulation of cosmic ray electrons and positrons \citep[e.g.,][]{Potgieter2004}, 
%i.e., different behavior of the positron fraction $e^+/(e^++e^-)$ for solar cycles 
%with different polarities. This effect is clearly seen when comparing the measurements
%done at different epochs \citep{Adriani2009}.
Though the e$^+$   fraction  inside the heliosphere is small ($\sim$0.1), the e$^+$  flux 
in the ISM below $\sim$1 GeV 
is estimated to be comparable to CR electron flux 
at the same energies \cite{2004ApJ...613..962S}.

Direct information about
the CR  fluxes and spectra in distant locations is provided by the Galactic 
diffuse $\gamma$-rays.
Continuum diffuse emission  is expected in the 
hard X-ray and $\gamma$-ray regime from the physical processes of positron 
annihilation (through formation of positronium), 
inverse Compton scattering and bremsstrahlung from CR electrons and positrons, and
via decay of neutral pions produced by interactions of CR nuclei with the 
interstellar gas \citep{2000ApJ...537..763S,2004ApJ...613..962S,Porter2008}.  
Positron annihilation in flight (continuum) may
contribute in the MeV range 
\citep{2000A&A...362..937A,Beacom-Yuksel:2006}. 
%(Aharonian & Atoyan 2000; Beacom \& Yuksel 2006).
%Together with CR electrons, secondary positrons contribute to the production
%of the diffuse X-ray and $\gamma$-ray emission via inverse Compton scattering
%of optical, IR, and CMB photons 
%\citep{2000ApJ...537..763S,2004ApJ...613..962S,Porter2008}.
%(Strong et al. 2000, 2004). 
That contribution 
%of secondary leptons (positrons and electrons) to the diffuse emission is non-negligible 
can be determined from
a comparison of model predictions to the data 
obtaied by  INTEGRAL, COMPTEL, and EGRET (and now by {\it Fermi}/LAT launched in June 2008,  \onlinecite{GLAST}).
%(Atwood et al. 2008).
The analysis of the INTEGRAL data shows that most of the emission 
between 50 keV and $\sim$1 MeV \cite{Bouchet+08} 
is produced via inverse Compton scattering of background photons 
off CR electrons \cite{Porter2008} and that
CR positrons  in  distant regions of the Galaxy 
(including the direction of the Galactic center)  are mostly secondary
(see also Sec. II.C.1).

%###################################################################
\subsubsection{Pulsars, milli-second pulsars and magnetars}

A large fraction of  high energy sources in the Milky Way consists of 
rapidly rotating magnetized neutron stars, which belong to several sub-classes:
rotation powered pulsars (including Crab-like and Vela-like pulsars),
accretion powered pulsars (including  milli-second pulsars or ms pulsars\footnote{ms pulsars spin-up by accreting
angular momentum from a companion star.}) and
strongly magnetized rotating objects (including soft $\gamma$-ray 
repeaters and magnetars). 
 All these objects have been the subject of intense observational 
and theoretical work (see  reviews by \onlinecite{Rudak01}; \onlinecite{SW04}; \onlinecite{Lorimer05}; \onlinecite{HL06}; and  references therein). 
Typical values of their main properties (magnetic field intensity, rotation period, activity lifetime, estimated
birthrate  and total number in the Milky Way) are provided in Table~\ref{TabPulsars}. 

The high energy radiation and/or a high magnetic 
field ($B_* \ge 10^{12}$ Gauss) of those objects are associated with intense e$^-$-e$^+$ pair creation. 
The pairs are further accelerated in parallel electric fields in the polar caps or in the outer gaps 
close to the light cylinder\footnote{The light cylinder is the surface inside which the closed field lines have a rotation velocity smaller than the speed of light; it separates the region of closed and open magnetic field lines.}. The $\gamma-\gamma$ interaction of secondary photons produced by
the primary particles yields a pair cascade; its particles can eventually escape into the pulsar wind.
Pair creation is accompanied by different high-energy photon production 
channels that directly contribute to the pair cascade: curvature radiation, magnetic inverse Compton scattering, synchrotron radiation, and photon splitting. 

It suffices of a few charged particles to be accelerated up to high Lorentz factors to initiate an 
 e$^-$-e$^+$ pair cascade either above the polar caps \citep{Harding81} or in a part of the outer magnetosphere
close to the frontier of the open magnetic field lines region called the outer gap \citep{CHR86}.
In the outer gap model of  \citet{ZC97}, primary  e$^-$-e$^+$ pairs have a typical energy $E_{\rm p} 
\simeq 5  \ 10^{6} \, P^{1/3}$ MeV (where $P$ is the pulsar's period in seconds). 
Photons with energy $E>E_{\rm crit} \simeq$ 
3 B$_{\rm 12}^{-1}$ MeV (see Sec. IV.B.1b) will generate a pair cascade involving a total of 
$N_{e^{\pm}}$=$E_{\rm p}/E_{\rm crit}$     pairs, 
most of which will be reflected by the magnetic mirror effect and then move towards the light cylinder.

%The averaged number of positrons produced per each primary at the end of the cascade is:
%\begin{equation}
%n_{\rm e^{\pm}} = {E_{\rm p} \over E_{\rm crit}} \simeq 1.9 \times 10^6 \ B_{\rm d, 12} \times P^{1/3}  \ ,
%\end{equation}
%The total rate of pair production is $\dot{n}_{\rm e^{\pm}} = f \times \dot{n}_{\rm GJ} \times n_{\rm e^{\pm}}$.
%The quantity $f < 1$ is the fractional volume of the outer magnetosphere occupied by the outer gap and is evaluated to 
%$f \simeq 5.5 \, P^{26/21} \, B_{\rm 12}^{-4/7}$. Finally, $\dot{N}_{\rm GJ}= 2.7 \times 10^{30} \, \rm{s^{-1}} P^{-2} 
%\, B_{\rm d,12}$ is the Goldreich-Julian charge density (Goldreich \& Julian 1969). 
In terms of the surface magnetic field and 
the pulsar period the total e$^+$ production rate of the cascade is:
\begin{equation}
\dot{n}_{\rm e^{\pm}}  \simeq 2.8  \ 10^{37}  \ B_{\rm d,12}^{10/7} \ P^{-8/21} \  \rm{s^{-1}}
\label{Eq:rateog}
\end{equation}
where $B_{\rm d,12}$ is the dipole magnetic  field in 10$^{12}$ Gauss.
%That rate is  about four orders of magnitude above the e$^+$ production rate in polar cap models \citep{CCY96}.
%It is clear that the outer gaps are the main sources of positrons  injected into the wind.
In the case of ms pulsars the dipole assumption at the stellar surface is not valid anymore and the magnetic field should be 
rescaled as $B_{\rm d} \rightarrow B_{\rm d} (R_*/\ell)^3$, where $R_*$ is the star radius and $\ell\sim$1 km  is the curvature radius 
of the magnetic field in the stellar surface (approximately equal to the stellar crust radius, \onlinecite{WPC06}). The effective rate of positrons injected into the pulsar wind is a fraction $\xi$ of $\dot{n}_{\rm e^{\pm}}$. 
This fraction is probably lower than one in normal pulsars and close to one in ms pulsars: due to the lower magnetic field  of the latter
($B \sim 10^{8-9}$ G) the light cylinder is much closer to the neutron star surface  and 
 particles are expected to  escape more easily \citep{WPC06}. Notice that in the extreme magnetic field conditions of magnetars
the production of most of the e$^+$-e$^-$ pairs 
is probably suppressed \citep{HL06}.

\begin{table}
\caption{Properties of magnetized neutron stars (data for pulsars
and ms pulsars are from \onlinecite{Lorimer05} and for magnetars from \onlinecite{GH07}, although \onlinecite{KK08} suggest
somewhat higher birthrates).
 \label{TabPulsars}}
\begin {center}
\begin{tabular}{llccc}
\hline \hline
   &    & Pulsars & ms Pulsars &  Magnetars \\
\hline

Magn. field & $\langle$B$\rangle$ (G)    & $10^{12}$   & $3\times10^{8}$  &  $3\times10^{14}$ \\

Period &  $\langle$P$\rangle$  (s)    & 0.5   &  $3\times10^{-3}$   & 10 \\

Birthrate & $R$ (yr$^{-1}$)   & $1.5 \times10^{-2}$  &   10$^{-5}$  & $2\times 10^{-3}$    \\

Lifetime &  $\langle\tau \rangle$ (yr)    & $10^{7}$   &  $3\times 10^{9}$  & $2\times 10^4$  \\

Total number   &  $N$ & $1.5\times 10^{5}$   &  $3\times 10^{4}$   &  40  \\

e$^+$ yield$^a$ &  $\dot n_{e^{\pm}} $ (s$^{-1}$) & $4\times 10^{37}$   & $5\times 10^{37}$   & $4\times 10^{40}$ \\

Total  e$^+$ yield$^b$ &  $\dot N_{e^{\pm}}$ (s$^{-1}$) & $5\times 10^{42}$   & $1.5\times 10^{42}$     &  1.6 $10^{42}$  \\
\hline \hline

\end{tabular}
\end{center}
a: Individual source yield  from Eq.~\ref{Eq:rateog}; b: Galactic yield from $\dot N_{e^{\pm}}$=$\dot n_{e^{\pm}} R \ \langle\tau\rangle$,
assuming $\xi$=1 (see text).
\end{table}

The total Galactic injection rate $\dot N_{e^{\pm}}$ of one particular class (normal pulsars, ms pulsars, magnetars) is 
$\dot N_{e^{\pm}}$=$\dot n_{e^{\pm}} N$, where
$\dot n_{e^{\pm}}$  is the average e$^+$ production rate of one source and $N$ =$R \ \langle\tau\rangle$ is
the number of sources in the Galaxy  (where $R$ is the  birthrate and $\langle\tau\rangle$ the typical lifetime of the sources, 
see Table~\ref{TabPulsars}).

%The total positron injection rate is thus $\simeq 6 \times 10^{42} \, \rm{e^+ \, s^{-1}}$. Several uncertainties may however question this result. First, the exact fraction of positron produced that leaves the light cylinder can be $\le 1$. Besides the uncertaintiesof the magnetic field/goemetrical properties of each object, the number of MSPs in the Galactic Center is probably an upper-limit since other type of sources can contribute to the EGRET gamma-ray background although any precise estimations of the MsPs birth rate is still missing.

In view of their young age, pulsars and magnetars are expected to have a radial distribution closely following the one of the
star formation rate (Fig.~\ref{SFRdens}), a small scaleheight ($\sim$100 pc) and an insignificant population in the bulge. Millisecond pulsars
are expected to have a different radial distribution, since they originate in binary systems of all ages. For that reason, their radial distribution should be closer to the  one of SNIa (Fig. ~\ref{SFRdens}), their scaleheight \footnote{\citet{SGH07} adopt 
a scaleheight of 200 pc for ms pulsars, but in view of the age of those systems - and the additional effect of a kick velocity
from the explosion - this should be $>$300 pc.} $>$300 pc and they should  have a substantial bulge component, albeit with a bulge/disk
ratio B/D$<0.5$.
 
The main problem with compact magnetized objects as candidate e$^+$ sources is the expected  high energy of the produced positrons
(E$\sim $30 MeV, \onlinecite{djD08}), which violates the constraint from the continuum MeV emission observed in the inner Galaxy. 

\subsubsection{X-ray binaries and micro-quasars}

X-ray binaries (XRBs) involve a compact object (neutron star or black hole, hereafter the primary) accreting matter from a normal star (the secondary) through an acretion disk. They are classified as high-mass (HMXRBs) or low-mass (LMXRBs)  depending on whether the mass of the secondary is of spectral type earlier or later than B (heavier or lighter than $\sim$4  \ms, respectively). About 300 XRBs are currently detected in the Galaxy, for a total X-ray luminosity of $\sim$10$^{39}$ erg s$^{-1}$; their luminosity function suggests, however, that  there are  more than a thousand of them. LMXRBs are $\sim$10 times brighter on average and slightly more numerous than HMXRBs. The corresponding scale heights are $\sim$410 pc in the former case and $\sim$150 pc in the latter  \citep{GGS02}.

Some XRBs exhibit radio emission, which is usually  attributed to synchrotron radiation emitted by leptons (electrons, and perhaps positrons); leptons are launched along diametrically opposite jets fuelled by the accretion energy.
If the jets are  confirmed by imaging, the system is called a microquasar
($\mu$Q), for it has a 
similar structure to quasars (the latter being a million times larger and brighter, i.e. \onlinecite{Mirabel08}, and references therein).  

The physics of microquasars is extremely complex and there is no generally accepted model at present. Empirical evidence suggests that at high accretion rates, X-ray emission peaks at $\sim$1 keV ({\it high-soft} state), whereas at low accretion rates X-ray emission appears at higher energies, with a power-law spectrum and an exponential  cut-off at $\sim$100 keV ({\it low-hard} state, e.g. \onlinecite{MR06}). It is expected that a persistent jet will be present in the low-hard state \citep{FBG04} and correlations between the radio and X-ray luminosity and the accretion activity have been proposed (e.g. \onlinecite{Corbel+03}).
It should be noticed that the issue of 
positron production in microquasars suffers from two major uncertainties:
i) while electrons are certainly responsible for the observed synchrotron emission,  it is not yet clear whether the positively charged component of the jets consists mainly of ions or  positrons; and ii) even if positrons are largely present, it is not yet known whether the e$^-$-e$^+$ pairs of the jets are ejected at ultra-relativistic velocities or not: current wisdom is that the jets are mildly relativistic \citep{GFP03} but alternative views have been expressed (e.g. \onlinecite{FPFH08}). This has obvious implications for the inflight e$^+$ annihilation and the production of $>$1 MeV $\gamma$-ray continuum.

Positrons can be pair-created in the vicinity of the compact object, either in the hot inner accretion disk, 
in the X-ray corona surrounding the disk, or at the base of the jets; the latter may 
channel a fraction of the e$^-$-e$^+$ pairs out of the system. 
Alternatively, if the jets consist mainly of relatively cold e$^-$-e$^+$ or e$^-$-p plasma, they may create new  e$^-$-e$^+$ pairs  at the termination shock with the ISM.  \citet{HS02} noticed that the total kinetic luminosity of microquasar jets in the Galaxy, evaluated at $L_{kin}\sim3 \ 10^{38}$ erg s$^{-1}$, can  produce up to $4 \ 10^{43}$ e$^+$ s$^{-1}$, i.e. more than required from observations. This estimate requires $\sim$5\% of  the kinetic power to be converted to e$^-$-e$^+$ pairs, i.e. a reasonable conversion efficiency.

Soon after the first data release of the 511 keV image by SPI/INTEGRAL, \citet{Prantzos04} noticed that 
i) the observed  distribution of LMXRBs in the Galaxy is strongly
peaked towards the central regions \citep{GGS02},  similar to
that of the 511 keV emission, and ii)
their total X-ray luminosity is $\sim$10$^{39}$ erg s$^{-1}$, a hundred times larger than the corresponding mass-energy of the observed 10$^{43}$ e$^+$ s$^{-1}$  in the Galaxy.
He also noticed, however, that most of
the strongest sources (accounting for 80\% of the total Galactic X-ray flux)
are evenly distributed in the Galactic plane, with no preference for 
the bulge; he concluded  that, if the positron emissivity scales with their X-ray
flux, then LMXRBs cannot be  the origin of the bulge Galactic positrons. The argument is invalid, however,
if the timescale for the variability of the X-ray flux is much smaller than the slowing down timescale of
the positrons, and/or if positrons annihilate far away from their sources. 

Various features of the scenario of microquasars as positron producers were studied in \citet{GJP06}, on the basis of
i) existing theoretical models (e.g. \onlinecite{Beloborodov99}; \onlinecite{YTK99})
and ii) global energetic considerations (of XRB luminosities correlated to jet power and 
to positron ejection rates). They found that
such considerations lead to average values up to 10$^{41}$ e$^+$ s$^{-1}$ for a jet.
Interestingly enough, this is roughly the current  upper limit for 
e$^+$ production rates in XRBs with SPI/INTEGRAL (see Table 1 in \onlinecite{GJP06}).
If a hundred microquasars exist in the Milky Way (not an  unreasonable extrapolation from their currently known population
of two dozen), then these objects may contribute substantially to the observed 511 keV emission. 
The distribution of the  known microquasars
shows indeed some clustering towards the inner Galaxy, but the data are insufficient for statistically significant conclusions.

A similar investigation is performed in \citet{BSTM09}, who consider hadronic jets (also containing e$^-$-e$^+$ pairs) 
launched by all LMXRBs, down to the lowest X-ray luminosities. Extrapolating from the results of recent deep X-ray surveys
of the central bulge, they estimate that a bulge population of 300-3000 LMXRBs would inject mass in  jets at a rate
of $10^{17}-10^{18}$ g s$^{-1}$; the observed e$^+$ production (=annihilation) rate of $2\times 10^{43}$ s$^{-1}$
requires then a yield of 40--400 e$^-$-e$^+$ pairs per proton. 
As an example the authors discuss observations of the giant galaxy M87, and they
argue that such a high ratio is justified by the observational finding that the jet plasma in that
galaxy is cold, i.e. the energy spectrum of the electrons has no measurable low energy cut-off and it is thus dominated by low energy particles (electrons and  positrons). They also argue that this effect  allows one to satisfy the observational constraint in the MeV range (Sec. II.B.2). However, an analysis of the M87 jet by \citet{DFC06}
concludes that the jet is e$^-$-e$^+$ dominated {\it only under the assumption of a low-energy cut-off}  E$_{min}\sim$0.5 MeV,  but higher E$_{min}$ (implying smaller pair fractions per proton) cannot be excluded and are even suggested by polarization measurements. It is clear then that neither the energies nor the abundance of positrons in the jets are known at present.

Note that in both   \citet{GJP06} and \citet{BSTM09} the bulge/disk ratio can only be reproduced if it is assumed that a fraction
of the disk positrons do not annihilate in the disk, but leave it altogether.  These arguments will be discussed in Sec. IV.D along
with the claim of \citet{Weidenspointner+08a} that their recent finding of asymmetric 511 keV emission from the inner disk (Sec. II.B.3) favours LMXRBs in the hard state as the e$^+$ sources.

\subsubsection{\label{sec:BlackHole}Positron production by the Galactic black hole} 

As already discussed in Sec. II.B, the supermassive black hole (SMBH) in the Galactic center (GC) had been
suggested already in the 1980s as a e$^+$  source, on the basis
of the variability found in the HEAO-3 data by \citet{Riegler:1981}. Variability is not an issue anymore, but the
first year data of SPI/INTEGRAL revived the idea of the SMBH as e$^+$ source because of the difficulty met by
other candidate sources to explain the 511 keV image.

Compared to the situation in the 1980s, the SMBH models have to cope with two new requirements:

i) The emission does not originate from a point source in the GC, but from a region extended over the whole bulge.
This implies that positrons from the central source have to travel distances comparable to the bulge radius, i.e.
$\sim$2 kpc.

ii) Sgr A$^*$, the multi-wavelength emission source at the GC, is notoriously weak (e.g. \onlinecite{Eckart+08}): its
X-ray luminosity is $\sim$10$^{35}$ erg s$^{-1}$ and its bolometric luminosity is estimated to 
$\sim$10$^{36}$ erg s$^{-1}$, i.e. $\sim$3  10$^{-9}$ times the corresponding Eddington luminosity\footnote{The
Eddington luminosity is the limiting value for which gravitational attraction of a point source of mass
$M$ (accreting from surrounding material) is matched by repulsive radiation pressure due to accretion luminosity $L$. 
It is given numerically by $L_{Edd}\sim$1.3 $\times$ 10$^{38}$
$(M/M_{\odot})$ erg s$^{-1}$ (assuming  spherical symmetry and Thompson scattering of radiation).}.
The X-ray emissivity of Sgr A$^*$ is $\sim$10$^4$ times weaker than the combined emissivity of the population of Galactic XRBs (see Sec.
IV.B.4) or even some individual XRBs;  if positron production is correlated to X-ray emissivity, 
it is difficult to conceive Sgr A$^*$ as important e$^+$ source.

\begin{figure}
\includegraphics[width=0.49\textwidth]{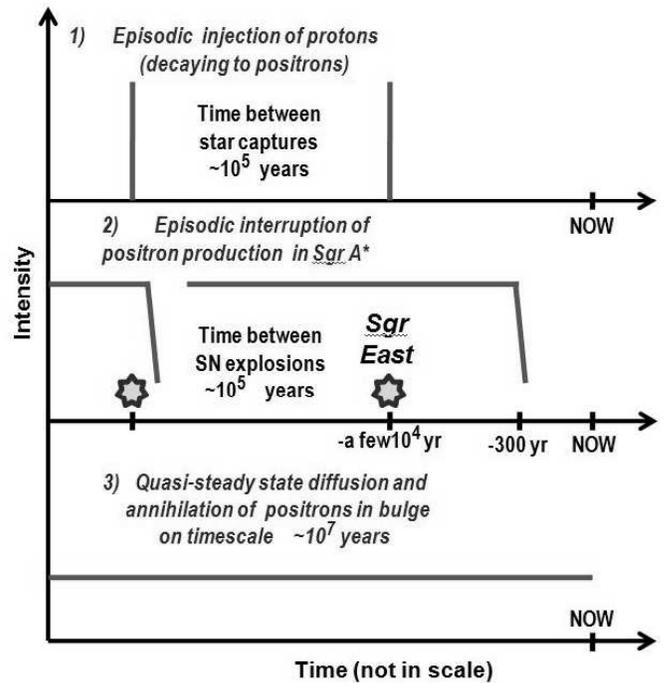}% Here is how to import EPS art
\caption{\label{fig:epsart} Illustration of the two schemes conceived for the positron production activity of Sgr A$^*$.
In (1), protons are injected/accelerated every $\sim$10$^5$ yr following  the tidal disruption of a star, and  their collisions
with the ISM produce positrons (after $\pi^+$ decay). In (2), Sgr A$^*$ produces quasi-continuously positrons at high rate,
except when its accretion flow is interrupted by the passage of the shock front of a nearby SN explosion (expected to occur
on a timescale of $\sim$10$^5$ yr); the last SN explosion created the SN remnant Sgr East a few 10$^4$ yr ago and its
expansion interrupted the accretion flow 300 yr ago (according to Totani 2006). Despite the discontinuous e$^+$ injection
in both  cases (1) and (2), the annihilation of positrons and the resulting 511 keV emission are in a quasi-steady 
steady state (3) {\it if} positrons diffuse in the bulge, because of the long timescale of the latter process ($\sim$10$^7$ yr).
}
\label{fig:SMBH_activ}
\end{figure}

\begin{figure*}
\includegraphics[width=0.99\textwidth]{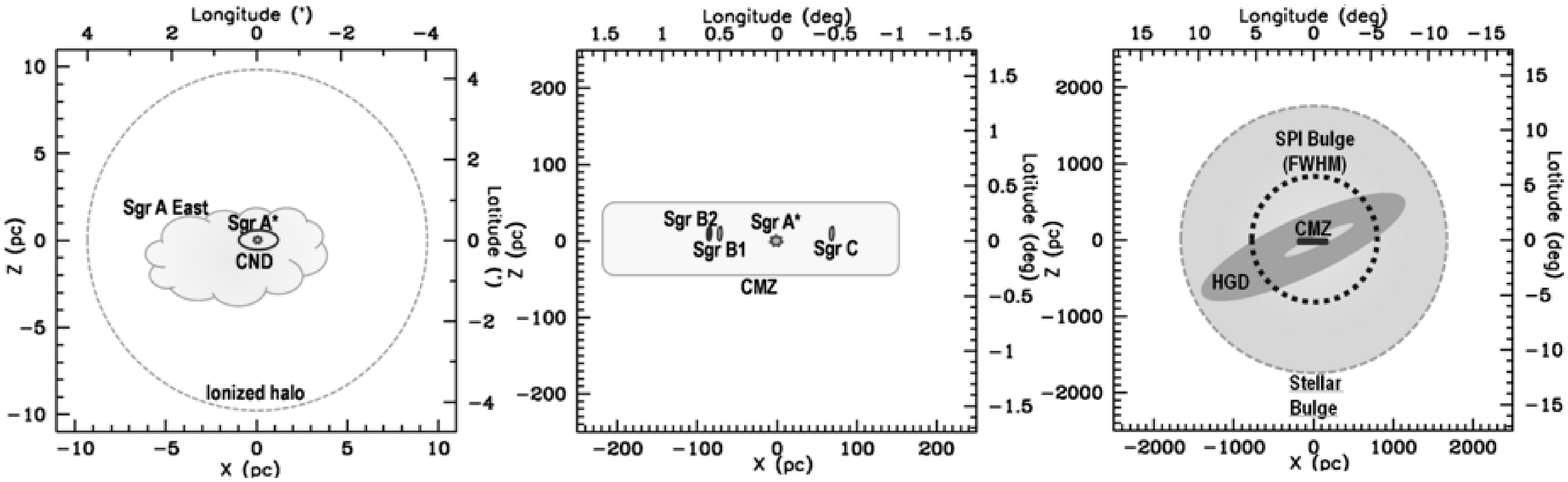}% Here is how to import EPS art
\caption{\label{fig:epsart} Schematic representation of the inner Galaxy at different linear ({\it bottom and left axis}) and angular ({\it top and right axis}) scales. The main features discussed in Sec. IV.B.4 (SMBH scenario) appear on the figures, as
e.g. CND (Circumnuclear disk, {\it left}), CMZ (Central Molecular Zone, {\it middle}) and the tilted HGD (Holed Gaseous Disk, containing atomic and molecular gas, {\it right}). If indeed Sgr A$^*$ is the main positron source, (most of) its positrons have to diffuse {\it through} the CMZ into the bulge. Notice that the
size of the CMZ corresponds to the size (FWHM) of the inner bulge in the first model of SPI data fitting (Table I). The size of
the outer SPI bulge in that same model is indicated by a {\it dotted circle} in the rightmost panel.
  }
\label{Bulge_SgrA}
\end{figure*}

It turns out that a viable solution may emerge, satisfying both constraints, if one drops the assumption of steady
state (i.e. equality between e$^+$ production and annihilation rates): since positrons need time to slow down 
{\it and} to fill the bulge, one may invoke a much higher activity of Sgr A$^*$ in the past. 
The question is then whether that high activity was due to a rare event (in which case the current
 low activity  represents the
normal state of Sgr A$^*$) or whether the past high activity was the norm (in which case today's low activity is of low probability).
Proposed models explore both possibilities (Fig.~\ref{fig:SMBH_activ}):

a) Tidal disruption of nearby stars and subsequent accretion of their material may boost the activity of Sgr A$^*$ for timescales
of 10--100 yr (\onlinecite{Rees88}). In  view of the star density in the vicinity of Sgr A$^*$ such events may occur 
every $\tau_{cap}\sim10^4-10^5$ yr (e.g. \onlinecite{SU99}). \citet{FMR01}, \citet{CCD06} and \citet{CCD07} suggest that such events are at the origin of past high activity of Sgr A$^*$.

b) Quasi steady-state accretion of surrounding gas was $\sim$10$^4$ times higher in the past, but it was interrupted by some external
factor which destroyed the accretion flow $\sim$300 years ago (\onlinecite{Totani06}). This timescale results from the X-ray emission observed from Sgr B and Sgr C (both at distances of $\sim$75 pc from Sgr A$^*$, see Fig.~\ref{Bulge_SgrA})  which has been interpreted as the
delayed reflexion of (a much stronger) past emission from Sgr A$^*$ (e.g. \onlinecite{MSP93}, \onlinecite{KHI06}, \onlinecite{Murakami+00},
\onlinecite{Revnivtsev+04}). Building on those ideas, \citet{Totani06}  suggested that 
it is the expansion of the SN remnant Sgr East which destroyed the high accretion flow.
The age of that remnant (a few $10^4$ yr) is smaller than the estimated timescale between two SN explosions in the GC vicinity 
($\sim$10$^5$ yr), but   the probability that we observed Sgr A$^*$ just 300 yr after its crossing by the expanding shell of Sgr East is rather small ($\sim$1\%,  \onlinecite{Totani06}).

It should be noted that the timescale for variability of 511 keV emission is the longer of the two timescales $\tau_{prod}$ (for
variability of e$^+$ production in the above models) and $\tau_{ann}$ (for e$^+$ slow down and annihilation). The former is $\sim$10$^5$ yr, while the latter depends strongly on the physical conditions of the ISM (Sec. V). However, {\it if} the positrons survive annihilation in the GC vicinity and manage to fill the bulge (and this is a big {\it if}), 
the corresponding diffusion timescale can be estimated by quasi-linear diffusion theory as $\tau_{diff}\sim$10$^7$ yr (e.g. Jean et al. 2006). It is this long timescale that determines variability of 511 keV emission and makes it essentially constant
(Fig.~\ref{fig:SMBH_activ}), despite the variability of the e$^+$ production rate in both cases (a) and (b).
We shall discuss the problem  of e$^+$ propagation in the bulge (a crucial issue for all models producing bulge positrons in the SMBH) in Sec. VI.

% the existence of a hot ($\sim$8 keV) and dense (10$^3$ cm$^{-3}$) 
%ionized halo within 10 pc of Sgr A$^{\star}$ seems to require much more 
%energy ($\sim$ 10$^{40}$ ergs/s) for its heating than is currently 
%available (Maeda et al. 2002), and this hot plasma -- Totani (2006) 
%suggests -- can be explained as the result of shock-heating by the wind. 
%Also, one should note that Maeda et al. (2002) have proposed that the past 
%higher activity could have been the result of the accretion from the 
%dense supernova shell as it was expanding into the ionized halo.

Production of positrons can be envisioned by either
a) collisions of protons accelerated by the SMBH with the ISM and subsequent $\pi^+$ decay
or,  b) direct pair-production (Sec.  IV.B.1). 
We  briefly present below a few  models which constitute specific realisations of the aforementioned ideas.
%We note that in all cases it has to be demonstrated that positrons from the central source can indeed
%fill most of the bulge without excessive annihilation in the dense inner zones (e.g. the Central Molecular Zone or CMZ),
%because in that case a point source would have been seen by SPI/INTEGRAL.

%the positrons being then somehow evacuated by jets (e.g. Yamasaki, 
%Takahara \& Kusunose 1999), some soft photon wind (Beloborodov 1999), 
%or any other mechanism. The existence of jets, which are often 
%invoked as necessary components of such models has been argued for by 
%Markoff \& Falcke (2003), Yuan et al. (2002 \& 2005) and Yuan (2007) 
%among others, based on similarities with M81$^{\star}$, on Chandra-observed 
%X-ray features in the vicinity of Sgr A$^{\star}$, etc.
%Let us describe the main models that have been proposed in the two 
%categories of processes. One of the on-going avenues of research here 
%is the investigation of the extent to which the positrons produced in 
%each case can propagate and at least fill the bulge region in order 
%to have a chance to reproduce the annihilation map. We shall come 
%back to this issue later and comment on it.

\medskip

\noindent{\it a. Production of positrons via p-p collisions}

\smallskip

Models of this class usually try to reproduce simultaneously observations of 511 keV emission from the bulge {\it and}
of higher energy $\gamma$-rays from the central regions. Such emission has been detected at E$_{\gamma}>$500 MeV
with EGRET/CGRO (\onlinecite{MayerHasselwander+98}) and at TeV energies with various instruments
 (\onlinecite{Aharonian+04}; \onlinecite{Aharonian+06}, and references therein). 
 %Comparing observations to theoretical predictions \citet{Aharonian+06} conclude  that the TeV emission cannot result from dark mater annihilation.

The generally large timescale for e$^+$ thermalization/annihilation (see Sec. V) creates problems to any scheme
trying to explain simultaneously current observations of 511 keV and higher energy $\gamma$-rays, {\it if}
accelerated protons are assumed to be at the origin of both emissions. This was realized in \citet{FMR01},
who explored the fate of positrons produced by energetic protons, required to explain the GeV
emission detected by EGRET from the source 2EG J1746-2852; the latter is located in the inner arcmin of the Galaxy
and its $\gamma$-ray emissivity is associated to proton acceleration in Sgr East by \citet{MYF98}.
\citet{FMR01} find that in the physical conditions of Sgr East, the e$^+$ thermalization timescale is much
longer than the age of Sgr East (or that the current e$^+$ thermalization/annihilation rate is much lower than the
e$^+$ production rate, which is inferred from the high-energy $\gamma$-ray emissivity of 2EG J1746-2852).
They conclude that recently accelerated protons in Sgr A East cannot be the  source of the observed annihilation radiation,  
unless some more efficient e$^+$ cooling mechanism is at work. Alternatively, they suggest that  more positrons 
from previous episodes of activity in the Galactic center (i.e. from tidal disruption of stars) 
have been "stockpiled" in that region and are annihilating now.
 
%Fatuzzo et al. (2001) then explore such a scenario and calculate the 
%rate of positron production, basing themselves on one of the variants 
%of the Melia et al. (1998) model. They take into account the cooling 
%of the positrons they produce (the rate and energy distribution of 
%which is given in their Figure 1), considering synchrotron, Compton 
%scattering, bremsstrahlung, and Coulomb losses. They find that 
%positrons which energies are less than 100 MeV, and majority of them 
%fulfill this condition, thermalize to the same temperature as the 
%ambient medium (on a timescale equal to the inverse of their initial 
%cooling rate) before they annihilate. They also find, however, that 
%with the Sgr A East physical conditions the current production rate 
%of positrons would be much higher than the annihilation one, 
%resulting in their ``stockpiling''. 

\citet{CCD06} explore further those ideas, but they consider in more detail the propagation of protons and positrons in the 
Galactic bulge. They find that in the case of an energetic but rare event (such as the disruption of a 50 \ms \ star, releasing
10$^{54}$ erg in energetic protons every $\sim$10$^7$ yr)  it is impossible to explain simultaneously the 511 keV and 
high energy $\gamma$-ray emissions: although positrons take a long time to diffuse and annihilate, protons interact
rapidly and the corresponding $\gamma$-ray emission fades out in less than 10$^5$ yr. In a subsequent version of that work,
\citet{CCD07} return then to the idea proposed by  \citet{FMR01}, namely that the observed 511 keV emission results from the annihilation of positrons produced and stockpiled in the bulge by dozens of tidally disrupted low mass stars in the past
$\sim$10$^7$ years: each one of those events is less energetic ($\sim$10$^{52}$ erg) but more frequent 
(corresponding timescale: $\sim$10$^5$ yr) than the rare massive star disruption. Because of the short timescale of
e$^+$ production with respect to the timescale of e$^+$ annihilation, the resulting 511 keV emission is quasi-steady
(see Fig.~\ref{fig:SMBH_activ}). Thus, although the invoked e$^+$ production event is of low probability (small frequency), 
the observed intensity of the 511 keV emission is not.

 Obviously, most of the positrons 
are produced in the high-density region where protons interact. \citet{CCD07}
argue that  positrons do not annihilate in those regions, but they avoid them 
(because of the screening effect produced by magnetohydrodynamic waves excited near 
molecular clouds) and  they propagate through the intercloud medium. 
%They further heuristically argue  that the injection of  (high energy) positrons in the low-density region 
%is equivalent to their constant injection with energies less than 
%10 MeV; they claim that this helps avoiding the emission of MeV photons (produced via direct 
%annihilation of relativistic positrons in flight), which constitutes one of the main constraints on the e$^+$ sources. 
In general, models where positrons result from the 
production and decay of $\pi^+$ will have typical  energies of 
$\sim$30~MeV, resulting in too much emission at $>$MeV energies from in-flight annihilation.  
\citet{Chernyshov+09} find that this difficulty
may be circumvented if the magnetic field in the galactic bulge is high enough ($>$0.4 mG), because in that case
positrons lose their energy rapidly (before significant in-flight annihilation) through synchrotron emission; this cooling
process takes place within the timescale of e$^+$ production, i.e. long before
e$^+$ annihilation,  and should be currently undetectable in radio. 
However, as discussed in Sec.~\ref{sec1:GalaxyMagnField}, observations do not favour, a present, such high  values for the magnetic field of the inner Galaxy.

\medskip 

\noindent{\it b. e$^+$-e$^-$ pair production by photons}

\smallskip

As in the case of XRBs, pair production around the Galactic supermassive black hole may occur either in the inner hot accretion disk,
in a corona above it or in the jets; the latter case has not been considered up to now, in the absence of relevant observational
evidence in Sgr A$^*$.

\citet{Beloborodov99} studied a detailed model of pair production, resulting from collisions of
$\gamma$-rays from the hot inner disk with X-rays from the outer disk; the resulting e$^-$-e$^+$ pairs are blown outwards
from the radiation pressure of the disk, in a mildly relativistic wind. The model is not specifically designed for the case of 
Sgr A$^*$ but rather for extragalactic black holes in Active Galactic Nuclei (AGN), but its results can be extrapolated to the case
of Sgr A$^*$. The maximum pair production  is obtained for a dense, optically thick, wind. In that case positrons annihilate
mostly {\it inside} the wind flow and produce a very broad annihilation line, unlike the one observed with SPI/INTEGRAL.
Positrons can escape and annihilate in the ISM in the case of a less dense,  optically thin, wind but in that case their production rate is substantially smaller.

Totani (2006) considered pair production in the hot, inner accretion disk during past phases of higher activity in Sgr A$^*$.
The invoked past accretion rate, $\dot{m}\sim$10$^{-4}$\ms/yr, is not extravagant and could easily result from the material
released by the regular tidal disruption of nearby low mass stars (as in Gheng et al. 2007) or from winds of nearby massive stars \citep{Quataert04}, although Totani (2006) assumes that it originates from the
ionized "halo" surrounding Sgr A$^*$ (see Fig. \ref{Bulge_SgrA}, left). Totani (2006)  considers
pair production in the framework of the so-called Radiatively Inefficient Advection Flow (RIAF) model for
accretion disks; this model, decoupling accretion from emerging luminosity, has been applied with considerable success to the case of Sgr A$^*$  (e.g. \onlinecite{YQN04}; \onlinecite{Xu+06}). Totani (2006) finds that
the very high temperatures of the inner disk (T$\sim$10$^{11}$ K) implied by  the RIAF models are essential for a high
rate of e$^+$ production, which he evaluates up to $\sim$10$^{43}$ e$^+$ s$^{-1}$, i.e. close to
the observationally inferred annihilation rate. 

In all models of e$^+$ production from Sgr A$^*$,
e$^+$ annihilation occurs on much longer timescales than e$^+$ production and varies much less in time than the latter.
An advantage of direct pair production models with respect to those involving energetic proton collisions is
the low energy of the positrons produced, allowing them to satisfy the constraint of the observed MeV continuum.

In all cases, it has to be demonstrated that positrons may diffuse from Sgr A$^*$ throughout the bulge without
excessive annihilation in the dense inner regions (the Circumnuclear Disk), which would give a strong, point-like emission.
On the other hand, the latest analysis of SPI data suggests a narrow bulge component (Table I), the size of which
(3\degree ~FWHM) corresponds to the size of the CMZ: $\sim$1/3 of the bulge positrons may annihilate there and the remaining 2/3 may diffuse in the outer bulge (11\degree FWHM in Table I, see also Fig.~\ref{Bulge_SgrA}). This picture may have difficulties, however,
with the results of the spectroscopic analysis of SPI (Sec. V.E) which finds that bulge positrons annihilate mostly in a
warm medium (neutral or ionized), not in a molecular one.

%%%%%%%%%%%%%%%%%%%%%%%%%%%%%%%%%
%\input{Celine4.tex}
%%%%%%%%%%%%%%%%%%%%%%%%%%%%%%%%%

\subsection{Dark matter and "non-standard" models}
\label{sec1:DM}

\subsubsection{General properties of dark matter particles}
\label{sec:dmg}

%Astrophysical evidence on galactic and cosmological scales suggests that, while gravitating matter
%amounts for $\sim$28\% of the critical density of the Universe, only $\sim$4\% of this density corresponds to the common,
%baryonic, form \citep{Bartelmann09}. The remaining $\sim$24\% would correspond to a new type of matter, called dark matter because
%it does not emit visible light.  

In the past 40 years or so, particle physicists searched for possible dark matter (DM) candidates
meeting  three basic requirements,  namely: stability (at least on timescales comparable to the age of the Universe), charge neutrality (to avoid electromagnetic - ELM - interactions and prevent DM to shine) 
and with a non negligible mass (so that it can contribute gravitationally)\footnote{These criteria are in fact supported by recent observations (assuming a Friedman-Robertson-Walker Universe). For example, extrapolation of the physics of ordinary baryons to DM suggests  
that DM ELM interactions should  damp the DM primordial fluctuations on a cosmological scale, 
and prevent the formation  of small scale structures (smaller than a Milky Way size galaxy).  
%However, the existence of small size galaxies 
%by 2dF and SDSS ($109 M_{\odot}$) 
%and Lyman-$\alpha$ clouds ($10^6 M_{\odot}$) 
%by weak lensing experiments now 
%shows  that DM cannot be electrically charged.
}.
The absence of electric charge favours DM models with weak  interactions. 
However weakly interacting particles may 
%also experience damping. Indeed they may 
suffer from a prohibitive "free-streaming" effect\footnote{Free streaming refers to the motion of non-interacting particles endowed with
some initial velocity across the Universe. It has the effect of erasing  irregularities on scales smaller than the free-streaming length.
}, depending on their mass: for example, if DM is composed of massless neutrinos, the formation of Milky Way size galaxies is strongly suppressed. Hence there is a lower limit on the mass of weakly interacting DM candidates to explain the formation of the smallest ojects  observed in the Universe, which is of $\sim$a few keV. This leads to the notion of {\it weakly interacting massive particles} (WIMPs) and the idea of {\it collisionless dark matter}. The existence of WIMPs with a mass in the GeV-TeV range is  compatible with the absence of signal in present DM direct detection experiments. However, the fact that none of these experiments has observed a positive signal yet \citep{Baudis07}\footnote{Apart, perhaps,  from DAMA/LIBRA,  \citep{Bernabei:2003wy}.} may lead to different interpretations, as discussed below.

%Although dark matter may be weakly interacting, it is likely to have non vanishing interactions. These are expected to play a crucial role and explain the dark matter disappearance, despite the symmetries which may prevent it to decay quickly.
%The dark matter relic density as measured by WMAP indicates that the  number density $n_{\rm{dm}}$ of DM particles 
%is heavily suppressed with respect to the number density of relativistic particles: indeed, i
If the DM number density today $n_{dm}$ were similar to the relativistic particle density  
($n_{\gamma}  \sim$ 400 cm$^{-3}$ for photons), the DM mass-energy density $\rho_{dm} \simeq m_{dm} n_{dm}$ would exceed the critical density $\rho_c$ by several orders of magnitude, for DM particle masses $m_{dm} >$1 keV.
% thus leading to a DM cosmological parameter $\Omega_{dm} \gg 1$. 
Given that particles lighter than keV are forbidden by the free-streaming argument, DM particles which have been in thermal equilibrium with radiation at some stage of the cosmic evolution should subsequently disappear. This may occur in two ways: either through an extremely small decay rate, which ensures a lifetime  comparable to the age of the Universe (and thus implies quasi-stability) or through annihilation processes. 

The latter mechanism has  received a lot of attention in the last three decades.
% because it enables to make predictions.  
The requirement  $\rho_{dm} \sim \rho_c$  implies that the annihilation cross-section of DM
particles should be comparable to the weak interaction cross section,  see  e.g. \citet{1977PhLA...69...85H},
\citet{1980BAAS...12..861P}, \citet{1983PhLB..120..133A}
\footnote{The abundance of DM particles today $\Omega_X$ is fixed in the
so-called "freeze-out" epoch, when the expansion rate of the Universe $H$ (a function of $\Omega_X$) equals the DM annihilation rate
$\Gamma=n_X <\sigma v>$, where $n_X$ is the DM particle abundance and $<\sigma v>$ their annihilation cross-section; this leads to a "relic" abundance of DM particles of $\Omega_X=f(<\sigma v>)$ and for the observed $\Omega_X\sim$0.25 one obtains $<\sigma v> \sim$ 10$^{-26}$ cm$^3$ s$^{-1}$, which is close to the value of weak interaction cross sections (the actual value depends  on the nature of the DM candidate and its mass, see e.g. \onlinecite{BHS04}).}.
In some cases, this requirement can also be used to constrain the DM mass.
As was pointed out by \citet{Lee:1977ua} and \citet{1977PhLA...69...85H}, the annihilation rate ($<\sigma v> \propto v^2$) of
fermionic particles with typical weak interactions has a square dependence on DM mass $m_{dm}^2$.  
The observed $\rho_{dm}$  implies then that DM particles should be heavier than a few GeV. 
This constitutes an extra motivation for considering WIMPs as DM candidates.

About twenty five years ago, such properties (weak but non negligible interactions and $m_{dm}>$1 GeV)
suggested that  direct detection of DM would be relatively easy. However:
%the absence of a positive signal in direct or indirect experiments up to now opens up new possibilities. Indeed: 

1) The DM spin independent interactions with matter  (as measured by direct detection experiments) 
is at least eight orders of magnitude weaker than the weak interactions (\onlinecite{Lemrani:2006ec,Angle:2007uj}) 

2) Indirect detection experiments find no "smoking gun "evidence (i.e. the emission of a monochromatic line at an energy $E = m_{dm}$)  allowing for a clear identification of $m_{dm}$ \cite{Abdo:2009zk,Aharonian:2009ah}.

3) No signature of new physics, which would indirectly validate the existence of DM particles, was found
at LEP or TEVATRON \cite{Piper:2009ne,Abbiendi:1998ea}.

The aforementioned facts could either mean that DM has much weaker interactions than the Standard Model or point towards very heavy or very light DM particles\footnote{If DM is lighter than a few GeV, its interaction with matter would be essentially "invisible" for current detectors (MeV particles would require for example detectors with eV energy threshold, while they are currently in the keV range). If DM particles are too heavy, their number density in our Galaxy is  too small to generate a significant number of events in a detector.}.
Besides, the absence of ELM interactions could also imply strong interactions but at different energy scale than  previously considered \cite{Boehm:1900zz}.

The aforementionned constraints, individually taken, 
can be easily accommodated in existing DM models (e.g. WIMPs).
However, when combined together, they actually eliminate many proposed models.
For example, to reduce the tension between direct detection experiments, which are now
sensitive to elastic cross sections of the order of $10^{-43}
\rm{cm^2}$ (corresponding to $<\sigma v> \sim$ 10$^{-33}$ cm$^3$ s$^{-1}$),
and the annihilation rate $\sigma v \sim$ 3 \ 10$^{-26}$ cm$^3$ s$^{-1}$ 
imposed by the "relic density" criterion (corresponding to $\sigma \sim$ 10$^{-36}$ cm$^2$),
one often has to ``decouple'' the corresponding processes \cite{Ellis:2000ds}\footnote{
This decoupling can be made by invoking either
"co-annihilation", i.e.  annihilation of DM  with another particle, present during
the dark matter transition  to the non-relativistc regime \cite{Griest:1990kh}, or
by fine tuning the DM parameters so that
the annihilation cross section is enhanced but the elastic scattering cross section remains very small.}.
%(in supersymmetry, the latest possibility happens in the so-called the focus point or Higgs pole regions \cite{focuspoints,higgspole}).
The lack of evidence could also mean that $m_{dm}<<$1 GeV  or  $m_{dm}>>$1 TeV.
In the case of sub-GeV DM,  one may naturally circumvent  the Hut-Lee-Weinberg limit 
if DM has been produced out of thermal
equilibrium or if it is made of scalar particles with non chiral couplings to Standard Model particles. Non-thermal production also
helps very heavy DM particles to avoid conflict with the relic density criterion. 
Thus,  the acceptable range for $m_{dm}$ still lies from the sub-eV range (with axions) 
to several TeV  (such as ``excited'' DM, Kaluza-Klein particles) and, in fact in some case, to even higher masses.
Hence, keV, MeV and sub TeV candidates (such as, respectively, sterile neutrinos, light dark matter, and neutralinos)
remain potential solutions (see, e.g. \onlinecite{Bertone07}).

Progress in the  field of DM may come from indirect detection.
Indeed, cosmic ray spectra ``anomalies'' (with respect to ``standard'' astrophysical predictions)
appear puzzling enough  \citep{Picozza:2006nm} to open up new particle physics scenarios
\cite{Adriani2009,2009PhRvL.102e1101A,2008Natur.456..362C}. 
For example, 
recent results from the PAMELA satellite indicate an excess of CR positrons above 10 GeV over the background expected from CR interactions with interstellar matter \citep{MS1998}, but no
corresponding excess in antiprotons\footnote{The balloon-borne experiment ATIC reported an excess of 
electrons plus positrons in the 300-800 GeV range, but the excess was not confirmed
by Fermi-LAT \citep{Abdo:2009zk} and
by HESS \citep{Aharonian:2009ah}}. However, the all lepton
spectrum as measured by the Fermi-LAT came out flatter than previously
thought \citep{Abdo:2009zk}. The sources of these particles (or electrons and positrons)
should be less  than $\sim$1 kpc away,
since electrons and positrons in this energy range suffer heavy energy 
losses through inverse Compton and synchrotron
processes. Conventional astrophysical sources (e.g. nearby pulsars, see Sec. IV.B.3) 
could explain these excesses, but the
possibility of the first indirect detection of DM annihilation created a lot of excitement in the particle physics community (see \onlinecite{ESS09} and references therein).
%\footnote{Particle physics enhancement mechanisms  are required to explain the
%positron excess and the lack of anomaly in anti proton through DM. Among these enhancement mechanisms there is a resonance effect, which only happens for a very specific value of the DM mass and which is called the "Sommerfeld effect".}
%This also assumes that DM would be rather lepto-philic 
%\cite{cirelli,weiner,boehm}.} 
%(although strong interactions in the quark sector would not create more anti protons \cite{boehm}).}

The recent excitement illustrates the new trends in the DM particle physics community and shows how far we are from the determination of the nature of DM.  This also demonstrates that considering only neutralinos or Kaluza-Klein particles as DM may be too restrictive.
In this review, we shall focus only on the DM candidates which have been explicitly invoked to explain
the Galactic 511 keV emission.

\subsubsection{Specific dark matter candidates for e$^+$ production }
\label{sec:dmm}

Positrons produced by DM annihilation or decay may annihilate {\it in flight}, before losing a large fraction of their energy. 
Inflight e$^+$ annihilation would provide an additional source of continuum $\gamma$-ray emission towards the Galactic centre (Sec II.C.1) which can be used to constrain the e $^+$ energy at injection and, consequently, the mass of annihilating or decaying DM particles.
Assuming that the
contribution of known astrophysical sources to the
measured continuum in this energy range is well understood, \citet{Beacom-Yuksel:2006} and
\citet{Sizun:2006} obtained a mass upper limit of $m_{dm}\sim$ a few MeV
(see Fig.~\ref{fig:sizun}). 
Among  the many proposed DM scenarii, those  that may  satisfy such a constraint
can be classified in two main categories:

i) Light DM particles of $\sim$MeV mass, either
{\it annihilating} \cite{511,Gunion:2005rw}, or
{\it decaying} \cite{Hooper:2004qf,Picciotto:2004rp} %   Conlon:2007gk,Cembranos:2008bw,Khalil:2008kp}
or even both \cite{Pospelov:2007mp}.

ii) Heavy DM particles in the $\sim$GeV - TeV range, {\it de-exciting} (or decaying into another particle) with a mass
difference of a few MeV between initial and final states (\onlinecite{Finkbeiner:2007kk};
\onlinecite{Pospelov:2007xh}).

Other, more intricate possibilities, involving dark matter, cosmic strings, primordial black holes and
other exotica, will be briefly presented in Sec.~\ref{subsec:Exotica}.

\begin{figure}
\centering
\includegraphics{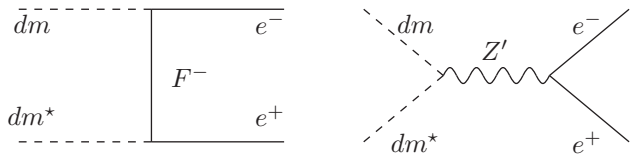}
\caption{Possible Feynman diagrams for light annihilating dark matter particles.}
\label{fig:Fey}
\end{figure}

Low mass {\it annihilating} DM particles were initially proposed
to illustrate a new damping effect
\footnote{The mixed damping effect is in fact analogous to the "Silk damping", with
dark matter playing the role of baryons and neutrinos replacing the photons \cite{Boehm:2000gq,bs}.}
but also as a counter example of the "Hut-Lee-Weinberg limit": \citet{bens} pointed out and
\citet{bf} showed that this limit is only valid in
the case of fermionic DM candidates interacting with Fermi (i.e. weak)
interactions. But, if one or both of those assumptions are relaxed, a very different conclusion may be obtained:
for instance, if the DM particle  is a scalar (spin-0) which
annihilates into a e$^-$-e$^+$ pair  via the exchange of a
fermionic particle $F$, then the relic density criterion constrains
the characteristics (mass and couplings) of the $F$ particle
instead of the  mass of the DM particle.  Particles substantially
lighter than a few GeV (and down to the MeV range)  may then account for the
observed dark matter relic density in that case.
Such particles are expected to annihilate into e$^-$-e$^+$  pairs (Fig.~\ref{fig:Fey}) either via a
heavy charged particle exchange or a new neutral gauge boson
\cite{bf,bens}.   \citet{boehmascasibar} and \citet{boehmsilk}
found that the properties of such particles, if they are at the origin of the 511 keV line, should
affect the value of the fine structure constant. In fact, using this very argument, \citet{boehmascasibar} and \citet{boehmsilk}
could exclude DM particles heavier than 7 MeV (assuming a NFW dark matter halo and the corresponding best fit cross section).

The light DM particle idea, at least in it simplest form, has been challenged
by an analysis of the explosion of supernova SN1987A \citep{FHS06},
which puts a lower limit of $\sim$10 MeV to the mass of the
particle. 
%assuming however a different model than that which provides the best fit to the data and
%the same couplings to quarks as that for the electrons. 
Such a limit is very close to the upper limit allowed by the observed $\sim$MeV continuum.
%  and may indicates, in principle, some tension.
% subsequent work showed, however, that all available constraints can still be accomodated by the model.

Low mass ($<$100 MeV) {\it decaying} DM candidates were proposed by several groups,
after the release of the 511 keV map by the SPI collaboration, in order
to explain the large amount of low energy positrons  in the Galactic bulge.
%(see \onlinecite{Pospelov:2007xh}, and references therein).
Decay into e$^+$-e$^-$ pairs would be one of the dominant decay modes
of such particles since,  %with a decay rate of about $10^{-26} \ s^{-1}$
apart from the neutrino and photon channels, electrons would be the only
other kinematically accessible channel. Depending on the model, these particles
may (the axinos in \onlinecite{Hooper:2004qf}) or may not (sterile neutrinos in \onlinecite{Picciotto:2004rp}, moduli
in  \onlinecite{Kasuya:2006kj} or in \onlinecite{Craig:2009zv})
be the major contributors to cosmic DM density.

The second category of DM particle candidates   invokes electroweak scale WIMPs, with masses
in the 100 GeV - 1 TeV range and possessing almost mass degenerate excited states, i.e. the difference
between the excited and ground states should be of the order of $\sim$MeV. 
\citet{Finkbeiner:2007kk} noticed that
the velocity dispersion of DM particles in the gravitational potential of the inner Galaxy is of the order of a few
100 km s$^{-1}$, endowing a 500 GeV WIMP with  a kinetic energy $>$511 keV. Inelastic scattering between WIMPs could
raise one or both of them in their excited state(s) and de-excitation to the ground state could proceed via emission
of a e$^-$-e$^+$ pair.  This scenario has also been invoked to  explain other
observables such as the DAMA/LIBRA signal or the WMAP "haze"\footnote{Observations with the Wilkinson Microwave Anistropy Probe (WMAP)
revealed an excess of microwave emission in the inner 20\degree of the Galaxy, which cannot be accounted for by conventional
astrophysical explanations, such as thermal Bremsstrahlung from hot gas, synchrotron emission, etc.} (\onlinecite{HFD07}, \onlinecite{FPW08})
and the PAMELA e$^+$ excess \cite{ArkaniHamed:2008qp}. A recent investigation of this idea suggests, however, that the model parameters
have to be pushed to their extreme values for the e$^+$ production rate to agree with observations \citep{Chen+09}.

%The electrons and positrons produced in those  scenarios  are predicted to have an
%initial (injection) energy of about $E = m_{dm}/2$ (for annihilating
%candidates), $E= m_{dm}$ (for decaying particles) or $E = m_{dm,2} -
%m_{dm,1}$ (for excited dark matter). This injection energy should be
%smaller than a few MeV in oredr to avoid overproduction of the  X
%and MeV $\gamma$-ray backgrounds
%\cite{Beacom-Yuksel:2006,Cordier:2004hf}.

Independently of their "naturalness" (or lack of) as extensions of the standard model, the various proposed
scenarios for DM particles at the origin of Galactic 511 keV line also differ  as to the predicted
spatial profile of the resulting e$^+$  population. Assuming that positrons annihilate close to their production sites,
this may constrain and discriminate between the various models.

%\subsubsection{Spatial profiles of dark matter positrons }
%\label{sec:dme}

The rate of positrons produced locally by annihilation/decay/de-excitation of DM particles is given by:
$\dot{n}_{e^+}(r) \propto$  $\Gamma_{X \longrightarrow e^-e^+} \ n_X(r)$, where
$n_X(r)$ is the number density of DM particles at distance $r$ from the Galactic center\footnote{Generally,
$n_X(r)$ is assumed to be spherically symmetric; deviations
from spherical symmetry, due to triaxiality of the DM halo, are negligible with respect to other uncertainties of the problem.}.
The interaction rate  $\Gamma_{X \longrightarrow e^-e^+}$
in the case of decaying DM particles has a maximum value corresponding to the inverse
of the age of the Universe. In the case of annihilating or collisionally excited DM particles,
$\Gamma_{X \longrightarrow e^-e^+}\ = \ <\sigma v> n_X$, where  $<\sigma v>$ is the annihilation or excitation
cross-section folded with the velocity distribution of the DM particles.

In the case of decaying DM particles, $\dot{n}(r)_{e^+} \propto \ n(r)_X$, i.e. the positron production profile
follows closely the DM density profile; notice that this also the case for all "conventional"
astrophysical sources studied in the previous sections. On the other hand, in the case of annihilating or de-excited
DM,  $\dot{n}_{e^+}(r) \propto <\sigma v> n_X^2(r)$, i.e. the positron profile is generically more centrally concentrated (because
of the  $n_X^2(r)$ term), but this shape can be also modulated by the possible velocity dependence of the interaction cross-section
$\sigma(v)$, since the typical particle velocities depend on the gravitational potential $\Phi(r)$ of the DM halo.
In the latter case, comparison of the model to the data requires a more elaborate analysis;
an example of such an analysis will be given in Sec. IV.D.3.

\subsubsection{Other "exotica"}
\label{subsec:Exotica}

Several rather "exotic" objects have been
invoked as sources of the Galactic positrons producing the 511 keV
emisssion. Here we provide a non-exhaustive liste of them.

For example,  \citet{Huh:2007zw} suggested 
%that the 511 keV line could originate from 
MeV milli-charged (fermionic) particles. As in the first version of MeV annihilating dark matter, this scenario assumes a new light ($U(1)$) boson exchange, called exphoton. Due to the presence of kinetic terms, the dark matter would be millicharged, while the rest of the scenario ressembles the fermionic candidates introduced in \cite{bf}).  
%The claim is that such kind of scenario can be illustrated in string theory and 
The required parameter range to explain the 511 keV line appears to be 
compatible with  the constraints from the relic density requirement and collider experiments.

The idea  of  Q-balls in gauge mediated supersymmetry breaking scenarios was proposed in \citet{Kasuya:2005ay}. 
Q balls can be depicted as  stable localized field configurations, their stability being
 guaranteed by a conserved charge $Q$ associated to a
$U(1)$ symmetry. For example, $Q$  could be the electric charge; in \citet{Kasuya:2005ay}, $Q$ 
is in fact the lepton number. These objects may have a long enough lifetime and yet a small energy
density, possibly  enabling them to be present in our Galaxy and to explain the 511 keV emission.

Macroscopic objects, like superconducting dark matter
\cite{Oaknin:2004mn} or compact composite objects
\cite{Zhitnitsky:2006vt} were also proposed. These objects are hypohesized  to
form during the QCD phase transition and could be schematically
depicted as "quark"-balls. They would  introduce a link between the
dark matter and baryonic energy densities nowadays and eventually
explain why these two quantities are of the same order of
magnitude. Positronium formation would happen through  electrons or baryonic matter
 interactions (annihilations) inside the compact composite objects.
 However, the ability of such a scenario to explain the 511 keV emission
has been criticized in \citet{Cumberbatch:2006bj}, who find that positronium formation
is hardly possible at all in such objects.

If a tangle of light superconducting strings exist in the Milky Way, it may also 
act as a low energy positron source \cite{Ferrer:2005xva,Ferrer:2006pf}. 
If the string curvature radius is smaller than a 
characteristic scale, the string can move with respect to the magnetized plasma in the galaxy. 
The possible interaction of the string with the magnetic field can then generate a current 
composed of zero modes of charged particles. Owing to this mechanism, positrons could propagate along the string.  
They eventually leave the string if they become non relativistic and scatter with counter-propagating particles 
in the string. Depending on the string curvature radius, the positrons would then have an energy of  $E < \rm{MeV}$,
 although this energy could reach the GeV scale if there are superconducting strings at the TeV scale.

Finally, Titarchuk \& Chardonnet (2006) proposed that
X-rays from the SMBH collide with 10-MeV $\gamma$-rays from 
small-mass black holes (10$^{17}$ g) to give pairs; 
this can produce about 10$^{42}$ e$^+$/s, about an order of magnitude less 
than needed. This rate is obtained by taking the total X-ray and 
$\gamma$-ray luminosity of the inner GC regions: $L_{X} \sim 2 \
10^{39}$ ergs/s and $L_{\gamma} \sim 4  \ 10^{38}$ ergs/s 
\citep{2000ApJ...537..763S}, then assuming the 
gamma radiation comes from an optically thick medium and that its 
spectrum is therefore a blackbody one with a temperature of T$_{\gamma}$ 
= 10~MeV; the two energy distributions (X and $\gamma$) are 
convolved to compute the pair production rate. A simpler scenario, involving evaporating
primordial black holes was proposed by \citet{BDP08}.

\subsection{Assessment of sources }

In this subsection we summarize the pros and cons of each one of the candidate positron sources presented so far, in
the light of the observational constraints of Sec. II.B, namely: i) the total e$^+$ annihilation rate ($\gsim$2  10$^{43}$ s$^{-1}$),
ii) the typical energy of the injected positrons, or the equivalent mass of annihilation DM particles ($<$3--7 MeV) and
(perhaps, most significantly) iii) the morphology of the 511 keV emission (with a bulge/disk ratio B/D$>$1 in the case 
of a thin disk emission). A fourth constraint, namely the longitudinally asymmetric disk emission,
should be added to this list, once robustly established  by further data and analysis. 

\subsubsection{Positron annihilation rate}

Assuming a steady state regime, the e$^+$ annihilation rate has to be  equal to the {\it average {\rm e}$^+$ production rate  during the lifetime of {\rm e}$^+$ in the ISM}. 

The only source definitely known to provide substantial amounts of e$^+$ at a well constrained rate is the radioactive decay of $^{26}$Al: 0.4 10$^{43}$ e$^+$ s$^{-1}$. 
The decay of $^{44}$Ti probably provides another 0.3 10$^{43}$e$^+$  s$^{-1}$. 
GCRs probably provide 0.1 10$^{43}$ e$^+$ s$^{-1}$. Nova models (as constrained against several observables such as ejecta abundances, velocities etc.) may provide a  e$^+$ yield from $^{22}$Na decay not be much below the reported value of 10$^{41}$ e$^+$ s$^{-1}$. The e$^+$ of all other candidate sources is entirely speculative at present. 
Values discussed in previous sections should be considered as optimistic rather than typical values. 
Observed upper limits of individual sources (see Table 4 in \onlinecite{Knodlseder:2005} and Table 1 in \onlinecite{GJP06}) are of little help
 to constrain positron sources. No useful observational constraints exist up to now on the e$^+$ yields of hypernovae/GRBs, pulsars, ms pulsars, magnetars, LMXRBs, microquasars, the SMBH at the Galactic center or dark matter annihilation. 
%each of those {\it may} add some fraction to above values, thus those are constrained from below by these sources.
SNIa remain an intriguing, but  serious candidate, with a potential Galactic yield of  2~10$^{43}$ e$^+$ s$^{-1}$.

%The case of SNIa (once the most serious candidate e$^+$ sources) is quite intriguing in that respect. Both the typical  $^{56}$Ni yield of a SNIa and the Galactic SNIa rate are rather well constrained, resulting in 5 10$^{44}$ e$^+$ s$^{-1}$ produced {\it inside} SNIa. If only f$_{esc}\sim$4\% of them escape the supernova to annihilate in the ISM, the observed e$^+$ annihilation rate can be readily explained. Recent detailed observations of three SNIa, interpreted in the framework of 1-D (stratified) models, suggest that f$_{esc}\sim$0 {\it at late times}. 
%However, the late lightcurves ay be (at least partially) powered by electrons from $^{57}$Co decay; moreover, 
%both observations of early spectra and 3-D models of SNIa suggest that a sizeable fraction of 
%$^{56}$Ni  is found at high velocity (close to the surface), making easier the escape of the $^{56}$Co positrons. For that reason, we feel that 
%SNIa remain a serious candidate, with a potential Galactic yield of  2  10$^{43}$ e$^+$ s$^{-1}$.
% Nikos: Why do we need to repeat those "details"again here, diluting the nice summary above; it has been said. ROD 18Oct09

\subsubsection{Positron energy}

Radioactive decay produces positrons of E$\leq$1 MeV, naturally fulfilling the observational constraint on continuum $\gamma$ rays from in flight annihilation. The same applies to pair creation through $\gamma-\gamma$ collisions in the inner accretion disk or at the base of the jets of LMXRBs, microquasars
and the SMBH at the Galactic center. Conversely, pair creation involving very high energy photons, as in e.g. pulsars or magnetars, will produce positrons of too high energy. The same holds for energetic p-p collisions
in Galactic cosmic rays or in the baryonic jets of LMXRBs, microquasars and the Galactic SMBH. Those processes produce
e$^+$ of $>$30 MeV, thus may be discarded as major e$^+$ sources in the Milky Way. 
Also, that same constraint  limits the mass of putative decaying or annihilating DM particles to $<$10 MeV, while it does 
not constrain the mass of de-exciting DM particles.

\begin{figure}
\includegraphics[width=0.49\textwidth]{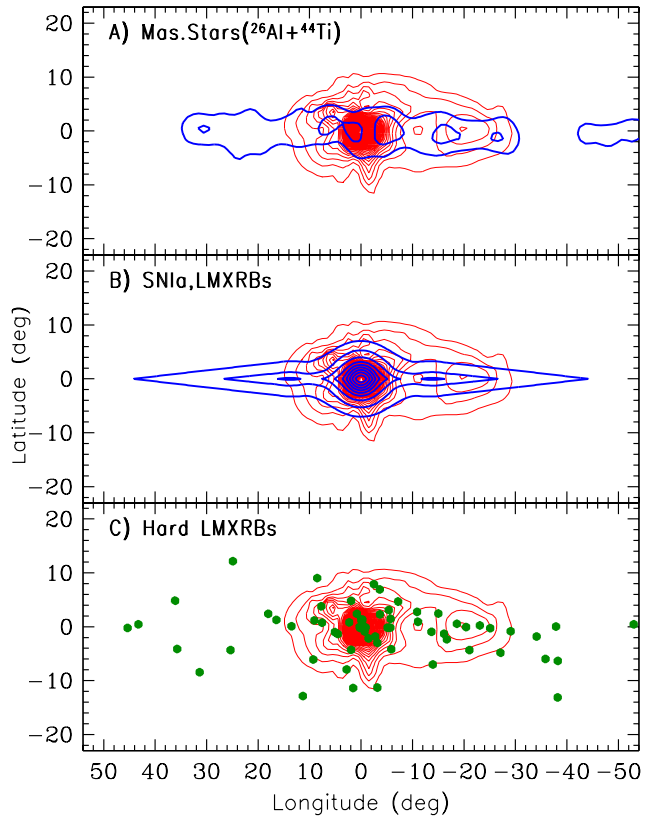}% Here is how to import EPS art
\caption{\label{fig:epsart} Maps of the Galactic 511 keV emission (flux in cm$^{-2}$ s$^{-1}$ sterad$^{-1}$), 
as observed from SPI (in all panels, {\it thin isocontours}
from \onlinecite{Weidenspointner+08a}) and from observationally based or theoretical estimates. A) Observed 
\Al \ (and, presumaby,  $^{44}$Ti) map (from \onlinecite{Pluschke+01} ) ; 
B) Accreting binary systems (SNIa and, presumably, LMXRBs, see text); C) 
Observed Hard LMXRBs (from \onlinecite{Bird+07}). The robustly expected e$^+$ annihilation from radioactivity in the disk 
(upper panel) is not yet fully  seen by SPI.
}
\label{Flux_Image}
\end{figure}

\begin{figure}
\includegraphics[width=0.49\textwidth]{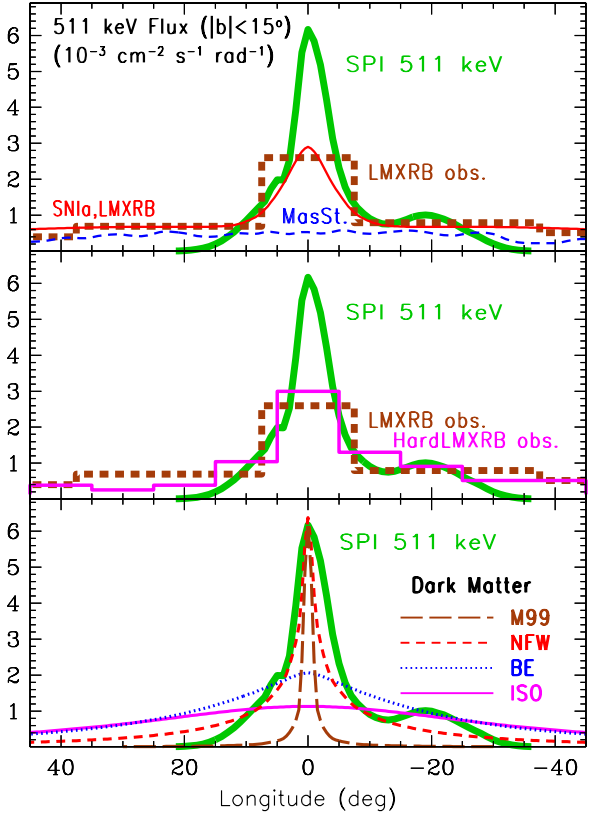} % Here is how to import EPS art
\caption{\label{fig:epsart} Intensity of 511 keV emission as a function of Galactic longitude.
All fluxes  are integrated for latitudes $|b| <$15$^o$. In all panels, the {\it thick solid curve}
corresponds to  SPI observations, i.e. the map of Fig.~\ref{Flux_Image}
({\it Note}: We emphasize that SPI maps and fluxes are
provided here for illustration purposes only; quantitative comparison of model predictions to data should only be made through
convolution with SPI response matrix.). 
The {\it thick dotted histogram} ({\it top} and {\it middle}) is the observed longitude distribution of LMXRBs (from Grimm et al. 2002);  the latter resembles  closely the theoretically estimated  longitude distribution of SNIa
({\it thin solid curve} in the {\it upper panel}), which  has been normalised to
a total emissivity of 1.6 10$^{43}$ e$^+$ s$^{-1}$, with Bulge/Disk=0.45 (maximum Bulge/Disk ratio for SNIa from Table~\ref{tab:SNrates}). Also, in the upper panel, the
{\it lower dashed  curve} 
corresponds to the expected contribution of the $^{26}$Al and $^{44}$Ti $\beta^+$-decay from massive stars. 
The {\it thin solid histogram} in the {\it middle panel} is the observed longitude distribution of Hard LMXRBs (from 
\onlinecite{Bird+07}) and it has the same normalization as the thick histogram.
In the {\it bottom} panel, the SPI 511 keV profile is compared to profiles expected from dark matter
annihilation (after Table~\ref{tab:DMHaloProfile}). 
%Notice that for {\it divergent} DM density profiles ($\gamma\geq$1) the resulting flux
%profiles depend strongly on the adopted inner cut-off $r_{in}$ in radius; $r_{in}$=50 pc is adopted here. 
%making the NFW profile to fit better the SPI observations. However, for higher $r_{in}$ values, the M99 profile may fit better, 
%whereas for lower values {\it none} of the divergent pofiles provides a satisfactory fit.}
}
\label{Flux_Lon}
\end{figure}

\subsubsection{Morphology}

None of the e$^+$ sources studied in this section reproduces the large B/D$\geq$1 ratio inferred from SPI data, as can be seen in Figs. \ref{Flux_Image} and \ref{Flux_Lon}, where
we present flux skymaps and longitude profiles, respectively, 
comparing the SPI data of  \citet{Weidenspointner+08a} to various expected  source profiles,
either theoretical or observed ones. The comparison is made under the explicit
assumption that positrons produced from the corresponding sources annihilate close
to them. 

The best-established
e$^+$ sources, $\beta^+$-decay from $^{26}$Al and $^{44}$Ti produced in massive stars, yield B/D$\leq$0.2, as derived from the observed distribution of the 1.8 MeV line
%. The corresponding skymap %\footnote{Photon flux (in photons cm$^{-2}$ s$^{-1}$ sterad$^{-1}$) as a function of Galactic longitude $l$ and latitude $b$  is calculated by
%$$ dF(l,b)=\frac{1}{4 \pi} \int_0^{\infty} \rho^k(l,b,S) \ dS \ dsinb \ dl$$
%where $S$ is the distance along the line of sight. The heliocentric coordinates ($l,b,S$) are related to Galactocentric
%cylindrical coordinates $R$ (distance from Gal. center) and $Z$ (distance from plane of the disk) through $Z/S=sinb$ and
%$R^2=R_{\odot}^2+r^2-2rR_{\odot}cosl$, with $R_{\odot}$=8 kpc and $r$ being the distance from the Sun of the projection
%of the source onto the Galactic plane. $\rho$ is the local photon emissivity (in $\gamma$ cm$^{-3}$ s$^{-1}$), resulting from the
%corresponding local density of positrons (provided they annihilate near their sources);  
%for all positron sources $k$=1, except for dark matter annihilation and de-excitation, for which $k$=2.}
%ROD Oct09: this is unnecessary technical burdon at this stage
% and   longitude profile 
 (normalised here to a total disk emissivity of 0.7 10$^{43}$ e$^+$ s$^{-1}$ , Sec. IV.D.1)  (Figs. \ref{Flux_Image} and \ref{Flux_Lon}, top). 
Notice that similar profiles are expected for pulsars, magnetars, hypernovae
and gamma-ray bursts (albeit with different normalizations).

Binaries involving low mass stars, such as SNIa, novae and LMXRBs,  are expected to have a steeper longitude profile, with a maximal B/D$\leq$0.5 (assuming the bulge and disk masses of Sec. III). Using  data from Fig.~\ref{SFRdens}  and Table~\ref{tab:SNrates} and adopting
an exponential density profile for the bulge (with scalelength of 400 pc and normalised to  1.4 10$^{10}$ \ms) one may estimate an expected sky distribution and corresponding longitude profile of 
SNIa, also displayed in  Figs. \ref{Flux_Image} (middle)  and \ref{Flux_Lon} (top), where it is assumed that
the e$^+$ escape fraction from SNIa is 3\%. 
Fig. \ref{Flux_Lon} (top) indicates that the theoretically
expected SNIa longitude profile ressembles the observed profile of LMXRBs (from \onlinecite{GGS02}). 
This similarity  reflects the fact that both classes of sources involve
an old stellar population, which is proportionally  more abundant in the inner Galaxy and the bulge than in the 
rest of the disk. 
Novae are  also expected, on those grounds,  to have  a similar distribution (albeit with a much lower
normalization constant).

The upper panel of Fig. \ref{Flux_Lon} clearly shows that 
%\begin{itemize}

a) The expected outer disk ($l>$20$^o$) contribution of massive star radioactivity
($^{26}$Al and $^{44}$Ti) is not yet detected, due to insufficient sensitivity;
SPI/INTEGRAL or a future instrument  should reveal that component, or else
it should be concluded that positrons diffuse far away from their sources.

b) SNIa or LMXRBs or microquasars can explain only about
half of the strong 511 keV emission from the inner Galaxy,  assuming they produce as many positrons as in Table~\ref{tab:Pos_Sources})  

c)  Any remaining annihilation $\gamma$-ray emission requires a supplementary source (dark matter or the central SMBH, provided its positrons can diffuse to kpc distances  and fill  the bulge); alternatively, it may be assumed that
SNIa or LMXRBs or microquasars produce twice as many positrons as assumed in Fig. \ref{Flux_Lon}, but half
of the disk positrons are transported to annihilate outside the Galactic disk (see Sec. VI.C).

In the middle panel of Fig. \ref{Flux_Lon} we compare the longitude profile of the observed 511 keV emission
to the one of the hard LMXRBs (emitting in the 20-100 keV range), as seen with IBIS/INTEGRAL (from \onlinecite{Bird+07})
\citet{Weidenspointner+08a} notice that the latter distribution exhibits a pronounced asymmetry, with 
source number ratio $N(l<20^o)/N(l>20^o)$=1.7, which matches well the asymmetry in the 511 keV flux reported in the same paper.
They suggest then that hard LMXRBs may be  at the origin of the disk emission.

%%%%%%%%%%%%%%%%%%%%%%%%%%%%%%%%%%%
\begin{table*}
\caption{Properties of candidate  positron sources in the Milky Way
 \label{TabSources}}
\begin {center}
\begin{tabular}{lcccccc}
\hline \hline

Source & Process  & E(e$^+$)$^a$ & e$^+$  rate$^b$ &  Bulge/Disk$^c$ & Comments \\

 &  & (MeV)  & $\dot N_{e^+}$(10$^{43}$ s$^{-1})$ & $B/D$  &       \\
\hline

Massive stars: $^{26}$Al  & $\beta^+$-decay & $\sim$1 & 0.4 &  $<$0.2& $\dot N,B/D$: Observationally inferred \\

Supernovae: $^{24}$Ti  & $\beta^+$-decay & $\sim$1 & 0.3 &   $<$0.2 & $\dot N$: Robust estimate \\

SNIa: $^{56}$Ni        &  $\beta^+$-decay & $\sim$1 &{\it 2} &  $<$0.5 & Assuming $f_{e^+,esc}$=0.04 \\

Novae             & $\beta^+$-decay & $\sim$1 &{\it 0.02} &  $<$0.5   & Insufficent e$^+$ production \\

Hypernovae/GRB: $^{56}$Ni & $\beta^+$-decay  & $\sim$1&  ? &    $<$0.2  & Improbable in inner MW \\

Cosmic rays & p-p  & $\sim$30 & 0.1  &  $<$0.2 & Too high e$^+$ energy \\

LMXRBs  & $\gamma-\gamma$ & $\sim$1  & {\it 2}  & $<$0.5 & Assuming $L_{e^+}\sim$0.01 $L_{obs,X}$  \\

Microquasars ($\mu$Qs) & $\gamma-\gamma$  &$\sim$ 1 &  {\it 1}  &  $<$0.5 &  e$^+$ load of jets uncertain\\

Pulsars  & $\gamma-\gamma$ / $\gamma-\gamma_B$ & $>$30 & {\it 0.5} & $<$0.2   &  Too high e$^+$ energy \\

ms pulsars &  $\gamma-\gamma$ / $\gamma-\gamma_B$ & $>$30 & {\it 0.15}  &$<$0.5 &  Too high e$^+$ energy \\

Magnetars & $\gamma-\gamma$ / $\gamma-\gamma_B$  & $>$30 & {\it 0.16}  &$<$0.2 & Too high e$^+$ energy  \\

Central black hole &  p-p & High &  ? &  &Too high e$^+$ energy, unless $B>$0.4 mG \\

                           & $\gamma-\gamma$  & 1& ?   &  & Requires e$^+$ diffusion to $\sim$1 kpc\\

Dark matter  & Annihilation & 1 (?) & ?  &  & Requires light scalar particle, cuspy DM profile \\

   & Deexcitation & 1 & ? &  & Only cuspy DM profiles allowed\\
   & Decay & 1 & ? &  & Ruled out for all  DM profiles\\

\hline

Observational constraints & & $<$7 & 2 &  $>$1.4 &  \\

\hline \hline

\end{tabular}
\end{center}
$a$: typical values are given.
$b$: e$^+$ rates: in roman: observationally deduced or reasonable estimates; 
in italic: speculative (and rather closer to upper limits).  
$c$: sources are simply classified as belonging to either young ($B/D<$0.2) or old($<$0.5) stellar populations.\\
\label{tab:Pos_Sources}
\end{table*}
%%%%%%%%%%%%%%%%%%%%%%%%%%%%%%%%%%%

We note that the study  of the same SPI/INTEGRAL data by \citet{Bouchet+08,BRJ10}  with different methods does not find significant disk asymmetry. Obviously, the important (and intriguing) observational result of \citet{Weidenspointner+08a} needs confirmation by further observations/analysis.  Assuming the asymmetry is real, what might the implications be? The interpretation of \citet{Weidenspointner+08a} implicitly assumes that:

i) Positrons annihilate relatively close to their sources, such that the  annihilation morphology reflects the source morphology. 

ii) Among all LMXRBs only the bright and hard LMXRBs of the IBIS/INTEGRAL catalogue  \citep{Bird+07}  are important e$^+$ contributors.
 
 iii) Those X-ray bright and hard-spectrum LMXRBs have the same average positron yields, which are therefore not correlated to their currently observed (but known to be widely varying) X-ray luminosities; in that way, the collective e$^+$ production of hard-spectrum LMXRBs  is just proportional to their total number, not to their total X-ray brightness.

Assumption (i) underlies  all efforts to match the observed 511 keV morphology with some particular class of sources. However,
if it is adopted, and if it is assumed that the observed disk emission is due to 0.7 10$^{43}$ e$^+$ s$^{-1}$ released by hard LMXRBs, one has to explain why the robustly established e$^+$ production of $^{26}$Al and $^{44}$Ti is not detected by SPI. Indeed, the corresponding e$^+$ production rate is quite high (0.7 10$^{43}$ s$^{-1}$, with small uncertainty) and positrons are released in the 
 dense environment of massive stars and CCSN. In comparison, positrons from LMXRBs are released away from the disk (in view of the $\sim$400 pc scaleheight), i.e. in less dense environments,  and could travel and annihilate further away from their sources than those of massive star radioactivity. If both radioactivity and LMXRBs release 0.7~10$^{43}$ e$^+$ s$^{-1}$, the former should dominate the observed 511 keV emission (the latter having a lower surface brightness), and no significant asymmetry should be seen \citep{Prantzos08}.

Assumption (ii) has been criticized in \citet{BSTM09}, who note that a lower sensitivity cut-off than the one of IBIS
would   lead to a different spatial distribution of the hard LMXRBs, in view of the steeply rising luminosity function of
those sources. Besides, in view of the time variability of LMXRBs, the present day asymmetric profile (merely a snapshot in time) does not guarantee that the same morphology characterizes the total number of hard LMXRBs that may contribute to e$^+$ production during the e$^+$ lifetime ($\sim$10$^6$ yr).  Notice also that in the 4th IBIS source catalogue \citep{Bird+10}
there is no strong evidence for a LMXB distribution asymmetry in the Galactic plane.

Finally, assumption (iii) is far from obvious. This assumption certainly applies to e.g. SNIa, which are assessed to have an average $^{56}$Ni yield of 0.7 \ms \  and to constitute a relatively homogeneous class of objects. One may certainly imagine that LMXRBs also produce, on average, the same yield of positrons, at least on timescales comparable to the positron annihilation timescale. However, if LMXRB positrons
are produced in the inner accretion disks by processes depending on parameters of the binary system
(e.g. temperature, depending on black hole mass) then only a few of those systems may be important e$^+$ producers; 
their spatial distribution may not be represented at all by the one of {\it all } hard LMXRBs. 

The morphology of the observed 511 keV emission provides also some interesting constraints
in the case of dark matter  particles as positron sources (under the assumption of negligible  e$^+$ propagation).
An illustration of such an analysis is provided in the work of \citet{ascasibar}, who
convolved the  positron maps
predicted for various light DM particle scenarios and types of DM halo
profiles with the response function of SPI. Comparison to the data showed that:
i) Particle candidates with velocity dependent cross section  are excluded as the main
source of 511 keV emission,   ii) Fermionic DM candidates
are also excluded, since they would need to exchange too light charged
particles, and iii)  Decaying dark matter cannot be the main source
of low energy positrons, because the resulting flux profile is too flat, compared to SPI data. Notice that this
latter feature is a generic property of all models involving decaying particles, where the positron production (and annihilation)
rate is proportional to the DM density profile: even "cuspy" profiles, such as the NFW (see Fig.\ref{fig:Rot_Curve})
do not provide a $\gamma$-ray flux profile sufficiently peaked towards the inner Galaxy.
Annihilating or de-exciting DM produces positrons at a rate proportional to the square of the DM density profile (Sec. IV.C.2) and leads to 
a much more peaked $\gamma$-ray profile.
\citet{ascasibar} found that light scalar annihilating
particles remain a possible candidate, provided the DM halo is at least 
as cuspy as the NFW profile with $\gamma \sim 1$  (see bottom panel of Fig.~\ref{Flux_Lon}); 
however, as stressed in Sec.~\ref{sec:GalaxyDM},
astrophysical evidence favors flatter DM halo profiles.
 
 The proximity of the Galactic center and the expected high density of DM particles there, make it the
 prime target for the detection of all kinds of  radiation emitted indirectly by DM (either decaying, annihilating or de-exciting).
 However, because of  the uncertainties presently affecting the  density profile of DM haloes (see Sec. III.E) and the possible 
 contamination  of the signal by more conventional astrophysical sources, other potential targets have been seeked for.
 The dwarf spheroidal (dSph) satellites of the Milky Way, with their high mass/light ratio and relative proximity, may constitute such targets.
 \citet{Hooper:2003sh} suggested that the light DM hypothesis could be tested on the nearby (25 kpc) dSph galaxy Sagittarius, which 
 appears to be dominated by dark matter. A search for the expected annihilation signal  at 511 keV \citep{Cordier+04}
was unsuccessful. 

\subsubsection{Summary of candidate sources}

The main features of the candidate e$^+$ sources discussed in this section are summarized in Table~\ref{tab:Pos_Sources}.
 As already emphasized, e$^+$ production rates of all those sources are extremely uncertain (except those of $^{26}$Al, $^{44}$Ti and GCRs) and the values listed above should be considered as optimistic  rather than typical ones. Only in the case of novae may the estimated production value be used to eliminate those sources as important e$^+$ producers.
 Source morphology and high energy of produced positrons appear to exclude pulsars, magnetars and GCRs as major
contributors to the observed 511 keV emission from the bulge. 
Source morphology alone would exclude hypernovae and GRBs. The high energy of produced positrons disfavors ms pulsars, as well as  p-p collisions from any source (micro-quasars, LMXRB jets, the central SMBH). This still leaves several potentially important e$^+$ contributors, but none of them has  the observed morphology of 511 keV emission.

Thus, {\it assuming that positrons annihilate near their sources}, one has to conclude that

i) Either an unknown class of sources dominates e$^+$ production, or

ii) Positrons are produced by a combination of the sources of Table~\ref{tab:Pos_Sources}, e.g. (a) \Al+$^{44}$Ti for the disk and dark matter for the bulge, or (b) \Al+$^{44}$Ti+LMXRBs(or microquasars) for the disk {\it and } the bulge plus a contribution from the central SMBH for the inner bulge, or (c) some other combination.

Alternative (ii) bears an interesting "philosophical" issue: how is it possible that two widely different classes of sources have so similar e$^+$ yields (to within a factor of a few), such as required to fit the observations? However, such "coincidences" are not unusual in astronomy\footnote{Among the most famous "coincidences" are: (i) the contributions of baryonic and non-baryonic matter, as well as those of dark matter and dark energy, to the cosmic density; (ii) the solar 
abundances of s- and r- nuclei (both of $\sim$10$^{-6}$ by mass fraction); (iii) the approximately equal contributions
of CCSN and SNIa to the solar Fe content.} these days.

A more important issue  arises in solutions involving the central SMBH as the main e$^+$ producer within the Galaxy's bulge: If its positrons can diffuse to kpc scales  in the dense environment of the inner bulge, then positrons {\it should}  diffuse to even larger scales in the less dense environment of the disk; this would be even more true outside the spiral arms, where most of the SNIa/LMXRB/$\mu$Q positrons are expected to be released. Positron escape from the disk to the halo would alleviate the morphology problem, by reducing the disk 511 keV emissivity
and thus increasing the B/D ratio of those classes of sources. 
However, although some of the basic physical processes underlying e$^+$ propagation are well understood, there is no clear global picture of how far positrons can propagate in the magnetized, turbulent ISM of the Galaxy. We turn to those two issues in the next two sections.

% Sec. V    Annihilation  .......................................................................................................................................................

%
% Edited by Nidhal May 27, 2010
%
% Corrected by Guessoum & Jean --  29 july, 2009
%
\section{ Positron interactions with matter and annihilation}
\label{sec:Interactions}

\begin{figure}
\includegraphics[width=0.49\textwidth]{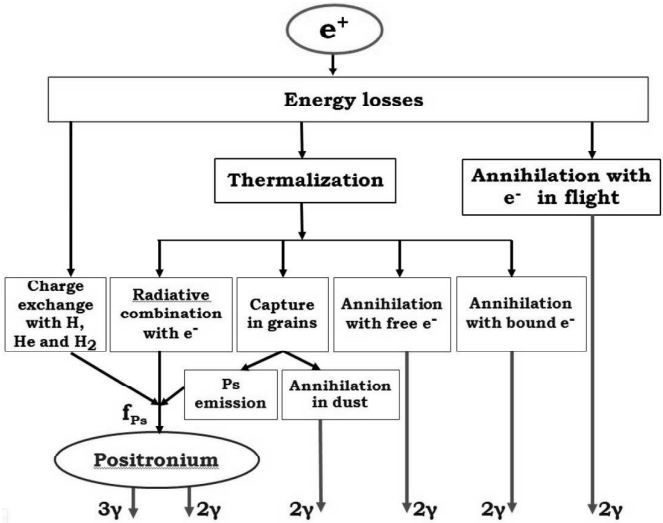}
\caption{\label{fig:sumproc} The processes leading to gamma-ray 
production from positron annihilation \citep[taken 
from][slightly modified to account for the 
annihilation in flight of relativistic positrons]{Guessoum:1991}.}
\end{figure}

Positrons are initially produced with kinetic energies higher than
those of the ISM (see previous  section). 
Most of them slow down to low energy before they annihilate with 
bounded or free electrons of the ISM. However, when their initial 
kinetic energy 
is above a tenth of MeV, a significant fraction of them may 
annihilate 
in-flight. Figure Fig.~\ref{fig:sumproc} summarizes the 
processes that lead to gamma-ray production from positron 
annihilation. 
The following sections list the interactions that are responsible 
for the energy losses of positrons in the ISM and present the 
different ways in which
they annihilate with electrons. 

%%%%%%%%%%% Energy losses and annihilation processes %%%%%%%%%%%%%%%
%%%%%%%%%%%%%%%%%%%%%%%%%%%%%%%%%%%%%%%%%%%%%%%%%%%%%%%%%%%%%%%%%%%%

\subsection{Energy losses}

\begin{figure}
\includegraphics[width=0.49\textwidth]{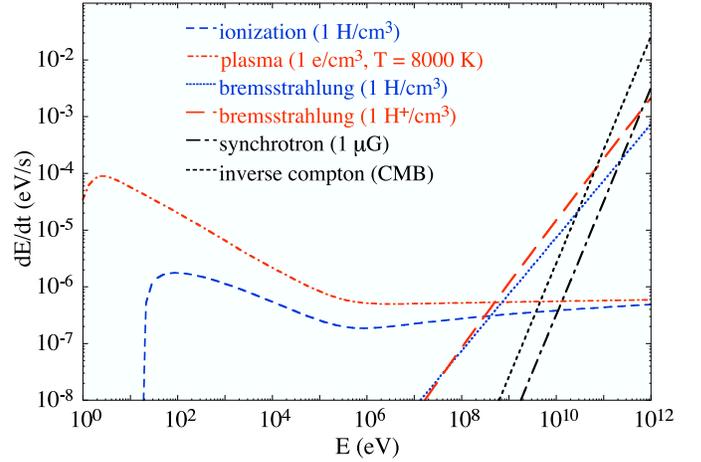}
\caption{\label{fig:desdt} Energy loss rate for positrons in ISM 
conditions. For synchrotron losses, the pitch angle is here taken as 
$\pi$/2.}
\end{figure}

As charged leptons, positrons interact via the electromagnetic force 
with all basic constituents of the ISM, namely: electrons, ions, atoms, 
molecules, solid dust grains, photons and magnetic fields. Since 
their initial kinetic energy is generally larger than 
the kinetic energy of the targets in the ISM, positrons lose energy 
in these interactions. The energy loss rate and the kind of interaction depend 
on the energy of positrons and the density of target particles. 
Fig.~\ref{fig:desdt} presents the 
energy loss rate as a function of positron energy; the contributions 
of each type of interaction are shown separately.

%%%%%%%%%%%%%%%%% Energy loss rate vs. Energy %%%%%%%%%%%%%%%%%%%%%%
%%%%%%%%%%%%%%%%%%%%%%%%%%%%%%%%%%%%%%%%%%%%%%%%%%%%%%%%%%%%%%%%%%%%

Ultra-relativistic positrons (E$>$10 GeV) lose their energy mainly by 
inverse Compton scattering with cosmic microwave background (CMB) photons and 
interstellar radiation fields. 
%(The process is called inverse because the  positrons 
%lose energy rather than the photons, the opposite of the standard  Compton effect.) 
When the interaction occurs with an isotropic photon 
gas in the Thomson scattering regime (i.e. $ h \nu \ll m_{e} 
c^{2}$),  the energy loss rate (in eV/s) can be calculated from 
\citep{Blumenthal-Gould:1970}:

\begin{equation}
 \left(\frac{dE}{dt}\right)_{IC} = - 2.6 \times 10^{-14} \; u_{\rm{rad}} \; \gamma^{2} 
\; \beta^{2} \; ,
 \label{eq:dedtic}
\end{equation}

\noindent where $u_{\rm{rad}}$ is the radiation energy density 
(eV/cm$^{3}$),  and $\beta=\upsilon/c$ 
its velocity relative to the speed of light. The radiation energy density 
depends on the position of the positron in the Galaxy; it ranges from 
0.26 eV/cm$^{3}$ (CMB) to 11.4 eV/cm$^{3}$ in the Galactic Centre 
region \citep{Moskalenko-Porter-Strong:2006}.

Ultrarelativistic positrons also lose their energy by emitting 
synchrotron radiation when they spiral along a magnetic field line. The energy 
loss rate depends on the magnetic field intensity, the positron's 
kinetic energy and the {\it pitch angle}\footnote{The pitch angle of a charged particle is the angle between the vectors of the particle's velocity and of
the local magnetic field}. Its expression (in eV/s) is 
\citep{Blumenthal-Gould:1970}:

\begin{equation}
\left( \frac{dE}{dt}\right)_{SY} = - 9.9 \times 10^{-16} \; B^{2} \; \gamma^{2} \; 
\beta^{2} \; sin^{2}(\alpha) \; ,
 \label{eq:dedtsy}
\end{equation}

\noindent where $B$ is the magnetic field (in $\mu$G) and $\alpha$ is 
the pitch angle. For a positron moving through a randomly oriented 
magnetic field, the mean energy loss rate is obtained by replacing 
$sin^{2}(\alpha)$ by its average value 2/3. The synchrotron energy 
loss rate is proportional to the square of the magnetic field while the inverse-Compton 
one is proportional to the radiation energy density. Therefore, 
synchrotron losses dominate for ultrarelativistic positrons in an environment where 
$B > 6.3 \mu$G $\sqrt{u_{\rm{rad}}}$, e.g. near Pulsars, the Galactic 
Centre region, etc.

In the 1-10 GeV energy range, positrons lose their energy mainly by 
emitting bremsstrahlung radiation in interactions with ions, electrons and 
atoms. The energy loss rate depends on the target mass, charge, and density. For 
relativistic positrons, it is equivalent to the electron one. Methods 
for calculating the average energy loss rate and the bremsstrahlung 
energy spectrum are described in \citet{Hayakawa:1969}, \citet{Evans:1975} and 
\citet{Blumenthal-Gould:1970}. 
A recent update, with a more accurate treatment of the differential 
cross section for electrons with midly relativistic energies in a fully ionized plasma, 
was presented in \citet{Haug:2004}. A good approximation of the bremsstrahlung 
energy loss rate (in eV/s) of relativistic positrons in a fully ionized gas is 
\citep[][p 408]{Ginzburg:1979}:

\begin{equation}
\left( \frac{dE}{dt}\right)_{BR} = - 3.6 \times 10^{-11} \; Z(Z+1)  \; n \; \gamma \; 
\left[ ln(2\gamma) - \frac{1}{3} \right] \; ,
 \label{eq:dedtbi}
\end{equation}

\noindent where $Z$ and $n$ are the nuclear charge and the number 
density (in cm$^{-3}$) of the ion, respectively. In a neutral hydrogen gas, the 
energy loss rate can be estimated via \citep[][p. 386]{Ginzburg:1979}:

\begin{equation}
 \left(\frac{dE}{dt}\right)_{BR} = A \  n \ \gamma   ,
 \label{eq:dedtba}
\end{equation}

\noindent where A= -4.1$\times$10$^{-10}$ for hydrogen and A=-1.1$\times$10$^{-9}$
 for helium, respectively.

Below 1 GeV, positrons lose their energy mainly via Coulomb 
scatterings with free electrons and/or inelastic interactions with 
atoms and molecules. The former process is a continous energy loss, 
whatever the energy of positrons. At high energy, the target 
electrons can be considered at rest and the energy loss rate depends 
mostly on their density. \citet{Dermer:1985} calculated the rate of e-e
Coulomb collisions in relativistic thermal plasmas. He 
also treated the case of collisions in cold plasmas.
This energy loss can be approximated by \citep[][p 361]{Ginzburg:1979}:

\begin{equation}
\left( \frac{dE}{dt}\right)_{COU} = - 7.7 \times 10^{-9} \; \frac{n_{e}}{\beta} \; 
\left[ ln \left(\frac{\gamma}{n_{e}} \right) + 73.6 \right] \; ,
 \label{eq:dedtco}
\end{equation}

\noindent where $n_{e}$ is the electron density. 

At low positron energy (E$\lesssim$10kT, where T is the ambient 
temperature), the electrons of the ISM cannot be considered at 
rest anymore. The energy loss rate depends on their 
temperature and density in the plasma and is given by 
\citet{Book-Ali:1975} and Huba (2006) \citep[see also][]{Murphy:2005}.

Inelastic collisions of positrons with atoms and molecules can be 
considered as a continous process for a positron energy $>$1 keV, and 
the energy loss rate can be evaluated using the Bethe-Bloch formula. 
The ionization loss is larger than the excitation loss.
This energy loss can be approximated by \citep[][p 360]{Ginzburg:1979}:

\begin{equation}
\left( \frac{dE}{dt}\right)_{ION} = - 7.7 \times 10^{-9} \; \frac{nZ}{\beta} \; 
%\left[ ln \left(\frac{(\gamma - 1) \gamma^2 \beta^2 m^2 c^4}{2 I^2} \right) 
\left[ ln \left(\frac{(\gamma - 1) (\gamma \beta m c^2)^2}{2 I^2} \right) 
+ \frac{1}{8} \right] \; ,
 \label{eq:dedtion}
\end{equation}

\noindent where $n$ is the neutral atom density, $Z$ is the number of 
electron of the atom and $I$ is its ionization potential 
(e.g. 13.6 eV for H and 24.6 eV for He). 
Below 1 keV the interaction between positrons and atoms/molecules 
should be estimated via Monte Carlo simulations since positrons release a large 
fraction of their energy in one interaction and, below $\sim$ 100 eV, they can 
pick-up an electron from an atom or a molecule to form a positronium
``in flight'' (see Table~\ref{tab:thres} and the next section). In this case, 
the cross sections for such collisions should be used in evaluating 
the energy loss rate \citep[see][]{Guessoum:2005}. The Monte-Carlo 
method to calculate the interaction probability as a function of the 
positron's energy was 
presented by \citet{Bussard:1979}.

% e$^{+}$e$^{-}$ 
%%%%%%%%%%%%%%%%%%%%%%%% Table - thresholds %%%%%%%%%%%%%%%%%%%%%%%
\begin{table}
     \caption{Energy thresholds of inelastic reactions produced by 
positrons.} 
\label{tab:thres}
 \begin{tabular}[b]{lc}
\noalign{\smallskip}
\hline
\hline
\noalign{\smallskip}
\mbox{Process}  & \mbox{Threshold (eV)} \\
\hline
\noalign{\smallskip}
\mbox{e$^{+}$ + H $\rightarrow$ Ps + H$^{+}$} & \mbox{6.8} \\
\mbox{e$^{+}$ + H $\rightarrow$ e$^{+}$ + e$^{-}$ + H$^{+}$} & \mbox{13.6} \\
\mbox{e$^{+}$ + H $\rightarrow$ e$^{+}$ + H$^{*}$} & \mbox{10.2} \\
\mbox{e$^{+}$ + H $\rightarrow$ e$^{+}$ + H$^{**}$} & \mbox{12.1} \\
\mbox{e$^{+}$ + He $\rightarrow$ Ps + He$^{+}$} & \mbox{17.8} \\
\mbox{e$^{+}$ + He $\rightarrow$ e$^{+}$ + e$^{-}$ + He$^{+}$} & \mbox{24.6} \\
\mbox{e$^{+}$ + He $\rightarrow$ e$^{+}$ + He$^{*}$} & \mbox{21.2} \\
\mbox{e$^{+}$ + H$_{2}$ $\rightarrow$ Ps + H$_{2}^{+}$} & \mbox{8.6} \\
\mbox{e$^{+}$ + H$_{2}$ $\rightarrow$ e$^{+}$ + e$^{-}$ + H$_{2}^{+}$} & \mbox{15.4} \\
\mbox{e$^{+}$ + H$_{2}$ $\rightarrow$ e$^{+}$ + H$_{2}^{*}$} & \mbox{12.0} \\
\noalign{\smallskip}
  \hline
 \end{tabular}
\end{table}
%%%%%%%%%%%%%%%%%%%%%%%%%%%%%%%%%%%%%%%%%%%%%%%%%%%%%%%%%%%%%%%%%%%%
Equations \ref{eq:dedtic}-\ref{eq:dedtion} allow one to estimate to a good
aproximation energy losses of positrons. More accurate
expressions, valid to energies $>$ 1 GeV, are given in \citet{SM98}.

\subsection{Annihilation in flight}

In this section, we present two kinds of annihilation in flight: (1) 
direct annihilation of relativistic positrons with electrons 
and (2) annihilation via positronium produced in interactions 
of non relativistic positrons with atoms/molecules.

\smallskip

\smallskip

\subsubsection{Direct annihilation in flight}

\smallskip

When high-energy positrons ($\gtrsim$ 10 keV) slow down, 
they may annihilate in flight with 
free or bound electrons. The energies of the two photons emitted
in this process are strongly shifted as per the Doppler effect. This 
produces a continuous spectrum in the energy interval $mc^{2}/2 \lesssim 
E_{\gamma} \lesssim E + mc^{2}/2 $ with $E$ the total energy of the 
positron. 

%%%%%%% Fraction of annihilation inflight vs. Energy %%%%%%%%%%%%%%%
\begin{figure}
\includegraphics[width=0.49\textwidth]{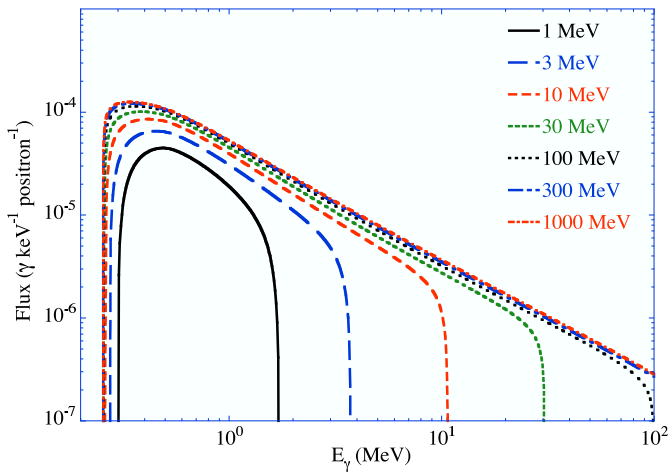}
\caption{\label{fig:specaif} Gamma-ray spectra from the annihilation 
in flight in the ISM for various initial kinetic energies of positrons.}
\end{figure}
%%%%%%%%%%%%%%%%%%%%%%%%%%%%%%%%%%%%%%%%%%%%%%%%%%%%%%%%%%%%%%%%%%%%

\begin{figure}
\includegraphics[width=0.33\textwidth]{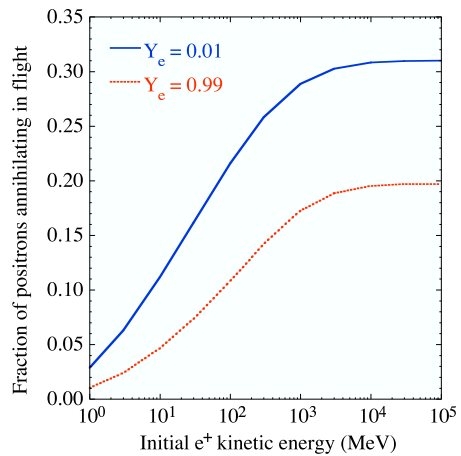}

\caption{\label{fig:probann} Probability for a positron to annihilate 
in flight as a function of its initial kinetic energy, in a neutral 
medium (solid line) and an ionized medium (dashed line). $Y_{e}$ 
represents the ionization fraction.}
\end{figure}

The probability that a positron with an initial kinetic energy $E_{0}$ 
annihilates in flight before reaching an energy $E$ is:

\begin{equation}
 P(E_{0}, E) = 1 \; - \; exp\left( - n_{e} \int_{E}^{E_{0}} 
 \frac{v(E') \sigma_{a}(E') dE'}{|dE'/dt|} \right) \; ,
 \label{eq:probann}
\end{equation}

\noindent where $v$ and $dE'/dt$ are the positron velocity and the 
energy loss rate, respectively; $n_{e}$ is the density of target electrons, 
and $\sigma_{a}$ the annihilation cross section, which can be 
estimated for kinetic energies larger than 75 keV via \citep{Dirac:1930}:

\begin{equation}
 \sigma_{a} = \frac{\pi r_{e}^2}{\gamma + 1} \left[ 
 \frac{\gamma^{2}+4\gamma+1}{\gamma^{2}-1} 
 ln(\gamma+\sqrt{\gamma^2-1}) - \frac{\gamma+3}{\sqrt{\gamma^2-1}} 
\right]
 \label{eq:sigann}
\end{equation}

\noindent where $r_{e}$ is the classical electron radius ($r_{e} = 
e^{2}/m_{e}c^{2}$); for an evaluation of this cross section below 75 
keV, see Gould (1989).
The probability $P(E_{0}, E)$ depends on the energy loss rate and 
consequently on the physical conditions of the interstellar medium in 
which the positron propagates. 
Fig.~\ref{fig:specaif} presents the spectra of gamma-rays 
emitted by in-flight annihilation of relativistic positrons slowing 
down in the interstellar medium, for several initial kinetic energies.
Fig.~\ref{fig:probann} shows the fraction 
of positrons annihilating in flight as a function of the inital 
kinetic energy of positrons, in both a neutral and a fully ionized medium. It is 
negligible ($\lesssim$4\%) for energies lower than 1 MeV.
Above $\sim$ 1 GeV, the fraction does not change because the energy 
loss rate is so large (see Fig.~\ref{fig:desdt}) that positrons do 
not  have enough time to annihilate in flight at these energies.

The implications of e$^+$ annihilation in flight for the observed Galactic
MeV emission are further analyzed in Sec. IV.E. 
%%%%%%% Fraction of annihilation inflight vs. Energy %%%%%%%%%%%%%%%
%%%%%%%%%%%%%%%%%%%%%%%%%%%%%%%%%%%%%%%%%%%%%%%%%%%

\subsubsection{ Positronium formation in flight}

When positrons have kinetic energies lower than $\sim$~100 eV, they 
can pick up an electron from an atom or a molecule to form a 
positronium in flight while they slow down. This reaction, also called 
charge exchange, is endoenergetic. It can happen as long as the 
kinetic energy of the positron is larger than the charge exchange 
threshold energy of Ps formation with the given atom or molecule 
(see Table.~\ref{tab:thres}). This threshold energy is equal to the 
ionization potential of the atom reduced by the binding energy of the 
Ps (6.8 eV). The cross sections for positronium formation by charge 
exchange were measured mostly in rare gases, particularly in helium, 
and in molecular and atomic hydrogen 
\citep[see references in][]{Guessoum:2005}. 

\begin{figure}
\includegraphics[width=0.33\textwidth]{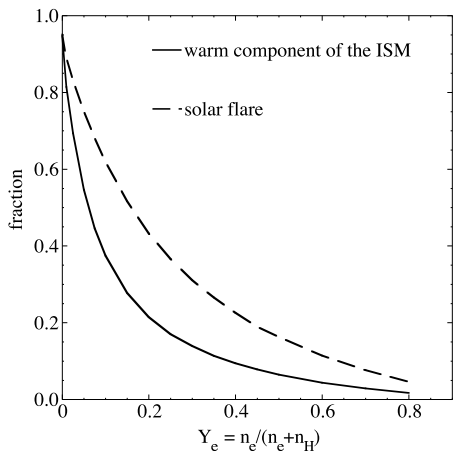}
\caption{\label{fig:f1} Fraction of positrons forming positronium in 
flight by charge exchange with atomic hydrogen as a function of the 
ionization fraction (Y$_e$) in a warm component of the interstellar 
medium (electron density: n$_{e}$ = 0.1 cm$^{-3}$, electron temperature: 
T$_{e}$ = 8000 K) and in solar flare (n$_{e}$ = 5 $\times$ 10$^{13}$ 
cm$^{-3}$, T$_{e}$ = 1.16 $\times$ 10$^{4}$ K).}
\end{figure}

The fraction of positrons that form a positronium in flight is 
obtained by Monte Carlo methods which consist in simulating the interactions 
of positrons with atoms and molecules on the basis of the cross sections and 
energy loss mechanisms presented in the previous section. This 
fraction depends on the density, on the temperature, and strongly on 
the ionization fraction of the medium in which positrons slow down. 
Fig.~\ref{fig:f1} shows the fraction of positrons forming positronium 
in flight as a function of the ionized fraction in two types of 
media \citep{Murphy:2005}. The fraction of positrons that 
form Ps decreases quickly with increasing values of the ionized 
fraction not only due to the reduction of the density of neutral H but also 
because the energy loss rate increases quickly with the ionized 
fraction  (see Fig.~\ref{fig:desdt}). It makes the positrons slow down so 
rapidly that they do not have time to exchange charge with H.
Measured \citep{Brown:1984, Brown:1986} and calculated values of 
the fraction of positrons forming positronium in flight in H, He and 
H$_{2}$ are summarized in Table \ref{tab:f1}.

%%%%%%%% Table - comparaison des differents f1 obtenus %%%%%%%%%%
\begin{table}[b]
\caption{Fraction (in \%) of positrons forming positronium in flight, 
in totally neutral media.}
\label{tab:f1}
\begin{tabular}{lccc}
\noalign{\smallskip}
\hline
\hline
\noalign{\smallskip}
\mbox{References}  & \mbox{H} & \mbox{H$_{2}$} & \mbox{He} \\
\hline
\noalign{\smallskip}
\mbox{Bussard et al. (1979)} & 95 & 93 & - \\
\mbox{Brown \& Leventhal (1986)} \qquad & - & \quad 89.7 $\pm$ 0.3 \quad & 80.7 $\pm$ 0.5 \\
\mbox{\citet{Wallyn:1994}} & 98 & 90 & - \\
\mbox{\citet{Chapuis:1994}} \qquad & - & - & 78 \\
\mbox{Guessoum et al. (2005)} & 95.5 & 89.6 & 81.7 \\
\noalign{\smallskip}
  \hline
\end{tabular}
\end{table}%%%%%%%%%%%%%%%%%%%%

%%%%% Fraction of Ps formed inflight vs. ionization fraction %%%%%%%
%%%%%%%%%%%%%%%%%%%%%%%%%%%%%%%%%%%%%%%%%%%%%%

%Depending on whether the positron and the electron have parallel or 
%anti-parallel spins, the positronium will be in the long-lifetime 
%($\tau$ = 140 ns) ortho-positronium (ortho-Ps, $^3$Ps) state or the 
%short-lifetime ($\tau$  = 0.125 ns) para-positronium (para-Ps, $^1$Ps) state. It is expected 
%from spin  statistics that positronium is formed with a population ratio of 
%$^3$Ps/$^1$Ps equal to 3:1. At rest, the para-Ps decays mainly 
%into 2 photons of 511 keV, while the ortho-Ps decays mainly into 3 
%photons  of energies totaling 1022 keV. Decays into more than 3 photons are 
%not forbidden  but are very rare. The distribution of photon energies resulting from 
%the  decay of ortho-Ps was first calculated by \citet{Ore-Powell:1949} and 
%more recently by \citet{Adkins:1983}.

When the Ps is produced in flight, its kinetic energy is
equal to the energy of the positron minus the threshold of the charge 
exchange reaction. Consequently, the energy of the photons emitted in 
the annihilation are shifted as per the Doppler effect.
The spectral shape of the 511~keV line emitted in the annihilation of 
para-Ps was also derived from Monte Carlo simulations by extracting 
the kinetic energy distribution of the produced Ps. The most recent 
calculations \citep{Guessoum:2005} yield widths of 5.8~keV, 6.4~keV and 
7.4~keV for H, H$_{2}$ and He, respectively. The calculated line widths for He 
and H$_{2}$ are in good agreement with previous measurements \citep{Brown:1984}.

For large gas density, the time scale for a Ps to collide with an 
ambient particle X (atom, electron, photon) may be comparable to or 
lower than the lifetime of the ortho-Ps. In that case, the ortho-Ps 
can be destroyed ($^3$Ps + X $\rightarrow$ X + e$^+$ + e$^-$) or converted 
into a para-Ps (spin flip -- $^3$Ps + X $\rightarrow$ X + $^1$Ps). These processes 
tend to reduce the contribution of the annihilation via the ortho-Ps state. They are 
expected to occur in the solar atmosphere during flares 
\citep[see][]{Guessoum:1997, Murphy:2005} and in nova 
envelopes \citep{Leising-Clayton:1987}. In such media, the interactions 
of Ps with components (atoms, electrons, photons) of the plasma must be taken 
into account in evaluating the proportion of annihilations through the ortho-Ps 
and the para-Ps states and is known as ``ortho-Ps quenching".

\subsection{Thermalisation}

%As explained above, the positrons lose energy down from their initial 
%kinetic energies of $\sim 1$ MeV (more or less, depending on the production 
%process) before the interactions which lead to their annihilation start to take 
%place. Depending on the ionization state of the medium, and to some extent on its 
%temperature, the more important energy-loss processes can be Coulomb interactions with 
%free electrons, losses to the plasma through collective (wave) mechanisms, ionization 
%and excitation of atoms (H, He) and molecules (H$_2$).

Once the positrons have come down to energies similar to those of the 
ambient medium, they start to ``thermalize", i.e.  their energy distribution 
relaxes to the Maxwellian function which characterizes the 
interstellar gas (or plasma)\footnote{It is understood 
that unless the background medium is in thermodynamical 
non-equilibrium (occurrence of irregular heating, sudden energy losses, 
etc.), it will have adopted a Maxwellian energy distribution 
characterized by a temperature T.}. The ISM is usually 
considered to consist of a few phases, each with rather well defined physical characteristics 
(temperature, density, ionization fraction) -- see Table~\ref{tab:TableISM}. The timescale 
needed for the energetic positrons to relax to 
the ISM Maxwellian distribution is to be compared to the timescale 
for subsequent annihilation processes; if the
former timescale is longer than the latter, it would be 
incorrect to assume a Maxwellian distribution for both the 
positrons and the ISM when calculating  the e$^+$ annihilation rates.

To tackle this question one may simply estimate the 
relaxation  timescale or perform a full statistical-physics treatment. The 
former can be simply done by using the energy loss rate: 
$\tau = - ({1 \over E} { {dE} \over {dt} })^{-1}$ 
or (a simpler but cruder evaluation) $\tau^{-1} = R = <n \sigma v>$, 
taking for the cross section some typical inelastic scattering value. The 
first formula (using values in Fig.~\ref{fig:desdt} for the energy loss rate) gives 
$\tau \sim 6 \times 10^{7}$ seconds for 1 keV positrons, while using the cruder 
approach gives $\sim 2 \times 10^{7}$ seconds (taking $\sigma \sim 10^{-16}$ cm$^2$, E$_+ \sim 
1$ keV, and n $\sim 1$ cm$^{-3}$); estimates for the relaxation of positrons from their initial energies of $\sim$ MeV , is more complicated, both because the particles undergo many different processes and the overall cross section is difficult to estimates, but roughly the two simplistic approaches give timescales $\lesssim 10^{12}$ seconds. It must then be noted that all of these values are much less than typical annihilation 
timescales in  the ISM \citep[see][]{Guessoum:2005}, which range around $10^{12}-10^{14}$ seconds. 

The sophisticated statistical-physics approach is much more rigorous and conclusive, though complicated. 
A few authors have attempted such treatments, ranging from very broad, highly theoretical \citep{Wolfe-Melia:2006} to others covering a specific area of application \citep{Crannell:1976},  the latter focusing on low-energy (50 eV) positrons produced and annihilated in solar flares. \citet{Dermer:1989} and \citet{NM98} have performed thorough statistical-physics treatments for the thermalization of high-energy electrons (and secondarily of positrons), %\citet{Dermer:1989} 
assuming the interaction of the injected particles takes place with a  relativistic electron-proton plasma. \citet{Baring:1987}, modifying a simple treatment by \citet{Spitzer:1965},  showed how to calculate the relaxation time $\tau(E_e)$ of an electron for any temperature (i.e. relativistic or not). 
The statistical-physics method uses a Fokker-Planck equation to follow the energetic particles from their injection (at a given time and energy) to their thermalization either with the background plasma/gas or with each other, taking into account the various inelastic scattering and loss processes; in this approach, one evaluates the energy and dispersion coefficients (${1 \over n_e} { {dE} \over {dt} }$ and ${1 \over n_e} { {d(\Delta E)^2} \over {dt} }$, respectively). 
 \citet{CCD06} performed such a treatment for high-energy (30 MeV) positrons,
 presumably  produced by proton-proton collisions at the Galactic center
 and  approaching thermalization with the ISM gas and plasma conditions (T $\sim 10^4$ K and $10^6$ K, respectively).
 % the goal is then to calculate the rate of gamma-ray production as a function of time. 

These various approaches, heuristic or elaborate, have shown that the positrons do indeed thermalize before the annihilation processes become important, which makes valid and legitimate the usage of Maxwellian distribution functions in the calculations of the rates of annihilation and other processes that the positrons undergo.

\subsection{Annihilation}

Positrons annihilate by various processes during the two stages of their ``lives'': 
a) during the slowing-down time (the process that is referred to as 
``in-flight'' annihilation, described in section V.B); b) after thermalization with the ISM.
In this section we consider the latter stage and its processes. As 
explained in sections V.A and V.B, while positrons lose the bulk of their energies 
from $\sim 1$ MeV to $\simeq 100 $ eV, their probability of undergoing ``charge exchange'' 
(``picking up'' an electron from an atom or molecule and 
forming a positronium) increases steadily. Thus, by the time they have 
thermalized with the medium, the probability $f_{1}$ that they will have 
formed a Ps can be as high as 95 \%, depending 
on the ionization state of the medium (see Fig.~\ref{fig:f1}).
%(the lower the free electron abundance, the higher the probability).

Depending on the physical conditions of the medium (ionization state, 
temperature, composition), positrons may undergo a variety of annihilation 
processes, which are listed here by order of decreasing strength of their cross sections (when 
the physical conditions allow them to occur): 1) charge exchange (with atoms and molecules); 
2) radiative (re)combination with (free) electrons; 3) direct annihilation with 
free electrons; 4) direct annihilation with bound electrons (of atoms and molecules). 
For dust grains, the cross section for collision and annihilation (see 
below), may be larger than those of some of the above processes; however, when the 
abundance of dust grains is taken into account, the reaction rate for this process turns out to 
be smaller than the others', except in rare cases.

%%%%%%%%%%%%%%%%%%%%% cross-sections e+ + X %%%%%%%%%%%%%%%%%%%%%%%%
\begin{figure}
\includegraphics[width=0.49\textwidth]{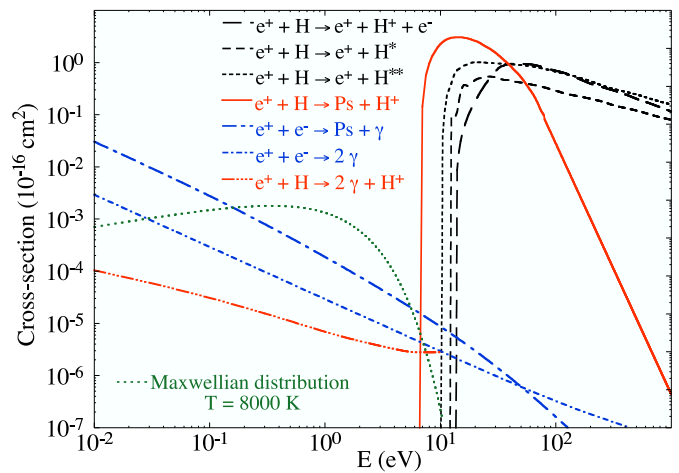}
\caption{\label{fig:cspos} Cross-sections for ionization, excitation, 
charge exchange, radiative recombination and direct annihilation 
interactions of positrons with atomic hydrogen and free electrons. Also shown is 
the Maxwellian distribution for a temperature of 8000 K (in arbitrary units).}
\end{figure}
%%%%%%%%%%%%%%%%%%%%%%%%%%%%%%%%%%%%%%%%%%%%%%%%%%%%%%%%%%%%%%%%%%%%
The key features of  these  annihilation processes (see Fig.~\ref{fig:cspos})
are the following:

{\it a) Charge exchange} (with H, He, and H$_2$): although difficult to 
measure, the cross section 
for this(ese) process(es) has been obtained by several experimental 
groups \citep[see][ for references]{Guessoum:2005}; the important feature
of that process is its threshold energy (6.8, 17.8, and 8.6 eV, respectively), 
implying that this reaction (which has by far the highest cross section of all positron 
annihilation processes) can only occur at temperatures larger than several thousand K; 
hence this cannot occur in the cold media of the ISM, and of course it cannot take place in the 
hot ISM phase either (because the medium is completely ionized).

The width of the line resulting from the annihilation of thermal 
positrons by charge exchange (decay of Ps) with H or He can be 
derived from the kinetic energy distribution of the produced Ps. This 
distribution is obtained simply by computing the charge exchange rate 
for a population of positrons that follows a Maxwellian distribution. 
The calculated widths in a warm medium (T = 8000 K) are 1.16 keV and 
1.22 keV, for H and He, respectively \citep{Guessoum:2005}.

%The widths of the lines resulting from the annihilation by charge 
%exchange (decay of Ps) with H, 
%He, and H$_2$ can be measured in the laboratory (e.g. Brown \& 
%Leventhal 1984, 
%Brown \& Leventhal 1986) as well as calculated by Monte Carlo 
%simulation (Bussard, Ramaty, and 
%Drachman 1979, Guessoum, Jean, and Gillard 2005). The latest 
%results give 5.8 keV, 7.4 keV, 
%and 6.4 keV, for H, He, and H$_2$, respectively, agreeing well 
%with past measurements and 
%calculations; we consider the new results as more accurate, 
%however, for they 
%were based on newer measurements of the cross sections for 
%these processes.

{\it b) Radiative (re)combination  with free electrons}: the cross section 
for this process is too small ($\sim 10^{-20}$ cm$^2$ at 1 eV) to be measured 
experimentally;  one then has to rely on theoretical calculations, 
such as the determinations (by different approaches) 
made by \citet{Crannell:1976} and by \citet{Gould:1989}.

{\it c) Direct annihilation with free electrons}: this process has an even 
smaller cross section (about an order of magnitude less than the previous one at 
temperatures less than about $10^5$ K), hence it is only important in the hot 
phase of the ISM; the cross section has also been estimated by \citet{Crannell:1976} 
and \citet{Bussard:1979} from the early theoretical work of 
\citet{Heitler:1954}.

The width of the line resulting from both radiative combination and 
direct annihilation of positrons and electrons was calculated by \citet{Crannell:1976} 
using a simple argument of thermal broadening due to the pair's 
center-of-mass motion; they obtained the simple expression $\Gamma_{rc, dae} = 1.1 \times 
(T/10^4)^{1/2}$ keV, which applies to both processes.

{\it d) Direct annihilation with bound electrons}: the cross sections (for 
H, He, and H$_2$) are the weakest of all (see Fig.~\ref{fig:cspos}), but they become 
important by default at very low temperatures (in the cold phases of the ISM) where free 
electrons do not exist and the charge exchange (with atoms and molecules) 
cannot take place due to the threshold energies, which are (at about 10 eV) much 
too large compared to the particles' average thermal energies of $\sim 0.01$ 
eV. The first work which has performed a calculation of this cross 
section is that of \citet{Bhatia:1977}; 
more detailled calculations, taking into account short range  
interactions between the positron and the target electron (e.g.  
virtual formation of positronium), were performed by \citet{2002PhRvL..89l3201I}
 and \citet{1990JPhB...23.3057A}, for positrons colliding with H and H$_2$, respectively.
%; it has not been updated.

The widths of the lines resulting from these processes have been 
measured by Brown and Leventhal (1986) for H and by \citet{Iwata97}
for He and H$_2$; the values obtained were: 1.56 keV, and 2.50 
keV, and 1.71 keV for H, He, and H$_2$, respectively. These values have 
a very weak dependence on the temperature.

{\it e) Annihilation on dust grains}: the importance of this process was first pointed
out by \citet{Zurek:1985}, who stressed the important effect this process would
have on the Ps formation fraction, which is a quantity that can be inferred from
observational data (see section II.A) and thus represent an important constraint
on models;\citet{Guessoum:1991} then refined the calculation of the rate, adding
electric-charge and positron-grain reflection effects, on the one hand, and spectral
considerations (line width and effect on the overall calculated spectrum).
\citet{Guessoum:2005} have done the most extensive astrophysical treatment of this
process to date, despite the dearth of some crucial information on the processes,
considering the materials that constitute the dust grains; in particular, the widths
of the lines resulting from the annihilation of positrons inside the grain (after capture)
and from the decay of the positronium which is formed in/on the grain and ejected out
have been evaluated to $\Gamma_{Ps, in} \approx 2.0 $ keV and $\Gamma_{Ps, out} \approx 1.4 $ keV,
respectively; this is highly important in that it affects the amount of dust one will infer
from the galactic positron annihilation line spectra (see the discussion in \onlinecite{Guessoum:2004}
 and \onlinecite{Guessoum:2005}); note, incidentally, that due to the fact
that Ps inside the grain undergoes ``pick-off'' annihilation, it always gives two photons, never three. 

The corresponding  reaction rates,  taking into account  the (Maxwellian) 
energy distribution of the particles (positrons, electrons, atoms, or molecules) and 
the abundance and density of each species, are given by: 

\begin{equation}
  r_p = n < \sigma v > = \int_{E_T}^{\infty}{2 \over \sqrt{\pi}} 
{{\sqrt{E}\over {(kT)}^{3/2}} e^{-E/kT} \sigma(E) v dE}  \; .
 \label{eq:annrate}
\end{equation}

\noindent The reaction rates then allow one to determine the fraction 
of positrons which annihilate through each process, $f_p = r_p / \Sigma r_p$ (the index p 
generically referring to a process), and these fractions are then used to determine the spectrum 
of emission in a given physical medium.

\begin{figure}
\includegraphics[width=0.49\textwidth]{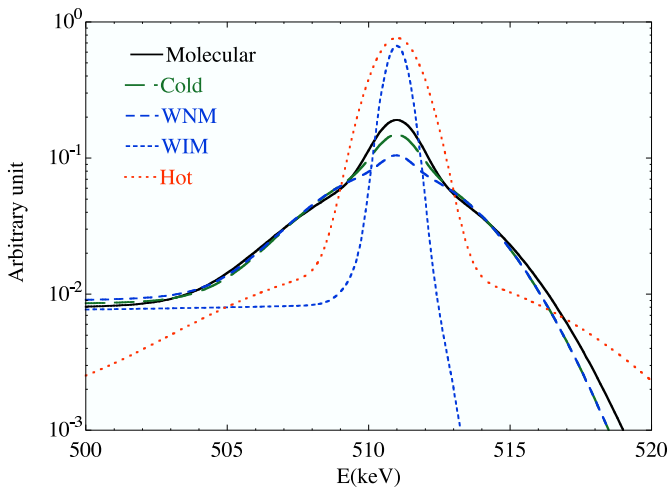}
\caption{\label{fig:specphase} Annihilation spectra for the five ISM 
phases.  Adopted temperatures (see Table \ref{tab:TableISM}) 
are 10 K (Molecular), 80 K (Cold), 8000 K (WNM and WIM) and 10$^6$ K (Hot) and
ionization fractions are 0 for the neutral phases (Molecular, Cold and WNM) and 1 for the ionized 
     phases (WIM and Hot).}
\end{figure}

The spectrum of gamma-ray emission  includes contributions 
from various processes: each one of them consists of either a Gaussian 
function describing the line emission at 511 keV with a given line width (FWHM denoted by 
$\Gamma$) or a Gaussian (denoted below by $G(E, E', \Gamma)$, of variable E' and 
center E) and an ortho-positronium continuum (at $ 0 < E  < 511$ keV), the latter 
given by the \citet{Ore-Powell:1949} function $P_t (E)$. The spectrum
is then given by:

\begin{eqnarray}
S(E) &=& \int dE' \left[ 3 \times {3 \over 4} P_t(E') + 2 \times {1 
\over 4} \delta(E'-E_0) \right] \nonumber \\
     &\{&  X \times f_{1, H/H_2} G(E, E', \Gamma_{if, H/H_2}) \nonumber \\
     &+& Y \times f_{1, He} G(E, E', \Gamma_{if, He}) \nonumber \\
     &+& \left( 1 - X \; f_{1, H/H_2} - Y \; f_{1, He} \right) \nonumber \\  
     &\times& [ f_{ce, H/H_2} \; G(E, E', \Gamma_{ce, H/H_2}) \nonumber \\  
     &+& f_{ce, He} \; G(E, E', \Gamma_{ce, He})  \nonumber \\
     &+& f_{rce} \; G(E, E', \Gamma_{rce})  \nonumber \\
     &+& f_{gr, out} \; G(E, E', \Gamma_{gr, out}) \, ]  \; \}\;  \nonumber \\
     &+& 2 \left( 1- X \; f_{1, H/H_2} - Y \; f_{1, He} \right) [ f_{dae} \; G(E, E_0, \Gamma_{dae}) \nonumber \\
     &+& f_{da, H/H_{2}} \; G(E, E_0, \Gamma_{da, H/H_2}) \nonumber \\
     &+& f_{da, He} \; G(E, E_0, \Gamma_{da, He}) \nonumber \\
     &+& f_{gr, in} \; G(E, E_0, \Gamma_{gr, in})] \; ,
    \label{eq:spec}
\end{eqnarray}
%%%%%%%%%%%%%%%%%%%%%%%%%%%%%%%%%%%%%%%%%%%%%%%%%%

\noindent where X and Y are the relative abundances of H(H$_2$) and He (90 \% and 
10\% respectively, by number) and $f_{1}$ is the fraction of positrons forming 
positronium in flight. The spectra are presented in 
Fig.~\ref{fig:specphase} for each phase of the ISM. In neutral media 
 the line is broad due to the annihilation of Ps formed in flight. 
The width of the line is $\sim$1 keV in the warm ionized phase where positrons 
annihilate mainly by radiative recombination with electrons.  
The contribution of the annihilation of positrons in grains is 
negligible in all the media except in the hot phase where it produces the $\sim$ 
2~keV width line superimposed on the broad line ($\sim$ 11 keV). The 
latter results from positrons that annihilate directly or via the radiative 
recombination process with electrons at a T $\sim$10$^6$ K.

%%%%%%%%%%%%%%%%%%%%% spectra vs. phases %%%%%%%%%%%%%%%%%%%%%%%%
%%%%%%%%%%%%%%%%%%%%%%%%%%%%%%%%%%%%%%%%%%%%%%%%%%%%%%%%%%%%%%%%%%%%

The ``global'' spectrum of annihilation of positrons in the ISM is then constructed 
by combining the spectra for each phase, considering the relative contributions 
(densities and filling factors) of each. For this, a model of the ISM, like
those briefly sketched in Sec. III 
\citep[e.g.][]{McKee-Ostriker:1977, ferriere98, Ferriere:1999, ferriere01} is needed.
The resulting ``global'' spectra can then be compared with the observational 
data.  Of course, it would be at least as interesting and useful to compare the 
individual phase spectra with observational data, but this has been impossible so far,
because of insufficient spatial resolution and sensitivity of  the detctors.

\subsection{Spectral analysis of observed emission}
\label{SecV_spect_anal}

The previous sections described the annihilation processes and the 
resulting characteristics of the possible annihilation emissions. One 
must distinguish the emission produced by the in-flight annihilation of 
relativistic positrons, which is characterized by a continuous 
spectrum in the MeV domain (see section V.B.1) from the 511~keV line and 
ortho-Ps continuum emissions produced in the annihilation of 
low-energy positrons (sections V.B.2 and V.D).

%%%%%%%%%%%%%%%%%%%%% spectra vs. phases %%%%%%%%%%%%%%%%%%%%%%%%
%\begin{figure}
%\includegraphics[width=0.49\textwidth]{sizun-spect.eps}
%\caption{\label{fig:sizun} From Sizun et al....       .}
%\end{figure}
%%%%%%%%%%%%%%%%%%%%%%%%%%%%%%%%%%%%%%%%%%%%%%%%%%%%%%%%%%%%%%%%%%

As already discussed in Sec. II.B.2, the observed MeV continuum in the direction of
the inner Galaxy significantly constrains the energies of injected positrons: they have to
be lower than a few MeV, otherwise the continuum emission would be much higher than
observed by COMPTEL (see Fig. \ref{fig:specaif} and \ref{fig:sizun}). 
This allows one to eliminate several classes of sources {\it in the steady-state regime}, such as
pulsars, ms pulsars, magnetars, or energetic proton collisions (either from
cosmic rays or from the central black hole), as major positron sources.
The same argument was used by \citet{Beacom-Yuksel:2006} 
and \citet{Sizun:2006} to constraint the mass of the decaying or annihilating 
dark matter particles  which could be the sources of positrons in the spheroid 
(see section IV.D) to lower than a few MeV. Note that the MeV continuum 
does not constrain the mass of de-exciting dark matter particles, as discussed in Sec. IV.D.
Moreover, \citet{Chernyshov+09} showed that the injection of positrons with initial kinetic energy higher than several GeV is allowed if the injection is non-stationary (e.g. through intermittent emission
from the central black hole) and if the magnetic field is higher than 0.4 mG in this region (see
also Sec. IV.5.a).

%%%%%%%%%%%%%%%%%%%%% spectra vs. phases %%%%%%%%%%%%%%%%%%%%%%%%
\begin{figure}
\includegraphics[angle=90, width=0.49\textwidth]{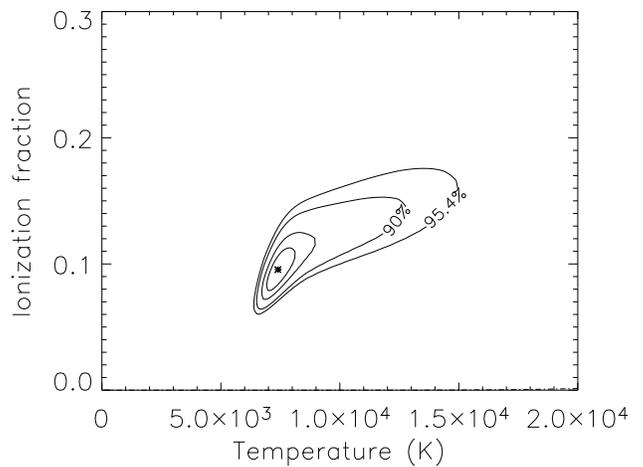}
\caption{\label{fig:specTYe} Confidence regions of the fit of the 
temperature and ionization fraction to the SPI data obtained after one 
year of mission. The best fit values are T = 7800 K and y$_{e}$=0.1.}
\end{figure}
%%%%%%%%%%%%%%%%%%%%%%%%%%%%%%%%%%%%%%%%%%%%%%%%%%%%%%%%%%%%%%%%%%%

Concerning the annihilation of low-energy positrons, the shape of the 
annihilation line and the relative intensity of the ortho-Ps 
continuum are closely related to the abundances and thermodynamical conditions (density, 
temperature, ionization fraction) of the plasma in which positrons annihilate. 
The broadening of the 511~keV line induced by bulk motions of the gas 
in which positrons annihilate is not taken into account since we do not expect a 
significant Doppler shift due to Galactic rotation and/or turbulence in the 
Galactic Centre region 
($\Delta \upsilon \lesssim$ 100 km/s, $\Delta$E $\lesssim$ 0.17 keV). Consequently, 
the spectral characteristics of the annihilation emission offer valuable 
information on the physical conditions of the ISM where positrons annihilate. 

The observed spectrum can be simply characterized by the sum of its 
independent components: the Gaussians (to describe the 511~keV 
lines) and the ortho-positronium continuum. The fraction of positrons 
annihilating via Ps ($f_{Ps}$) is derived from the ratio of the 
ortho-positronium flux ($I_{3\gamma}$) to the total line flux 
($I_{2\gamma}$) using Eq.~\ref{eq:fPos}.
The measured characteristics (widths, $f_{Ps}$) can then be compared 
with what we expect from the physics of the annihilation of positrons 
in the ISM (section V.D).

From the OSSE data, \citet{Kinzer:1996} inferred a positronium 
fraction $f_{Ps} \approx$0.97$\pm$0.03 in the Galactic Centre 
region. Measurements with the Ge detector 
TGRS onboard the WIND mission (1995-1997) gave a compatible 
value of 0.94$\pm$0.04 \citep{Harris:1998}. From the line width -- 
(1.8$\pm$0.5)~keV -- and the positronium fraction measurements, 
\citet{Harris:1998} concluded that a scenario in which annihilation does 
not occur either in cold molecular clouds or in the hot phase of the ISM is 
favored. Using preliminary SPI data of \citet{Jean:2003} and TGRS data of 
\citet{Harris:1998}, \citet{Guessoum:2004} showed that the bulk of the annihilation 
occurs in warm gas. However, the two groups did not exclude that a significant 
fraction of the annihilation may occur in hot gas and in interstellar dust. 

%%%%%%%%%%%%%%%%%%%%% spectra vs. phases %%%%%%%%%%%%%%%%%%%%%%%%
\begin{figure}
\includegraphics[width=0.49\textwidth]{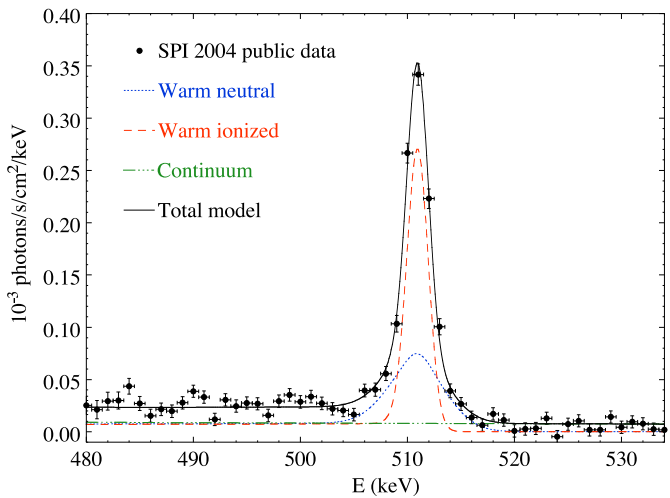}
\caption{\label{fig:fitphase} Best fit of the spectrum measured by 
SPI using the warm components of the ISM and the Galactic continuum. Contributions 
from the molecular, cold and hot components are not needed to explain the data 
\citep{Jean:2006}.}
\end{figure}
%%%%%%%%%%%%%%%%%%%%%%%%%%%%%%%%%%%%%%%%%%%%%%%%%%%%%%%%%%%%%%%%%%%%

Another approach consists in fitting annihilation models to the 
observed spectrum. Churazov et al. (2005) fitted the temperature and 
the ionized fraction of the gas where the annihilation occurs with a 
measured spectrum based on SPI observations of the Galactic centre 
region. They inferred from their analysis that the spectral parameters 
of the emission can be explained by positrons annihilating in a warm 
gas with a temperature ranging from 7000 to 40000 K and an ionized 
fraction $>$ 1 $\%$. However, they did not exclude a combination of warm 
and cold gases. When a similar analysis is performed using a spectrum 
measured with three times more exposure, a temperature T = 
(7.8$^{+0.8}_{-0.5}$) $\times 10^{3}$ K and an ionized fraction of 
(10$\pm$2)$\%$ (see Fig.~\ref{fig:specTYe}) are obtained (Jean, unpublished).

Instead of fitting the temperature and the ionized fraction to the 
SPI data, \citet{Jean:2006} adopted the spectral models for the 
different ISM phases (see Fig.~\ref{fig:fitphase}) and adjusted 
the phase fractions $f_{i}$ (with $i$=$\{$molecular, cold, warm neutral, 
warm ionized, hot$\}$) so as to obtain the best fit to the spectrum measured 
by SPI. This model is described by:

%%%%%%%%%%%%%%%%%%%%%%%%%%%%%%%%%%%%%%%%%%%%%%%%%%%%%%%%%%%%%%%%%%%%%%%%%%

\begin{equation}
    S_{ISM}(E) \, = \, I_{e^{+}e^{-}} \times \sum_{i=1}^{5} f_{i} 
\times 
    S_{i}(E,x_{gr}) \; ,
    \label{eq:spism}
\end{equation}
%%%%%%%%%%%%%%%%%%%%%%%%%%%%%%%%%%%%%%%%%%%%%%%%%%%%%%%%%%%%%%%%%%%%%%%%%%

\noindent
where $S_{i}(E,x_{gr})$ is the normalized spectral distribution 
(in keV$^{-1}$) of the annihilation photons in phase $i$, $I_{e^{+}e^{-}}$ is 
the annihilation flux (photons s$^{-1}$ cm$^{-2}$), and $x_{gr}$ 
represents the  fraction of dust grains and  allows for uncertainties 
in dust abundance and positron-grain reaction rates 
\citep[$x_{gr}$ = 1 in the standard grain model of][]{Guessoum:2005}.
With this method, \citet{Jean:2006} found that 49$^{+2}_{-23}$ \% of the annihilation 
emission come from the warm neutral phase and 51$^{+3}_{-2}$\% from the warm 
ionized phase. While they may not exclude 
that up to 23\% of the emission might come from cold gas, they have 
constrained the fraction of annihilation emission from molecular 
clouds and hot gas to be less than 8\% and 0.5\%, respectively, and the 
contribution of grains to be less than 1.2$\%$.

Finally, \citet{GJG10} examined the annihilation of
positrons on polycyclic aromatic hydrocarbon (PAH) molecules in ISM
conditions. They showed that PAHs would play a significant role only if their
abundances are higher than about $10^{-6}$ (by number), the low abundances
being compensated by the large enhancement in the PAHs' (resonant)
annihilation cross sections. They used the 511 keV spectrum measured by SPI
(Fig.~{\ref{fig:fitphase}) to constrain the PAH abundance in
the bulge and find an upper limit of $4.6 \times 10^{-7}$, consistent with
results obtained  from IR observations on other Galactic regions.

% Sec. VI Propagation..........................................................................................................................................................
%\input{SecVI_v5}

\section{ Positron propagation in the ISM}
\label{sec:Propagation}

\begin{figure*}[t]
\includegraphics*[width=0.99\textwidth,viewport= 20 30 600 235]{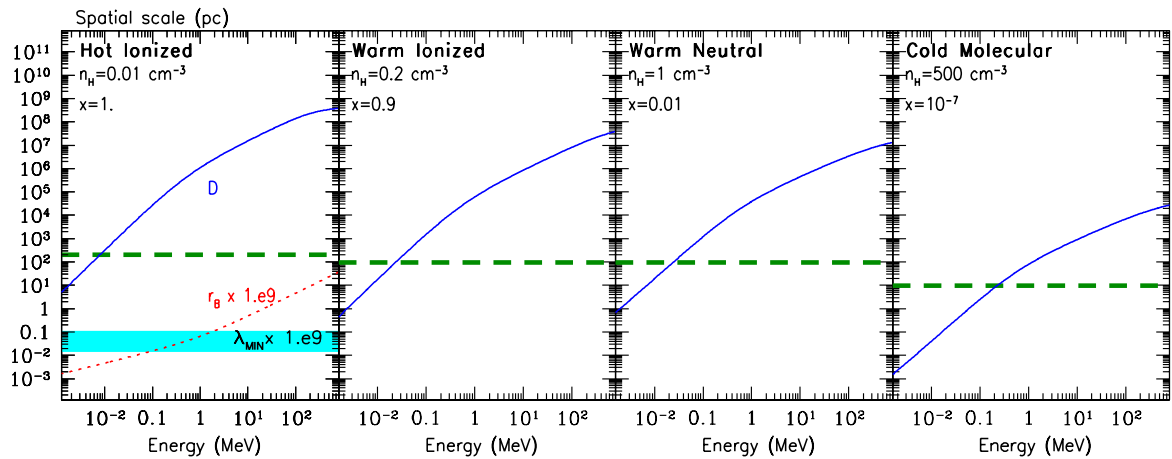}
%\includegraphics*[viewport= 52 168 52 380]{RMP_Pos_StopDist.eps}
%\includegraphics*[scale=0.9]{reaccel.eps}
%\plotone{fig3a.eps}
\caption{Positron stopping distance $D$ %({\it solid}) {\bf and}  Coulomb mean free path
%$\lambda_C$ ({\it dotted})
%and effective propagation distance $S\sim \sqrt{D \lambda_C}$ ({\it dashed}, assuming
%isotropic e$^+$ scattering, i.e. unmagnetized plasma)
in various phases of the
ISM (characterized by typical densities $n_H$ and ionization fractions $x$)
as a function of positron energy $E$. Horizontal dashed lines display typical sizes of the corresponding ISM phases.
In left panel, the lower {\it dashed} curve  displays positron gyroradius $r_g$ in a  magnetic field of $B=5$ $\mu$G, and
shaded aerea displays minimum scalelength $\lambda_{MIN}$ of MHD Alfven waves (the latter being calculated in \citet{JGMF09}; both $r_g$ and $\lambda_{MIN}$ are enhanced by a factor of 10$^9$.
}
\label{Fig:Pos_StopDist}
\end{figure*}

%Positron transport in the ISM plays a central role in understanding of the origin of Galactic 511 keV emission.
While most astrophysical sources are producing  relativistic
positrons (Sec. \ref{sec0:Processes}), the results of the spectral
analysis of the Galactic 511 keV emission (line width $\Delta
E/E<$0.01 and positronium fraction $f_{Ps}\sim$97\%) imply that
positrons annihilate at very low energies (Sec. \ref{sec:obs} and
V.E). This implies significant deceleration of positrons
on their way to the annihilation sites, through the various energy
loss processes summarized in Sec. \ref{sec:Interactions}. Furthermore,
various plasma processes - such as  advection due to the Galactic wind, adiabatic deceleration, 
energy gain due to stochastic acceleration and diffusion -  may affect
significantly the propagation of positrons and the extent of the
region of e$^+$ annihilation emission.
 
%The fate  of  the Galactic low energy positrons (deccelerated mainly through Coulomb collisions and, perhaps,  re-accelerated by turbulence or locally converging flows) depends on the level of the interstellar turbulence at small scales (i.e.  comparable to the e$^+$ gyroradii), which is  poorly constrained by current observations.
%The presence of small scale magnetic fluctuations is expected in the vicinity of the localized
%energy/momentum sources in the ISM like supernova remnants and powerful stellar winds
%Moreover, the energy density of low-energy cosmic rays (mostly protons)  is comparable to the energy density of other components of the ISM, as e.g. the magnetic field. A local anisotropy in the proton distribution  could then create magnetic turbulence in the ISM and set up preferential directions for the motion of positively charged positrons; therefore, transport of low-energy positrons may be somewhat different from transport of electrons.

There are two distinctly different regimes of positron transport in the ISM: {\it collisional} and {\it collisionless}. The former implies Coulomb  interactions of positrons with particles
in various gas phases of the ISM (in the presence of  radiation and magnetic fields), while the latter is essentially due to the
scattering off magnetic turbulence in the interstellar plasma. Observations of the
solar energetic particles in the heliosphere reveal that their
transport is indeed dominated by
 the scattering off  fluctuating magnetic fields (see Sec. VI.B.1).

In the collisional regime, positrons change their momentum and lose their energy through collisions with gas particles while
propagating along magnetic field lines in the ISM. Fig.~\ref{Fig:Pos_StopDist} shows the {\it stopping distances} $D=\int_0^E v(E') dE'/\dot{E'}$
(where $v(E)$ is positron's velocity and $\dot{E'}$ represents the
sum of energy losses of Sec. V.A)
%and the Coulomb mean free path lengths $\lambda_C=v(E) E/\dot{E}_{COU}$
in the various phases of the ISM. For MeV
positrons, $D$ %and $\lambda_C$ 
is much larger than the typical size of HIM,
WIM and WNM phases, and comparable to  the size of
CMM phase (horizontal lines in Fig.~\ref{Fig:Pos_StopDist})\footnote{
For a complete description of the ISM gas phases see Section \ref{sec1:GalaxyISM}}.
Therefore, in the collisional regime, only the CMM phase may be  efficient in stopping MeV positrons.

In the case of a magnetized  plasma,
positrons are spiraling along the magnetic
field lines. The gyroradius of a positron with  Lorentz
factor $\gamma$ is $r_g \sim$ 1.7 10$^9$ $B_{\mu G}^{-1}  (\gamma^2 - 1)^{1/2}$
cm, where the local mean magnetic field $B_{\mu G}$ is expressed in
$\mu$G. For typical values of the Galactic $B$-field (1-10 $\mu$G, see Sec. III.D), $r_g$ is
many orders of magnitude smaller than $\lambda_C$ in all ISM phases.
In a magnetized, turbulent, plasma,
the most efficient of collisionless processes is scattering  off  magnetic
fluctuations of size $r_B\simeq r_g$,
which induce {\it resonant} pitch angle scattering of positrons
(e.g. \onlinecite{kulsr05}, and references therein), or
{\it non-resonant} interactions with fluctuations on scales just above $r_g$  \citep[see
e.g.][]{topt85,ragot06}.  The transport of energetic ($>$GeV) GCR is driven by such collisionless
processes (Sec. IV.B.2), but in the case of MeV positrons the situation is not clear,
because there is no observational evidence on the level of
intestellar turbulence at such small scales (although one may reasonably expect  that
it can be quite high in the vicinity of  some positron sources, e.g. supernovae).

Because of the lack of observational evidence on the dominant regime of
propagation of low energy positrons, in this Section we consider both regimes (Sec.~\ref{S:Coltran} and
Sec.~\ref{N:Coll}, respectively). Some implications of this analysis for the spatial morphology of the Galactic
511 keV emission are presented in Sec.~\ref{sec1:Implications}.
%In the collisionless regime, we distinguish two cases,

%We shall consider below both collisional ( \S \ref{S:Coltran}) and collisionless regimes
%of the positron transport. In the latter case,  Sec.
%\ref{S:Loctran} evaluates the transport in the presence of
%small scale fluctuations, close to the positro sources; sec.
%\ref{S:Glotran} is devoted to the transport by large scale
%structures or fluctuations far from the sources, in the interstellar
%medium.

\subsection{Transport by collisions}
\label{S:Coltran}

The propagation of positrons in different gas phases in the collisional regime
is thoroughly studied in the recent work of  \citet{JGMF09}, with a  Monte Carlo code describing
e$^+$ "cooling" and annihilation\footnote{The process of e$^+$ cooling encompasses two distinctly different phases:
slowing down and thermalization. During the slowing-down period, the e$^+$ kinetic
energy is above the charge-exchange threshold and the
particles cool through binary or Coulomb interactions. Once the
e$^+$ kinetic energy is between the charge-exchange
energy and the thermal energy of the gas, the particles enter into a
thermalization phase where the elastic and/or Coulomb interactions
dominate. The thermalization ends once
the e$^+$ kinetic energy becomes close to the gas thermal energy
and positrons annihilate. \citet{JGMF09} found that the slowing-down process is always
the longer of the two.}.
The Monte Carlo simulation consists in calculating the trajectory of positrons (injected through a point source)
 along magnetic field lines, taking into account  collisions that change their pitch angle and energy. Their initial velocity distribution is assumed to be isotropic, so the cosine of their initial pitch-angle is distributed uniformly between -1 and 1. The initial kinetic energy of positrons is in the range 10 keV $ <$ E $<$ 10 MeV (appropriate for radioactivity emitted positrons). The simulation provide the spatial position of positrons at the end of their life. Analysis of a large number of test particles leads to the spatial distribution of the annihilation sites. 

The simulations show that the spatial field-aligned distributions of positrons at the end of their life are nearly uniform out to the maximum distance traveled along the field lines (corresponding to initial zero pitch angle). This implies that the pitch angle does not change significantly during the slowing-down period. For a given density, the most significant effect on positron propagation is found to be due to the variation of the ionization fraction in the neutral gas (warm neutral, cold neutral and molecular): as the ionization fraction increases, Coulomb scattering quickly becomes the dominant process of e$^+$ energy losses. 

Fig.~\ref{Fig:PosDistrib} shows the spatial extents of the final (before annihilation)
distribution of positrons along and perpendicular to the regular magnetic field, respectively, 
as a function of their initial kinetic energy, for each ISM phase. The extent along the regular magnetic field corresponds to approximately twice the maximum distance travelled by positrons and gives the size of the annihilation site
(even though positrons originate from a point source).
A "realistic"  Galactic magnetic field was assumed, composed of an average magnetic
uniform field $B$ plus a random field of intensity  $\delta B \simeq$ B, due to
the turbulent motions of the gas (Sec. \ref{sec1:GalaxyMagnField}).
The effective distances traveled by
positrons (i.e. in a straight line from their initial to final position) following the chaotic field lines
are smaller than in the case of a simple, uniform magnetic field, but only by 25\%.
The reason for such a surprisingly
small effect is that, because of the assumed injection scale
of turbulence from supernovae (10-100 pc) and the
adopted turbulence power-spectrum (Fig.~\ref{fig:Turb_spec}),
the number of small scale fluctuations interacting with positrons is quite small.
For that reason, the distances calculated by Monte-Carlo for a realistic Galactic
magnetic field in Fig. \ref{Fig:PosDistrib}
are quite similar to the stopping distances $D$  calculated semi-analytically
for a uniform average magnetic field in Fig. \ref{Fig:Pos_StopDist}.
In other words, the magnetic field rapidly orients the motion of
 randomly injected positrons
along its direction and MeV positrons travel an effective distance
$\sim S \sim$10 $n^{-1}$ kpc (where $n$ (cm$^{-3}$) is the ISM density)
from their origin before stopping by Coulomb collisions.
As already stressed, these distances are larger than typical sizes of
the warm ISM phases where Galactic positrons annihilate, according to the spectral analysis 
of 511 keV emission (Sec. \ref{SecV_spect_anal}).
This may imply either that the adopted model for turbulence is inadequate
(for instance, turbulence can show anisotropic or intermittent behaviour, instead of the
isotropic turbulence cascade assumed in \onlinecite{JGMF09}) or
 that positron transport (at least, in the warm phases) is dominated by
 collisionless processes. The latter issue is discussed in the next section.

\begin{figure}
{\includegraphics*[width=0.49\textwidth,]{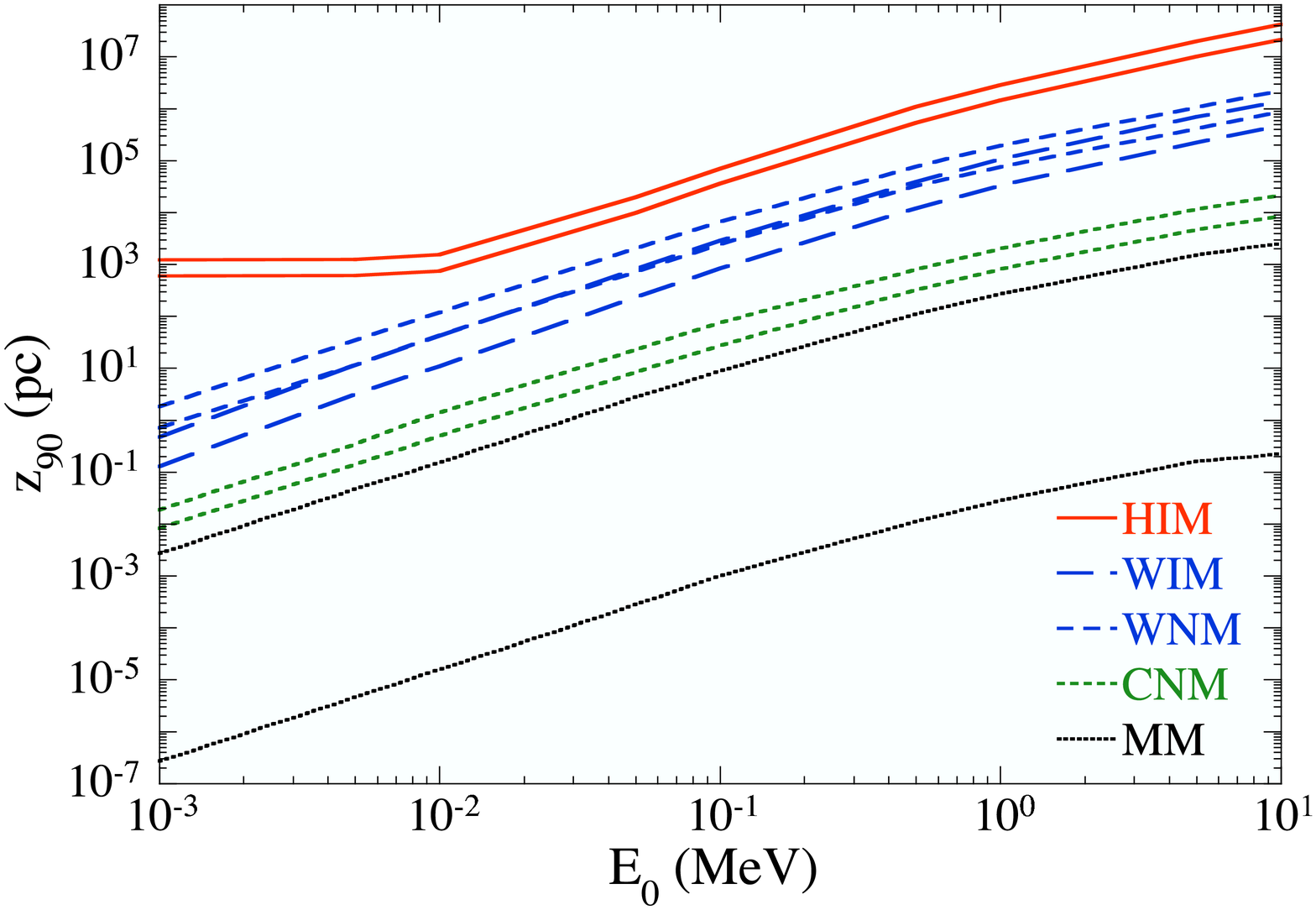}}
{\includegraphics*[width=0.49\textwidth,]{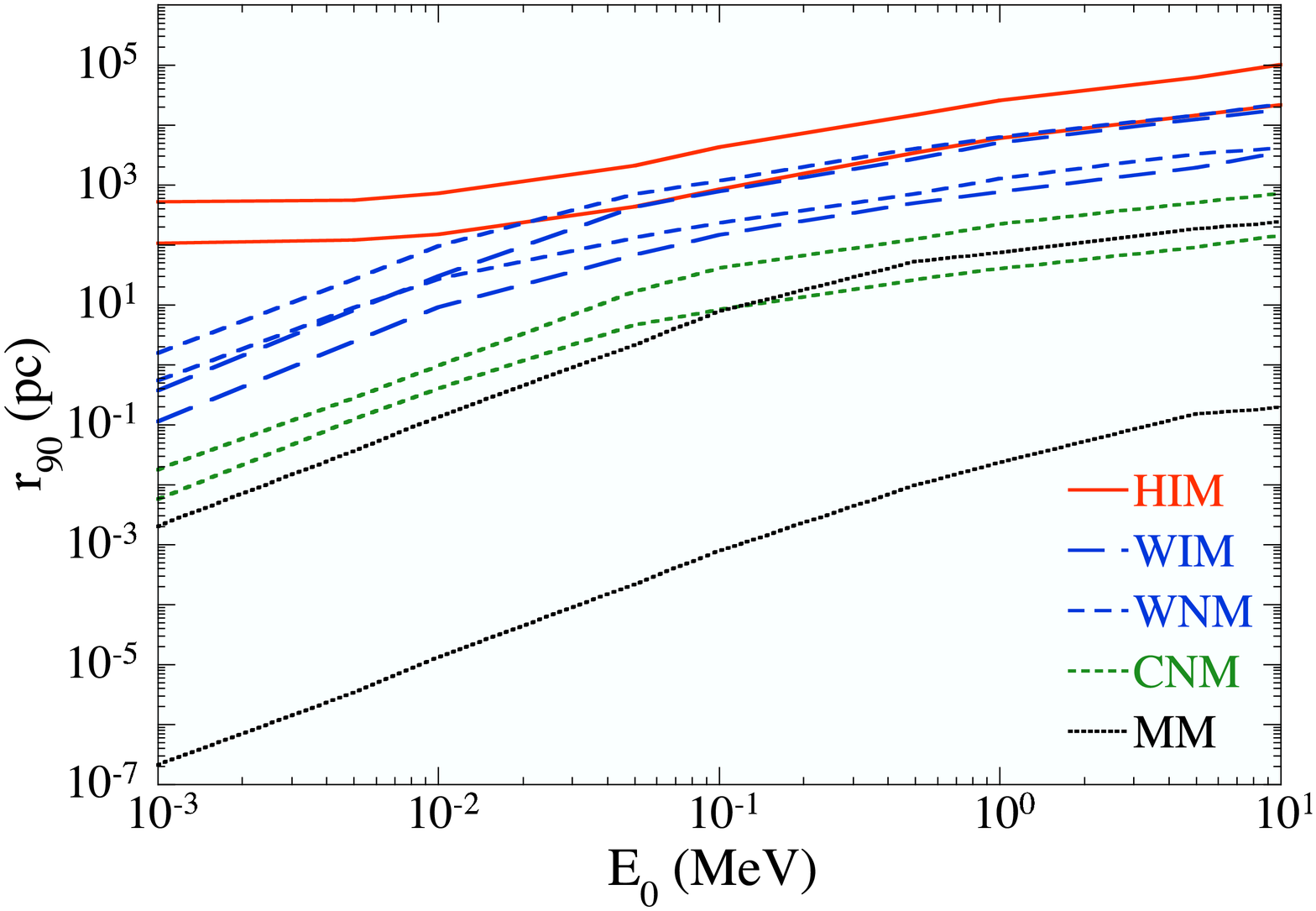}}
\caption{Minimum and maximum extents of the spatial distributions of 1 MeV
positrons slowing down to  100 eV, along (top) and perpendicular (bottom) to the regular Galactic magnetic field, taking
into account the turbulent behavior of the field lines as well as
realistic values for the density in each ISM phase (from \onlinecite{JGMF09}.}
\label{Fig:PosDistrib}
\end{figure}

\subsection{Wave-particle interactions and collisionless transport }
\label{N:Coll}

In the case of weak homogeneous  turbulence --
electromagnetic fluctuations over several orders of magnitude in
wavelength, homogenously distributed in space -- the wave-particle
interactions and collisionless charged particle transport are rather
well understood.
%AB: "fully developed" usually contradicts to weak, so I remove it...
 Since the gyro-radii of particles in interstellar
magnetic fields are usually much smaller than the relevant spatial
scales, one may use the so-called "gyro-phase averaged distribution"
of particles, which depends on four variables: time $t$, spatial
coordinate $s$ along the field lines, momentum $p$, and pitch angle
$\mu$.
% (the angle between the particle velocity and the local magnetic field).
The evolution of
the particle distribution, $f(t, s, p, \mu)$, can be described by
a Fokker-Planck equation \citep{melrose80}, while the injected spectrum of positrons
and its spatial distribution depend on the nature of the e$^+$ source
(see Sec.~\ref{sec0:Processes}).

Positrons undergo pitch-angle scattering described by the angular
diffusion coefficient $D_{\mu\mu}$ and stochastic acceleration by
interaction with plasma turbulence, described by the momentum
diffusion coefficient $D_{pp}$. In collisionless turbulent
interstellar plasmas, the kinetic coefficients in the Fokker-Planck
equation
%can be expressed through the correlation functions of stochastic electromagnetic fields. They
are dominated by resonant wave-positron interactions involving  both
{\it cyclotron}  and {\it Cherenkov} resonances\footnote{The {\it
electron cyclotron resonance} involves the gyro-motion of electrons
(or positrons) perpendicularly to the magnetic field: the transverse
electric field associated with the wave rotates at the same velocity
and in the same direction with the particles, which absorb its
energy and accelerate. The {\it Cherenkov resonance} involves
particle motion along the magnetic field lines.} \citep{melrose80}
and they can be evaluated in the "quasi-linear" theory, valid in the
weak turbulence regime \footnote{To calculate the diffusion
coefficients beyond the quasi-linear approach (e.g. for particle
transport and acceleration by strong turbulent fluctuations) the
re-normalization equations were developed by \citet{bt93} and
\citet{zankea04}, assuming that the particle propagation regime is
diffusive}.
%In the case of dense ISM phases with low level of magnetohydrodynamical (MHD) turbulence, Coulomb
%collisions may dominate  (see Sec. \ref{S:Glotran}).
%The effect of adiabatic focusing due to the mean magnetic field
%divergence is also accounted for in the Fokker-Plank equation. \\

Magnetohydrodynamical (MHD) waves\footnote{Under MHD
waves we mean shear Alfv\'en waves and fast
magnetosonic waves. Because the phase velocity of the magnetosonic mode is almost always 
larger than the Alfv\'en velocity $V_{\rm A}$, 
the magnetosonic wave is often called the "fast" hydromagnetic wave.
The dynamics of the third MHD mode, the
"slow" wave, has been shown to be entirely controlled by the Alfv\'en
wave cascade by \citet{lg01}; the slow wave spectrum is basically
the same as the Alfv\'en wave spectrum.}   are of
prime interest for the  transport of low energy positrons .
They can exist only at frequencies lower than the proton cyclotron frequency
$\Omega_{\rm cp} = (q B / m_{\rm p} c)$, where $q$ is the elementary charge,
and $m_p$ is the proton mass.
They are damped either by collisional effects (mainly viscous friction and ion-neutral collisions)
at low frequencies or
%, if they manage
%to survive collisional effects, then at frequencies approaching
%$\Omega_{\rm cp}$, they are heavily damped
by the Landau damping\footnote{Landau damping occurs due to the energy exchange between a wave with
phase velocity $v_P$ and particles  with velocity $v\simeq v_P$, which can interact strongly with the wave. Particles
with $v<v_P$  will be accelerated by the wave electric field, while those with $v >v_P$
will be decelerated, losing energy to the wave.}
(due to thermal protons) at frequencies approaching $\Omega_{\rm cp}$.
%In the latter case, they can induce {\it re-acceleration} of low energy positrons.
Higher-frequency waves are potentially important as well.
% as they can easily fulfill the resonance condition with  MeV positrons.
However, whistler waves produced by electron-proton plasma
instabilities, which are the most interesting waves in this frequency
domain, are right-handed polarized; therefore, they cannot be in
resonance with positrons, unless positron or proton flows generate
their own waves.
We shall restrict our analysis to  MHD waves with frequency
$\omega \ll \Omega_{\rm se}$, where  $\Omega_{\rm se}=\Omega_{\rm ce}/\gamma$ and
$\Omega_{\rm ce}=q B/m_e c$ are the synchrotron and the cyclotron frequencies of the positron,
respectively.

Collisionless processes may result in efficient deceleration  - or "cooling" -  of fast positrons
 (see, e.g. \onlinecite{pb08} and references therein).  The collisionless positron scattering due to
particle-wave interaction is efficient in warm ionized phases of the ISM
where the wave damping is not too strong (see eg.
\onlinecite{kulsr05}).
% As in the collisional regime, positrons suffer momentum losses due to Coulomb colisions, 
%but now the particles are coupled  to the magnetic field lines and 
%such collisions occur mostly during the propagation parallel to the mean magnetic field line. 
Moreover, {\it adiabatic deceleration} of positrons in jets or expanding shells (for example in SNRs) 
results in positron cooling,  even without Coulomb collisions; this occurs  
if the positron mean free path, which is  dominated by e$^+$ scattering by waves,  is
shorter than the typical scale of bulk plasma motion\footnote{
The adiabatic deceleration of a positron diffusing in an expanding shell has a
typical time scale  $\tau_{\rm ad} \sim {|\nabla \cdot \mathbf{\overline{U}}(\mathbf{r},t)|}^{-1}$, where
$\mathbf{\overline{U}}(\mathbf{r},t)$ is the bulk plasma velocity. The cooling time in  expanding wind, shell or jet (with $\nabla
\cdot \mathbf{\overline{U}}(\mathbf{r},t) >$ 0) can be much shorter than the Coulomb stopping time in a rarefied plasma.}.
Hence,  wave-particle interactions, both resonant and adiabatic non-resonant, could result in particle 
deceleration or re-acceleration, depending strongly on the local conditions.
The poor knowledge of those local conditions (concerning essentially the 
small-scale magnetic turbulence in the ISM)
imposes a case by case study and  precludes any generic conclusions  to be drawn.
Studies of wave-positron interactions have then to rely either on {\it in situ} measurements in the solar wind (Sec.~\ref{S:Loctran})
or on theoretical modeling of the properties of the MHD turbulence (Sec.~\ref{S:Glotran}) .

\subsubsection{Local e$^+$ transport in the ISM and re-acceleration}
\label{S:Loctran}

In the solar wind -- the only natural laboratory
for direct study of the propagation effects --
sub-MeV electrons and positrons
%of energy below MeV
resonantly interact with waves of high
enough frequencies that contain only a tiny fraction of energy of the
turbulent wave cascade (see e.g. \onlinecite{ragot05}).
%Their wavenumbers fall into the dissipative
%range of the solar wind turbulence; above the steepening observed at
%a fraction of the proton gyrofrequency \citep[see][and references
%therein for a discussion of the nature of the
%steepening]{leamonea98,alexandrovaea08}.
At 1 AU from the Sun  \footnote{1 Astronomical unit (AU)=1.5 10$^8$ km, the Earth-Sun distance.}
the solar wind turbulence steepens at a cut-off scale $V_{\rm A}/\Omega_{\rm ce}$
$\sim$10$^{7}$ cm, where the Alfv\'en velocity
$V_{\rm A} \sim$ 2.1 10$^5 B_{\mu G} \ n^{-1/2}$ cm s$^{-1}$
and $n$(cm$^{-3}$) is the local plasma density.
This means that for pitch-angles $\mu \le V_{\rm A}/\upsilon$ (where $\upsilon$ is
the positron velocity) sub-MeV particles  can no longer gyroresonate
%(either through synchrotron or Cherenkov resonance)
with waves in the inertial range of the solar wind turbulence.
%Low energy, sub-MeV particles are almost entirely out of the gyroresonance.
%We have already mentioned
%above that whistler waves cannot help much to scatter positrons in
%the standard scenario as they have an inverse polarization.
However, if non-resonant magnetosonic (i.e.  "fast") waves of lower frequencies
are present in the solar wind, they could
dominate the propagation of sub-MeV particles \citep[e.g.][]{topt85,ragot06}.
Depending on the level of non-resonant fast-mode wave turbulence and the
injection distribution of electrons, this can induce diffusion in energy space, i.e.
%invalidate the scatter-free assumption for the propagation
Fermi re-acceleration of sub-MeV electrons in the inner heliosphere \citep{ragot06}.
%Dynamics of strongly
%supra-Alfv\'enic sub-MeV electrons nonresonantly interacting with a
%spectrum of compressive waves in the solar wind were simulated by
%\citet{ragot06} using the equation for time variation of particle's
%pith-angle and applying some representation of the turbulence,
%including 10$^{15}$ turbulent modes. She found that the electrons
%can be reflected very quickly and even trapped along the magnetic
%field by the nonresonant waves. Reflection or trapping may involve a
%large fraction of the electrons with pitch angles in the range 40 -
%140 degrees. Electron reflection through tens of degrees of pitch
%angle occurs on timescales of the order of 1 s, providing a rather
%short transport length. If the magnetic fluctuations providing the
%particle pitch-angle scattering are due to MHD waves of non-zero
%phase velocity then electric fields of the waves necessarily provide
%also diffusion in the energy space i.e. Fermi re-acceleration.
The conditions for efficient re-acceleration can be estimated as follows:

\begin{figure}
\includegraphics[width=0.49\textwidth,]{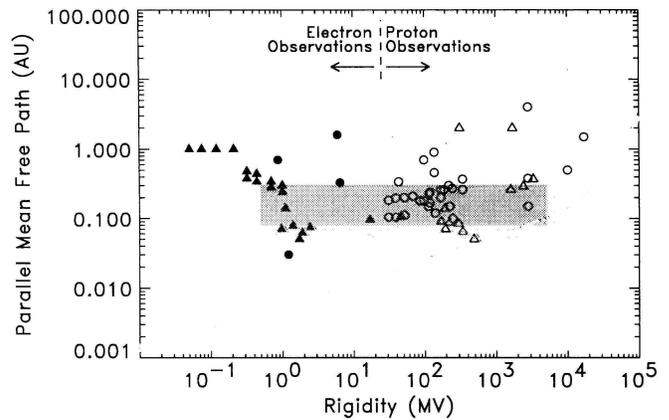}
%\includegraphics*[scale=0.9]{reaccel.eps}
%\plotone{fig3a.eps}
\caption{Mean free path of energetic particles parallel to magnetic field direction, as a function of
particle rigidity in the interplanetary medium;
circles and upward pointing triangles denote the measured values and
upper limits,  respectively  (from \onlinecite{bieberea94}). }
\label{bieber94}
\end{figure}

If positron velocities are above the thermal electron velocities,
%and the Alfv\'en velocity of the background magnetized plasmas is $V_{\rm A}$,
the ratio of the energy diffusion rate $D_{pp}/p^2$ and the
positron pitch angle scattering rate $D_{\mu\mu}$ (both in s$^{-1}$) is
\begin{equation}\label{reaccel}
D_{pp}/p^2 \ \ D_{\mu\mu}^{-1} \approx (V_{\rm A}/v)^2 \ll 1,
\end{equation}
\citep[e.g.][ and references therein]{pb08}.
%Even though the energy diffusion rate
%$D_{pp}/p^2 \approx (V_{\rm A}/v)^2\,
%D_{\mu\mu}$
%is typically slower than the pitch-angle scattering
%rate,
The re-acceleration effect is non-negligible when  the re-acceleration timescale is shorter
%related to the energy diffusion could be important if the particle propagation time
than  the energy loss timescale (or the e$^+$ propagation timescale).
% are longer than (or comparable with) the
%re-acceleration time. The Coulomb energy loss rate dominates the total
%positron energy losses at
%mildly relativistic
%MeV energies.
%(at energies $\sim m_e c^2$)
%in a medium of a local number density $n$ (measured in cm$^{-3}$).
%In that case
Stated differently,
the acceleration rate $D_{pp}/p^2$ must exceed the momentum loss rate
$\dot p_L/p = [dE/dt]/(m_e v^2 \gamma)$,  where - apart from fastly expanding regions - the energy loss rate of sub-MeV
particles is dominated by Coulomb losses (Eq.~(\ref{eq:dedtco}).
Therefore, for re-acceleration to be important,
the positron scattering rate by MHD waves $D_{\mu\mu}$ must satisfy the condition
\begin{equation}\label{reaccel1}
D_{\mu\mu} > 0.02\, \beta^{-1}\,\gamma^{-1}\, \frac{n_e^2}{B_{\mu G}^2}~ (s^{-1}),
\end{equation}
%where $\beta = v/c$.
which corresponds to a particle mean free path $\lambda (= 1/3 \ v D_{\mu \mu}^{-1})$ satisfying
%\footnote{{\bf To obtain this result we have used the equality in (Eq~(\ref{reaccel1})) and
%fixed the acceleration rate $D_{pp}/p^2$ to the Coulomb loss rate $\dot p_L/p$. Hence we have used the definition of the %mean free path $\lambda$ produced
%by the pitch-angle scattering at a rate $D^{-1}_{\mu\mu}$; i.e. $\lambda = 1/3 v/D_{\mu\mu}$.} This definition inserted into Eq.~({reaccel1}) gives Eq.~({Eq:lambda}).}
\begin{equation}
\label{Eq:lambda}
\lambda < 0.1\, \beta^2 \gamma  \, \frac{B_{\mu{\rm G}}^2}{n_e^2}\ {\rm (AU)}.
\end{equation}

In the solar wind, the electron mean free path $\lambda$ reaches a plateau
at energies $\sim$1 MeV with $\lambda$ about $0.01 - 1$ AU
\citep{ragot99}, depending on the parameters chosen for the interplanetary
medium (IPM). This is comparable to typical values of $\lambda$
estimated from observations of
electrons propagating in the inner heliosphere  (see Fig.~\ref{bieber94}). Interestingly,
those observations show no dependence
on electrons' rigidity (or momentum).
%adopted from \citet{bieberea94}.
%This implies potentially important role of
This can be interpreted as an indication that re-acceleration of low energy particles
is important in the case of the solar wind
(for further discussion see \onlinecite{droge00} and \onlinecite{shalchiea06}).

In the ISM, Eq. \ref{Eq:lambda} implies that the effect of re-acceleration is expected to be important
in the tenuous phases filled with strong MHD turbulence.
Fig.~\ref{figreaccel} shows the ISM parameter space (number
densities vs. mean magnetic fields) where MeV positrons may be affected by
re-acceleration, depending on their mean free path $\lambda$, which depends in its turn on the level
of MHD turbulence.
%In the shaded
%domains bounded by the lines corresponding to different particular
%values of the scattering mean free paths $\lambda$, positron
%re-acceleration is efficient.
Re-acceleration  becomes progressively important in regions
of low density and/or high magnetic field (such as those of the inner Bulge), resulting
%The mean free path
%is determined by the local magnetic turbulence. We parameterized
%the MeV positron scattering mean free paths to be in the
%range from 1 to 1000 AU to illustrate the effect.
in longer thermalization timescales for positrons
%and correspondingly lower annihilation rate.
%In turn, this
%Thus the re-acceleration effect
and increasing the  sizes of corresponding e$^+$ annihilation regions.
%Thus, positrons might escape the ISM phases where they are produced, even if first-order estimates (Fig. \ref{Fig:Pos_StopDist})
%suggest that they should be confined there.
%even in the case of small scattering mean free paths of the
%MeV positrons $\lambda <$ 0.01 pc.
%Notice hat in the  case of large
%$\lambda >$ 10 pc, the slowing down lengths also become long
%enough, typically exceeding 1 kpc for $n < 0.1 \cmc$ \citep{HLR09}.

\begin{figure}
\includegraphics[width=0.47\textwidth]{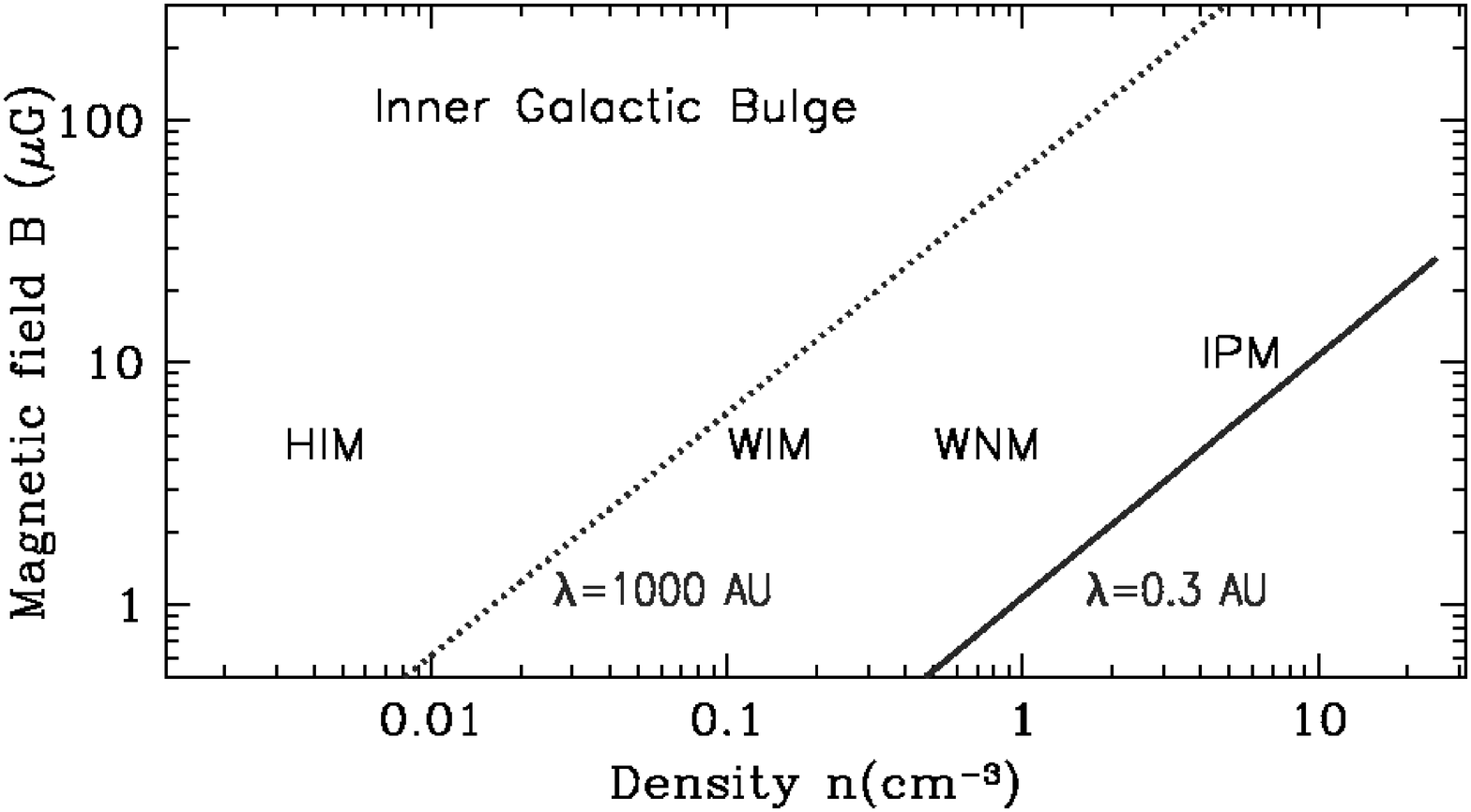}
%\includegraphics*[scale=0.9]{reaccel.eps}
%\plotone{fig3a.eps}
\caption{
%The ISM parameter space for regions of efficient
%in-situ re-acceleration of MeV positrons ($\gamma_{e^{+}}$ = 2.1) by
%small scale MHD waves. Shaded regions are bound by the lines corresponding to different
%values of the positron mean free path $\lambda = 1, 10, 100, 1000$ AU.  \label{figreaccel}}
Regions of the ISM in the $B$ vs $n$ (magnetic field vs  number density) plane. Physical conditions for
HIM, WIM, WNM and Inner Bulge are discussed in Sec. III, while IPM stands for the interplanetary medium.
The two lines correspond to Eq.~\ref{Eq:lambda} for MeV positrons, with the particle mean free path
$\lambda$=0.3 AU ({\it solid}, appropriate for the
 interplanetary medium from Fig.~\ref{bieber94}), and  $\lambda$=1000 AU ({\it dotted},
 provided for illustration purposes only, since its value is unknown in the ISM).
In regions of the ISM found to the left of those lines, re-acceleration effects may be important
 for positrons (provided there is a sufficient level of small-scale turbulence), 
 as suggested from observations of the IPM.}
  \label{figreaccel}
\end{figure}

Unfortunately, our knowledge of the small-scale ISM turbulence is limited at present.
While the quasi-linear theory provides a simple
analytical description of charged particle transport, recent
test particle simulations revealed some problems of this
approach \citep[e.g.][]{tautzea06}. In particular, while non-linear effects
were shown to be essential for particle transport in MHD turbulence
\citep[e.g.][]{matthaeusea03,zankea04,lerouxwebb07},
test-particle simulations of the coefficient $D_{pp}$
do not exist at present. Therefore, one has to  rely on  simulations of
$D_{\mu\mu}$ and use that quantity in order to estimate $D_{pp}$ from
Eq.~(\ref{reaccel}); in their turn, simulations of  $D_{\mu\mu}$
require assumptions on the polarization and spectral properties of MHD waves.

 The aforementioned uncertainties affect any
attempt to evaluate the propagation of MeV positrons in the Galaxy without
detailed information on the magnetic fluctuation spectra in the
propagation region. In a recent work, \citet{HLR09} adopted a particular
model calculation of the charged particle mean free path in the
interplanetary medium (in the quasi-linear regime), which
they extrapolated to the conditions of the ISM (see Sec. VI.C for
details of their model). However, in view of the different
conditions  - density, magnetic field and unknown level of
small-scale turbulence - in the Bulge, such a calculation appears
rather arbitrary (although not necessarily wrong) . Their model can be considered as a quantitative
illustration of a possible scenario, the plausibility of which remains to be shown.

Notice that small-scale MHD waves can be injected into the ISM in-situ by a
variety of {\it kinetic instabilities.}
%These waves can be directly injected into the collisionless regime.
%Each case requires a
%dedicated investigation of the wave growth and damping processes.
A kinetic instability may be locally triggered, for instance, by the
streaming (through the ISM) of cosmic rays with  bulk velocity larger
than a few times  the local Alfv\'en speed \citep{wentzel74}. Such
streaming instability is expected to develop in the intercloud
medium and may provide particle re-acceleration, compensating for the strong ionization losses
inside molecular clouds. In other terms, low-energy cosmic rays scatter off their
self-generated waves and are, therefore, pushed outside molecular
clouds \citep{ss76}. The waves thus generated can help to confine the positrons in the warm phases
around the molecular clouds. This mechanism was adapted to the transport of
cosmic-ray electrons by \citet{morfill82}, but
%A preliminary study was performed by \citet{Jean2009}. The authors concluded that
it operates over a limited range of positron momenta
and its ability to confine positrons within the ionized phases
of the ISM is questionable \citep{JGMF09}. \citet{HLR09} proposed
an alternative mechanism in which positrons scatter
off their own self-generated waves.  It is possible that kinetic instabilities
 play a key role in the observed  e$^+$ 
 annihilation (mostly) in the warm phases of the ISM.

\subsubsection{Global positron transport}
\label{S:Glotran}

Description of positron transport on a small scale $r_g<L<\lambda$
(where $\lambda \sim v\,D_{\mu\mu}^{-1}$ is the
positron mean free path)
requires the solution of the Fokker-Planck equation.
On much larger spatial scales (L$>  \lambda )$) the pitch-angle distribution is nearly isotropic. This justifies
the use of a simplified Fokker-Planck equation, in the
so-called {\sl diffusion approximation};
the corresponding equation is called the {\sl diffusion-advection equation}
%The Fokker-Planck equation is needed to describe the
%positron propagation on the spatial scale $r_g < l < v\,
%D_{\mu\mu}^{-1}$.  Since the coefficient $D_{\mu\mu}$ is a
%pitch-angle scattering rate, the pitch-angle distribution will be
%nearly-isotropic at spatial scales larger than the positron mean
%free path $\lambda \sim v\, D_{\mu\mu}^{-1}$. Greatly simplified
%transport equations can be obtained from the general Fokker-Planck
%equation to describe the positron propagation on the scales $l >>
%\lambda$. One can use in that case the {\sl diffusion approximation}
%which leads to the {\sl diffusion-convection equation}.
%\bigskip
%\noindent{\it The diffusion-advection transport}
%\medskip
%The large scale plasma motions have %the mean
% spatial scales $L$ well in excess of the
%particle mean free path $L \gg \lambda$. Therefore,
%particle transport can be well described by the
%diffusion-advection equation.
%Spatial diffusion $\chi_{\alpha \beta}$ and the momentum diffusion $D$
%coefficients are expressed in terms of spectral functions that
%describe correlation properties of large scale turbulent motions by
%renormalization equations
\citep{bt93,bykov01}.
%The large scale velocity field is described by the spectral
%correlation function of velocity $\langle u_{\alpha} u_{\beta}\rangle$
%and the magnetic field
%$\langle B_{\alpha} B_{\beta}\rangle$ that determine the transport of particles
%on the spatial scales $>L$.
If the amplitude of velocity
fluctuations is $u > v\, \lambda /L$, the particle transport is dominated by the turbulent
advection.

It is important to note that only compressible large scale
turbulent motions provide  efficient re-acceleration
\citep[][]{bt93}. In the regime dominated by turbulent advection
(TA),
%(i.e. for $u > v\, \lambda /L$),
the re-acceleration rate can be approximated as $D_{\rm pp-TA}
\approx (u/9L)p^2$, e.g.  \citet{bt93,bt01}.
%(Bykov and Toptygin 1993, Bykov and Toptygin 2001).
The re-acceleration effect is important if $D_{\rm
pp-TA}/p^2
> \dot{p}_L/p $, where the stopping rate due to Coulomb collisions
$\dot{p}_L/p$ is discussed in Sec.~\ref{S:Loctran}. Therefore,
particle re-acceleration would significantly affect the global
positron propagation in the bulge if the amplitude of compressible
turbulence $u_{50}$ (measured in units of 50 km/s) and the energy
containing scale $L_{50}$ (measured in units of 50 pc) satisfy the
condition\footnote{In case of important energy losses, other than
from Coulomb collisions, higher values of $u_{50}/L_{50}$ than given in Eq.~\ref{reaccel2}
are required.}
\begin{equation}\label{reaccel2}
u_{50}/L_{50} > 300\, \frac{n}{ \beta^3\, \gamma}.
\end{equation}

A word of caution is in order here. The re-acceleration effect, in
principle, could boost a substantial amount of mildly relativistic
MeV-positrons to ultra-relativistic energies. It might therefore lead to
violation of  the limit on $\gamma$-ray continuum emission from
in-flight annihilation of positrons imposed by Beacom and
Y\"{u}ksel (2006), which  was discussed  in Section IV. In
the GC vicinity and in the presence of a strong magnetic field,
 the severe synchrotron-Compton losses of
 relativistic positrons can make it impossible to accelerate them
 to energies above a few MeV by the Fermi-type mechanism. 
In the absence of synchrotron-Compton losses the differential
spectral index of the Fermi re-accelerated positron momentum
distribution $N(p) \propto p^{-a}$ can be approximated as $a =
-1/2 + \sqrt{9/4 + 36 (L/R)/2}$, where $R$ is the minimal size of the
confinement region (if anisotropic);   in the bulge $R$ is likely
below 1 kpc and it is much less in the disk. Exact estimates of the
positron spectra can be done only if one knows the energy (or momentum)
dependence of the positron mean free path in the MeV regime, which 
is governed by yet uncertain small scale magnetic fluctuations
in the regions of interest.

The question of whether MHD cascades may extend over several
orders of magnitude on spatial scales is still a matter of active debate.
Electron density power spectra in the local ISM
have been measured down to scales of 10$^{10}$ cm
%by means of analysis of angular broadening and decorrelation bandwidth
\citep[e.g.][]{ars95}. It was established that radio
scattering in the interstellar plasma implies widespread
inhomogeneities in density with a power law spatial spectrum and
suggests a turbulent origin. Observations suggest that the small
scale plasma turbulence is often highly anisotropic and clumpy \citep{Shebalin1983,Desaj2001,Brisken2010} ; its fluctuations
%The anisotropic plasma fluctuations
are likely to concentrate in
filamentary density structures aligned by the local magnetic field
\citep[e.g.][]{higdon84}. Since the scattering would become
isotropic if the irregularities were uniformly distributed over many
length scales of the magnetic field, \citet{rc04} concluded that the
plasma turbulence is distributed  intermittently. %{\bf SO WHAT ???}
%The idea is also supported by the success of the screen model for pulsar scattering
%over paths up to a kiloparsec ({\bf imos: what's this?}).

To estimate the positron scattering
effect from the electron density measurements one has to assume a
relationship between the amplitudes of density and magnetic field
fluctuations. Fluctuations  due to magnetosonic waves could
contribute to non-resonant scattering of MeV particles or  - if they  survive down
to small scales - to resonant scattering. However, if  the observed electron
density fluctuations are simply due to entropy-type ("isobaric")
%\footnote{\bf vortex entropy fluctuations solutions of the equation of hydrodynamics.
%Entropy fluctuations have density perturbations but are isobaric}
fluctuations, they are inefficient in e$^+$ scattering.

Depending on the ISM phase, the MHD waves can suffer from collisionless or collisional
damping.  \citet{JGMF09} concluded that in the neutral atomic and molecular
phases of the ISM the Alfv\'en and fast magnetosonic wave cascades
are both cut off by ion-neutral collisions,  on
scales considerably larger than the gyroradius of MeV
positrons; therefore, MHD waves cannot resonantly
interact with positrons. The situation is  different in
the ionized phases of the ISM, where the Alfv\'en wave cascade
suffers insignificant (collisional) damping down to the
thermal proton Coulomb mean free-path
\footnote{On smaller scales,  the cascade enters the collisionless regime.
and cuts off by  linear
Landau damping around the proton inertial length  $c/\omega_{\rm pp}$ (where  $\omega_{\rm pp} =
4\pi n_{\rm H} e^2/ m_p$).}. 
This leaves some room for possible resonant interactions
%of the extended inertial range
of the Alfv\'en wave cascade with positrons\footnote{
Notice that, theoretical considerations show that the energy transfer in the
Alfv\'en cascade proceeds mostly through the interaction of
oppositely propagating wave packets. The distortion of the wave
packets during the interaction produces anisotropic fluctuations
elongated along the mean magnetic field \citep{lg01}. Using
numerical simulation, the  magnetosonic cascade has been found
to keep its isotropy along the whole range of spatial scales \citep{yl04}.
In case of strong anisotropy of the cascade, the scattering efficiency of
MeV positrons by small scale Alfv\'en waves could be considerably reduced while the effect
of magnetosonic waves could be enhanced.} but the effect has not been properly investigated up to now.

 In summary, contrary to the case of the solar wind (where
 nonresonant compressible perturbations  may control the
 positron mean free path),  the cascade of MHD waves in the ISM
(injected  at scales $L_{\rm inj} \sim 10-100$ pc by
 supernovae)  is expected to be damped at scales well  above
  $r_{\rm g}$. Small scale MHD (and whistler) waves could be  generated
 by anisotropic distributions of energetic particles, through streaming instabilities,
 but the waves could be damped by ion-neutral collisions as well \citep{HLR09}; in any case,
 the effects have  not been studied in detail up to now. Streaming instabilities
 could help to confine  positrons at the border of the molecular clouds, thus
enhancing the fraction of positrons that annihilate in warm phases.
 In ISM regions of low density (and/or high magnetic field)
  with even a moderate level of MHD
 turbulence (see Fig.~\ref{figreaccel}) particle re-acceleration
  could substantially suppress the positron stopping by Coulomb
 collisions (even in the absence of any other cooling effect, e.g. adiabatic losses), thus providing a
 possibility to extend largely the volume filled with annihilating positrons. Most of the
 current models rely on quasi-linear wave treatment, but nonlinear
 effects could substantially change some estimates. Notice that, in general, a quasi-homogeneous
 distribution of the  sources of turbulence is assumed, although the
 intermittency effects (due to e.g. large scale ISM shocks) can modify
 the analysis. In Table \ref{TabSum} we list  processes that
 are potentially important for positron propagation in the ISM (see
 also \onlinecite{JGMF09})  and which require further investigation.
 Our current understanding of MeV positron propagation does not allow one to
 conclude whether such positrons undergo strong diffusion or essentially free propagation.

 \begin{table}[t]
 \caption{Collionless and collisional transport processes
 in  ISM phases (see Table IV).}
% \begin{tabular}{|c|c|c|c|c|c|}
 \begin{tabular}{c c c c c c }
 %\hline
 \hline
  &\multicolumn{5}{ c }{ISM phase} \\
 {\rm Transport processes}   & CM & CN & WN & WI & HI \\
 \hline
 {\rm NR MHD modes: Large scale} & N & N & N & Y & Y \\
 {\rm R MHD modes} {\rm Alfv\`en} & N & N & N & Y & Y \\
 {\rm R MHD modes} {\rm Fast MS} & N & N & N & N (Y) & N (Y) \\
 {\rm Streaming modes} & Y? & Y? & Y? & Y & Y \\
 {\rm Collisions} & Y & Y & Y & Y & Y  \\
 \hline
 %\hline
 \end{tabular}
 Y (N) means that the process can (cannot) take
 place. NR (R) stands for nonresonant (resonant) MHD motions
 at large scales.  Streaming
 modes are generated through the streaming instability produced by
 the CRs propagating away from their sources 
 %(the cosmic ray sources like SNRs, superbubbles, X-ray binaries could be distributed around  molecular cloud complexes.) 
 The streaming processes not yet studied in  depth are indicated  with question marks.
  \label{TabSum}
 \end{table}

\subsection{Implications of e$^+$ propagation for 511 keV emission}
\label{sec1:Implications}

The implications of low energy positron transport for the Galactic
511 keV emission were  raised by \citet{Prantzos06}, who
pointed out that the morphology of the 511 keV emission does not
necessarily reflect the morphology of the underlying e$^+$ source
distribution. He noticed that e$^+$ from SNIa are released in the
hot and rarefied ionized medium, since the scaleheight of SNIa is
considerably larger than the scaleheight of the disk of cool,
dense,  gas (see Fig.~\ref{SFRdens}). The e$^+$ propagation
distances are then quite large (Fig.~\ref{Fig:Pos_StopDist}),
allowing e$^+$  from the disk to annihilate far away from their
sources (perhaps in the halo, where a low surface brightness
emission should  be expected); this fact may considerably reduce
the bulge/disk ratio of 511 keV emission of any class of
astrophysical e$^+$ sources, thus alleviating the morphology
problem discussed in Sec. IV. D. Models of cosmic ray propagation in the Galaxy
(like those described in Sec.~\ref{subsubsec:GalaxyCR})
usually adopt an isotropic diffusion coefficient. \citet{GdB09} introduced recently
an anisotropic diffusion coefficient
in the GALPROP code to simulate the advection of cosmic rays by a Galactic wind. They find
that, by adopting observationally derived wind velocities, their scheme naturally produces large escape
fractions ($>$50\%) of positrons from the disk.

Furthermore, \citet{Prantzos06}
suggested that {\it if} the halo magnetic field of the Milky Way
has a strong poloidal component, as suggested by several authors
(e.g. \onlinecite{Han04}; \onlinecite{Pv03} and Fig.~\ref{Fig:MagnFields}, {\it top}) then some positrons
escaping the disk may be channeled into the bulge and annihilate
there, enhancing even more the bulge/disk e$^+$ annihilation
ratio; he noticed that, in that case, positrons from SNIa may
suffice to explain {\it quantitatively both} the total observed
e$^+$ annihilation rate ($\sim$2 10$^{43}$ e$^+$ s$^{-1}$) {\it
and} the corresponding bulge/disk ratio,  {\it provided} that the
escaping e$^+$ fraction from SNIa is $\sim$4\%  .
As discussed in  Sec.~\ref{sec1:GalaxyMagnField}, it is rather unlikely  
- although it cannot yet be ruled out -
that  the  poloidal component of the regular Galactic magnetic field
is close to a dipole. Observations of external spirals
suggest rather an X-shaped halo field (e.g. Fig.~\ref{Fig:MagnFields}, {\it bottom}),
in which case it would be difficult for disk positrons to find their way into the bulge.
Still, the issue is of considerable interest and urgently calls for a better assessment of the poorly known
global configuration of the Galactic magnetic field.

 %\begin{figure}
 %\includegraphics[width=0.49\textwidth]{MagnFieldX.eps}
%  \caption{Radio continuum emission of the edge-on spiral galaxy NGC~4631 at
%$\lambda3.6$cm (8.35~GHz) with the 100-m Effelsberg telescope with $84 \arcsec~HPBW $.
%The contours in give the total intensities, the vectors the intrinsic magnetic field orientation.
%The radio map is overlayed on an optical image of NGC~4631 taken with the Misti
%Mountain Observatory. (Copyright: MPIfR Bonn)
%}
%\label{n4631}
%\end{figure}

\begin{figure}
\includegraphics[width=0.49\textwidth]{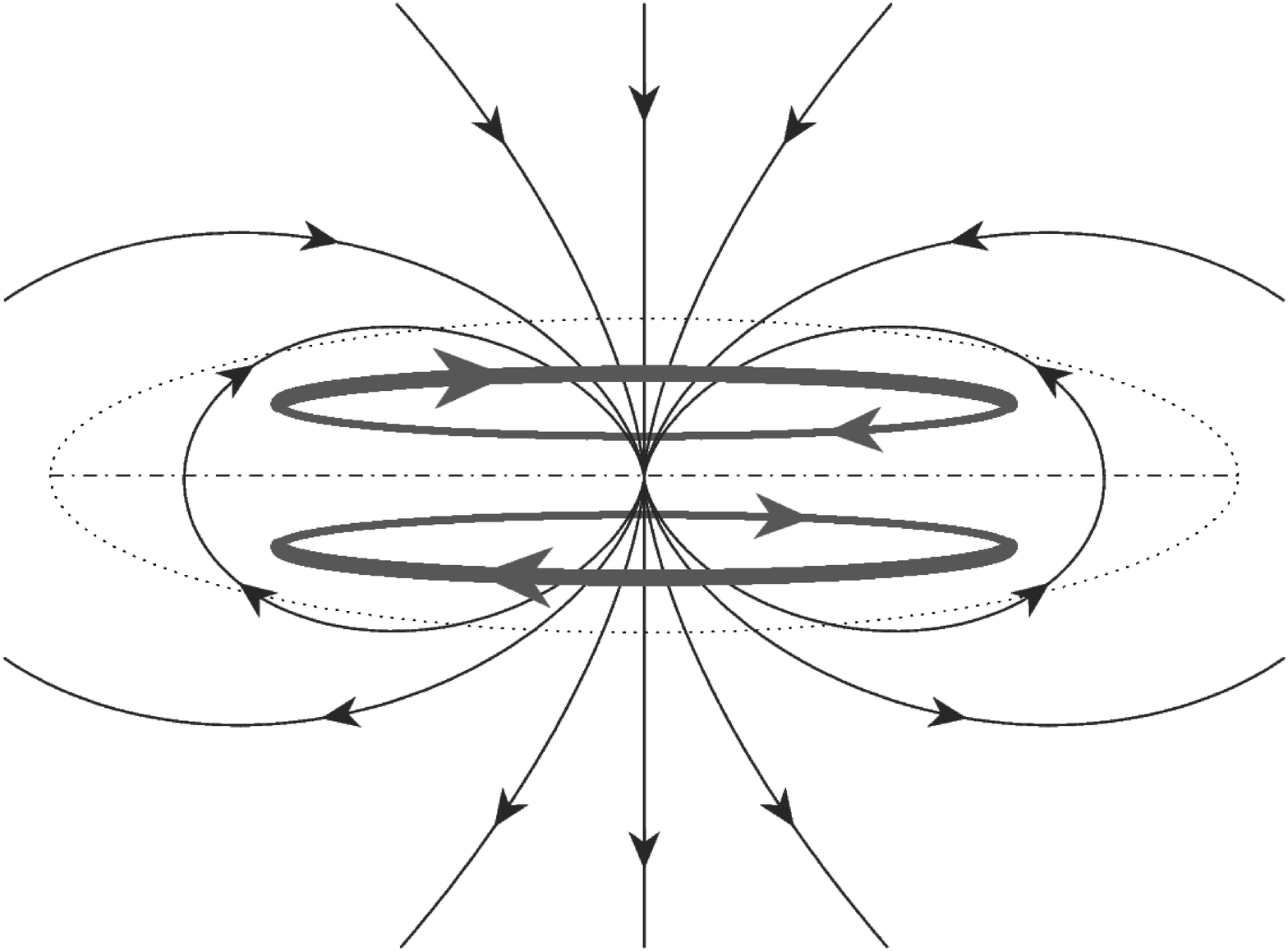}
\includegraphics[width=0.49\textwidth]{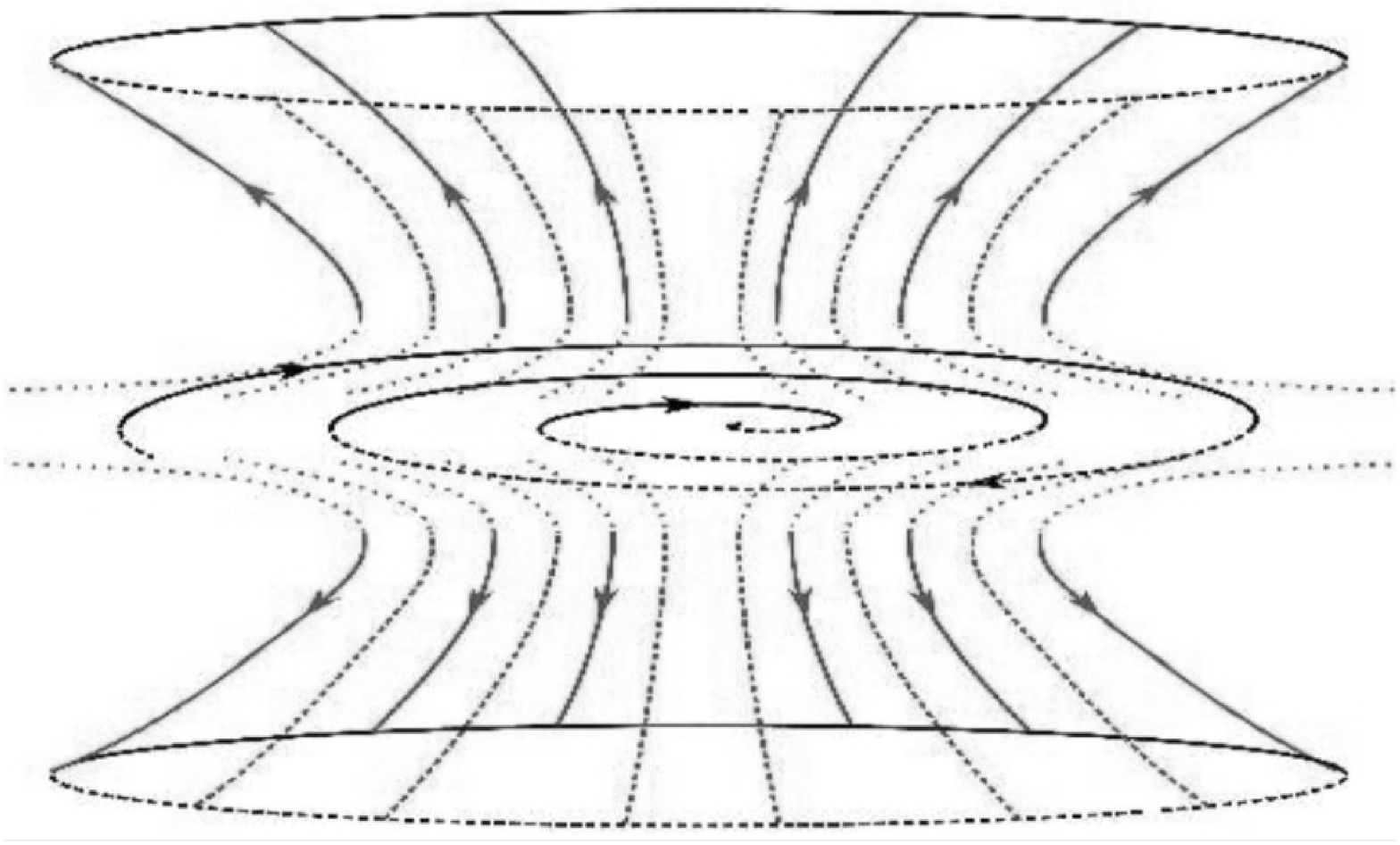}
\caption{Possible configurations of the large scale magnetic field of the Milky Way. {\it Top}: As derived from
Faraday polarization measurements of the MW, according to \citet{Han04}. {\it Bottom}: Scetch of the observable components
of the large scale magnetic field of the disk galaxy NGC253 (which shows, however, signs of starburst activity, unlike the Milky Way); the halo magnetic field is even and pointing outwards, whereas the dotted parts are not observed (from \onlinecite{HBKD09}).
}
\label{Fig:MagnFields}
\end{figure}

%A number of models discussed above were recently proposed to
%understand the origin and the  observed properties of the 511 keV emission.
\citet{HLR09} suggested that positron propagation may be
the key for understanding not only the spatial morphology of the 511
keV emission, but also its spectral properties.  They made the bold
assumption that radioactivity (from $^{26}$Al, $^{44}$Ti and,
mostly, from $^{56}$Co, see Sec. IV.A) is the sole e$^+$ source in
the Galaxy. They considered (i) a fairly detailed description of the
various phases of the ISM (Sec. III.B) and (ii) a particular phenomenological model
of collisionless scattering of MeV positrons by turbulent
fluctuations of the ISM, that was used to describe energetic
particle (electron or proton) propagation in the interplanetary medium.   Depending on the
nature of the medium, positrons are assumed to propagate either by
diffusion along magnetic flux tubes (in ionized media, where
turbulence cascades down to the gyroradius $r_g$) or by streaming
with mean velocity $v=\beta c$ (in neutral media, where turbulence is
quenched by ion-neutral collisions at scales $>> r_g$).

Putting together  the aforementioned ingredients, \citet{HLR09}
proceed then in an impressive calculation of positron
production rates from radioactivity along the Galaxy, typical
distances of e$^+$ propagation in the corresponding ISM phases and
probabilities that positrons will finally annihilate in one or
another of those phases. In the end, they find excellent agreement
with each and every observable of the 511 keV emission available
so far (Sec. II.D): spatial morphology, i.e. bulge/disk
annihilation ratio, spectral futures (including a narrow and a
broad 511 keV line) and even the claimed asymmetry between fluxes
from negative and positive longitudes;  they explain the latter as
due to the corresponding overall asymmetry of the spiral arm
pattern of the Milky Way disk, as viewed from the Sun.

The work of \citet{HLR09} constitutes the first study of Galactic
e$^+$ production, propagation and annihilation in a "global"
framework, trying to include all (or most of) the various aspects
of this complex topic and to account for all the available
observational data.
%They emphasized the role of positron
%propagation effects to account for the apparent extended structure
%of the annihilation source.
However, its extremely precise
"predictions" for the various properties of the 511 keV emission
(which fit extremely well - to better than 10\% - each and every
observable), concealed  the various uncertainties of the problem.
For instance, it is assumed that most Galactic positrons result
from $^{56}$Co produced in SNIa: but, as discussed in Sec. IV.A.4,
recent observations suggest that the e$^+$ escape fraction from
such objects is very small, at least  at late times. Only an early
e$^+$ escape (due, perhaps, to 3D effects, not yet theoretically
treated and unobservable at present) could make SNIa plausible
e$^+$ sources again. And even if that were the case, the poorly
known SFR of the bulge does not allow one to estimate the SNIa rate of
that region to better than a factor of two (see
Table~\ref{tab:SNrates}).
The same criticism applies to their treatment of e$^+$
propagation. Their prescription for deriving the e$^+$ mean free path is,  strictly speaking,
valid in the framework of interplanetary plasma, while the properties
of the ISM and magnetic fields in the inner Galactic regions are
too poorly understood to allow for any strong conclusions on the nature of  turbulence there.
%Moreover, a key role to the quantitative success of their model is played by the zone of galactocentric radius 1.5$<$r$<$3.5 kpc, called "outer bulge" in their work (although it lies clearly outside the stellar bulge). Positrons produced there are claimed to end up mostly in the inner bulge, enhancing considerably the final bulge/disk e$^+$ annihilation ratio. However, . \citet{HLR09} overlooked the effect 
For instance,  reacceleration - which extends the e$^+$ annihilation zone - is expected to be particularly
efficient in the ionized, low density, medium of the bulge (Sec. VI.B.1).
%The main consequence of reacceleration  is to produce a delay in the annihilation process and to extend the annihilation zone. 
As for the claim that the observed  spatial asymmetry of
the 511 keV emission is simply due to the corresponding asymmetry
of the spiral arms as viewed by the Sun, it is apparently in
contradiction with the fact that no such strong asymmetry is
observed in the 1.8 MeV emission of $^{26}$Al (a product of
massive stars), which {\it should} be a direct tracer of spiral
arms, since $^{26}$Al nuclei travel much slower than positrons.

One of the most interesting results of the work of \citet{HLR09} is,
perhaps, their prediction of a spatial differentiation between the
broad and narrow 511 keV line components, revealed by the spectral
analysis of SPI data (Sec. II.C). The former should occur in the
"middle bulge" (the region of radius 0.5 $<$r$<$1.5 kpc, according
to the authors' terminology), through in-flight positronium
formation, while the latter in the "inner bulge" (r$<$0.5 kpc)
through positronium formation at thermal energies. An analogous
spatial differentiation is predicted to exist between the bulge and
disk 511 keV line components. It is unlikely that such predictions
will be checked with INTEGRAL/SPI observations, but they certainly
hold important clues to positron propagation issues.

In a companion paper, \citet{LHR09} argued that, since the \citet{HLR09} model (with propagation of radioactively produced positrons from SNIa) can fully account for the observations, dark matter should be excluded as a major positron source.
However, such a conclusion is premature, since it is not yet clear whether positrons from SNIa escape at all.
It is true that, since the confirmation of the disk 511 keV emission, DM has lost a lot of its "appeal" as positron source,
but it cannot yet be excluded as such, at least for the bulge: indeed, in that case e$^+$ propagation merely smears out the spatial profile
of the 511 keV emission, and even decaying DM cannot be excluded then (contrary to the arguments of Sec. IV.D, which
neglect positron propagation).

\section{Summary and perspectives}
\label{sec:Conclusions}

The Galactic 511 keV emission from e$^+$ annihilation is the first
$\gamma$-ray line detected from outside the solar system. Its
unambiguous identification came soon after its detection, with
high resolution Ge-detectors aboard balloon experiments. However,
its spatial morphology remained elusive for almost three decades
after its first detection. Only long running experiments aboard
satellites could tackle this issue, in view of the importance of
the treatment of the background  at those energies (mostly created
inside the detectors by cosmic ray interactions). Observations in
the 1990s by  OSSE/CGRO offered the first hints for an
abnormally high bulge/disk ratio (compared to the situation at any
other wavelength). That property was firmly established only after
observations in the 2000s with SPI/INTEGRAL, which further
detected for the fist time an unambiguous disk emission. The
latter appears to be asymmetric, according to
\citet{Weidenspointner+08a}, with emission from negative
longitudes being 80\% brighter than from positive ones; however,
that claim is not supported by a different analysis
\citep{Bouchet+08,BRJ10} and has yet to be confirmed.

According to the latest  imaging analysis of SPI data
\citep{Weidenspointner+08a} the total Galactic e$^+$ annihilation
rate is at least $\dot{N}_{e^+} \sim$2 10$^{43}$ s$^{-1}$, with a
luminosity bulge/disk ratio B/D=1.4. This  model (Table I), is further
refined by considering  a narrow ($FWHM=3\degree$)
and a broad ($FWHM=11\degree$) bulge, the former contributing to
$\sim$35\% of the total bulge emission. However,  the data analysis also allows for other 
morphologies,
involving extended regions of low surface brightness but high total
emissivity, e.g. a "halo" of total $\dot{N}_{e^+} \sim$ 3 10$^{43}$
s$^{-1}$ and a thin disk of $\dot{N}_{e^+}\sim$5 10$^{42}$ s$^{-1}$,
leading to a high B/D$\sim$6 (Table I). Obviously, the  poorly known
configuration of the Galactic 511 keV emission precludes at present
the formulation of definitive statements about the origin of
annihilating positrons.

Information on the origin of those positrons is also obtained via the spectral analysis of the
511 keV emission: the observed flux at $\sim$MeV energies from the inner Galaxy constrains the initial
energy of the positrons to less than a few MeV (otherwise the emission from in-flight annihilation
would exceed the observed flux). Moreover, the spectral analysis provides important information
on the physical properties of the e$^+$ annihilation sites. The large positronium fraction
$f_{P_S}\sim$94-97 \% implies that positrons annihilate mostly at low energies, since direct
annihilation cross-sections are important only at high energies (Sec. V.D). The overall spectral
shape suggests that annihilation occurs mostly in warm (T$\sim$8 000 K) media, at about equal
amounts in neutral and ionized phases 
but it cannot be  excluded that less than 23\% of annihilation occurs in the Cold  
neutral medium (T $\sim$ 80 K; Sec. V. E); annihilation in the neutral  
media may account for the presence of a broad 511 keV line component  
(FWHM $\sim$5 keV) and the annihilation in the warm ionized medium for  
the narrow one (FWHM $\sim$1 keV).

Among the various astrophysical sources of positrons proposed so
far, the only one known with certainty to release e$^+$ in the ISM
is $\beta^+$ radioactivity of $^{26}$Al; the observed intensity of
its characteristic 1.8 MeV emission in the Galaxy corresponds to
$\sim$3-4 10$^{42}$ e$^+$ s$^{-1}$ (Sec. II.C.2). A similar amount
is expected from the decay of $^{44}$Ti, on the grounds of
nucleosynthesis arguments (Sec. IV.A.2). Both radionuclides are
produced mostly in massive stars and their positrons should be
released along the Galactic plane, as traced by the 1.8 MeV
emission; they could thus account for the observed disk 511 keV
emission.

Radioactivity of $^{56}$Co from SNIa was traditionally considered
to be the major e$^+$ producer in the Galaxy. Both the typical
$^{56}$Ni yield of a SNIa and the Galactic SNIa rate are rather
well constrained, resulting in 5 10$^{44}$ e$^+$ s$^{-1}$ produced
{\it inside} SNIa. If only f$_{esc}\sim$4\% of them escape the
supernova to annihilate in the ISM, the observed total e$^+$
annihilation rate can be readily explained. However, observations
of two SNIa, interpreted in the framework of 1-D (stratified)
models, suggest that the positron escape fraction is negligible
{\it at late times}. On the other hand, both observations of early
spectra and 3-D models of SNIa suggest that a sizeable fraction of
$^{56}$Ni is found at high velocity (close to the surface), making
- perhaps - easier the escape of $^{56}$Co positrons (Sec.
IV.A.4). In our opinion, SNIa remain a serious candidate, with a
potential Galactic yield of  2 10$^{43}$ e$^+$ s$^{-1}$. But the
expected spatial distribution of SNIa in the Galaxy  corresponds
to a much smaller B/D ratio than that of the observed 511 keV
profile.

Most of  the other astrophysical candidates   can be constrained to be only minor e$^+$ sources, on the basis of either
weak e$^+$ yields (novae, Galactic cosmic rays),  high e$^+$ energy (compact objects, like
pulsars or  magnetars), spatial morphology of sources (hypernovae, gamma ray bursts) or a combination
of those features (e.g. cosmic rays), as discussed in Sec. IV.E. Only two astrophysical candidates remain as
potentially important contributors: LMXRBs (or the microquasar variant of that class of sources, Sec. IV.B.3) and the supermassive
black hole at the Galactic center (Sec. IV.B.4). It should be stressed that there is no evidence that either of those sources
produces positrons and the e$^+$ yields evaluated by various authors are close to upper limits rather than typical
values (Sec. IV.D.1). Furthermore, because of the current low activity of the central MBH (much lower than that  of LMXRBs) it has to be assumed that the source was much more active in the past, thus dropping the
assumption of "steady state" between e$^+$ production and annihilation,  which is likely in all other cases.

The observed spatial distribution of LMXRBs is similar to the theoretically derived one for SNIa (Sec. IV.D.3),
as expected, since both classes of sources have old and young stellar components; however, none of them has
the large B/D ratio of the observed 511 keV emission. The only hint that LMXRBs may contribute to, at least, the disk
511 keV emission stems from the asymmetric distribution of the hard LMXRBs in the 3rd IBIS catalogue, of similar
magnitude to the detected 511 keV emission; however, the former asymmetry is not confirmed 
(the 4th IBIS catalogue shows no strong evidence
for such an asymmetry) and even if the  511 keV asymmetry is confirmed by future analysis, there is
considerable debate on whether such a similarity has a causal origin or is just fortuitous (Sec. IV.D.3).

Dark matter (DM) has been proposed as an alternative e$^+$ source,
at least for the bulge 511 keV emission; in principle, it could
complement disk emission originating from radioactivity of $^{26}$Al
and $^{44}$Ti or $^{56}$Co. Observations of the MeV continuum from the inner
Galaxy constrain the large phase space of DM properties. The mass of
annihilating or decaying DM particles should be smaller than a few
MeV, otherwise their in-flight annihilation would overproduce the
MeV continuum (Sec. II.C.1 and IV.B.1).  Scalar light DM particles
with fermionic interactions still appear as a possible candidate;
alternatively, the collisional de-excitation of heavy (100 GeV) DM
particles could provide the required positrons, provided the energy
separation between their excited levels is in the MeV range (Sec.
IV.C). On the other hand, the observed spatial profile of the 511
keV emission constrains the production mode of DM positrons, {\it if
it is assumed that they annihilate close to their production
region}: only "cuspy" profiles are allowed in the case of
annihilating or de-exciting DM particles (for which
$\rho_{\gamma}\propto \rho_{DM}^2$), while decaying DM particles
(for which $\rho_{\gamma}\propto \rho_{DM}$) are excluded (Sec.
IV.D.3); the problem is that observations of external galaxies
suggest rather flat, not cuspy, DM profiles (Sec III.E).

Positrons produced in the hot, tenuous plasma filling the bulge
(either from SNIa, LMXRBs or DM), have to travel long distances
before slowing down and annihilating. This is corroborated by the
spectral analysis, which suggests that positrons annihilate in warm
gas: such gas is filling mostly the inner bulge. Positron
propagation appears then unavoidable, undermining the assumption
that the e$^+$ production and annihilation profiles are correlated,
at least in the bulge. A similar situation should hold for positrons
produced away from the plane of the disk (i.e. from SNIa or LMXRBs),
which is also dominated by hot, tenuous gas. The situation is less
clear for positrons produced by massive star radioactivity, in the
plane of the disk and inside spiral arms: although some of them may
fill hot bubbles and cavities created by the SN explosions and
ultimately escape from the disk, another fraction may annihilate in
closeby dense molecular clouds. Propagation of MeV positrons in the
ISM may then hold the key to understanding the 511 keV emission. It
depends on the physical properties of the ISM (density, ionization,
Sec. V) but also on the properties of turbulence and magnetic field
configuration (Sec. VI). Preliminary attempts to evaluate the extent
of positron propagation (Sec. VI.A) and their implications for the
Galactic 511 keV emission (Sec. VI.D) are promising in that respect,
but the situation is far from clear at present: the entanglement
between the various uncertainties (concerning e$^+$ sources, e$^+$
propagation and annihilation sites) does not allow any strong
conclusions to be drawn.

More than 30 years after its discovery, the origin of the first extra-solar $\gamma$-ray line remains unknown.
Progress in the field will require advances in several directions:

{\it (i) Observations of 511 keV emission}: what is the true spatial distribution of the emission? how far do the spheroid
and disk extend ? are there yet undetected regions of low surface brightness? is the disk emission asymmetric indeed?
how do the 1.8 MeV and 511 keV disk emissions compare to each other? A much deeper exposure of the Galaxy and a better understanding of the backgrounds will be required to tackle those issues. Even if  INTEGRAL's mission is extended
to 2012, it may not provide the answers; unfortunately,  no other mission of similar scope
is on the horizon.

{\it (ii) Physics of e$^+$ sources}: what is the e$^+$ escaping fraction in SNIa ?   what is the SNIa rate in the inner
(star forming)  and in the outer (inactive) bulge? what are the e$^+$ yields, activity timescales, and spatial distribution
in the inner bulge of LMXRBs or microquasars? how can the past level of activity of the central supermassive black hole be reliably inferred?

{\it (iii) Positron propagation:} what is the large scale
configuration of the Galactic magnetic field? what are the
properties of interstellar plasma turbulence and how do they affect the
positron transport? what are the dominant propagation modes of
positrons and what is the role of re-acceleration?

The many facets of the Galactic 511 keV emission make this problem one of the most intriguing problems  in high energy
astrophysics today and for many  years to come.

%.........................................................................................................................................................................................

%%  \input{SecVII}
\acknowledgements{This work was largely performed in the framework of the ISSI
 (International Space Science Institute at Bern) programme for International Team work (ID \# 110). We are grateful to ISSI for
generous support and for the hospitality of its staff. I.V.M. acknowledges support from NASA grant NNX09AC15G. 
A.M.B. was supported by RBRF grant 09-02-12080 and the RAS Presidium
Programm. He also acknowledge  Joint Supercomputing
Centre (JSCC RAS), the Supercomputing Centre at Ioffe Institute and
the program "Particle Acceleration in Astrophysical Plasmas" at KITP.
We  are grateful to Martin Pohl and Peter Kalberla, 
who kindly provided us with their data files.  We acknowledge constructive reports by three anonymous
referees, which greatly improved the manuscript.}

%\bibliography{RMP_bibNP,RMP_bibPJ,RMP_bibAB,RMP_bibCB,RMP_bibKF,RMP_bibRD,RMP_bibIM,RMP_bibJK}% Produces the bibliography via BibTeX.
%\include{RMP_Pos_Rev.bbl}

\end{document}